# Waves, modes, communications and optics

David A. B. Miller

*Stanford University*

Modes generally provide an economical description of waves, reducing complicated wave functions to finite numbers of mode amplitudes, as in propagating fiber modes and ideal laser beams. But finding a corresponding mode description for counting the best orthogonal channels for communicating between surfaces or volumes, or for optimally describing the inputs and outputs of a complicated optical system or wave scatterer, requires a different approach. The singular-value decomposition approach we describe here gives the necessary optimal source and receiver "communication modes" pairs and device or scatterer input and output "mode-converter basis function" pairs. These define the best communication or input/output channels, allowing precise counting and straightforward calculations. Here we introduce all the mathematics and physics of this approach, which works for acoustic, radio-frequency and optical waves, including full vector electromagnetic behavior, and is valid from nanophotonic scales to large systems. We show several general behaviors of communications modes, including various heuristic results. We also establish a new "M-gauge" for electromagnetism that clarifies the number of vector wave channels and allows a simple and general quantization. This approach also gives a new modal "M-coefficient" version of Einstein's A&B coefficient argument and revised versions of Kirchhoff's radiation laws. The article is written in a tutorial style to introduce the approach and its consequences.

## 1. Introduction

The idea of modes is very common in world of waves, especially in optics. Modes are very useful in simplifying many problems. But, there is much confusion about them. Are modes "resonances"? Are they "beams"? Do they have to stay the same "shape"? Are they "communication channels"? How do we "count" modes? Are they properties of space or of objects like scatterers? Just what is the definition of a mode? The purpose of this paper is to sort out the answers to questions like these, and to clarify and extend the idea of "modes". In particular, we want to use them for describing waves in communications and in describing sophisticated optical devices. Such applications are increasingly important: communications may require mode- or space-division multiplexing to increase capacity, and we are able to fabricate progressively more complex optical devices with modern micro- and nano-fabrication.

### 1.1. Modes and waves

At their simplest, modes can be different shapes of waves. Some such modes arise naturally in waveguides and resonators; these modes are well understood and are taught in standard texts (see, e.g., [1 - 4]). A key benefit of modes is that, when we choose the right ones, problems simplify; instead of describing waves directly as their values at each a large number of points, we can just use the amplitudes of some relatively small number of modes. But when we want to use modes to understand communications with waves more generally, or when we want to describe some linear optical device or object economically using modes, we need to move beyond the ideas of just resonator or waveguide modes. Specifically, we can introduce the ideas of *communications modes* in communicating with waves [5] and *mode-converter basis sets* [6, 7] in describing devices. These modes are not yet part of standard texts, nor is there even any broad and deep introduction to them. Further, many of their details and applications are not yet discussed in the literature.

The reason for writing this paper is to provide exactly such an introduction. In addition to sorting out the ideas of modes generally, we explain the physics of these additional forms of modes, which bring clearer answers to our opening questions above. We show how these ideas are supported by powerful and ultimately straightforward mathematics. We introduce novel, useful and fundamental results that follow. This approach resolves many confusions. It reveals powerful concepts and methods, general limits, new physical laws, and some simple and even surprising results. It works over a broad range of waves, from acoustics, through classical microwave electromagnetism, to quantum mechanical descriptions of light.

## 1.2. The idea of modes

One subtle point about modes is that it can be difficult to find a definition or even a clear statement of what they are. We should clarify this now.

Modes are particularly common in describing oscillations of physical objects and systems. Simple examples include a mass on a spring, or waves on a string, especially one with fixed ends. In these cases, an informal definition of an oscillating mode is that it is a way of oscillating in which everything that is oscillating is oscillating at the same frequency. This is a sense in which a "mode" is a "way" or "manner" of oscillation. Musical instruments offer many other examples of such modes, as in standing waves in a pipe, or resonances in the vibrations of plates or hollow bodies. Such a mode will have a specific frequency of oscillation, and the amplitude of the vibration will take a specific physical form – it can be a function of position along the string or pipe or on the surface of some plate or body.

The underlying mathematical idea of modes is associated with eigenfunctions or eigenvectors in linear physical systems; in oscillating systems or resonators, the function that gives the amplitude of oscillation at each position is the eigenfunction and the frequency (or often the square of the frequency) is the eigenvalue. Indeed, we can state a useful, general definition of a mode [8 - 10]:

A mode is an eigenfunction of an eigen problem describing a physical system. (1)

Conventional resonator and waveguide modes are each the eigenfunctions of a single eigen problem. The fixed "shape" of this oscillation amplitude inside the resonator is often thought of as the "mode" or eigenfunction in this sense. Waveguide modes use the same mathematics, but the concept here is that the transverse shape of the mode does not change as it propagates. An analogous informal definition of a propagating mode is that everything that propagates with the same wavevector, which also implies that the (transverse) shape does not change as it propagates. That transverse shape is the eigenfunction. Though such waveguide modes may well be modes of a specific frequency that we have chosen, the eigenvalue is typically a propagation constant or wavevector magnitude (or, again, often the square of this quantity).

Before going any further, to support these ideas of modes, we need good notations; they should be general enough to handle everything we need, but they should suppress unnecessary detail. Wherever possible, we use a Dirac "bra-ket" notation, which operates at just such a useful level of abstraction. We introduce this notation progressively (see also [9]). In this notation a function can be represented by a "ket" or "ket vector", written as $|\psi_S\rangle$ or $|\phi_R\rangle$, for example. Linear operators, such as Green's functions or scattering operators, are represented by a letter, and here we will mostly use "sans serif" capital letters such as $\mathsf{G}$ and $\mathsf{D}$. Most simply, we can think of kets as column vectors of numbers and the linear operators as matrices. Dirac notation implements a convenient version of linear algebra equivalent to matrix-vector operations with complex numbers, and indeed such a matrix-vector view can be the simplest way to think about Dirac notation.

## 1.3. Modes as pairs of functions

To handle communications and complex optical devices, however, we need to go beyond just resonator or waveguide modes; fortunately, though, we can use much of the same mathematics. The key mathematical difference between resonator and waveguide modes on the one hand and our new modes on the other is that communications modes and mode-converter basis sets each result from solving a *singular-value decomposition* (SVD) problem, which corresponds to solving *two* eigen problems.

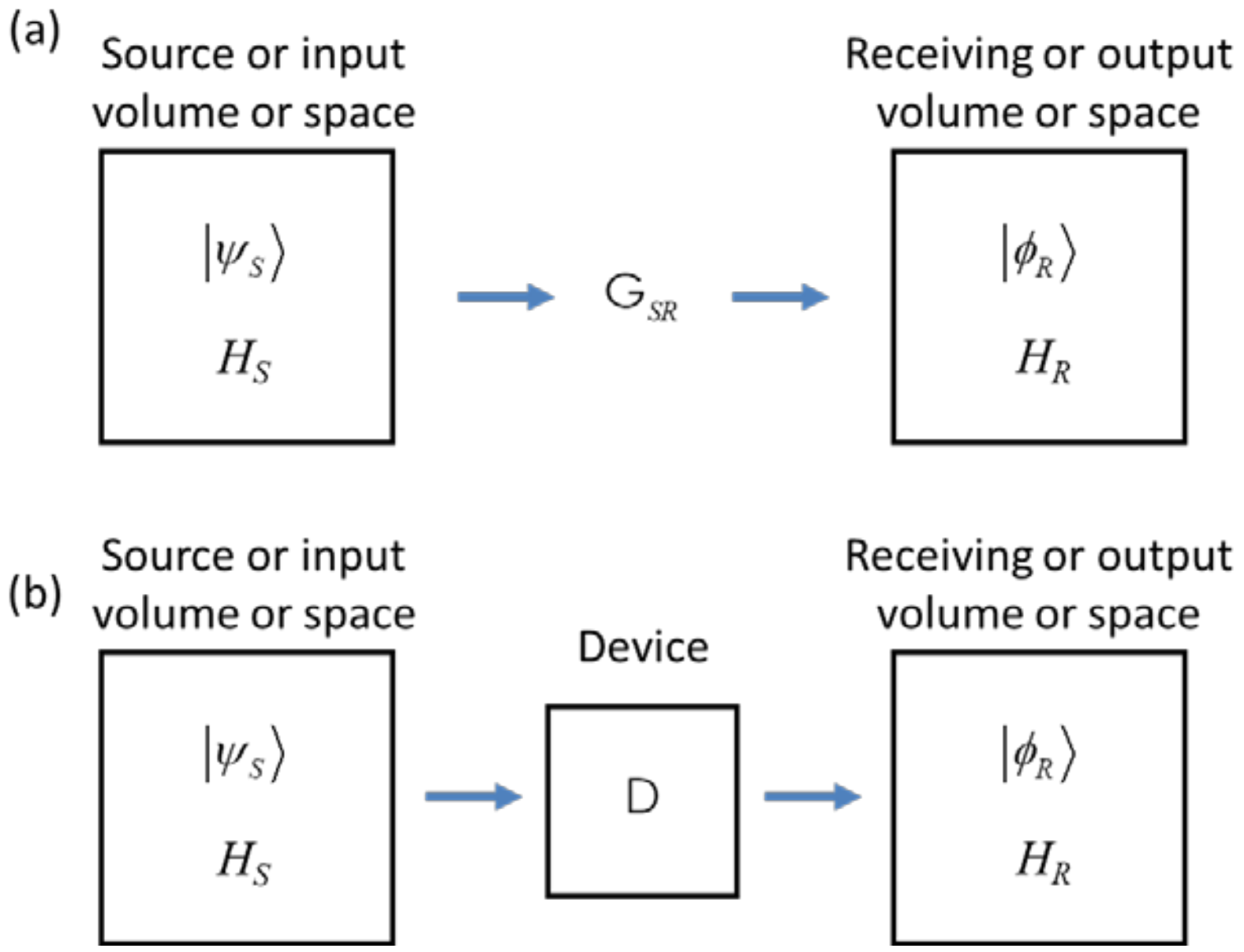

Fig. 1. Conceptual view for (a) communications modes and (b) mode-converter basis sets. In both cases a source function $|\psi_S\rangle$ in a source or input volume $V_S$, or more generally in a mathematical (Hilbert) space $H_S$, results in a wave function $|\phi_R\rangle$ in a receiving or output volume $V_R$, or more generally in a mathematical (Hilbert) space $H_R$. In the communications mode case (a) the coupling is through a Green's function operator $G_{SR}$ as appropriate for the intervening medium between the spaces. In the mode-converter case (b) , the coupling is through the action of a device (or scatttering) operator $D$.

The physical reason for having two such eigenproblems is because we are defining optimum mappings between two different spaces.

For example, in communications, we may have sources or transmitters in one "source" volume and resulting waves communicated into another "receiving" volume (Fig. 1 (a)). The solutions to our problem are then the set of optimum source functions in the source or input volume that couple, one by one, to the resulting optimal waves in the receiving or output volume; SVD solves for both of those sets of functions, and it is these *two* sets of functions that are the communications modes. So, a given communications mode is not one function but two.

In practice we may only need to solve one of these two SVD eigenproblems, and we can then deduce the solutions to the other. But because we can view this through two eigen problems, each of these sets of functions, one in the source space and one in the receiving space, therefore has all the useful mathematical properties of eigenfunctions. This mathematics of eigenfunctions has profound consequences for the physical interpretation and the mathematics that follows.

### 1.3.1. Communications modes

Note immediately that, in this view,

> the communications mode is *not* the propagating wave (or what we will call the *beam*) between the source volume and receiver volume.

Indeed, in general the beam will change shape as it propagates, and it is not itself the "eigenfunction" of the mathematical problem (though it is easily deduced from the actual eigenfunctions in simple communication problems). In this SVD way of looking at communications, the *j*th communications mode is a pair of functions – $|\psi_{Sj}\rangle$ in the source or input space, and $|\phi_{Rj}\rangle$ in the receiving or output space. Explicitly, therefore,

> communications modes are pairs of functions – one in the source space and one in the receiving space.

They are a set of *communications mode pairs* of functions – a pair $|\psi_{S1}\rangle$ and $|\phi_{R1}\rangle$, a pair $|\psi_{S2}\rangle$ and $|\phi_{R2}\rangle$, and so on. To find these functions, we perform the SVD of the coupling operator $G_{SR}$ between the volumes or spaces. For the communications problems we consider first, $G_{SR}$ is effectively the free-space Green's function for our wave equation.

### 1.3.2. Mode-converter basis sets

When we change from thinking just about waves in free space to trying to describe a linear optical device, we can consider how it scatters input waves to output waves (Fig. 1 (b)). By analyzing this also as an SVD problem, in this case of a device (or scattering) operator $D$, we can similarly deduce a set [11] of input source functions $\{|\psi_{Sj}\rangle\}$ that couple one by one to a set of output wave functions $\{|\phi_{Rj}\rangle\}$; these two sets of functions are the mode-converter basis sets.

In this second case, we want to describe the device as one that converts from a specific input mode $|\psi_{Sj}\rangle$ to the corresponding output mode $|\phi_{Rj}\rangle$, and so on, for all such mode pairs; again, as in the case of communications modes, we think in terms of pairs of functions here, one in the source or input space, and one in the receiving or output space. We can consider these as *mode converter pairs* - a pair $|\psi_{S1}\rangle$ and $|\phi_{R1}\rangle$, a pair $|\psi_{S2}\rangle$ and $|\phi_{R2}\rangle$, and so on, just as in the communications modes. In this way of looking at a linear optical device [6],

> any linear optical device can be viewed as a mode converter, converting from specific sets of functions in the input space one-by-one to specific corresponding functions in the output space, giving the mode-converter pairs of functions.

The device converts input mode $|\psi_{S1}\rangle$ to output mode $|\phi_{R1}\rangle$, input mode $|\psi_{S2}\rangle$ to output mode $|\phi_{R2}\rangle$, and so on. In this case, though the mathematics is similar to the communications modes, this is more a way of describing the device, whereas the communications modes are a way of describing the communications channels from sources to receivers. For the device case, we may not have anything like a simple beam between the sources and receivers, but we do have these well-defined functions or "modes" inside the source space or volume and inside the receiving space or volume. We could also view the mode-converter basis sets as describing the communications modes "through" the device.

In an actual physical problem for a device, there are ways in principle in which we could deduce the mode-converter pairs of functions by experiment [7, 12] without ever knowing exactly what the wave field is inside the device. Then we could know the mode converter pairs as eigenfunctions without knowing the "beam"; this point emphasizes that it can be more useful and meaningful to use the pairs of functions in the source and receiving spaces as the modes of the system rather than attempting to use the beam through the whole system as the way to describe it.

## 1.4. Usefulness of this approach

There are several practical and fundamental reasons why these pairs of functions are useful.

### 1.4.1. Using communications modes

In communications, we continually want larger amounts of useful bandwidth. This need is strong for wireless radio-frequency transmission [13], for optical signals in fibers [14 - 17] or free space [17 - 20], and even for acoustic information transmission [21- 23]. Recent progress in novel optical ways to separate different [16] and even arbitrary modes [24 -29], including automatic methods [24 - 29], gives additional motivation to consider the use of different modes (or "spatial degrees of freedom") in communications.

Increasingly, therefore, we need to understand the spatial degrees of freedom in such communications and the limits in their use; a natural way to describe and quantify those is in terms of communications modes. Specifically,

> we can understand how to count the number of useful available spatial channels.

Essentially, this can also be viewed as a generalization of the ideas of diffraction limits, and we will develop these ideas below.

A key novel result is that

> this SVD approach gives a sum rule that bounds the number and strength of those channels.

As we solve the problem this way, we can also unambiguously establish just exactly what the best channels are; we do not need to presume any particular form of these modes to start with. So, specifically, we do not need to analyze in terms of plane wave "modes", Hermite-Gaussian or Laguerre-Gaussian beams, optical "orbital" angular momentum [19, 20, 30 - 32] "modes", prolate spheroidals [33], arrays of spots, or any other specific family of functions; specifically,

> the SVD solution will tell us the best answers for the transmitting and receiving functions – the communications modes – and those will in general be none of the standard mathematical families of functions or beams.

### 1.4.2. Using mode-converter basis sets

In analyzing linear optical devices or scatterers,

> if we establish the mode-converter basis sets by solving the SVD problem, we will have the most economical and complete description of a device or scatterer.

Essentially, we establish the "best" functions to use here, starting with the most important and progressing to those of decreasing importance. An incidental and universal consequence of this approach is that we realize that

> there is a set of independent channels through any linear scatterer

(which are the mode-converter basis sets), and that we can describe the device completely using those. The implications of the mode-converter basis sets go beyond simple mathematical economy.

> Mode-converter basis functions have basic physical meaning and implications, giving fundamental results that can be economically and uniquely expressed using them.

They allow us, for example, to write new versions and extensions of Kirchhoff's radiation laws [7], including ones that apply specifically and only to the mode-converter pairs, and to derive a novel modal version of Einstein's "A & B" coefficient argument on spontaneous and stimulated emission (section **11.2**). Such results suggest that this mode-converter basis set approach is deeply meaningful as a way to describe optical systems. These mode-converter basis functions can also be identified in principle for a given linear object through physical experiments [7], independent of the mathematics.

### 1.5. Approach of this paper

Because the ideas here go beyond conventional textbook discussions, and because we are combining concepts and techniques that cross several different fields, the approach of this article is quite tutorial. Most algebra steps are written explicitly, and many "toy" examples illustrate the key steps and points. I have tried to write the main text so that it is readable, and with a progressive flow of ideas. I introduce core mathematical ideas in the main text, but relegate most other derivations and mathematics to appendices.

This article has been written to be accessible to readers with a good basic undergraduate knowledge of mathematics and some physical science, such as would be acquired in a subject like electrical engineering or physics or a discipline like optics (for specific presumed background, see [34]), but I explicitly introduce all other required advanced mathematics and electromagnetism. Wherever possible, I take a direct approach in derivations, working from fundamental results, like Maxwell's equations or core mathematical definitions and principles, without invoking intermediate results or methods.

## 2. Organization of this paper

The mathematics of SVD is relatively straightforward for finite matrices; such matrices arise, for example, if we have a finite number of small sources communicating to a finite number of small receivers. The mathematics is particularly simple if we also initially consider just scalar waves, like acoustic waves in air, for example. Such scalar waves allow a good tutorial introduction to communications modes and more generally to the ideas of this SVD approach. We start with the mathematics of such sources, receivers and waves in section **3**. In section **4**, we go through a simple "toy" example explicitly, showing both the mathematical results and physical systems that would implement them.

Many quite general physical and mathematical behaviors emerge as we look at wave systems this way, only some of which are currently well known. Though these behaviors are relatively simple and even intuitive, only some have simple analytic solutions. On the other hand, numerical "experiments" and examples are straightforward, at least for finite numbers of "point" sources and receivers. Then the main calculation is just finding eigenvalues and eigenvectors of finite matrices. So, we introduce these behaviors informally through a sequence of further numerical examples in section **5** (supported by additional heuristic arguments in **Appendix A**, **Appendix B**, and **Appendix C**). Pretending we can approximate any set of "smooth" source and receiver functions with sufficiently many such point sources and receivers, we can reveal much of the behavior of the more general case and many of the results.

To be general enough for real problems in optics and electromagnetism, we need two major sophistications. First, we need to expand the mathematics to handle sources and received waves that are continuous functions of space, and to consider possibly infinite sets of source and or wave functions. A key point is that we will be able to show that

> even with continuous source and wave functions, and with possibly infinite sets of them, we end up with finite numbers of useful communications channels or mode-converter basis functions for describing devices.

Furthermore, this gives a general statement of diffraction limits for any volumes and any form of waves.

The mathematics has to go beyond that of finite matrices, and cannot be deduced from it [35]. Fortunately, that mathematics – functional analysis – exists. Unfortunately, this field is often impenetrable to the casual reader; necessarily it has to introduce ideas of convergence of functions, and that involves an additional set of concepts, mathematical tools and terminology. The important results can, however, be stated relatively simply; in section 6, I summarize key mathematical results, deferring some detail to **Appendix D** and **Appendix E**. I have also written a separate (and hopefully accessible) introduction [36] to this functional analysis mathematics, including all required proofs. With the results from functional analysis, continuous sources and waves for the simple scalar case can then be understood quite simply. In section 7, we relate these mathematical results to known families of functions in the "paraxial" case often encountered in optics.

The second major sophistication we require is the extension to electromagnetic waves, and we summarize the key results in section 8. Scalar waves are often a good first model in optics, and much of their behavior carries over into the full electromagnetic case; dealing with electromagnetic waves properly is, however, more complicated. Not only are electromagnetic waves vectors rather than scalars, but, on the face of it, we have two kinds of fields to deal with – electric and magnetic. Maxwell's equations relate these two kinds of fields, of course. Existing sophisticated approaches to electromagnetism, such as the use of scalar and vector potentials, are helpful here in understanding just how many independent field components or "degrees of freedom" there really are, but standard approaches are not quite sufficient for clarifying this number. This difficulty can be resolved by proposing a new "gauge" (the "M-gauge") for the electromagnetic field. We provide a full explanation and derivation of the necessary electromagnetism in **Appendix F**, supported with a derivation in **Appendix G** and additional notation and identities in **Appendix H**.

This new M-gauge, together with the results of the functional analysis and the SVD approach, allows a revised quantization of the electromagnetic field, summarized in section 9 and in more detail in Appendix I. This resolves several difficulties. In particular, we can avoid artificial "boxes" and "running waves" in quantizing radiation fields, and they can be quantized for any shape of volume. This quantization means that our results here are generally valid and meaningful for both classical and quantum-mechanical radiation fields.

In section 10, we describe how to apply this same mathematics and physics in considering mode-converter basis sets for devices or scatterers. Section 11 includes discussion of fundamental aspects of such mode-converter basis sets, including new radiation laws and a revised and simplified "Einstein's A & B" coefficient argument (with a full derivation in **Appendix J**) in modal form. Finally, in section 12, I draw some conclusions.

Length constraints here mean some relevant topics are omitted. First, in discussing waves, mostly I consider just the monochromatic case, but the underlying mathematics and electromagnetism supports general time-dependent fields [37] (and I give those results explicitly for electromagnetic fields), and hence "temporal modes" [25, 38, 39]. Second, though the communication channels are well-suited for adding information theory to calculate capacities (e.g., [18, 40]), I have to omit that discussion here.

To improve the narrative flow of the paper, I have avoided extensive historical and research review in the main body of the text, but I have included these important discussions in **Appendix K**. The subject is easier to explain without the constraint of the historical order and way in which the concepts arose, and the history and other research and connections are easier to explain once we understand the concepts.

Though some aspects of this material are well-known in the literature, and our treatment of those aspects is therefore purely a tutorial, for some other aspects, we have to present some original material. To clarify this, and to allow the reader to make their own judgements of the approaches and validity of any new results, I have listed what I believe to be novel results in **Appendix L**.

This work is quite long overall, and that length might be daunting. I suggest that the reader starts with sections 3, 4, and 5 – that will convey much of this new way of thinking about waves – followed by section 10 to understand how this approach describes optical devices and scatterers. Sections 6, 7, and 8 add depth and rigor to the wave discussion, and sections 9 and 11 add discussion of fundamental physical results from this approach.

# 3. An introduction to SVD and waves – sets of point sources and receivers

We start here by introducing the main ideas of this SVD approach with the simple example of scalar waves with point sources and detectors.

## 3.1. Scalar wave equation and Green's functions

Suppose we have some uniform (and isotropic) medium, such as air, with a wave propagation velocity $v$. For simplicity, we presume we are interested in waves of only one (angular) frequency $\omega$. Then we could propose a simple Helmholtz wave equation; this would be appropriate, for example, for acoustic pressure waves in air [41], with $v$ being the velocity of sound in air. Then, for a spatial source function $\psi_{\omega S}(\mathbf{r})$ and a resulting wave $\phi_{\omega R}(\mathbf{r})$ , this Helmholtz wave equation would be

$$\nabla^2\phi_{\omega R}(\mathbf{r})+k^2\phi_{\omega R}(\mathbf{r})=\psi_{\omega S}(\mathbf{r}) \tag{2}$$

with

$$k^2=\omega^2/v^2 \tag{3}$$

Now, for such an equation, the Green's function (i.e., the wave that results from a "unit amplitude" point source $\delta(\mathbf{r}-\mathbf{r}')$ at position $\mathbf{r}'$) is

$$G_\omega(\mathbf{r};\mathbf{r}')=-\frac{1}{4\pi}\frac{\exp(ik|\mathbf{r}-\mathbf{r}'|)}{|\mathbf{r}-\mathbf{r}'|} \tag{4}$$

As usual with Green's functions (see e.g., [42] for an introduction), for an actual continuous source function $\psi_{\omega S}(\mathbf{r})$, the resulting wave would be

$$\phi_{\omega R}(\mathbf{r})=\int_{V_S} G_\omega(\mathbf{r};\mathbf{r}')\psi_{\omega S}(\mathbf{r}')d^3\mathbf{r}' \tag{5}$$

Such superposition of Green's functions (through the integral here) works because the medium (e.g., air) is presumed to be linear so that superpositions of solutions to the wave equation are also solutions to this (linear) wave equation (2).

## 3.2. Matrix-vector description of the coupling of point sources and receivers

We presume a set of $N_S$ point sources (Fig. 2) at positions $\mathbf{r}_{Sj}$ ( $j=1,\dots,N_S$ ) in the source volume, and with (complex [43]) amplitudes $h_j$ . These might be the (complex) amplitude of the drives to each of a set of $N_S$ small loudspeakers, for example, that we pretend we can approximate as point sources. Then the resulting wave at a point $\mathbf{r}_{Ri}$ in the receiving volume would be

$$\phi_{\omega R}(\mathbf{r}_R)=-\frac{1}{4\pi}\sum_{j=1}^{N_S}\frac{\exp(ik|\mathbf{r}_{Ri}-\mathbf{r}_{Sj}|)}{|\mathbf{r}_{Ri}-\mathbf{r}_{Sj}|}h_j=\sum_{j=1}^{N_S}g_{ij}h_j \tag{6}$$

where

$$g_{ij} = -\frac{1}{4\pi}\frac{\exp\left(ik\left|\mathbf{r}_{Ri} - \mathbf{r}_{Sj}\right|\right)}{\left|\mathbf{r}_R - \mathbf{r}_{Sj}\right|} \tag{7}$$

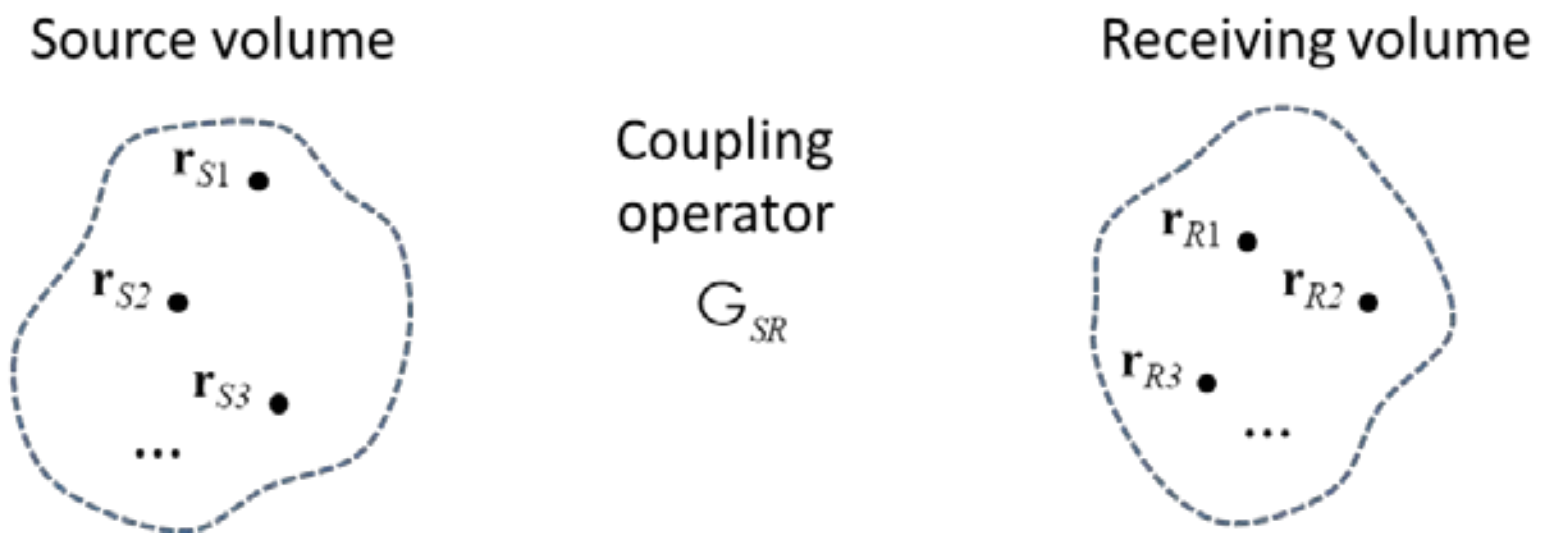

Fig. 2. A set of point sources at positions $\mathbf{r}_{Sj}$ in a source volume, and a set of point receivers at positions $\mathbf{r}_{Ri}$ in a receiving volume, coupled through the coupling operator $G_{SR}$ .

Suppose, then, that we had a set of $N_R$ small microphones at a set of positions $\mathbf{r}_{Ri}$ in the receiving volume (Fig. 2); we presume these are omnidirectional (so their response has no angular dependence). Then the received signal at one such microphone or point would be the sum of the waves from all the point sources, added up at the point $\mathbf{r}_{Ri}$ (as in Eq. (6))

$$f_i = \sum_{j=1}^{N_S} g_{ij} h_j \tag{8}$$

Equivalently, if we define the vectors $\left|\psi_S\right\rangle$ and $\left|\phi_R\right\rangle$ and the matrix $G_{SR}$ for such a problem as

$$\left|\psi_S\right\rangle = \begin{bmatrix} h_1 \\ h_2 \\ \vdots \\ h_{N_S} \end{bmatrix}, \left|\phi_R\right\rangle = \begin{bmatrix} f_1 \\ f_2 \\ \vdots \\ f_{N_R} \end{bmatrix}, \text{ and } G_{SR} = \begin{bmatrix} g_{11} & g_{12} & \cdots & g_{1N_S} \\ g_{21} & g_{22} & \cdots & g_{2N_S} \\ \vdots & \vdots & \ddots & \vdots \\ g_{N_R 1} & g_{N_R 2} & \cdots & g_{N_R N_S} \end{bmatrix} \tag{9}$$

then we can write the set of relations Eq. (8) for all *i* compactly as the matrix-vector expression

$$\left|\phi_R\right\rangle = G_{SR}\left|\psi_S\right\rangle \tag{10}$$

## 3.3. Hermitian adjoints and Dirac bra-ket notation

At this point, we can usefully introduce the final part of the Dirac notation, which involves the *Hermitian adjoint* (or *Hermitian conjugate* or *conjugate transpose*) [44]. Generally, this is notated with a superscript "dagger", written as "$\dagger$". The Hermitian adjoint of a matrix is formed by reflecting around the "top-left" to "bottom-right" diagonal of the matrix and taking the complex conjugate of the elements. For some matrix $G$ , with matrix elements $g_{ij}$ in the *i*th row and *j*th column, the corresponding "row-*i*, column-*j*" matrix element of the matrix $G^\dagger$ is the number $g_{ji}^*$ . The Hermitian adjoint of a column vector is, similarly, a row vector whose elements are the complex conjugates of the corresponding elements of the column vector. In Dirac notation, such a row vector is notated using the "bra" notation $\left\langle\phi\right|$ . So, explicitly, for our matrices and vectors here

$$\left(|\psi_S\rangle\right)^\dagger \equiv \begin{bmatrix} h_1 \\ h_2 \\ \vdots \\ h_{N_S} \end{bmatrix}^\dagger \equiv \begin{bmatrix} h_1^* & h_2^* & \cdots & h_{N_S}^* \end{bmatrix} \equiv \langle\psi_S| \tag{11}$$

and similarly for $|\phi_R\rangle$, and the Hermitian adjoint of the operator $\mathsf{G}_{SR}$ is

$$\mathsf{G}_{SR}^\dagger \equiv \begin{bmatrix} g_{11} & g_{12} & \cdots & g_{1N_S} \\ g_{21} & g_{22} & \cdots & g_{2N_S} \\ \vdots & \vdots & \ddots & \vdots \\ g_{N_R1} & g_{N_R2} & \cdots & g_{N_RN_S} \end{bmatrix}^\dagger \equiv \begin{bmatrix} g_{11}^* & g_{21}^* & \cdots & g_{N_R1}^* \\ g_{12}^* & g_{22}^* & \cdots & g_{N_R2}^* \\ \vdots & \vdots & \ddots & \vdots \\ g_{1NL}^* & g_{2N_S}^* & \cdots & g_{N_RN_S}^* \end{bmatrix} \tag{12}$$

Note too that the Hermitian adjoint of a product is the "flipped round" product of the Hermitian adjoints, i.e., for two operators $\mathsf{G}$ and $\mathsf{H}$

$$(\mathsf{GH})^\dagger = \mathsf{H}^\dagger\mathsf{G}^\dagger \tag{13}$$

(which is easily proved by writing such a product out explicitly using the elements of the matrices and summing them appropriately) and for matrix-vector products

$$\left(\mathsf{G}|\psi\rangle\right)^\dagger = \langle\psi|\mathsf{G}^\dagger \tag{14}$$

The Hermitian adjoint of a Hermitian adjoint just brings us back to where we started, i.e.,

$$\left(\mathsf{G}^\dagger\right)^\dagger = \mathsf{G} \tag{15}$$

and for some vector

$$\left[\left(|\phi\rangle\right)^\dagger\right]^\dagger = \left[\langle\phi|\right]^\dagger = |\phi\rangle \tag{16}$$

both of which results are obvious from the process of reflecting and complex conjugating matrices and vectors.

For a simple scalar wave, for an amplitude $f_i$ at a given receiving point (or microphone), the corresponding received power (in appropriate units) would typically be

$$P_i = f_i^* f_i \tag{17}$$

So the sum of all the detected powers would be

$$P = \sum_{i=1}^{N} f_i^* f_i = \langle\phi_R|\phi_R\rangle = \left(\langle\psi_S|\mathsf{G}_{SR}^\dagger\right)\left(\mathsf{G}_{SR}|\psi_S\rangle\right) = \langle\psi_S|\mathsf{G}_{SR}^\dagger\mathsf{G}_{SR}|\psi_S\rangle \tag{18}$$

where we have substituted from Eq. (10) and used the "bra-ket" shorthand notation for the "row-vector column-vector" product

$$\langle\alpha|\beta\rangle \equiv \langle\alpha||\beta\rangle \tag{19}$$

## 3.4. Orthogonality and inner products

In general, a "bra-ket" expression like $\langle\alpha|\beta\rangle$ is an example of an *inner product*, and one formed in this way, as the matrix product of a row vector on the left and a column vector on the right, is an example of a *Cartesian inner product*. Inner products are very important in our mathematics, and we will be expanding on this concept substantially. (One of the simplest common examples of an inner product is the usual "dot" product of two geometrical vectors; this Cartesian inner product can be thought of as a generalization of this idea to vectors of arbitrary dimensionality and with complex amplitudes.)

A key point about inner products is that they can define the concept of *orthogonality* of functions. Specifically, for two non-zero vectors $|\alpha\rangle$ and $|\beta\rangle$, if and only if their inner product is zero, then the functions are said to be *orthogonal*. (This is also a generalization of the concept of two (non-zero) geometrical vectors being at right angles or "orthogonal" if and only if their dot product is zero.)

An immediate consequence of the idea of orthogonality from the inner product is that, for a wave that is the sum of multiple different orthogonal components, then a power as in Eq. (18) is simply the sum of the powers of the individual components; all the "cross-terms" disappear. Explicitly, for a wave $|\phi\rangle$ that is a sum of set of (non-zero) waves $\{|\phi_q\rangle\}$

$$|\phi\rangle = \sum_q |\phi_q\rangle \tag{20}$$

where those waves are all orthogonal, which we can write as

$$\langle\phi_p|\phi_q\rangle = 0 \quad \text{if and only if} \quad p \neq q \tag{21}$$

then

$$P = \langle\phi|\phi\rangle = \left(\sum_p \langle\phi_p|\right)\left(\sum_q |\phi_q\rangle\right) = \sum_{p,q} \langle\phi_p|\phi_q\rangle = \sum_q \langle\phi_q|\phi_q\rangle = \sum_q P_q \tag{22}$$

where

$$P_q = \langle\phi_q|\phi_q\rangle \tag{23}$$

is the power in the wave $|\phi_q\rangle$.

Later, as we generalize the mathematics, we may formally define inner products that explicitly give the power or energy for electromagnetic waves; these will not just be the simple Cartesian products of wave functions, though they will still satisfy the more basic mathematical properties required of inner products.

A second important property that the inner product of a (non-zero) vector with itself is always positive; this is easy to see for the Cartesian inner product of a vector such as $|\phi_R\rangle$ as in Eq. (9); explicitly

$$\langle\phi_R|\phi_R\rangle = \begin{bmatrix} f_1^* & f_2^* & \cdots & f_{N_R}^* \end{bmatrix} \begin{bmatrix} f_1 \\ f_2 \\ \vdots \\ f_{N_R} \end{bmatrix} = \sum_{j=1}^{N_R} f_j^* f_j = \sum_{j=1}^{N_R} |f_j|^2 > 0 \tag{24}$$

because it is a sum of positive (or at least non-negative) quantities $|f_j|^2$, at least one of which must be greater than zero for a non-zero vector.

## 3.5. Orthonormal functions and vectors

Now, returning to Eq. (18), we presume we want to find the choices of source functions or vectors $\{|\psi_{Sj}\rangle\}$ that give the largest total powers in the set of receivers (or microphones). To make comparisons easier, we will presume that we *normalize* all the source functions of interest to us. Normalization means that we adjust the function or vector by some multiplicative (normalizing) factor so that the inner product of the function or vector with itself is unity, i.e.,

$$\langle \psi_{Sj} | \psi_{Sj} \rangle = 1 \tag{25}$$

A particularly convenient set of functions is one that is both normalized and in which all the different elements are orthogonal; this is called an *orthonormal* set of functions, and its elements would therefore satisfy

$$\langle \psi_{Sp} | \psi_{Sq} \rangle = \delta_{pq} \tag{26}$$

where the Kronecker delta is

$$\delta_{pq} = \begin{cases} 1 \text{ if } p = q \\ 0 \text{ if } p \neq q \end{cases} \tag{27}$$

## 3.6. Vector spaces, operators and Hilbert spaces

The mathematics here gives some very powerful tools and concepts. We are not going to prove these properties now for two reasons: first, for finite matrices, these properties are discussed in standard matrix algebra texts [45, 46]. Second, we will give these results for the more general (and difficult) cases of continuous functions and infinite matrices in section 6 (and with proofs in [36]); finite matrices are then a simple special case.

An operator's properties can only be completely described if we are specific about the mathematical "space" (often called a *vector space*) in which it operates. For example, the ordinary (geometrical) vector dot product is an operator that operates in the mathematical space based on ordinary three-dimensional geometric space. This mathematical space contains all vectors that can exist in a geometrical space that is three-dimensional, with the algebraic property of having a vector dot product – the "inner product" for this space.

The operators of interest to us act on vectors or functions in a *Hilbert space* (formally defined in section 6.4) Any given Hilbert space will have a specific dimensionality, which may be finite or infinite, and it must have an inner product. We can think of this mathematical Hilbert space as being analogous to the mathematical space of ordinary geometrical vectors, but allowing arbitrary dimensionality and complex coefficients (geometrical vectors can only be associated with real amplitudes or coefficients).

The possible source functions exist in one Hilbert space $H_S$, associated with the source volume $V_S$. In our example here, this space $H_S$ contains all possible $N_S$-dimensional mathematical vectors with finite complex elements that are the amplitudes of specific point sources. The possible wave functions exist in another Hilbert space $H_R$, associated with the receiving volume $V_R$. This space $H_R$ also contains all possible $N_R$-dimensional mathematical vectors with finite complex elements that are the possible amplitudes of specific waves at the point "microphones" (or the corresponding signals from those microphones). Each of these spaces $H_S$ and $H_R$ has a Cartesian inner product, though later we may use different "underlying" inner products in different spaces.

## 3.7. Eigen problems and singular value decomposition

Now we see that the operator $\mathsf{G}_{SR}$ is something that maps between these two spaces. Specifically, as in Eq. (10), it operates on the vector $|\psi_S\rangle$, which is in space $H_S$, to generate the vector $|\phi_R\rangle$, which is in space $H_R$. Now, we want to find some "best" choices of such source vectors $|\psi_S\rangle$ that will give us the "best" resulting waves $|\phi_R\rangle$.

For such best choices, our instinct might be to try to find eigenvectors of some operator. However, we cannot just find eigenvectors of $\mathsf{G}_{SR}$; we might be able mathematically to find eigenvectors of the matrix $\mathsf{G}_{SR}$, but these may have dubious physical meaning in our problem, because $\mathsf{G}_{SR}$ is an operator mapping between one space and another, not an operator within a space. So, $\mathsf{G}_{SR}$ does not map a function back onto a multiple of itself in the same space.

We could, of course, define a Green's function operating within a space, and we might do so for a resonator problem; we could even base that on the exactly the same kind of mathematical expression as in Eq. (4) for $G_\omega(\mathbf{r};\mathbf{r}')$, with $\mathbf{r}$ and $\mathbf{r}'$ being positions within the same volume. Here, however, we are effectively basing our operator $\mathsf{G}_{SR}$ on the mathematical operator (or *kernel* in the language of integral equations) $G_\omega(\mathbf{r}_R;\mathbf{r}_S)$, where $\mathbf{r}_R$ and $\mathbf{r}_S$ are definitely in different volumes.

The key to constructing the right eigen problems here is to look at those associated with the operator $\mathsf{G}_{SR}^\dagger \mathsf{G}_{SR}$ and with the complementary operator $\mathsf{G}_{SR}\mathsf{G}_{SR}^\dagger$. As we said, the operator $\mathsf{G}_{SR}$ maps a vector in $H_S$ to a vector in $H_R$. The operator $\mathsf{G}_{SR}^\dagger$, however, maps a vector in $H_R$ to a vector in $H_S$. So, overall, $\mathsf{G}_{SR}^\dagger \mathsf{G}_{SR}$ maps a vector in $H_S$ to a vector in $H_S$. Similarly, the operator $\mathsf{G}_{SR}\mathsf{G}_{SR}^\dagger$ maps a vector in $H_R$ to a vector in $H_R$. Hence, it is physically meaningful to consider eigen problems for each of these operators $\mathsf{G}_{SR}^\dagger \mathsf{G}_{SR}$ and $\mathsf{G}_{SR}\mathsf{G}_{SR}^\dagger$. It is the mathematics of such eigen problems that is at the core of SVD.

Here we can usefully introduce several more definitions and results (mostly without proofs for the moment). First, we note that the operators $\mathsf{G}_{SR}^\dagger \mathsf{G}_{SR}$ and $\mathsf{G}_{SR}\mathsf{G}_{SR}^\dagger$ are *Hermitian* – that is, each is its own Hermitian adjoint. Explicitly

$$\left(\mathsf{G}_{SR}^\dagger \mathsf{G}_{SR}\right)^\dagger = \mathsf{G}_{SR}^\dagger \left(\mathsf{G}_{SR}^\dagger\right)^\dagger = \mathsf{G}_{SR}^\dagger \mathsf{G}_{SR} \tag{28}$$

where we have used Eqs. (13) and (15), and similarly for $\mathsf{G}_{SR}\mathsf{G}_{SR}^\dagger$.

An operator like $\mathsf{G}_{SR}^\dagger \mathsf{G}_{SR}$ is also a *positive operator*, which means that an expression like $\langle \psi_S | \mathsf{G}_{SR}^\dagger \mathsf{G}_{SR} | \psi_S \rangle$ is always greater than or equal to zero. (Similarly the operator $\mathsf{G}_{SR}\mathsf{G}_{SR}^\dagger$ is also Hermitian and positive.)

Now, so-called "compact" [47] Hermitian operators (defined formally in section 6) have several properties (and all finite Hermitian matrices are compact Hermitian operators)

1) their eigenvalues are real
2) their eigenfunctions are orthogonal (or, at least formally, the ones corresponding to different eigenvalues are orthogonal, and different ones corresponding to the same eigenvalue can always be chosen to be orthogonal)
3) their eigenfunctions form *complete sets* for the Hilbert spaces in which they operate [48] – in other words, we can write any function in the space as some linear combination of these eigenfunctions

and if those operators are positive

4) their eigenvalues are greater than or equal to zero
5) their eigenfunctions and their corresponding eigenvalues satisfy maximization properties – specifically, if we set out to find the normalized "input" vector or function that led to the largest "output" vector (in terms of its inner product), then that is the eigenfunction with the largest eigenvalue, and we could find the eigenfunction with the second largest eigenvalue and corresponding eigenvector by repeating the maximization process to find a function orthogonal to the first one, and so on.

The formal proofs of all of these properties are given in [36] for finite and infinite dimensional spaces, and we discuss these topics further also in section 6.

Furthermore, any operator that can be approximated to any sufficient degree by a finite matrix is also effectively compact; indeed, this idea is beginning to get close to the idea of what a compact operator really is. So, certainly our finite matrix problem with positive operators $\mathsf{G}_{SR}^\dagger \mathsf{G}_{SR}$ or $\mathsf{G}_{SR}\mathsf{G}_{SR}^\dagger$ here has all of the properties (1) to (5) above, and it will retain these properties no matter how large we make the (finite) matrix.

Now, for specific choices of the numbers and positions of the point sources and receivers, we can simply write down the $N_R \times N_S$ matrix $\mathsf{G}_{SR}$ as in equation (9), using the formula Eq. (7) to work out the necessary matrix elements. So, we are ready to turn any such specific problem into a simple numerical problem to find the eigenvectors and eigenvalues. So, therefore, for any such problem, we can solve for the (orthonormal) eigenvectors $|\psi_{Sj}\rangle$ of the $N_S \times N_S$ matrix $\mathsf{G}_{SR}^{\dagger}\mathsf{G}_{SR}$. The eigenvalues are necessarily positive (because $\mathsf{G}_{SR}^{\dagger}\mathsf{G}_{SR}$ is a positive operator), and so we can write them in the form $|s_j|^2$. That is, explicitly,

$$\mathsf{G}_{SR}^{\dagger}\mathsf{G}_{SR}|\psi_{Sj}\rangle = |s_j|^2 |\psi_{Sj}\rangle \tag{29}$$

Similarly, we can solve for the (orthonormal) eigenvectors $|\phi_{Rj}\rangle$ of the $N_R \times N_R$ matrix $\mathsf{G}_{SR}\mathsf{G}_{SR}^{\dagger}$. It is not too surprising that these have the same [49] eigenvalues $|s_j|^2$. That is

$$\mathsf{G}_{SR}\mathsf{G}_{SR}^{\dagger}|\phi_{Rj}\rangle = |s_j|^2 |\phi_{Rj}\rangle \tag{30}$$

In fact, we can show (**Appendix D**) that

$$\mathsf{G}_{SR}|\psi_{Sj}\rangle = s_j |\phi_{Rj}\rangle \tag{31}$$

and

$$\mathsf{G}_{SR}^{\dagger}|\phi_{Rj}\rangle = s_j^* |\psi_{Sj}\rangle \tag{32}$$

Hence by solving two eigenvalue problems, one for $\mathsf{G}_{SR}^{\dagger}\mathsf{G}_{SR}$ and a second for $\mathsf{G}_{SR}\mathsf{G}_{SR}^{\dagger}$, we have established two sets of eigenfunctions, one, $\{|\psi_{Sj}\rangle\}$, for the source vectors or functions in $H_S$, and a second set $\{|\phi_{Rj}\rangle\}$ for the wave vectors or functions in $H_R$. Note, too, that these vectors or functions are paired: a source vector or function $|\psi_{Sj}\rangle$ in $H_S$ leads to the corresponding wave vector or function $|\phi_{Rj}\rangle$ in $H_R$, with an amplitude $s_j$. The numbers $s_j$ are called the *singular values* of the operator or matrix $\mathsf{G}_{SR}$.

Note that, in practice, we only actually have to solve one eigenvalue problem – that is, either (29) or (30). If we know the eigenfunctions $\{|\psi_{Sj}\rangle\}$ from solving Eq. (29), then we can deduce the eigenfunctions $\{|\phi_{Rj}\rangle\}$ from Eq. (31), and similarly if we know the eigenfunctions $\{|\phi_{Rj}\rangle\}$ from Eq. (30), we can deduce the $\{|\psi_{Sj}\rangle\}$ from Eq. (32), at least for all the cases where the singular value is not zero [49]. In practice, one of these eigen problems may be simpler or effectively "smaller" than the other, and we can conveniently choose that one if we prefer.

The fact that these two sets of functions $\{|\psi_{Sj}\rangle\}$ and $\{|\phi_{Rj}\rangle\}$ are each eigenfunctions of a Hermitian operator guarantees that each of these sets is orthogonal and complete [50] for its Hilbert space ($H_S$ or $H_R$ respectively).

From Eqs. (31) and (32), we can see that we can rewrite $\mathsf{G}_{SR}$ as

$$\mathsf{G}_{SR} = \sum_{j=1}^{N_m} s_j |\phi_{Rj}\rangle\langle\psi_{Sj}| \tag{33}$$

where $N_m$ is the smaller of $N_S$ and $N_R$ [49]. This expression Eq. (33) is called the *singular value decomposition* of the operator $\mathsf{G}_{SR}$. We can also similarly write

$$\mathsf{G}_{SR}^{\dagger} = \sum_{j=1}^{N_m} s_j^* |\psi_{Sj}\rangle\langle\phi_{Rj}| \tag{34}$$

Incidentally, a product of the form $|\phi_{Rj}\rangle\langle\psi_{Sj}|$, which has a column vector on the left and a row vector on the right, is sometimes called an *outer product*. Standard matrix manipulations show an outer product of two *N*-element vectors is an $N \times N$ matrix. So,

> this process of singular value decomposition, performed by solving the two related eigen problems, one for the matrix or operator $G_{SR}^{\dagger}G_{SR}$ and the second for the matrix or operator $G_{SR}G_{SR}^{\dagger}$, leads to our desired two sets of orthogonal vectors or functions.

These are source vectors or functions in $H_S$ and wave vectors or functions in $H_R$, and these are "paired up", with each source eigenvector in $H_S$ giving rise to its corresponding wave eigenvector in $H_R$ (with amplitude given by the corresponding singular value).

Furthermore, because these are eigenvectors or eigenfunctions of a positive Hermitian operator, by property (5) above, they are the "best" possible choices. Specifically, if we choose to order the eigenvectors by decreasing size of $|s_j|^2$, then (neglecting degeneracy of eigenvalues (i.e., more than one eigenvector for a given eigenvalue) for simplicity here at the moment) the source vector $|\psi_{S1}\rangle$ in $H_S$ gives rise to the largest possible magnitude of wave vector in $H_R$, and has the form $|\phi_{R1}\rangle$ the source vector $|\psi_{S2}\rangle$ is the source vector in $H_S$ that is orthogonal to $|\psi_{S1}\rangle$ and that gives rise to the second largest possible magnitude of wave vector in $H_R$, which has the form $|\phi_{R2}\rangle$ and is orthogonal to $|\phi_{R1}\rangle$ the source vector $|\psi_{S3}\rangle$ is the source vector in $H_S$ that is orthogonal to $|\psi_{S1}\rangle$ and $|\psi_{S2}\rangle$ and that gives rise to the third largest possible magnitude of wave vector in $H_R$, which has the form $|\phi_{R3}\rangle$ and is orthogonal to $|\phi_{R1}\rangle$ and $|\phi_{R2}\rangle$ and so on.

We have therefore established the best set of possible orthogonal (and therefore zero "crosstalk") "channels" between the two volumes (at least "best" as given by the magnitude of the inner product). Note explicitly that these channels are orthogonal to one another, both at the sources and at the receivers.

Incidentally, in matrix terms, the singular value decomposition as in Eq. (33) can also be written in the form

$$G_{SR} = VD_{diag}U^{\dagger} \tag{35}$$

where $D_{diag}$ is a diagonal matrix with the singular values $s_j$ as the diagonal elements, $V$ is a matrix whose columns are the vectors $|\phi_{Rj}\rangle$, and $U^{\dagger}$ is a matrix whose rows are the vectors $\langle\psi_{Sj}|$ (or equivalently $U$ is a matrix whose columns are the vectors $|\psi_{Sj}\rangle$). Technically, the matrices $V$ and $U^{\dagger}$ (and also $U$) are "unitary". (See **Appendix D**.)

## 3.8. A sum rule on coupling strengths

One very important point emerges from this algebra, which is a "sum rule" on the $|s_j|^2$. Specifically, we can show that, quite generally for finite numbers $N_S$ and $N_R$ of sources and receiver points

$$S = \sum_{q=1}^{N_m} |s_q|^2 = \sum_{i=1}^{N_R}\sum_{j=1}^{N_S} |g_{ij}|^2 \tag{36}$$

This mathematical result Eq. (36) is technically a consequence of the fact that the eigenfunctions have been written in normalized form, which in turn is why the matrices $\mathsf{V}$ and $\mathsf{U}^{\dagger}$ are "unitary", but this is really a deeper truth, and it has several important consequences that are central to the larger discussion here.

1) We can evaluate this sum *S* without even solving the eigenproblem.
2) Having evaluated *S*, we can know immediately that there is an upper bound on how many channels we could have that have at least some given coupling strength; there could only be a finite number of orthogonal channels with any given finite magnitude of coupling strength.
3) Suppose we solve the eigen problem by looking one by one, physically or mathematically, for the channels by some process, noting their coupling strengths. Then, when we find we have nearly exhausted the available sum rule *S*, we can stop looking for any more channels; there cannot be any more strongly coupled channels, because there is not sufficient sum rule *S* left.

It might seem that this kind of sum rule is only going to exist when we consider finite numbers of sources and receivers. However, we are going to find below in section **6** that the operators associated with wave equations are going to have a finite result for this sum rule *even as we consider continuous functions and infinite basis sets*. Indeed, this finiteness is the defining characteristic of so-called Hilbert-Schmidt operators (which incidentally, are necessarily also "compact"), and we return to this point in section **6**. So, though we continue for the moment with finite numbers of sources and receiver points, our results are generally going to survive even as we make the transition to such continuous source and receiver functions, with possibly infinite basis sets.

This finite sum rule can also be regarded as the source of diffraction limits with waves, as will become clearer below. Such limits apply both in conventional optical situations and more generally, and this result is a central benefit of this SVD approach.

## 3.9. A constraint on the choice of the coupling strengths of the channels

In looking at the sum rule, we might hope that we have the flexibility to choose how many channels are how strong – we might want to have several channels at a weaker coupling strength rather than a few strong ones, for example – while still "obeying" the sum rule. However, once we have chosen a given set of source and receiver points, or have done the equivalent for continuous functions in choosing the Hilbert spaces for source and receiver functions, we do not have that flexibility.

The eigenvalues, which are the coupling strengths, and any degeneracies they have, are uniquely specified when we solve the eigen problem. The eigenfunctions are also essentially unique, other than for the minor flexibility allowed in choosing the specific linear combinations for eigenfunctions associated with the same degenerate eigenvalue.

So, not only does the sum rule constrain the overall sum of connection strengths; the eigen solutions uniquely determine the coupling strengths of the channels if we want them to be orthogonal (with orthogonality determined by the inner products in the Hilbert spaces).

Suppose we write down the "power" coupling strengths $|s_j|^2$ in order, starting from the strongest and proceeding to weaker ones In this list, if a given eigenvalue is degenerate, with a degeneracy of *n*, we will write it down *n* times [51], and give each occurrence a separate value of the index *j*. So we would have a list of the form

$$|s_1|^2 \geq |s_2|^2 \geq \cdots \geq |s_j|^2 \geq \cdots \quad (37)$$

Then, provided we include all the occurrences of a given degenerate eigenvalue (so if we encounter degenerate coupling strengths in the list, our choice of *j* must be at the end of any such degenerate group of coupling strengths), we can state that The number of orthogonal channels (communications modes) with power coupling strength $|s|^2 \geq |s_j|^2$ is $j$ (38)

We cannot rearrange any linear combinations of communications modes to get more channels that are at least this strongly coupled. If we want more such channels, then we have to change the source and/or receiver points and/or Hilbert spaces.

# 4. An introductory example - 3 sources and 3 receivers

To understand how this approach works, both mathematically and physically, we can look at a simple example, one that is large enough to be meaningful, but small enough for explicit details. Fig. 3 shows the physical layout. We have $N_S = 3$ point sources, spaced by $2\lambda$ in the "vertical" $y$ direction, (with wavelength $\lambda = 2\pi / k$ as usual for a wavevector magnitude $k$). These are separated by a "horizontal" distance $L_z = 5\lambda$ from a similarly spaced set of $N_R = 3$ receiving points.

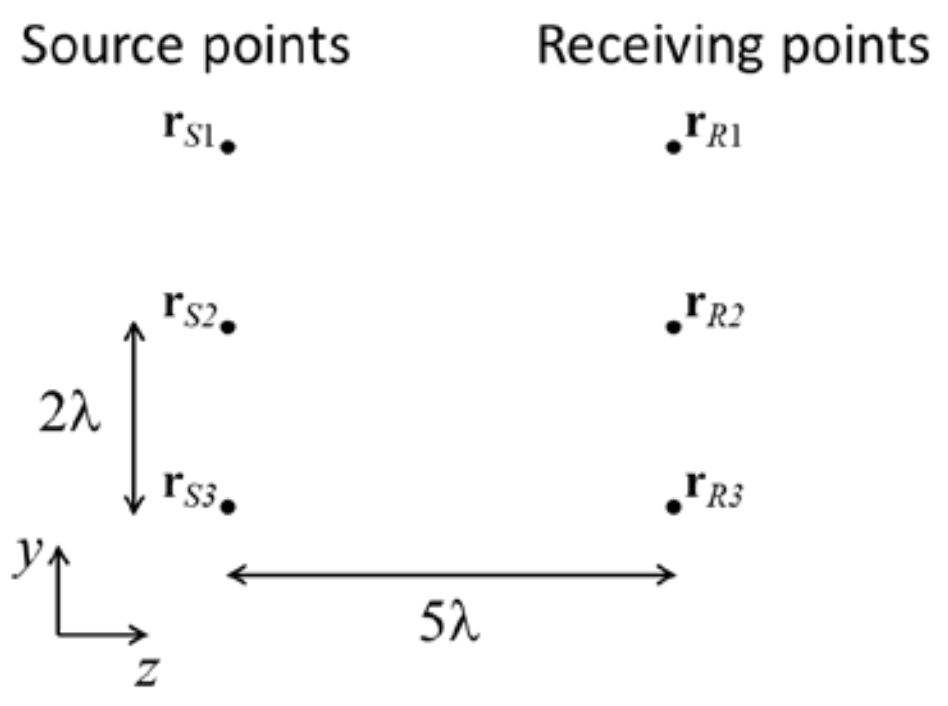

Fig. 3 A set of three sources, spaced by $2\lambda$ in the "vertical" $y$ direction, separated from three similarly spaced receiving points by a distance $5\lambda$ in the "horizontal" $z$ direction.

## 4.1. Mathematical solution

we use Eq. (7) to calculate the matrix elements $g_{ij}$ of the coupling operator $G_{SR}$. Explicitly, for example, for the matrix element $g_{13}$ that gives the wave amplitude at $\mathbf{r}_{R1}$ as a result of the source amplitude at $\mathbf{r}_{S3}$, noting first that

$$|\mathbf{r}_{R1} - \mathbf{r}_{S3}| = \sqrt{(y_{R1} - y_{S3})^2 + (z_{R1} - z_{S3})^2} = \lambda\sqrt{4^2 + 5^2} = \sqrt{41}\lambda \tag{39}$$

then

$$g_{13} = -\frac{1}{4\pi}\frac{\exp(ik|\mathbf{r}_{R1} - \mathbf{r}_{S3}|)}{|\mathbf{r}_{R1} - \mathbf{r}_{S3}|} = -\frac{1}{4\pi}\frac{\exp(2\pi i\sqrt{41})}{\sqrt{41}\lambda} \simeq \frac{0.01020 - 0.00711i}{\lambda} \tag{40}$$

Now we write distances in units of wavelengths for convenience (or equivalently, we set $\lambda = 1$). To get numbers of convenient sizes and signs, we multiply by a factor

$$g_{SR} = -4\pi L_z = -62.83 \tag{41}$$

So, then

$$g_{SR}g_{13} \simeq -62.83\times\left(0.01020-0.00711i\right)\simeq -0.64+0.45i \quad (42)$$

Proceeding similarly for the other matrix elements, we have

$$g_{SR}\mathsf{G}_{SR} \cong \begin{bmatrix} 1 & -0.7+0.6i & -0.64+0.45i \\ -0.7+0.6i & 1 & -0.7+0.6i \\ -0.64+0.45i & -0.7+0.6i & 1 \end{bmatrix} \quad (43)$$

The sum, Eq. (36), of the modulus squared of these matrix elements of $\mathsf{G}_{SR}$ as in Eq. (43) is the sum rule

$$S = 7.67 / g_{SR}^2 \quad (44)$$

The matrix $\mathsf{G}_{SR}^{\dagger}\mathsf{G}_{SR}$ can be written, with a convenient scaling factor $g_{SR}^2$,

$$g_{SR}^2\mathsf{G}_{SR}^{\dagger}\mathsf{G}_{SR} \cong \begin{bmatrix} 2.47 & -0.67-0.08i & -0.42 \\ -0.67+0.08i & 2.72 & -0.67+0.08i \\ -0.42 & -0.67-0.08i & 2.47 \end{bmatrix} \quad (45)$$

Note that this matrix is Hermitian, as required. Having established the matrix $\mathsf{G}_{SR}^{\dagger}\mathsf{G}_{SR}$, we can now use standard numerical routines to find eigenvalues and eigenvectors. The resulting eigenvalues of $\mathsf{G}_{SR}^{\dagger}\mathsf{G}_{SR}$ (and of $\mathsf{G}_{SR}\mathsf{G}_{SR}^{\dagger}$) are

$$\left|s_1\right|^2 = \frac{3.41}{g_{SR}^2}\ ,\ \left|s_2\right|^2 = \frac{2.89}{g_{SR}^2}\text{, and } \left|s_3\right|^2 = \frac{1.37}{g_{SR}^2} \quad (46)$$

Note that the sum of these is indeed also the sum rule $S$ [52], i.e.,

$$\left|s_1\right|^2+\left|s_2\right|^2+\left|s_3\right|^2 \cong \frac{3.41}{g_{SR}^2}+\frac{2.89}{g_{SR}^2}+\frac{1.37}{g_{SR}^2}=\frac{7.67}{g_{SR}^2}=S \quad (47)$$

The corresponding eigenvectors of $\mathsf{G}_{SR}^{\dagger}\mathsf{G}_{SR}$ are [53]

$$\left|\psi_{S1}\right\rangle = \begin{bmatrix} 0.41 \\ -0.81+0.1i \\ 0.41 \end{bmatrix},\ \left|\psi_{S2}\right\rangle = \begin{bmatrix} -0.71 \\ 0 \\ 0.71 \end{bmatrix}\text{, and } \left|\psi_{S3}\right\rangle = \begin{bmatrix} 0.58 \\ 0.57-0.07i \\ 0.58 \end{bmatrix} \quad (48)$$

which are all orthogonal to one another [54], and the corresponding eigenvectors of $\mathsf{G}_{SR}\mathsf{G}_{SR}^{\dagger}$ are

$$\left|\phi_{R1}\right\rangle \cong \begin{bmatrix} 0.41 \\ -0.81-0.1i \\ 0.41 \end{bmatrix},\ \left|\phi_{R2}\right\rangle \cong \begin{bmatrix} -0.71 \\ 0 \\ 0.71 \end{bmatrix}\text{, and } \left|\phi_{R3}\right\rangle \cong \begin{bmatrix} 0.58 \\ 0.57+0.07i \\ 0.58 \end{bmatrix} \quad (49)$$

For this very symmetrical problem, the receiving wave vectors and the source vectors are just complex conjugates of one another, though that is not generally the case. In this case, the complex conjugation means that the “phase” curvatures are equal and opposite. (The idea of phase curvature becomes clearer as we consider more source and receiver points, so we postpone that illustration.) If we wanted to construct the strongest “channels” between these sources and receivers at this wavelength (or frequency), we would

(a) choose to drive the point sources with relative amplitudes and phases given by one of the source eigenvectors in Eq. (48), and

(b) at the receiving points we would add up the signals from the different points weighted with the relative amplitudes and phase shifts given by the corresponding receiving eigenvector in Eq. (49).

We can also send and receive three separate channels at once through this system if we construct appropriate systems to create and to separate the necessary signals.

## 4.2. Physical implementation

To see how to create and detect the necessary signals, including running all three channels at once, we can look at example physical systems. These are not meant to be engineered solutions to a real problem, but they make the mathematics more concrete physically.

### 4.2.1. Acoustic and radio-frequency systems

First, for acoustic or radio-frequency signals, we can likely generate and measure the actual field directly. We will not consider the actual corresponding loudspeakers, microphones, or antennas for the moment, just approximating them instead by ideal point sources and receiving elements. (We are still postponing any consideration of vector electromagnetic fields, just considering scalar waves.)

We can then use appropriate electronic circuits that generate and collect the corresponding signals (Fig. 4). In these cases, we can imagine input voltage signals $V_{SIn1}$, $V_{SIn2}$, and $V_{SIn3}$ that represent the signals (such as three different binary bit streams, for example) that we want to send on the three different "channels" in the communication between the three sources and the three receivers. We would like these signals (e.g., the three binary bit streams) to appear as the three output voltage signals $V_{ROut1}$, $V_{ROut2}$, and $V_{ROut3}$ at corresponding electrical outputs at the far end of the system.

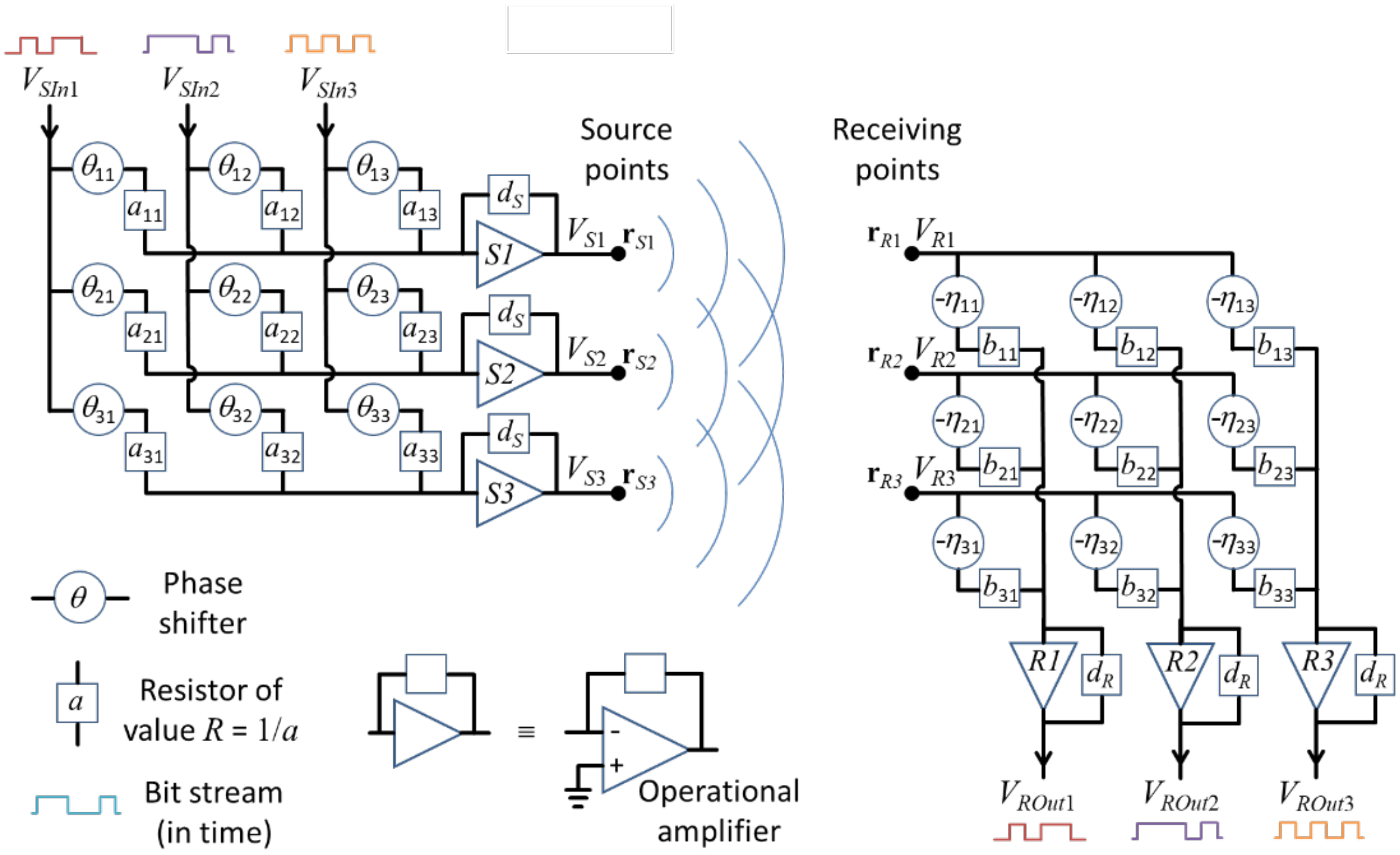

Fig. 4. Example electrical driving and receiving circuits to form the superposition of sources for transmission and to separate the channels again for reception. The input channels are the (voltage) bit streams $V_{SIn\,j}$ and the outputs are the corresponding (voltage) bit streams $V_{ROut\,j}$. The feedback conductances $d_S$ and $d_R$ set the overall electrical gain in transmission and reception in the operational amplifier circuits that sum the currents on their input "bus" lines as required to form the necessary linear superpositions.

Each such input signal voltage has to generate the corresponding vector of amplitudes to drive the sources. So $V_{SIn1}$ should generate a vector of voltage amplitudes $\propto V_{SIn1}|\psi_{S1}\rangle$, and similarly for the other two input voltage signals, and these three vectors should be added to generate the corresponding

set of output voltages $V_{S1}$, $V_{S2}$, and $V_{S3}$ that drive the sources (e.g., loudspeakers) at the corresponding positions $\mathbf{r}_{S1}$, $\mathbf{r}_{S2}$, and $\mathbf{r}_{S3}$.

We can perform this generation of the correct vectors of amplitudes and their summation by using the "analog crossbar" circuit on the left of Fig. 4. We presume we can make electrical phase shifters (circles in Fig. 4), whose phase delay is indicated inside the corresponding circle. The output of each such phase shifter is then passed as a voltage to drive a resistor (rectangular boxes in Fig. 4), whose conductance (i.e., 1/resistance) is given by the value inside the box (in some appropriate units). The other end of the resistors is connected in each case to a common "bus" line that is a "virtual ground" input to an operational amplifier (the triangles in in Fig. 4). The operational amplifiers each sum all the currents on their input bus line and each generate an output voltage proportional to this sum, giving the set of output drive voltages $V_{S1}$, $V_{S2}$, and $V_{S3}$.

At the receiving end of the system, we can construct a similar circuit. In this case the input voltages to the analog crossbar are the outputs $V_{R1}$, $V_{R2}$, and $V_{R3}$ from the measured fields at the three positions $\mathbf{r}_{R1}$, $\mathbf{r}_{R2}$, and $\mathbf{r}_{R3}$. If we have designed and set up our circuits correctly, the corresponding summed outputs, which become the voltage signals $V_{ROut1}$, $V_{ROut2}$, and $V_{ROut3}$, should each now be proportional to the original voltage signals $V_{SIn1}$, $V_{SIn2}$, and $V_{SIn3}$ that we wanted to send through this three-channel system (just with some propagation time delay). If we make the feedback conductances in the operational amplifier circuits all identical at some value $d_R$, then, for equal overall magnitudes of input voltage signals $V_{SIn1}$, $V_{SIn2}$, and $V_{SIn3}$, the relative sizes of the output voltage signals $V_{ROut1}$, $V_{ROut2}$, and $V_{ROut3}$ would be weighted by the corresponding singular values $s_j$ for the *j*th channel through the system. (Of course, we could compensate for the different singular values by using feedback conductances $\propto s_j$ in the circuits with the output operational amplifiers *R1*, *R2*, and *R3*.)

Writing each of the vectors in Eqs. (48) and (49) with matrix elements in "polar" form, i.e.,

$$\left|\psi_{Sq}\right\rangle = \begin{bmatrix} a_{1q}\exp\left(i\theta_{1q}\right) \\ a_{2q}\exp\left(i\theta_{2q}\right) \\ a_{3q}\exp\left(i\theta_{3q}\right) \end{bmatrix} \text{ and } \left|\phi_{Rq}\right\rangle \cong \begin{bmatrix} b_{1q}\exp\left(i\eta_{1q}\right) \\ b_{2q}\exp\left(i\eta_{2q}\right) \\ b_{3q}\exp\left(i\eta_{3q}\right) \end{bmatrix} \qquad (50)$$

where the *a*, *b*, $\theta$, and $\eta$ coefficients are all real, then we obtain the corresponding desired settings of the phase shifts and conductances in Fig. 4.

Note that, in the receiving analog crossbar circuit, the phases are set to minus the corresponding phases in the receiving vectors themselves. This is because, in summing the currents in this crossbar we are actually performing the inner product $\left\langle\phi_{Rq}\middle|V_{RIn}\right\rangle$, where $\left|V_{RIn}\right\rangle$ is the column vector of received voltages $\begin{bmatrix}V_{RIn1} & V_{RIn2} & V_{RIn3}\end{bmatrix}^T$, so as to extract the appropriate component of the signal. Since the $\left|\phi_{Rq}\right\rangle$ vector is in "bra" form $\left\langle\phi_{RIn}\right|$ in the inner product, we must take the complex conjugate of the phase factors.

Hence in this way, starting out with three quite separate signals, we are able to transmit them through the system and recover the original signals again. Note the three channels here have no crosstalk even though the waves from each source are mixed at the three receiving points. This remains true even if the three "channels" in the system have different coupling strengths (as given by the singular values $s_j$), as they do here, with relative power strengths of 3.41, 2.89 and 1.37 respectively. Taken together, these three channels use up all the available power coupling strength, as given by the sum rule (Eq. (47) or, more generally, Eq. (36)).

The mathematical function performed by each of the crossbar circuits is, of course, simply a (complex) matrix multiplication. By use of appropriate analog-to-digital and digital-to-analog conversion, such multiplications could be performed digitally instead.

### 4.2.2. Optical systems

At optical frequencies, we generally cannot measure the field directly (certainly not in real time), and circuit approaches as in Fig. 4 are not viable. Indeed, until recently, it was not generally understood in optics how to separate overlapping optical signals (without fundamental loss) to turn them into individual output beams, which is a process we require at the receiver in our scheme. Similarly, it was not clear how to losslessly generate arbitrary linear superpositions of inputs to give overlapping outputs, as required at the source. Indeed, without some apparatus to perform such functions, in optics the multiple-channel schemes presented here would remain mathematical curiosities.

Recently, however, specific schemes have been devised for both creating and separating arbitrary linear combinations of overlapping optical beams, and for emulating arbitrary linear optical components [24 - 27]. Indeed, these schemes constitute the first proof that arbitrary linear optics is possible [25]; the proof is entirely constructive because it shows specifically how to make the optical system, at least in principle.

These schemes rely on meshes of two-beam interferometers, in appropriate architectures and with associated algorithms [24 - 28] to allow the meshes to be set up. We will not review these in detail in this article, but some key aspects are important here. In some of the architectures, the setup of the mesh (and, if required, the calculation of the necessary settings of the interferometers) can be entirely progressive [24 - 28]. The setup of the mesh can also be accomplished by "training" the mesh with the beams of interest [24 - 28], based on a sequence of single-parameter power minimizations, without calculation or calibration.

For our present example, the electronic matrix multiplications implicit in the analog crossbar networks at the source and at the receiver can be implemented instead using the "triangular" source mesh and receiving mesh respectively, as in Fig. 5 (a). These meshes, which can also be viewed as analog crossbars, work directly by interfering beams in waveguides and waveguide couplers. The light in these meshes flows through them without fundamental loss, and they mathematically represent "unitary" (loss-less) matrix multiplications.

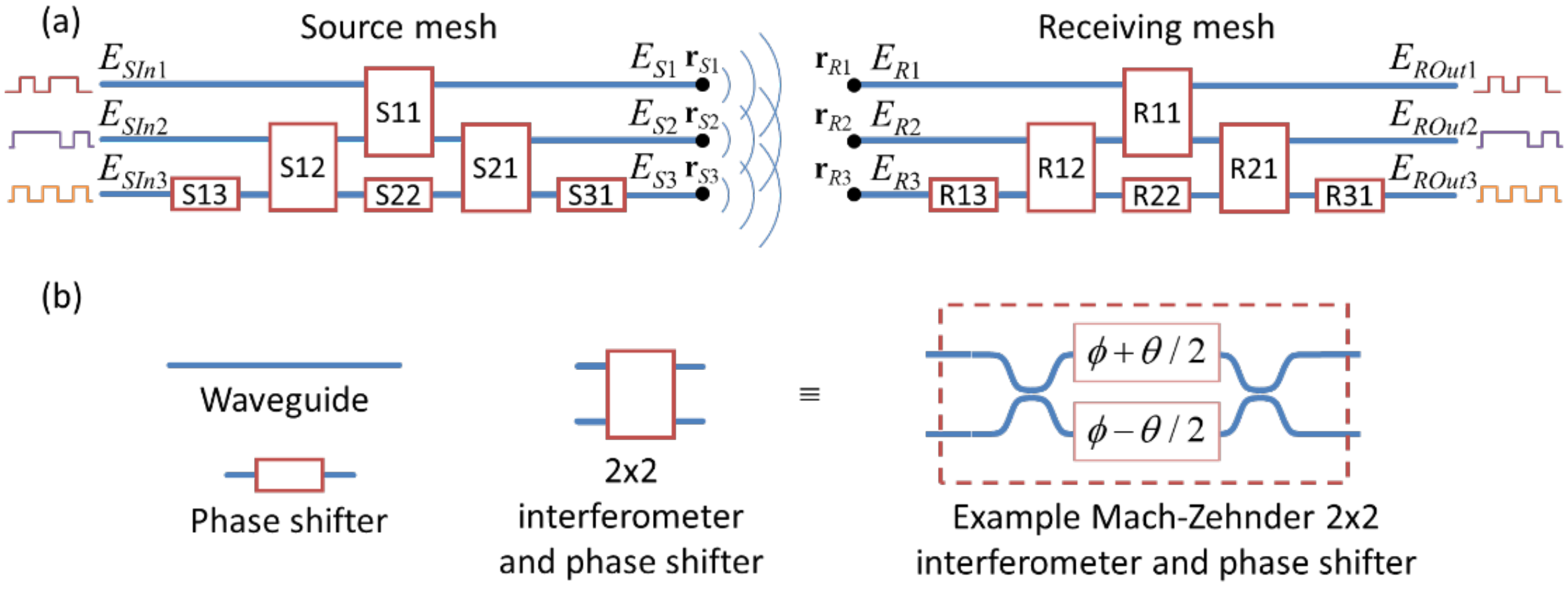

Fig. 5. (a) Interferometer mesh architectures to generate and superpose the necessary source communications mode source vectors from the separate input signals on the left, and to separate out the corresponding communications mode receiving vectors on the right to reconstruct the original channels of information. These processes work directly by interfering beams, and without fundamental loss in the meshes. (b) Key to the various elements in (a). One example form of Mach-Zehnder interferometer and phase shifter is shown that has the necessary functions for the 2x2 blocks.

In Fig. 5, we imagine that the input signals, instead of being voltages as in Fig. 4, are the amplitudes $E_{SIn1}$, $E_{SIn2}$, and $E_{SIn3}$ of the waves in single-(propagating)-mode input waveguides. By setting up the phase shifters and interferometers appropriately, the necessary superpositions are created as the amplitudes $E_{S1}$, $E_{S2}$, and $E_{S3}$ in the output waveguides. These amplitudes then feed the point sources at the corresponding positions $\mathbf{r}_{S1}$, $\mathbf{r}_{S2}$, and $\mathbf{r}_{S3}$.

In this example, we pretend that the outputs of those guides essentially represent point sources of waves that we can also approximate as being scalar (which is reasonable if we consider only one polarization for the moment). At the receiving end, we imagine that the inputs to receiving mesh waveguides, at points $\mathbf{r}_{R1}$, $\mathbf{r}_{R2}$, and $\mathbf{r}_{R3}$ are essentially "point" receiving elements to couple light into these waveguides, with corresponding waveguide amplitudes $E_{R1}$, $E_{R2}$, and $E_{R3}$ respectively. The receiving mesh then performs the appropriate matrix multiplication to separate out the signals (again without fundamental loss) to the output amplitudes $E_{ROut1}$, $E_{ROut2}$, and $E_{ROut3}$, recreating the input bit streams.

In Fig. 5 (b), we show one example way of making the necessary 2x2 interferometer block. This block should be able to control the "split ratio" of the interferometer – how the input power in, say, the top left waveguide is split between the two output waveguides on the right; this can be accomplished by controlling the phase angle $\theta$ (achieved by differentially driving the two phase shifters on the arms of the interferometer). The block should also independently be able to set one other phase on the outputs; in this example, this is achieved by the "common mode" drive of the phase shifters in the block, setting phase angle $\phi$ [55].

The setup of these unitary meshes is relatively straightforward [56]. Such a mesh has exactly the right number of independent real parameters (here, 9 phase shifters altogether) to construct an arbitrary 3x3 unitary matrix. Such design calculations are presented explicitly in [25] together with the self-configuring algorithms to allow direct training.

In a real system, rather than "bare" waveguide outputs and inputs for communications; likely one would add some optics, such as collimating lenses, in front of the waveguides, to avoid sending power in unnecessary directions. However, for our tutorial purposes at the moment, we will omit such optics (though it can ultimately also be handled by this approach by including the optics in the Green's function for the system).

The kind of architecture shown in Fig. 5 (a), with unitary meshes at both sides, has one other interesting property worth noting here. There is an iterative algorithm [12], based only on overall power maximization on one channel at a time, and working forwards and backwards through the entire system, that allows this system itself to find the best coupled channels. Essentially, by running such an algorithm, this system physically can find the SVD of the coupling operator $G_{SR}$ between the source points and the receiver points, without calculations. The results are then effectively stored as the settings of the various elements in the two meshes. This algorithm also still works even if there are other optics or scatterers between the source and receiver points, and so gives a way of finding the best orthogonal channels through any fixed linear optical system at a given frequency.

### 4.2.3. Larger systems

As we consider larger numbers of source and receiver points, the specific approaches in Fig. 4 and Fig. 5 would face technological limits of various kinds, especially for the purely optical approach of Fig. 5. However, practically, the ability to make large numbers of interferometers has been developing rapidly; working systems with up to 100's of interferometers [57 - 64] and small self-configuring systems [27, 29, 65] have both been demonstrated recently. Extensions to 1000's of interferometers may be feasible with current technology, and that number may not represent any particular fundamental limit. Indeed, these demonstrations and the potential for expanding to larger systems is one of the reasons why we consider these "modal" approaches. We need them both in wireless systems, where they can be viewed as extensions to MIMO (multiple input multiple output) antenna and communications systems, and in optical systems where we are now able to explore such configurable and optimizable multiple channel systems.

Whether we choose to make systems as in Fig. 4 or Fig. 5, these kinds of systems show the upper limits in terms of orthogonal channels and coupling strengths of what could be achieved through such linear processing and the resulting optimum channels. The scheme of Fig. 5 gives a method in principle of constructing arbitrary unitary (and hence nominally lossless) transforms of a given number of inputs and outputs (given ideal physical components). It operates with the minimum number of adjustable components and no loss in principle. No physical linear system can in principle do better than this scheme in constructing the communications modes for given numbers of source and receiver points. The existence of these approaches shows in principle that such systems could be made, both as actual physical systems up to some scale and as thought experiments at arbitrary scales for more basic discussions.

In what follows, we look at a variety of systems with larger numbers of source and receiver points in various different geometries. This progressively introduces many different behaviors. Some of these behaviors at large scales can relate to those seen in conventional optical systems, but some quite general behaviors have no particular well-known precedents. Though there are just a few results that can be expressed in analytic approximations for specific classes of systems (e.g., paraxial optics), there are several broader classes of behaviors that can be understood intuitively from these numerical simulations and some approximate heuristic results. These provide novel insights into communicating and sensing with waves in a wide range of systems, from acoustics, through radio and microwaves, to optics.

# 5. Scalar Wave examples with point sources and receivers

Now we continue to larger numbers and other geometries of source and receiver points to illustrate various behaviors.

## 5.1. 9 sources and 9 receivers in parallel lines

Now we space $N_S = 9$ point sources over the same total length of $4\lambda$ as in the 3 source case above (Fig. 3), and similarly for the $N_R = 9$ receiving points, which means the points are spaced by $\lambda / 2$ in each case [66].

### 5.1.1. Channels and coupling strengths

Using the same numerical approach as in the $3\times3$ case above, but now with a $9\times9$ matrix, there are 9 orthogonal source vectors and 9 corresponding orthogonal receiving vectors, and the sum rule $S$ is

$$S = 72.65 / g_{SR}^2 \qquad (51)$$

The results for the coupling strengths are summarized in Table 1.

**Table 1. Mode coupling strengths for 9 point sources and receivers**

| Mode number, $j$ | $\left|s_j\right|^2 / g_{SR}^2$ | % of $S$ | Cum. % of $S$ |
|---|---|---|---|
| 1 | 20.73 | 28.54 | 28.54 |
| 2 | 20.39 | 28.07 | 56.61 |
| 3 | 19.09 | 26.28 | 82.89 |
| 4 | 10.41 | 14.34 | 97.23 |
| 5 | 1.90 | 2.62 | 99.84 |
| 6 | 0.11 | 0.16 | ~100 |
| 7 | 0.0028 | 0.0038 | ~100 |
| 8 | 0.000027 | 0.000037 | ~100 |
| 9 | 0.000000065 | 0.000000089 | ~100 |

Though there are formally 9 orthogonal channels, there are only three strongly coupled channels of approximately the same coupling strength, one other channel about half as strong, one weak channel, one very weak channel, and three other extremely weak channels. Though we have 9 sources and receivers, we certainly do not have 9 practically usable channels.We see immediately that

> increasing the number of sources and/or receivers in given source and receiving volumes past a certain point does not increase the number of well-coupled channels.

The inability to form further well-coupled channels is being enforced by the sum rule *S*, and could be viewed as an effective "diffraction" limit.

### 5.1.2. Modes and beams

Once the eigenvectors $|\psi_{Sj}\rangle$ of amplitudes of the sources in each mode are calculated, it is straightforward to calculate the resulting wave or "beam" at any point **r** in space. Explicitly, with $N_S$ point sources, and writing out the *j*th (column) eigenvector as

$$|\psi_{Sj}\rangle \equiv \begin{bmatrix} h_{1j} & h_{2j} & \cdots & h_{N_S j} \end{bmatrix}^T \tag{52}$$

(with the superscript *T* indicating the transpose, used here just to save space) then the corresponding (complex) wave at point **r** is

$$\phi_j(\mathbf{r}) == -\frac{1}{4\pi}\sum_{q=1}^{N_S}\frac{\exp\left(ik\left|\mathbf{r}-\mathbf{r}_{Sq}\right|\right)}{\left|\mathbf{r}-\mathbf{r}_{Sq}\right|}h_{qj} \tag{53}$$

In Fig. 6, we have plotted the resulting amplitudes and phases of the first three (strongest) modes, together with the resulting waves or "beams". In Fig. 6, (a) is Mode 1, (b) is Mode 2, and (c) is Mode 3. The wave plotted is the real part of the wave amplitude, so it is essentially a snapshot of the wave at an arbitrary time, here for the values in the plane of the source and receiver points (i.e., the plane of the "paper").

In Fig. 6, we have also chosen the phase of all the source modes (source and receiving) to be zero in the middle of each mode [67]. Here also, because of the symmetry of this problem, the receiving vectors are just the complex conjugates of the transmitting vectors. We have deliberately used a scale to plot the phase of the source points so that $2\pi$ of phase corresponds to the same distance on the graph as one wavelength in the wave plot. This enables us to compare "phase curvatures" on the source vectors with the actual phase curvature as seen by eye on the wave plots.

With this larger number of sources and receivers, though without increasing the overall "size" of the source and receiving spaces, we can get intelligible behavior of the waves, including relatively clear "phase fronts" in the wave propagation. Mode 1 and Mode 2 have almost exactly the same coupling strength, with Mode 1 being (possibly surprisingly) just slightly stronger. Mode 2 is the simplest, corresponding to a "single-bumped" beam in the region between the source and receiver positions. Mode 1 is a "two-bumped" beam, with the upper and lower halves having opposite sign.

Note that the solution of these eigen problems has given the source vector a "phase curvature" that, for these first two Modes, leads to a beam "waist" approximately in the middle horizontally though, by eye, the beam in Mode 1 is wider "top to bottom" throughout than that of Mode 2. Again, the phase curvature of the source for these two Modes is such that, by eye, it is similar to the phase curvature of the wave. Mode 3 corresponds to a "three-bumped" beam, though the behavior overall is not so simple to describe as that of Modes 1 and 2. Note that the beam in Mode 3 in particular is nearly filling all the space between the source points and the receiver points.

The calculations also show that the corresponding receiving vector in each case has the opposite phase curvature to that of the source. In general the orthogonality of these three modes at both the source and receiver points is obvious graphically since they correspond to one, two, and three "bumped" beams over the lines of sources and receivers.

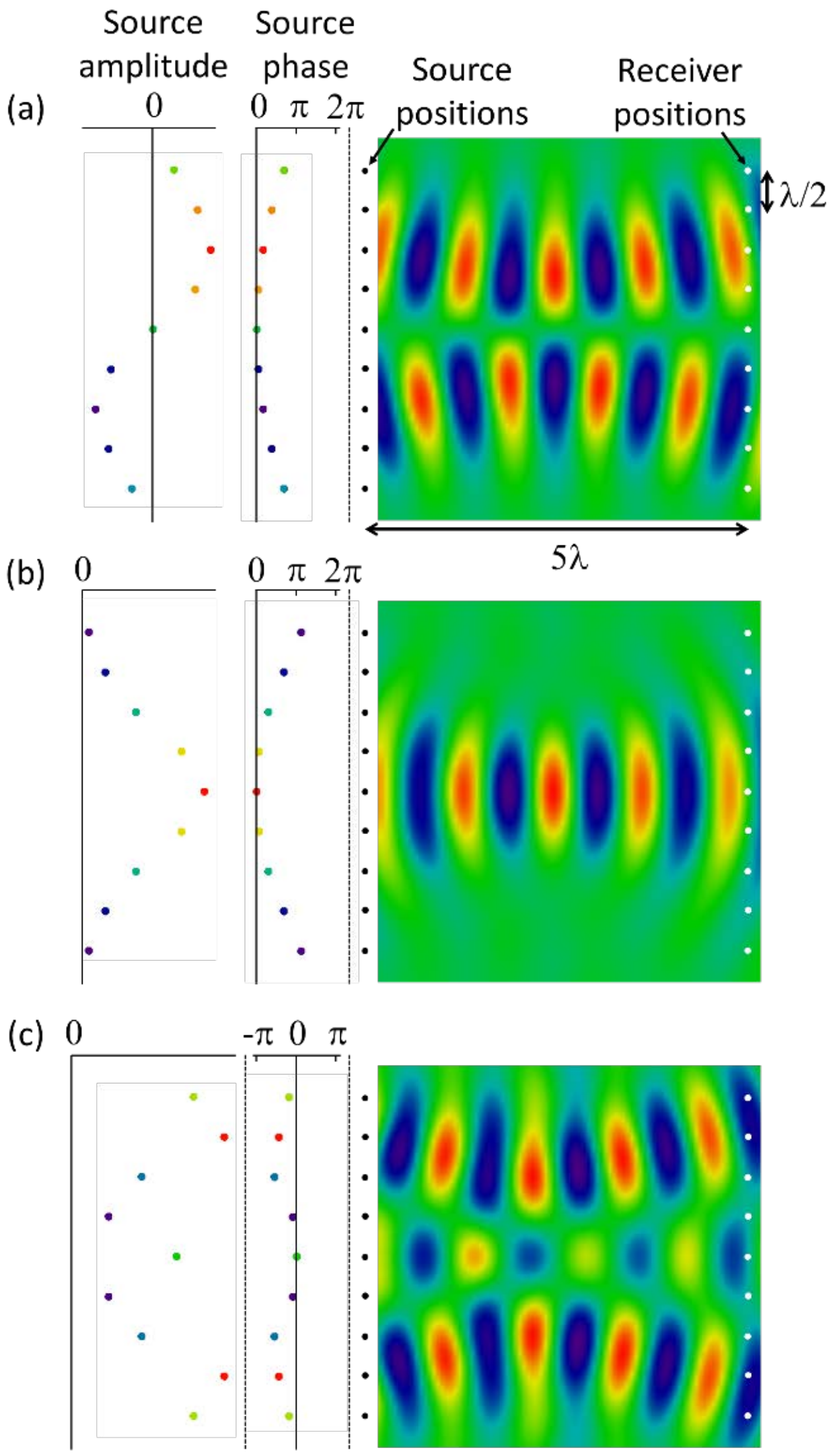

Fig. 6. Plots of the source relative amplitude, the source relative phases, and the resulting waves for the three most strongly coupled modes, (a), (b), and (c) respectively, for the nine point sources and receiving points shown. The phase of each source mode is chosen to be zero in the center of the mode. For graphic clarity, the wave is multiplied by $\sqrt{|z|}$ , where $z$ is the horizontal position relative to the source plane; the actual wave decays in amplitude from left to right, and the real part of the wave is plotted in false color. To avoid singularities, the waves just next to the source are not shown, so the positions of the sources, as shown, are just outside the graphed region on the left. The source amplitudes of the points in the "source amplitude" and "source phase" plots are also indicated using an amplitude false color of the points.

Note, incidentally, that

the solution of the singular value decomposition problem has "found" the necessary phase curvatures of the sources and corresponding receiver amplitudes so as to maximize the power coupling. These curvatures are not artificially put into the problem.

We can loosely interpret some of what is happening with these modes as diffraction limits. The beams associated with Modes 1, 2, and 3 are able to remain substantially within the space between the source points and the receiving points (at least in this two-dimensional plane). After these three Modes, however, attempts to create more orthogonal channels lead to waves that cannot be substantially contained in this way, and they start to “miss” the receiver points, as we will illustrate below for another case.

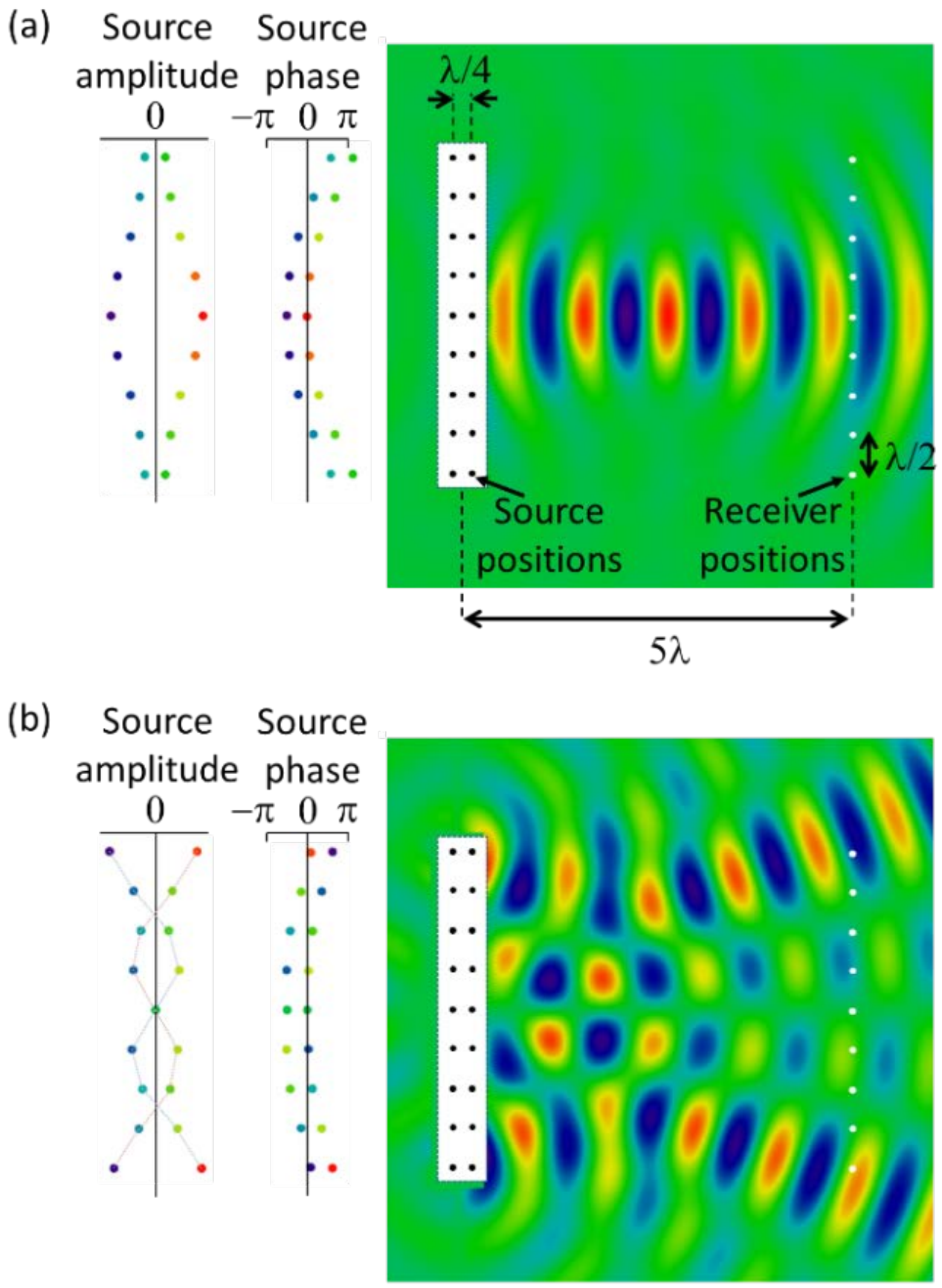

Fig. 7. Two example modes using a double line of sources. (a) A “single-bumped” mode. (b) A “four-bumped” mode that is only ~ half as well connected as the first three modes. (As can be seen, large parts of it miss the receiver positions entirely.) (As in Fig. 6, the real part of the wave is plotted in false color, and the wave is multiplied by $\sqrt{|z|}$ for graphic clarity.) The use of a double line of sources avoids substantial “left-propagating” waves for these modes, making the beam behavior clearer in this larger picture. To avoid singularities in the graphics, the wave is not plotted in the region of the white rectangle. The two lines of sources are spaced by a quarter wave. In the plots of the source phase, the “left” column of sources lag the phase of the right column of sources by approximately 90° (π/2), and the amplitudes are approximately equal and opposite. (In (b), we have joined the amplitude points in a given vertical column of sources by dotted lines to guide the eye.) The facts that the sources within a given “left-right” pair have opposite amplitudes and are phase-delayed in this way comes out of the numerical solution, and is not a starting constraint. This behavior is typical of “spatiotemporal dipoles” [68], and the calculations have “found” these as the best sources here.

Now, point sources at different points in space are in one sense automatically orthogonal since they do not overlap, and each different point source is mathematically a vector with one “one” in one unique position and so is orthogonal to the other point source vectors. However, a key point is that though the individual different point sources themselves can be orthogonal, the waves they generate in the receiving space are not necessarily orthogonal. To establish just what actual orthogonal sources give orthogonal waves in the receiving space, we need to perform the SVD, as we have done here.

We can usefully illustrate more behaviors of such modes, and this becomes somewhat clearer if we use a double line of sources as in Fig. 7. A single line of point sources like those in Fig. 3 and Fig. 6 actually broadcasts just as effectively "backwards" (i.e., to the left in these figures) as it does to the right [69]. To see more clearly what else is happening in the generated waves, we can first avoid such backwards waves. A double line of sources allows the sources to take the form of a "spatio-temporal dipole" which can suppress the backward (leftward) radiation [68, 70, 71].

In Fig. 7 (a) we show the "single-bumped" mode with such a "double line" set of sources, which is analogous to the similar mode in Fig. 6 (b) [72].

We also show, in Fig. 7 (b), the next most strongly coupled mode (after these first three), which has only 14.4% of the sum rule. We can see here that in this mode, significant parts of the beam are missing the receiver positions, in two angled parts that are just overlapping with the top and bottom receiver positions; this behavior is consistent with the power coupling strength here only being approximately half of that of the first three strongly coupled modes.

This fourth mode in Fig. 7 (b) illustrates another point. Note that, in this case, the beam is quite clearly not symmetrical about the center between the sources and receivers. (This asymmetry is not because we have different numbers of sources and receivers – similar behavior is seen with equal numbers of both.) Note that even with identical, symmetric source and receiver volumes and/or numbers, there is no requirement that the resulting beam for any given communications mode is similarly symmetric from "left" to "right".

With these first 4 modes, altogether ~97.16% of the available sum rule has been used up. The next ($5^{th}$) mode looks somewhat similar to Fig. 7 (b), though with one more "bump", and in this case, the upper and lower angled beam parts almost entirely miss the upper and lower receiver points. This $5^{th}$ mode consumes ~2.67% of the sum rule, which leaves sum rule therefore essentially exhausted, with only just over ~ 0.17% of the sum rule to be divided among the remaining 4 possible orthogonal modes. (We will return to look in more depth at such very weakly coupled modes in a later example.)

## 5.2. Two-dimensional arrays of sources and receivers

As another example, we can look at "planes" of source and receiver points, as in Fig. 8. Here we have arranged $17 \times 17 = 289$ source and receiver points (so $N_S = N_R = 289$), each spaced by the wavelength $\lambda$, to give parallel $16\lambda \times 16\lambda$ square source and receiver "surfaces" positioned $50\lambda$ apart. Now we plot, in false color, a snapshot (the "real part") of the amplitude of the wave on the receiver "plane" for each of the first 12 most strongly coupled modes (Fig. 8 (d)), in decreasing order of coupling strength. Also shown, Fig. 8 (c), is a graph of the relative magnitudes of the corresponding "power coupling strengths" [73] $|s_j|^2$, expressed as a percentage of the total available sum rule $S$ for this problem.

The most strongly coupled mode, Mode 1, is a simple "single-bumped" beam. An early heuristic approach to understanding how many independent "channels" there are between two surfaces is due to Gabor [74]. In Gabor's approach, he asks first what would be the minimum "diffraction limited" size of spot one would be able to form on the receiving surface using a source that is the size of the source surface. Then he asks how many such spots one could lay out on the receiving surface if those are to be approximately non-overlapping . If we interpret Mode 1 here as approximately a "diffraction-limited" spot, then by eye in Fig. 8 (b), we might reasonably expect that we could fit about 9 such spots within the square "receiver" region, roughly imagining a $3\times3$ array of such spots to fill that square region. In fact, this heuristic is a reasonable guess (and this correspondence is already known [5]). We will return to discuss such heuristics generally below.

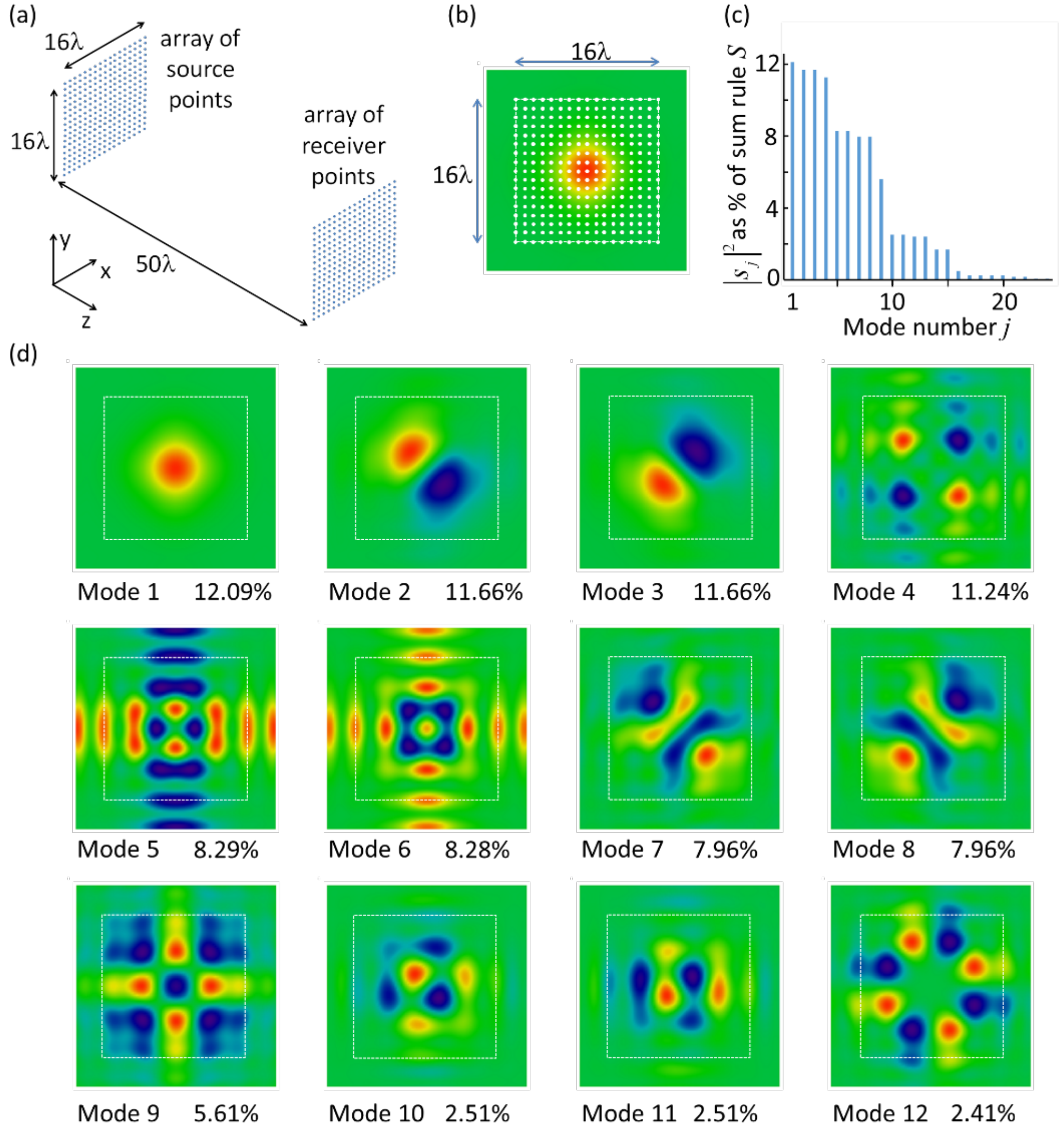

Fig. 8. (a) Layout of two-dimensional arrays of $17\times17=289$ point sources and receiver points, each on a square array, with points separated laterally by the wavelength $\lambda$, and the arrays separated, on the same axis, by $50\lambda$. (b) End view of the (real part of the) field amplitude for mode 1, with the receiver points superimposed. (c) In order, for the first 24 communications modes, the relative power coupling strengths ($|s_j|^2$), as a percentage of the power coupling strength sum rule $S$. (d) False color plots of the (real part of the) field amplitude at the receiver plane, together with the corresponding percentage of the sum rule. The dashed square represents the extent of the array of receiver points in each case, as in (b). Plots are relative to the maximum.

Looking at Fig. 8 (c), we see that there are roughly 9 Modes that are within just over a factor of 2 in relative power coupling strength (i.e., from 12.09% of $S$ down to 5.61% of $S$). After that, there are about 6 weakly coupled additional modes, and then the coupling strengths fall to very low values.

In all of these modes, the wave functions are orthogonal to one another within the square receiver region. This is even true as we go to the more weakly coupled modes (e.g., Modes 10, 11, and 12 in Fig. 8 (d)). From these views, it may not be obvious what is happening to the rest of the coupling. In part, for these more weakly coupled modes, there is significant power in the regions outside the areas plotted in Fig. 8 (d). Though the solutions are giving orthogonal channels into the receiver region, they are also broadcasting significant power elsewhere. With these sizes of source and receiver regions and this separation between them, it is not possible to keep the wave within the receiver region for these higher-numbered modes (as we will see explicitly below for another example.)

In this symmetric and "square" problem, we see behaviors in numbers and symmetries of beam forms that are typical in general of mode forms also found in related resonator problems. Because of the finite size of the region, however, the higher-numbered well-coupled modes, such as Modes 5 to 9, are showing strong influence of the finite size and shape of the receiver region, as also are the more weakly coupled Modes 10 – 12.

We can also guess (correctly) that Mode 1 here is essentially as well coupled as it could be. In this case (other from the backward wave from this single "sheet" of sources), there is no "wasted" wave. Essentially, all the ("forward") wave from the sources is indeed focused into the receiver plane. The corresponding 12.09% of $S$ in $|s_1|^2$ therefore corresponds to a well-coupled communications mode, and we could deduce immediately that we could not expect more than 100/12.09 ~ 8 or 9 modes that are this well coupled. In fact, we do see about 8 or 9 relatively well coupled modes, with strengths falling off somewhat with increasing mode number.

Not surprisingly with such square arrays of points, there are obvious degeneracies. Modes 2 and 3 are have the same coupling strength, as do the pairs Modes 7 and 8, and Modes10 and 11. The resulting beams in each pair also have shapes that are related by simple symmetries (e.g., rotating by 90°). We will call this particular kind of degeneracy a *symmetry degeneracy* (in part to distinguish from another degeneracy that emerges later).

## 5.3. Paraxial behavior

A common situation in optics is that the lateral sizes of the input and output spaces are relatively small compared to the separation between those spaces. Such situations are often called "paraxial" because much of the propagation is close to being parallel to the axis between the sources and receivers. Also, in optics often such input and output spaces are approximately surfaces (or lines) that are parallel to one another, and are both centered on the same axis that runs between them. The situations we have simulated so far (as in Fig. 3, Fig. 6, Fig. 7, and Fig. 8), are not well approximated as being paraxial (the lateral extent of the source and receiver spaces is somewhat too large for the separation between them). The situation of Fig. 9 is, however, approximately paraxial; the lateral extent here, 48 wavelengths, is relatively small (1/4) compared to the separation of 192 wavelengths. The paraxial case has a number of simplifying properties compared to the more general cases, and we can use it to illustrate several behaviors.

In Fig. 9(a) we show the positions of the sources and receivers, together with the beam for one calculated mode (actually, the "first" mode – the one with the largest magnitude of singular value). As in Fig. 7, we have used two lines of sources to allow suppression of the generation of "backwards" waves (i.e., to the left), and the sources in a line are spaced vertically by $\lambda / 2$ to suppress generation of spurious "higher order diffraction" waves at large angles. In Fig. 9 we now are plotting intensity (presumed $\propto |\phi(\mathbf{r})|^2$, where $\phi(\mathbf{r})$ is the calculated complex wave amplitude at each point $\mathbf{r}$). (Plotting the real part of the field would lead to a graph with too much structure on the very small wavelength scale for a graph of these dimensions.)

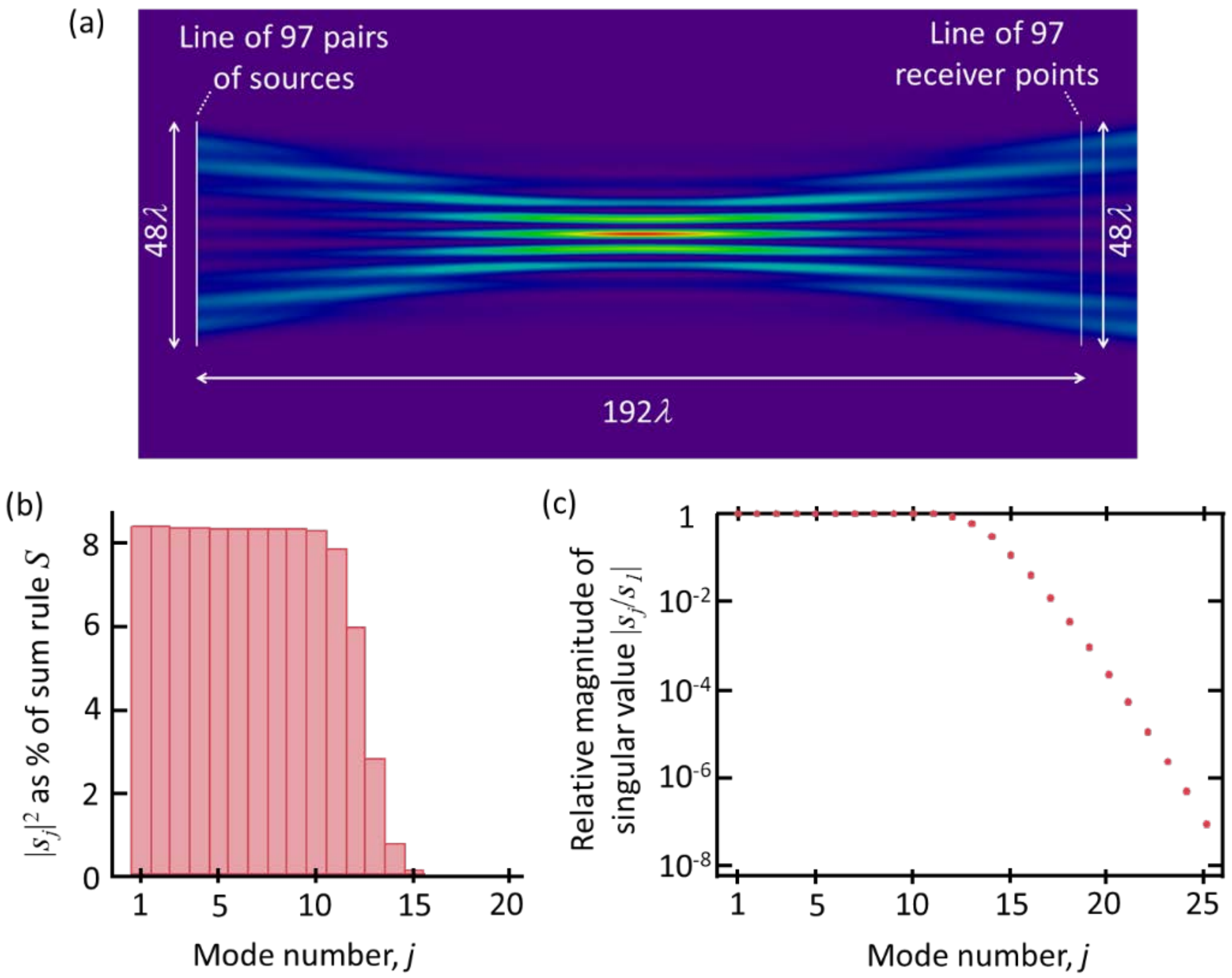

Fig. 9. (a) Positions of sources and receivers superimposed on the beam for Mode in the plane of the source and receiver points. Here the intensity of the mode multiplied by the horizontal distance $z$ from the source plane is shown in false color. This multiplication by $z$ compensates in the graphics for an underlying fall-off in intensity proportional to ~ $1/z$ with such lines of sources. A small region immediately adjacent to the sources is not plotted so as to avoid singularities and/or some very large amplitudes there in the graphics. The sources consist of 97 pairs of sources in two vertical lines. The sources in a vertical line are spaced by $\lambda/2$ ($\lambda$ is the wavelength), and the two lines of sources are spaced horizontally by $\lambda/4$ (similarly to those in Fig. 7). (b) Histogram of the modulus squared of the singular values $|s_j|^2$ for the different Modes (numbered in decreasing order of the singular value magnitude). These are shown as a percentage of the total sum rule $S$. (c) Relative magnitude of the singular values of each Mode, compared to the first (and largest) singular value, and plotted on a logarithmic scale.

### 5.3.1. Behavior of singular values

Fig. 9 (b) and (c) show the behavior of the magnitudes of the singular values. The mode numbers correspond to sorting the modes in decreasing order of the magnitude of the corresponding singular value. So, Mode 1 has the largest singular value, and subsequent modes have progressively smaller singular values. Fig. 9 (b) shows the modulus squared of the singular values for each Mode, $|s_j|^2$, shown as a percentage of the sum rule $S$ for this set of sources and receivers (as calculated using Eq. (36)). Fig. 9 (c) shows the relative magnitude of the singular values of the Modes, plotted on a logarithmic scale.

The relative sizes of the singular values in Fig. 9 (b) and (c) show two striking behaviors: (1) the singular values are nearly constant and equal up to a relatively abrupt threshold, here round about Mode 12 or so, after which they drop off rapidly with increasing Mode number; (2) for larger Mode numbers, in Fig. 9 (c) we see this drop-off becomes extremely rapid, with an apparently approximately exponential decrease in the singular value magnitude at large Mode numbers.

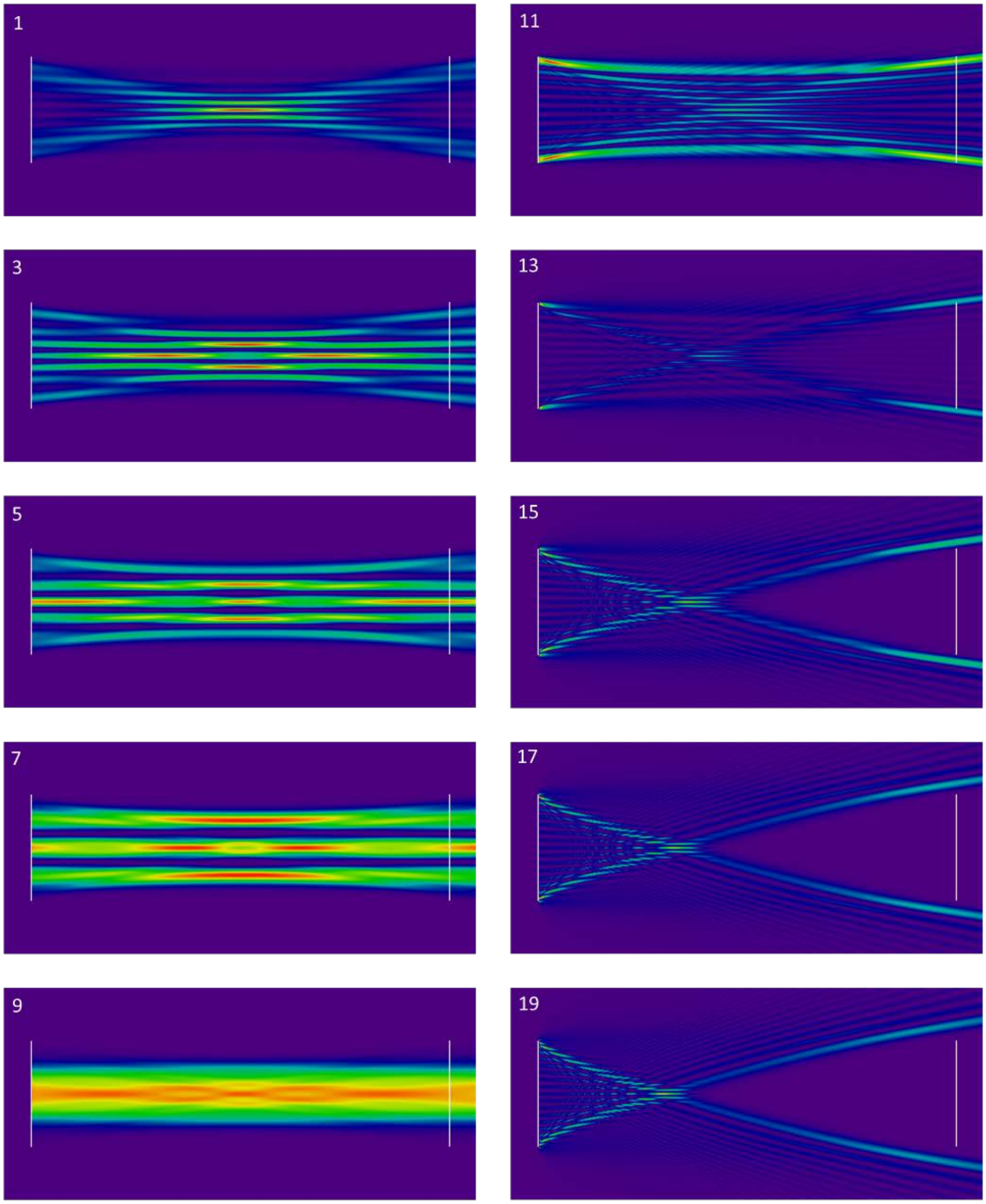

Fig. 10. Intensity graphs of the beams associated with the odd-numbered Modes 1 – 19 for the source and receiver points, as in Fig. 9 (a) (with intensities multiplied by the distance $z$ from the sources for graphic clarity). Relative intensities are rescaled for graphic clarity for each Mode plotted, so the absolute "brightness" has no meaning in comparing different Modes.

## 5.3.2. Forms of the communications modes

### *5.3.2.1. Strongly coupled modes*

In Fig. 10, we have plotted the beams for the odd-numbered Modes (in decreasing magnitude of singular value), as in Fig. 9 (a). At least for Modes 1 through 11, we can count the number of "bumps" relatively simply, which is an odd number in these cases. (The even-numbered Modes show similar behavior, though with even numbers of bumps.) The order of Modes 1 through 9 is somewhat surprising, being perhaps "backwards" compared to what we might expect. However, we should note that these modes have almost identical singular values, so this is a nearly degenerate eigenproblem, at least for Modes 1 through 10, so the order of these modes may have little physical importance.

##### 5.3.2.2. Weakly coupled modes

More strikingly, though, and more importantly for our discussion, once we pass Mode 11, the magnitude of the singular values starts to drop, and, consistent with that drop, the intensity starts to "miss" the line of receiving points.

It might seem that there is no intensity left at the line of receiving points for Modes 13 and above, especially in the "middle" vertically; certainly it is so small that after Mode 13 the intensity at the receiver points does not show up on the graphics in Fig. 10. However, if we plot the actual Modes at the sources and at the receiver points, we see that there is field at the receiver points, with well-defined if progressively very weak behavior for these higher numbered Modes.

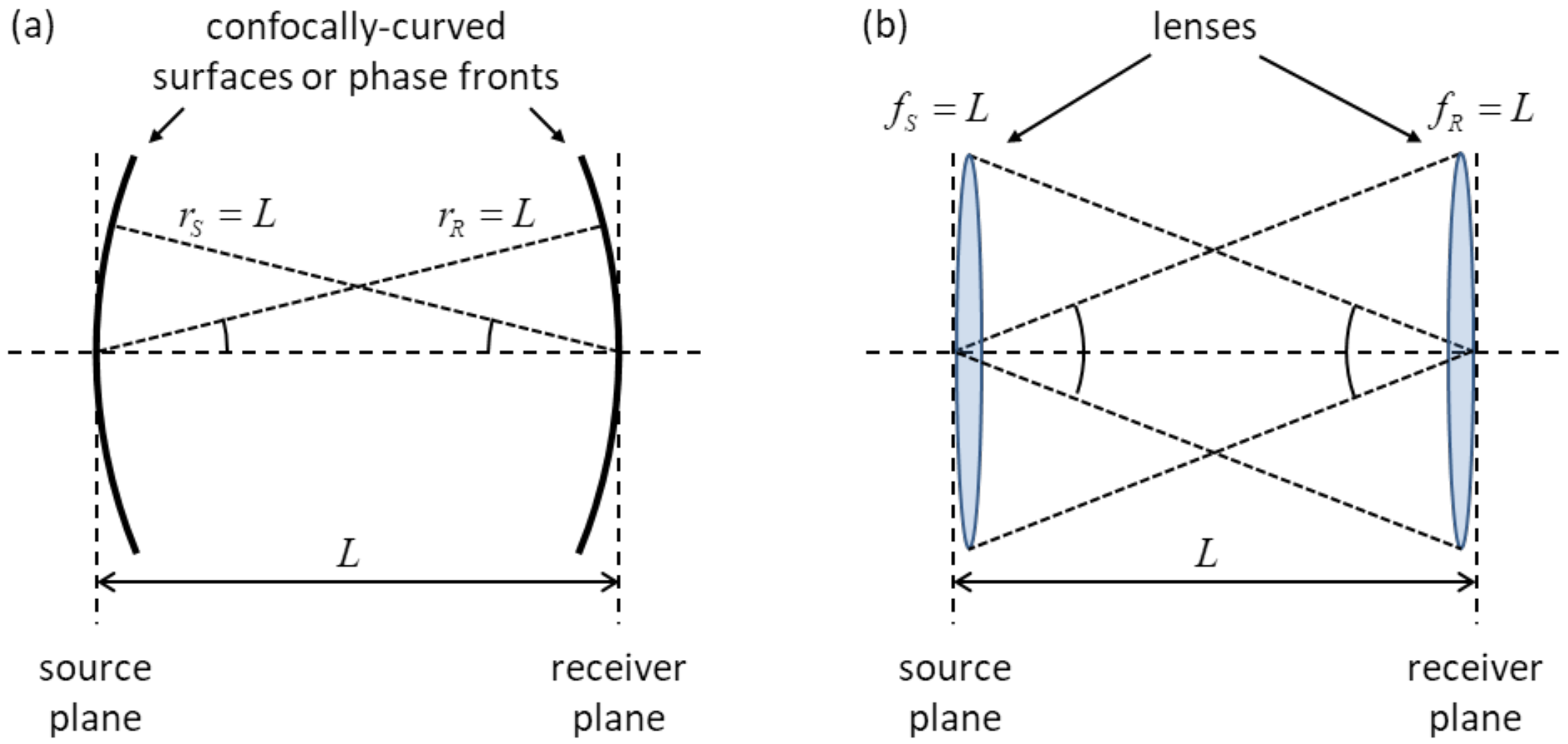

Fig. 11. Illustration of two surfaces or phase fronts that are "confocally curved" – each one is curved round about the center of the other surface: (a) from confocally curved surfaces; (b) using lenses with focal lengths $f_S$ and $f_R$ equal to the separation $L$.

Before plotting those higher numbered Modes, we should note the behavior of the phase curvature of the Modes. For the strongly coupled Modes 9, 8, 7, 6, 5, 4, 3, 2, 1, and then 10 and 11, the phase of the sources and of the waves at the receiver points goes (in this order) from being nearly a "flat" phase front for the approximately "single-bumped" Mode 9 to being progressively more curved, until by Mode 11 and for all the subsequent modes, the phases of both the sources and the receiver points show what could be described as approximately *confocal curvature*: for a pair of surfaces or phase fronts, the center of curvature of one surface or phase front is the center of the other surface (Fig. 11(a)) . We can formally describe the confocal curvature at the source and receiver planes using

$$c_S(x_S, y_S) = \exp\left[-ik\left(\sqrt{x_S^2 + y_S^2 + L^2} - L\right)\right] \tag{54}$$

and

$$c_R(x_R, y_R) = \exp\left[ik\left(\sqrt{x_R^2 + y_R^2 + L^2} - L\right)\right] \tag{55}$$

respectively, where $y_S$ is the vertical coordinate in the source plane in Fig. 11, and $x_S$ is similarly the coordinate in the direction into the "paper" in the source plane in Fig. 11 (a), and similarly for the coordinates $y_R$ and $x_R$ in the receiver plane. By taking out the confocal curvature (by multiplying by $c_S^*(x_S, y_S)$ at the sources and by $c_R^*(x_R, y_R)$ at the receiver points), and by choosing an overall phase factor that makes the resulting wave real in the middle of the line of receiver points vertically, we can conveniently plot (Fig. 12) both the source and the receiver amplitudes for these modes from Mode 11 onwards using just the real part of the source and receiver amplitudes [75].

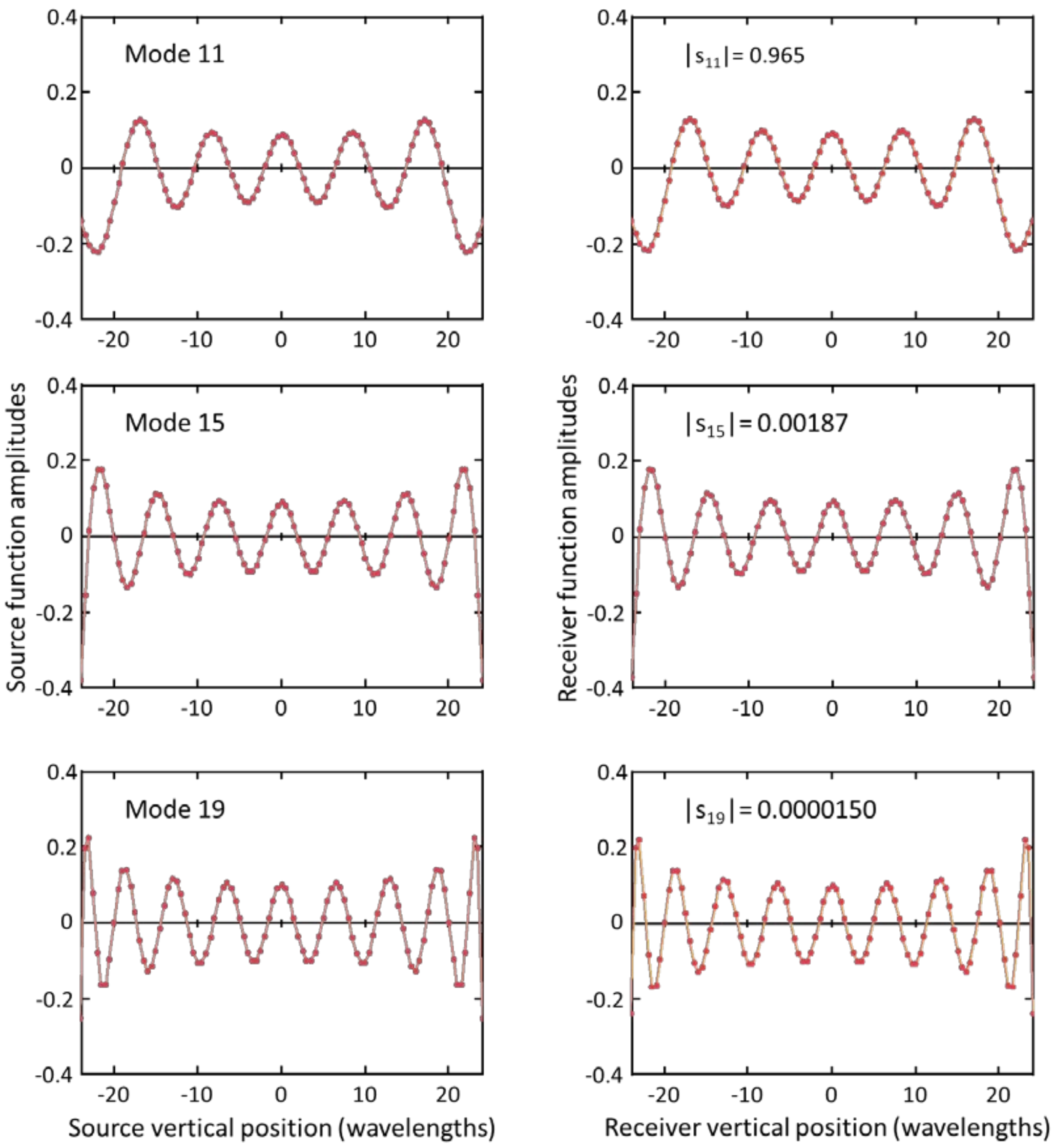

Fig. 12. Graphs for the source function amplitudes (left column) and the receiver function amplitudes (right column) for Modes 11, 13, and 19 for the sources and receiver points as in Fig. 9 and Fig. 10. The points are the amplitudes, and the lines join adjacent points to guide the eye. Underlying approximately confocal phase curvatures have been removed in each case, and the set of points in each graph has been multiplied by a constant phase factor to make the resulting points approximately real for graphic clarity, and only the real parts of the source and receiver function amplitudes are plotted. The receiver amplitudes are for the normalized vector of amplitudes. The vector of source amplitudes is also normalized, but because only the amplitudes of the "right" of the two vertical lines of sources is plotted here, this vector is additionally multiplied here by $\sqrt{2}$ to give the source function amplitudes plotted here for clarity in comparison since only half of the sources are plotted. The $|s_j|$ are the magnitudes of the singular values for each Mode.

In the plots of Fig. 12, we see that the actual relative amplitudes of the sources and of the waves at the receiver points are remarkably similar in form in each case (though possibly with very weak amplitudes at the receivers). This similarity is despite the fact that the total wave when we consider the region outside the receiver region is certainly not symmetric from left to right in Fig. 12 for the modes beyond Mode 11. The nominal behavior of the form of these source and receiver vectors is also relatively straightforward and similar for all of these higher numbered modes – we see a relatively sinusoidal behavior near the center, with correspondingly more "bumps" for the higher numbered modes, and with some increase in amplitude towards the edges (the top and the bottom of the source and receiver regions in Fig. 9 and Fig. 10) in all these modes.

The major difference between these higher-numbered modes, other than the proportionate increase in the number of "bumps" is, however, that the singular value is dropping rapidly with increasing mode number. To see the actual receiver amplitudes for the higher numbered modes for normalized source vectors, we should multiply these normalized receiver amplitudes by the magnitude of the singular values. Graphically, that makes the receiver amplitudes essentially invisible, as in Fig. 10 for these higher-numbered modes.

So, though the source and receiver eigenvectors behave in a straightforward and similar manner for these higher numbered modes, the coupling strength is vanishing very quickly with increasing mode number. The effect of this behavior for these higher numbered modes becomes very important when we consider below what happens as we try to pass the "diffraction limit", and we return to this point below.

### 5.3.3. An additional degeneracy of eigenvalues – paraxial degeneracy

We noticed above in considering a two-dimensional problem with square source and receiver "areas" (Fig. 8) that we had obvious 2-fold degeneracies that are associated with the symmetry of this problem. However, in this paraxial example (Fig. 9), we are seeing an additional behavior. Modes 1 to 10 are approximately degenerate – their coupling strengths or singular values are nearly the same. This approximate degeneracy does not obviously result from symmetry, and we it a *paraxial degeneracy* [76].

This paraxial degeneracy is common in paraxial problems with either parallel surfaces of parallel volumes of uniform thickness, as we will illustrate below with more examples. This paraxial degeneracy generally applies only to well-coupled modes. Unless the modes are also symmetry degenerate, the paraxially degenerate modes are generally only approximately degenerate – that is, they only approximately have the same coupling strength. However, the eigenvalues can be so nearly the same, typically up to some specific number, that from a physical point of view they can practically be thought of as degenerate. As a result,

> for the well-coupled modes in simple paraxial cases, there will also be many approximately equivalent ways of choosing the communications modes.

Again, just because there can be multiple approximately equivalent ways of choosing these modes does not change the counting of the modes or channels.

Generally, these two concepts of symmetry and paraxial degeneracy are different and overlapping. We can have symmetry degeneracy that is part of paraxial degeneracy, as in well-coupled modes between square apertures. We can have symmetry degeneracy that is not part of paraxial degeneracy, as in two very weakly coupled modes that may nonetheless have equal coupling strengths because of symmetry. We can have paraxially degenerate modes even if the "apertures" in the problem have no particular symmetry (as we will illustrate below); in such cases, there would be no symmetry degeneracy.

The number of geometrically degenerate modes in paraxial problems with large apertures is essentially the same thing as "space-bandwidth product" as used in Fourier optics, and can be described by a concept we will call the "paraxial heuristic number" $N_H$ , which we discuss next.

### 5.3.4. Paraxial degeneracy and paraxial heuristic numbers

To establish the intuitive idea of the paraxial heuristic number, suppose we have two point sources, on the surface of the source space, separated in the $y$ direction by a distance $d_s$. This separation $d_s$ is presumed small compared to the separation $L$ between the source plane and corresponding receiving plane on the incident surface of the receiving volume. (See Fig. 13.) Then, as in a "two-slit" diffraction, the resulting interference pattern on the receiving surface will approximately take the form, in the $y$ direction,

$$\phi(y) \propto \exp\left(ik\sqrt{\left(y-\frac{d_s}{2}\right)+L^2}\right)+\exp\left(ik\sqrt{\left(y+\frac{d_s}{2}\right)+L^2}\right) \tag{56}$$

where in all denominators we approximate the distance $r$ between points on the two planes by $L$.

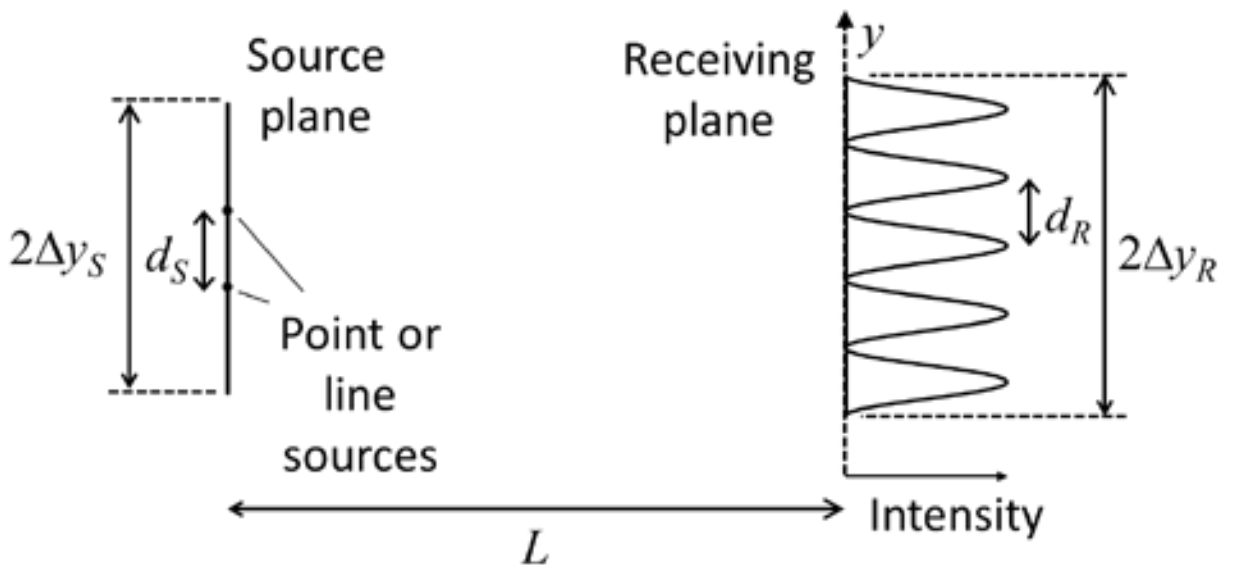

Fig. 13. Intensity pattern at the receiving space for two point sources on the source space.

Taking the approximation $\sqrt{1+\varepsilon} \simeq 1+\varepsilon/2$ for small $\varepsilon$, then we obtain the usual two-slit intensity pattern result for small $y$

$$|\phi(y)|^2 \propto \cos^2\left(\frac{\pi d_s y}{\lambda L}\right) \tag{57}$$

So the intensity "fringe spacing" or periodicity on the receiving plane is

$$d_r = \frac{\lambda L}{d_s} \tag{58}$$

So, if the source space extends from $-\Delta y_S$ to $\Delta y_S$ in the $y$ direction, and the receiving plane similarly extends from $-\Delta y_R$ to $\Delta y_R$, then the maximum number of intensity fringes we can form in the receiving space in the $y$ direction, with sources spaced as far as possible on the source plane, i.e., at distance $d_s = 2\Delta y_S$, is $N_{Hy} = 2\Delta y_R / d_r$, i.e.,

$$N_{Hy} = \frac{(2\Delta y_S)(2\Delta y_R)}{\lambda L} \tag{59}$$

This number $N_{Hy}$ is our paraxial heuristic number $N_H$ of well-coupled channels for source and receiver spaces that are one-dimensional "lines" in the $y$ direction.

It represents the maximum number of (intensity) "bumps" we could reasonably form in the receiver space from such "two-slit" interference from two points in the source space, for such "line" source and receiver spaces.

Note, for example that, for the situation in Fig. 9, we would have

$$N_{Hy} = \frac{48 \times 48}{192} = 12 \tag{60}$$

We could therefore regard is as not surprising intuitively that the number of strongly coupled channels, as indicated by the strengths of the singular values, is ~ 12 in the numerical results of Fig. 9.

Of course, considering a similar "two-slit" interference for sources spaced in the *x* direction instead would lead to a similar result there. With source and receiver space sizes in the *x* direction $2\Delta x_S$ and $2\Delta x_R$, we would obtain a similar paraxial heuristic number for that direction

$$N_{Hx} = \frac{(2\Delta x_S)(2\Delta x_R)}{\lambda L} \tag{61}$$

We can then take one more step, asserting we can reasonably postulate that for rectangular surfaces the corresponding paraxial heuristic number would be the product

$$N_H = N_{Hx} N_{Hy} \tag{62}$$

In this rectangular case, the areas of the surfaces are, respectively, for the source and receiver spaces

$$A_S = (2\Delta x_S)\times(2\Delta y_S) \text{ and } A_R = (2\Delta x_R)\times(2\Delta y_R) \tag{63}$$

Then we can write

$$N_H = \frac{A_S A_R}{\lambda^2 L^2} \tag{64}$$

This number $N_H$ is our paraxial heuristic number of well-coupled channels for planar source and receiver spaces.

Though we have provided a rationalization of this paraxial heuristic number $N_H$ only for the cases of rectangular source and receiver surfaces, we assert that can also use this as a useful characteristic number even when the source and receiver surfaces are not rectangular. The numerical calculation examples below will illustrate empirically the extent to which this number is useful in these cases.

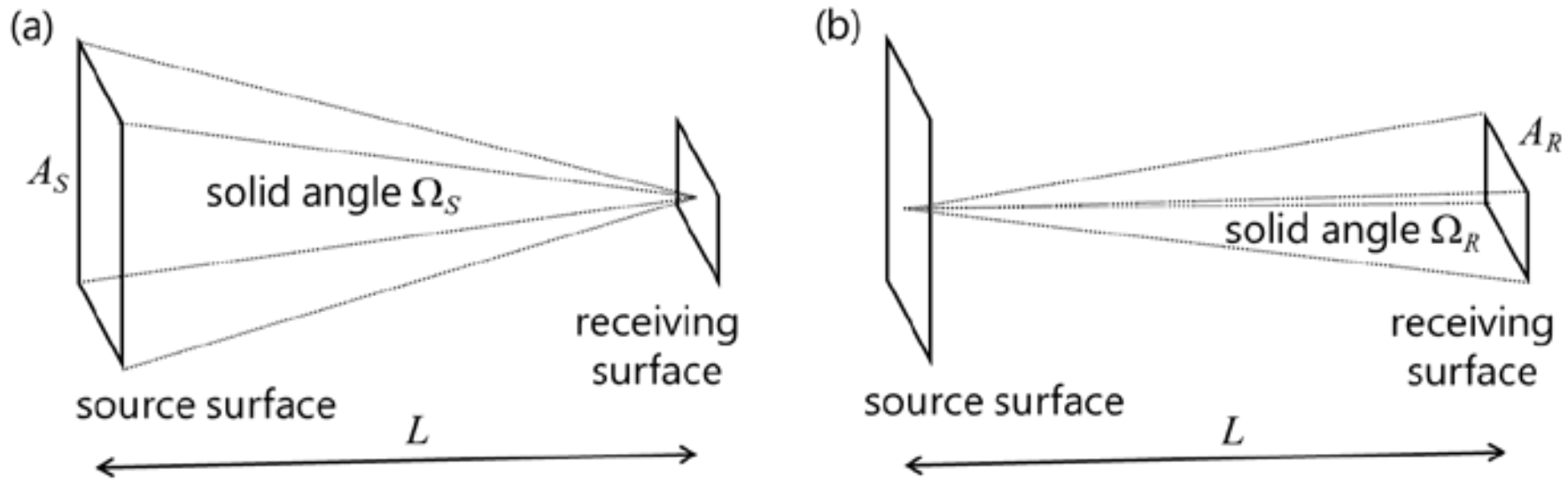

Fig. 14. Illustration of the solid angles (a) $\Omega_S$ subtended by the source surface (area $A_S$) at the receiving surface, and (b) $\Omega_R$ by the receiving surface (area $A_R$) at the source surface.

We can also usefully write Eq. (64) using solid angles. In a paraxial approximation, the solid angle subtended by one surface of area *A* at another surface or point at a (perpendicular) distance *L* is $\Omega \simeq A/L^2$. So, for the solid angles $\Omega_S$ and $\Omega_R$ subtended by the source and receiver surfaces respectively at the other surface, we have, as in Fig. 14,

$$\Omega_S \simeq A_S / L^2 \text{ and } \Omega_R \simeq A_R / L^2 \tag{65}$$

So we can write Eq. (64) in the alternative forms

$$N_H \simeq \Omega_S \frac{A_R}{\lambda^2} \simeq \Omega_R \frac{A_S}{\lambda^2} \tag{66}$$

We can also consider the source and receiver solid angles per "channel", $\Omega_{S1}$ and $\Omega_{R1}$ respectively, which are

$$\Omega_{S1} = \frac{\Omega_S}{N_H} = \frac{\lambda^2}{A_R} \quad \text{and} \quad \Omega_{R1} = \frac{\Omega_R}{N_H} = \frac{\lambda^2}{A_S} \tag{67}$$

which will be useful for a comparison later.

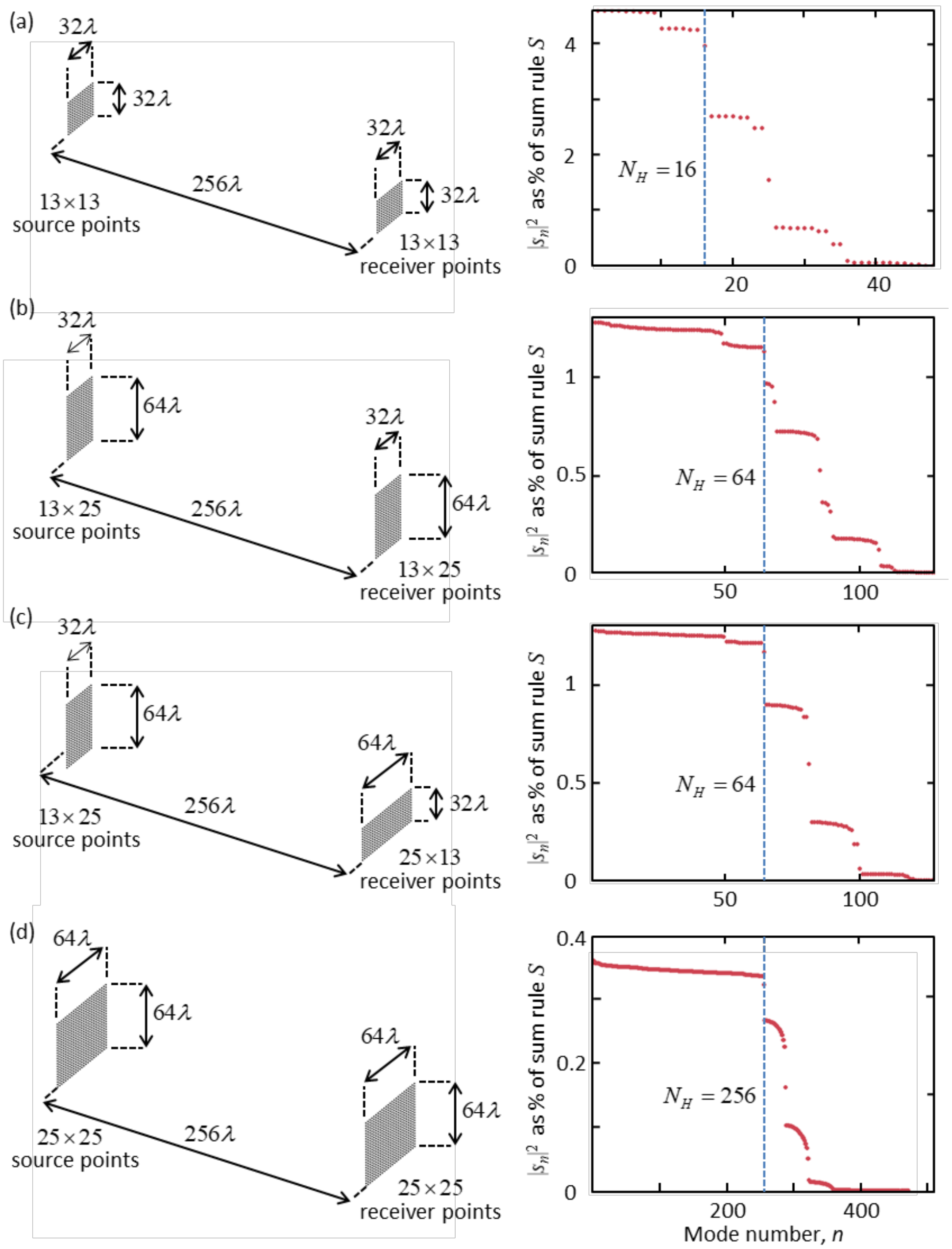

Fig. 15. Coupling strengths $|s_n|^2$ in decreasing order as a function of the mode number $n$, as calculated for the SVD of the scalar wave coupling between the sets of source points and receiver points as shown. $N_H$ is the paraxial heuristic number, here $A_S A_R / (\lambda^2 L^2)$, where $A_S$ and $A_R$ are the areas of the source and receiver surfaces, respectively, and $L$ ($= 256\lambda$ here) is the separation between the surfaces. Note that in each case $|s_n|^2$ is nearly constant up to $n = N_H$, after which it starts dropping rapidly.

#### 5.3.4.1. Paraxial heuristic numbers with rectangular surfaces

Fig. 15 shows the calculated coupling strengths $|s_n|^2$ for several situations with rectangular or square source and receiver surfaces under approximately paraxial conditions. In each case here, we have approximate paraxial degeneracy up to the paraxial heuristic number of modes $N_H$ , after which the coupling strengths drop rapidly.

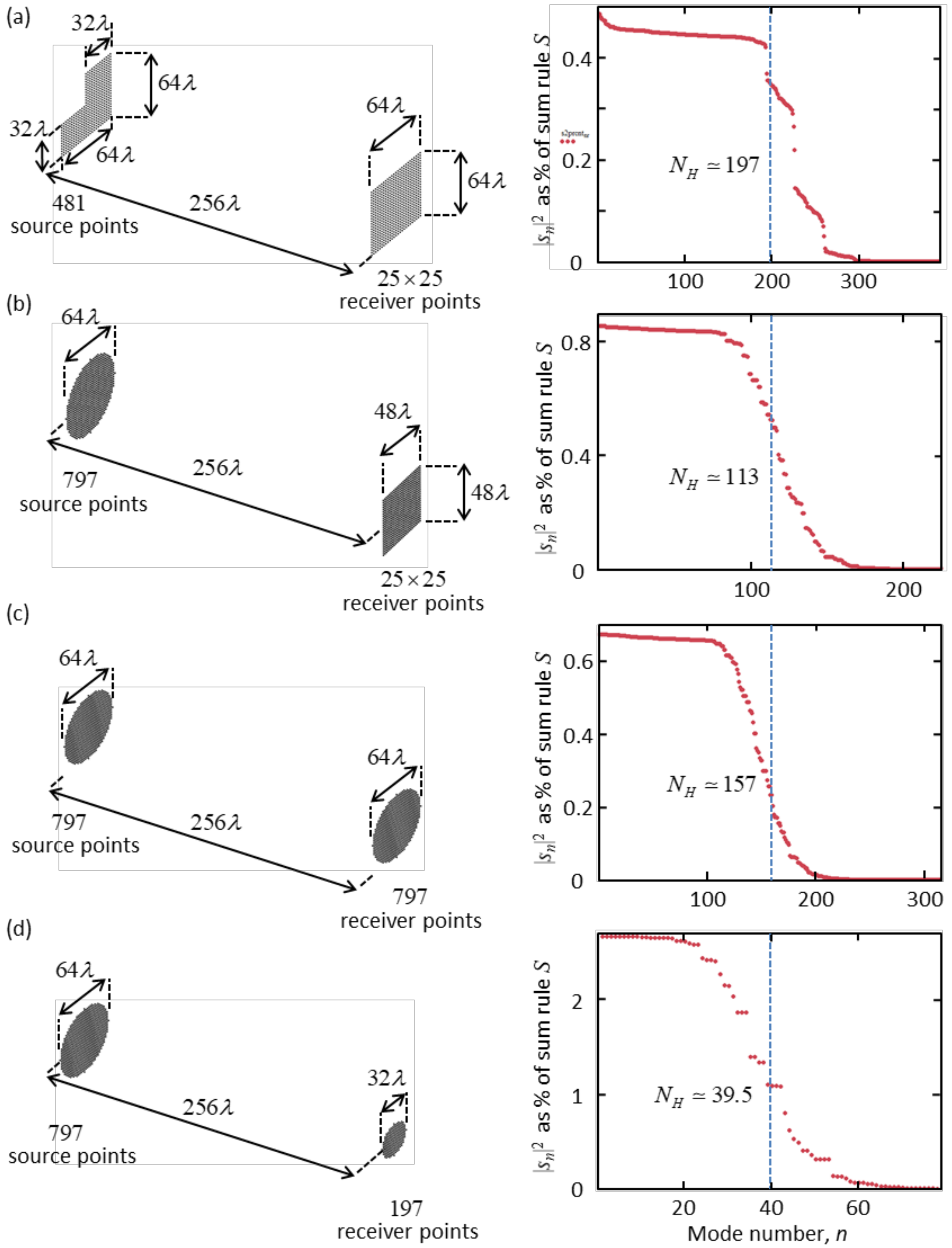

Fig. 16. Paraxial cases with source or receiver spaces with shapes other than rectangles. (a) An "L" shaped source space with a square receiving space. (b) A circular source space with a square receiving space. (c) and (d) – circular source and receiver spaces with (c) equal sizes and (d) different sizes. (All of these simulations were performed with arrays of source points and receiver points each spaced on a square grid of $2\lambda$ in both x and y directions, and truncated to fit within the corresponding shapes.)

Note that this behavior is similar (i) for both rectangular and square source and receiver spaces, (ii) for aligned rectangular spaces (Fig. 15 (b)) and (iii) where one is rotated by 90° (Fig. 15 (c)). For larger spaces, as in Fig. 15 (d), the paraxial (approximate) degeneracy of the $|s_n|^2$ is particularly smooth and relatively constant up to $N_H$ .

### 5.3.4.2. Other planar shapes of source and/or receiver spaces in paraxial problems

As we move to source and/or receiver spaces of different shapes, but still with plane-parallel surfaces, we see (Fig. 16 ) that the paraxial (approximate) degeneracy is retained, though possibly up to a number of modes somewhat smaller than the $N_H$ value calculated from the areas using Eq. (64). $N_H$ still remains a useful approximate guide to the number of strongly coupled modes, however. The case of an "L" shaped source space (Fig. 16 (a)) shows a slightly less uniform set of coupling strengths within the first $N_H$ modes than for the "rectangular" cases (Fig. 15), but still shows an abrupt drop starting at $N_H$ . Once we introduce a circular source area (Fig. 16 (b)), the initial drop in coupling strengths near $N_H$ is less abrupt, and we have fewer modes with the paraxial (approximate) degeneracy. This smoother drop in strengths is wider when both source and receiver spaces are circular (Fig. 16 (c) and (d)). (We find this less abrupt drop with the circular spaces is retained in other similar simulations with different sizes and separations.) So, empirically,

> the very abrupt drop in mode coupling strengths at $N_H$ appears to be a consequence of areas that are rectangular, but $N_H$ remains a useful guide to the number of strongly coupled modes for other shapes.

### 5.3.4.3. Effect of thickness on paraxial degeneracy

So far, we have considered only uniform "sheets" of sources (or pairs of sources) and receivers. It is, however, straightforward to simulate other situations. For example, we can use a cubic lattice of source points that fit within some other shape. Fig. 17 shows various cases where the two circular faces of the source and receiver spaces retain the same size and separation, so the paraxial heuristic number $N_H$ is the same in all these cases.

In Fig. 17, we see that changing from a circular set of sources (the red line and points) to a cylindrical one (the dashed grey line) makes little difference to the relative strengths of the various modes. The (approximate) paraxial degeneracy in both of these circular cross-section cases is good up to about mode 20, illustrating that a finite but uniform thickness of the source volume can retain the existing paraxial degeneracy. However, as we change to a non-uniform thickness of the source volume, given here by using ellipsoidal bounding volumes of increasing thickness ( $\rho_E$ , in units of the circle radius), the paraxial degeneracy is progressively lost. Little evidence of such approximate degeneracy is left by the $\rho_E = 0.45$ case, and there is essentially no such paraxial degeneracy by the $\rho_E = 1.5$ case [77]. These simulations illustrate, then, that

> paraxial degeneracy is a characteristic of paraxial systems of uniform thickness, but the paraxial heuristic number $N_H$ remains a good indicator of the number of strongly coupled modes, even if that coupling is not uniform between the different modes.

### 5.3.4.4. Strengths of weakly coupled modes

We have already seen in Table 1 for the case of 9 sources and receivers in a line, and in the similar case Fig. 9 (c) with 97 sources and receivers case, that, once we pass beyond the well-coupled modes, the singular values (or coupling strengths) drop off rapidly. For these "one-dimensional" (line) source and receiver cases, this drop-off becomes apparently exponential. This exponential behavior becomes quite general as we move to larger numbers of sources and receivers in these geometries. Three cases of progressively increasing $N_{Hy}$ are shown in Fig. 18.

Note first that all three of these case show clear paraxial degeneracy, with obviously nearly equal singular values for modes *n* up to very nearly $N_{Hy}$ . As *n* increases, the drop-off in singular values near $N_{Hy}$ is very abrupt, increasingly so on these graphs as we increase $N_{Hy}$ by using larger source and receiver spaces (though we have also increased their spacing to keep the relative geometry comparable).

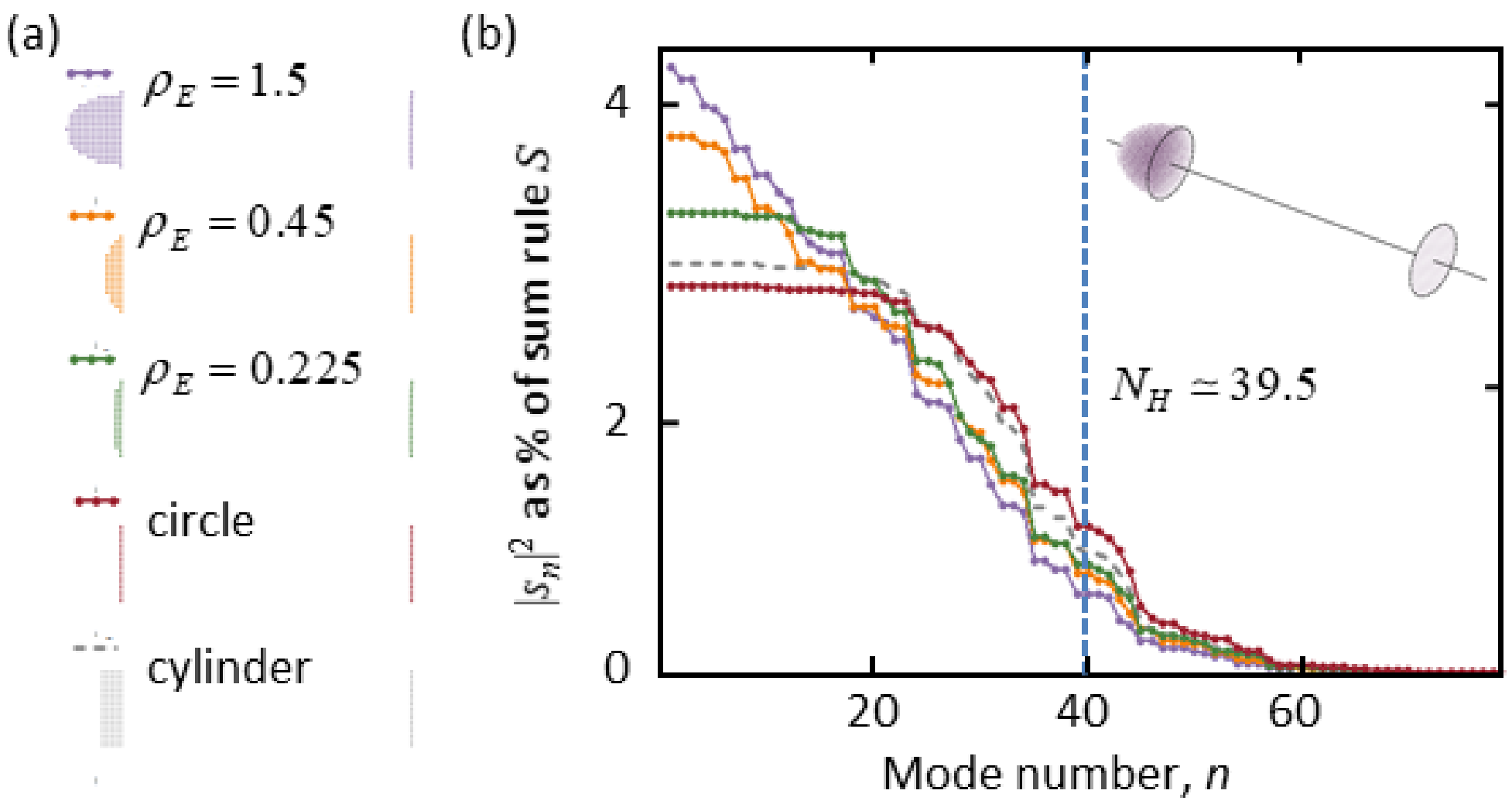

Fig. 17. Illustration of how non-uniform depth removes paraxial degeneracy. The receiving volume in all cases is a set of receiving points, on $2\lambda$ centers in both directions that fit with a circle of radius $16\lambda$. For the red points and line, the source is an identical circle of source points, at a distance $128\lambda$ away along their common axis. For the other sets of points and/or lines, the separation of the "faces" and the circular cross-section are retained, but the source consists of the points (also spaced on $2\lambda$ centers, now in all three directions) lying within a volume. For the grey dashed line, the volume is a cylinder of depth $8\lambda$. For the other three sets of points and lines, the bounding volume is half of an ellipsoid of revolution. $\rho_E$ is the ratio of the depth of the half ellipsoid compared to its cross-sectional radius (which is fixed at $16\lambda$) (so $\rho_E = 1$ would be a hemisphere). (a) shows the cross-sections of the source points and the receiving points in each case. The inset in (b) shows a perspective view, with the circular cross sections and the axis of rotation indicated, of the $\rho_E = 1.5$ case, which shows the ellipsoidal shape. The traces in (b) show the strengths of the various communications modes as a percentage of the sum rule.

Once the singular values start dropping off, they tend to decrease exponentially with increasing *n*. The dashed lines in Fig. 18 are exponentials given by

$$h(n) = 8\exp\left[-0.811\left(n - 0.985N_{Hy}\right)\right] \qquad (68)$$

(This formula is not derived, nor is it a fit to the calculations; it is simply heuristic, being judged by eye to represent the form of the decay. However, with one set of numerical coefficients, it approximately models the decaying exponential in all three cases.) The factor 0.811 in the formula Eq. (68) means that, between each successive mode in this decaying region very near to and after $n = N_{Hy}$, the singular value in all three cases decreases by a factor $\exp(0.811) \simeq 2.25$; we can see this behavior in the equal vertical spacing of the successive points on the right in the logarithmic-scale graph in Fig. 18. (In Eq. (68), the factor 0.985 accounts for the fact that the exponential decay starts slightly before $N_{Hy}$ in each case [78].)

As a practical matter, then, once $N_{Hy}$ becomes a significantly large number (e.g., in the 100's), there really will be essentially no usable modes beyond $N_{Hy}$ for this case of parallel one-dimensional source and receiver spaces.

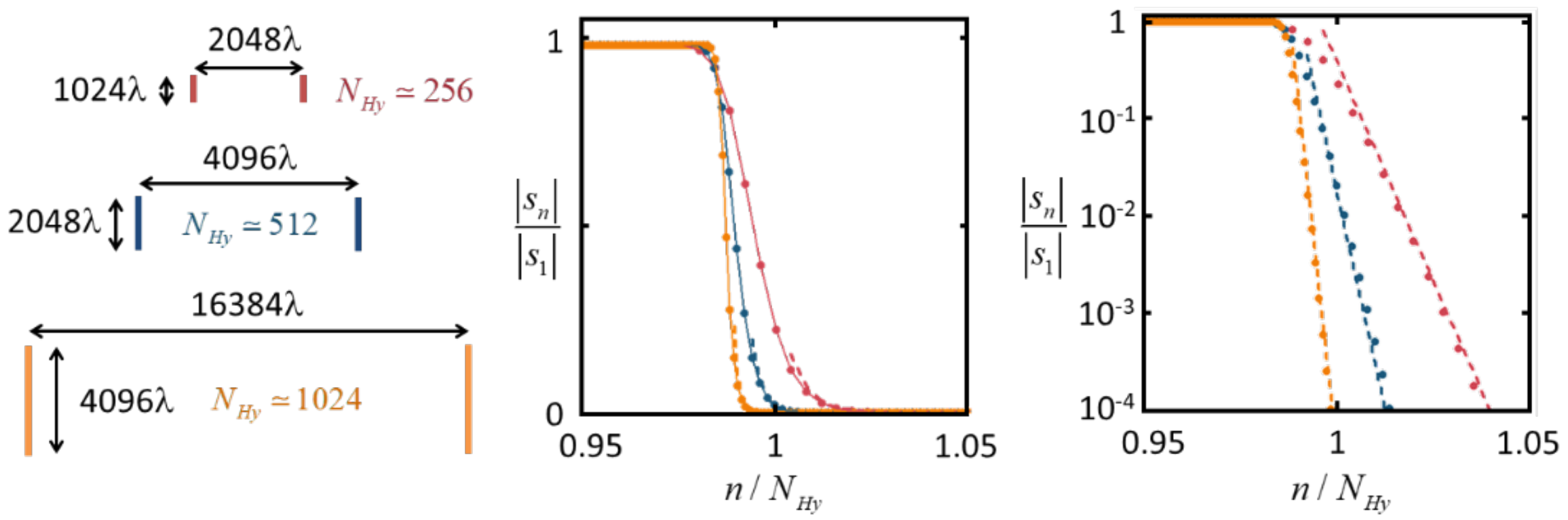

Fig. 18. Plots of the relative size of the singular values for several different approximately paraxial pairs of lines of source and receiver points, as a function of the mode number *n* compared to the paraxial heuristic number $N_{Hy}$ for each pair of source and receiver lines, on both linear (left graph) and a logarithmic (right graph) scales. The values are shown as points, with the solid lines in the linear graph joining them to aid the eye. The dashed lines are exponential functions given in the text. Note that, because the fall-off of the singular values is very rapid above $N_{Hy}$, the horizontal scale is expanded to show the behavior just round $N_{Hy}$. Three cases are plotted for different lengths of source ( $w_s$ ) and receiver ( $w_r$ ) lines and separation (*L*), as sketched on the left. Red points and lines (upper traces): $w_s = w_r = 1024\lambda$, $L = 4096\lambda$, $N_{Hy} \simeq 256$. Blue points and lines (middle traces): $w_s = w_r = 2048\lambda$, $L = 8192\lambda$, $N_{Hy} \simeq 512$. Orange points and lines (lower traces): $w_s = w_r = 4096\lambda$, $L = 16384\lambda$, $N_{Hy} \simeq 1024$. The number of source and receiver points used in the calculations was 513 (red), 1025 (blue) and 1036 (orange).

For two-dimensional source and receiver spaces, such as square arrays, Fig. 19 shows there is also a strong fall-off of the singular values above $N_H$, with an underlying exponentially decaying form, though in this case in a "stair-case" curve. The dashed lines in Fig. 19 are exponentials given by

$$h(r) = 4\exp\left(-\frac{3}{5}\frac{(n - N_H)}{\sqrt{N_H}}\right) \tag{69}$$

(Again, this formula is not derived, and is simply heuristic, judged by eye to show an approximate trend, though again only one set of coefficients (the numbers 4 and 3/5) is used for all three curves.) The "stair-case" behavior and the $\sqrt{N_H}$ in the denominator in the exponential can be rationalized [79].

In this two-dimensional case, then, the fall-off in the coupling strengths is not so abrupt as in the case of the one-dimensional source and receiver spaces, but it still has a strongly exponential underlying form. Again, once we pass significantly beyond $n \approx N_H$, the coupling strengths become very weak, with an underlying exponential fall-off.

In Fig. 19, in addition to plotting the relative strengths of the singular values between equal sized square source and receiver spaces, we have also plotted one additional set for a case where we have made the receiver "square" twice as large in linear dimension, but at double the distance (as shown by the grey points in the graph on the right). In this case, the solid angle subtended by the receiving space at the source space is retained; hence this additional case has the same paraxial heuristic number $N_H$ as the "orange" points and lines. We see that, indeed, the orange and grey relative singular values have very similar behavior. This similarity illustrates an important point. All of the modes we are showing here, including the ones past $N_H$ on these curves, are propagating modes; their amplitudes are falling off in an inverse square fashion. If they did not do that, then the orange and grey "curves" would have different forms. In particular, we can restate this point as follows:

These weakly coupled modes are *not* in general evanescent in the "far field".

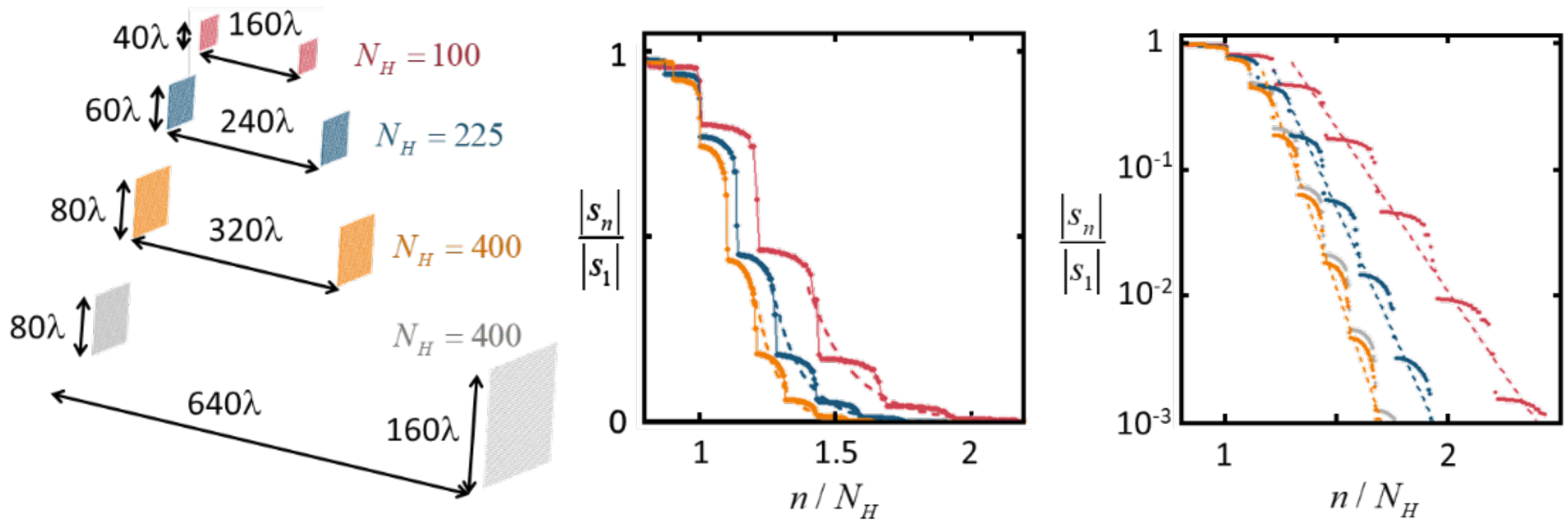

Fig. 19. Plots of the relative size of the singular values for several different approximately paraxial square arrays of source and receiver points, as a function of the mode number *n* compared to the paraxial heuristic number $N_H$ for each pair of source and receiver squares, on both linear (left graph) and a logarithmic (right graph) scales. The values are shown as points, with the solid lines in the linear graph joining them to aid the eye. The dashed lines are exponential functions given in the text. Note that the horizontal scale is displaced to show the behavior near to and above $N_H$. The square source and receiver spaces have linear dimension $w_s$ and $w_r$, respectively, and separation *L*, and the points are equally spaced on square lattices in each case, and with equal number of source and receiver points *N*. Red points and lines (upper traces): $w_s = w_r = 40\lambda$, $L = 160\lambda$, $N = 441$, $N_H = 100$. Blue points and lines (middle traces): $w_s = w_r = 60\lambda$, $L = 240\lambda$, $N = 961$, $N_H = 225$. Orange points and lines (lower traces): $w_s = w_r = 80\lambda$, $L = 320\lambda$, $N = 1681$, $N_H = 400$. Grey points (right graph only): $w_s = 40\lambda$, $w_r = 80\lambda$, $L = 640\lambda$, $N = 1681$, $N_H = 400$.

We might have thought that the exponential fall-off in the singular values was somehow associated with exponential fall-off in the field amplitudes with distance. In that case, the well-coupled modes with a "diffraction angle", which have ordinary inverse-square behavior, would remain similar in their coupling strengths as we increased the distance (while retaining solid angle), but the weakly coupled modes would not. However, the weakly coupled modes are showing very similar relative coupling strengths (as we increase the distance while retaining the solid angle) - in fact, slightly stronger in this numerical example for the case of the more distant but larger receiving surface.

There is some discussion in the literature of what is equivalent to our exponential fall-off in the singular values past our heuristic numbers, at least in the "Fourier transform" view, [80 - 82] for simple apertures [81] or some higher dimensional structures [82].

### 5.3.5. Use of point sources as approximations to sets of "patches"

So far, we have discussed point sources and point receivers because they enable a relatively straightforward set of "toy" problems to illustrate various behaviors. Below we will consider the mathematics of continuous sources more deeply, and to get that correct, we need several additional concepts. However, we can already argue that, at least under some circumstances, such point sources and receivers are some reasonable approximation to a situation where we have uniform source "patches" covering the source surface and similarly for a receiving surface. In other words, we can argue that a point source in the middle of a "patch" can be a good approximation to a uniform source that covers the patch, and that a similar approach can also work for uniform receiving "patches".

We can construct a heuristic argument, at least for the paraxial case, to get a characteristic maximum separation we need between our point sources if they are reasonably to approximate continuous source line segments on a "line" source (e.g., as in Fig. 9, Fig. 10, and Fig. 18) or a uniform "patch" source (e.g., as in Fig. 15, Fig. 16, and Fig. 19). This argument is based on keeping the variation in the path length to the receiving line or surface less than half a wavelength for two adjacent point sources (or, equivalently, between the extreme ends of the line segment or patch).

For a line source of total (lateral) length $w$ (e.g., the length of $48\lambda$ in Fig. 9) and separated from the receiver space by a distance $L$ (e.g., the distance $192\lambda$ in Fig. 9), and for a paraxial condition where $w \ll L$, we can argue (**Appendix A**) that the distance $d_p$ between the source points should obey

$$d_p < \lambda L / 2w \tag{70}$$

We can presume a similar constraint applies in each direction for a two-dimensional source, using different $w$ in each direction if the overall source has different sizes in the two directions.

For the paraxial situations above in Fig. 9, Fig. 10, Fig. 12, Fig. 15, Fig. 16, Fig. 17, Fig. 18, and Fig. 19 where $L / w \sim 4$, we should therefore have a separation between the point sources of $d_p < 2\lambda$ if those point sources are reasonably to approximate uniform patches. For the "line" sources in Fig. 9, Fig. 10, Fig. 12, and Fig. 18, we have $\lambda / 2$ source spacing, so we easily satisfy Eq. (70). In the various "area" examples, Fig. 15, Fig. 16, Fig. 17, and Fig. 19, where the source spacings used are $\sim 2\lambda$ or slightly larger, we are just on the edge of violating this simple criterion.

We could follow through similar arguments for the idea of replacing the point receivers with uniform patches of continuous "receivers", with a similar result. So,

> with some care in the spacing of point sources and receivers, such an approach can mimic the behavior we would have for uniform patches of sources and/or receivers with dimensions equal to the spacing between the point sources and/or receivers.

## 5.4. Non-paraxial behavior

### 5.4.1. Longitudinal heuristic angle

A set of sources can have directionality not only as a result of their transverse dimensions. Longitudinal sets of sources can also give rise to directional waves. This is routine in many designs of radio antennas, for example. This behavior is illustrated in the example in Fig. 20. Here we take horizontal lines of sources and of receivers and find the most strongly coupled communications mode.

We can rationalize this behavior with a heuristic argument based on interference of sources at the two extreme ends of the line of sources, and we construct this argument in **Appendix B**. This leads to a *longitudinal heuristic angle*

$$\theta_L = \sqrt{\frac{\lambda}{2\Delta z}} \tag{71}$$

that characterizes the "cone angle" of the resulting diffraction, which is a measure of the directionality we expect to be possible in the longitudinal direction from a line source of length $2\Delta z$. The calculated intensity pattern in Fig. 20 shows this angle $\theta_L$ gives a good approximate description of the resulting beam. Incidentally, in this case, there is only one communications mode; within numerical accuracy (at least to 5 significant figures), all of the sum rule $S$ is consumed by the coupling strength $|s_1|^2$ of this mode.

We can also evaluate the effective solid angle of this channel. The area of the disk of half-angle $\theta_L$, or equivalently, of radius given by the corresponding $\delta y$, is $A_L = \pi(\delta y)^2 = \pi z_o^2 \theta_L^2$, so the corresponding solid angle of this "disk" is

$$\Omega_L \simeq \frac{A_L}{z_o^2} = \pi\theta_L^2 = \frac{\pi\lambda}{2\Delta z} \tag{72}$$

which is $\pi$ divided by the length of the line source in wavelengths. We can compare this with the solid angle per channel for the two-dimensional surfaces, as in Eq. (67), which is 1 / (the area of the relevant surface in square wavelengths). This tells us that,

> for some object of comparable dimensions in all three directions, once the dimensions are some reasonable number of wavelengths, the solid angle per channel is determined more by the cross-sectional area than by the thickness.

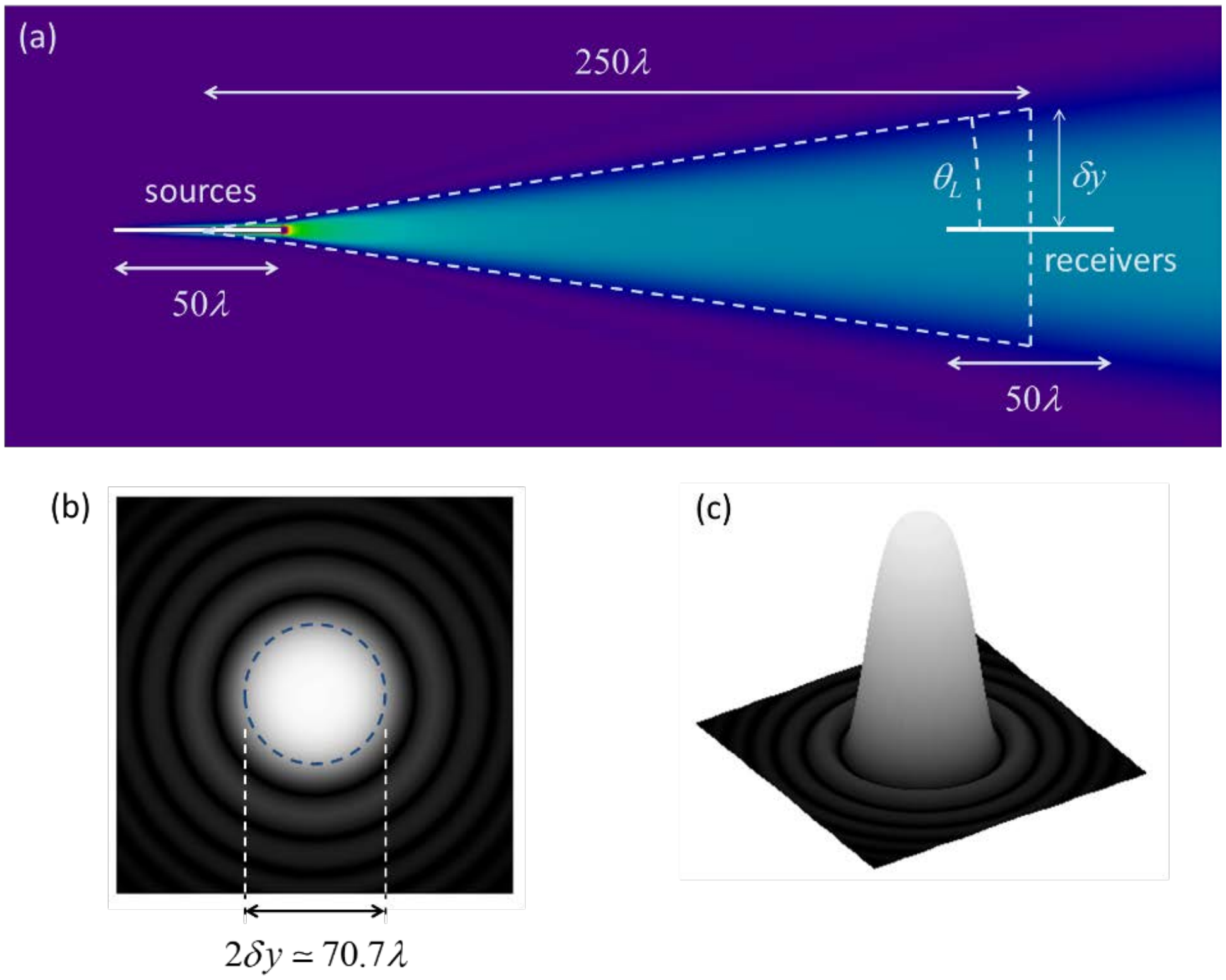

Fig. 20. Illustration of the beam resulting from finding the best-coupled mode between two horizontal lines of sources and receivers, showing the longitudinal heuristic angle $\theta_L$ . Both sources and receivers use 201 points spaced by $\lambda/4$, aligned in the $z$ horizontal axis, and with center-to-center spacing of $z_o = 250\lambda$ . From Eq. (71), $\theta_L \simeq 141\,\text{mrad} \simeq 8.1°$ and the corresponding $\delta y \simeq 35.35\lambda$ . (a) is a cross-section of the intensity. For graphic clarity, the magnitude is multiplied by $z^2$ once we leave the source region (technically, a factor $[\max(30\lambda, z)]^2$ ). The intensity in the region immediately around the source is omitted from the graphics to avoid singularities. (b) An *x-y* cross-section of the intensity in the middle of the receivers; (c) a perspective surface-plot view of the same data as in (b).

Equivalently, for some cuboid of dimensions $2\Delta x$, $2\Delta y$, and $2\Delta z$, the ratio of the solid angle per channel from the cross-sectional area $2\Delta x \times 2\Delta y$ to that from the length $2\Delta z$ alone is

$$\frac{\Omega_{S1}}{\Omega_L} \simeq \frac{2\Delta z\lambda}{\pi(2\Delta x)(2\Delta y)} \tag{73}$$

For example, for $2\Delta x = 2\Delta y = 2\lambda$, to get this ratio to be 1 (i.e., the solid angle from the length comparable to that from the cross-sectional area), the length $2\Delta z$ would need to be $4\times\pi \simeq 12.6$ wavelengths, much larger than any cross-sectional dimension. So, conventional “optical” situations with large cross-sectional dimensions in wavelengths have solid angles per channel roughly independent of the depth of the volume for any thickness comparable to the cross-sectional dimensions.

By contrast, if we consider transverse dimensions $2\Delta x = 2\Delta y = \lambda / 2$, then once the length $2\Delta z$ becomes significantly greater than a wavelength, the effect of the length will dominate in narrowing the solid angle of the channel. This is a typical situation in many multiple-element wireless or radio-frequency antennas (such as the classic "Yagi-Uda" antenna, for example), and could also occur in nanophotonic systems.

### 5.4.2. Spherical shell spaces

Concentric spherical "shell" source and receiver spaces (see Fig. 21) are a good example of a case that is very much not paraxial. Indeed, there is no preferred axis at all in this case. This case is also interesting from a fundamental point of view; it may allow us to deduce some limiting behavior for any and all waves emitted from some space because there is no way for the generated wave from a smaller "source" spherical shell to miss a larger spherical shell "receiving" space that surrounds it.

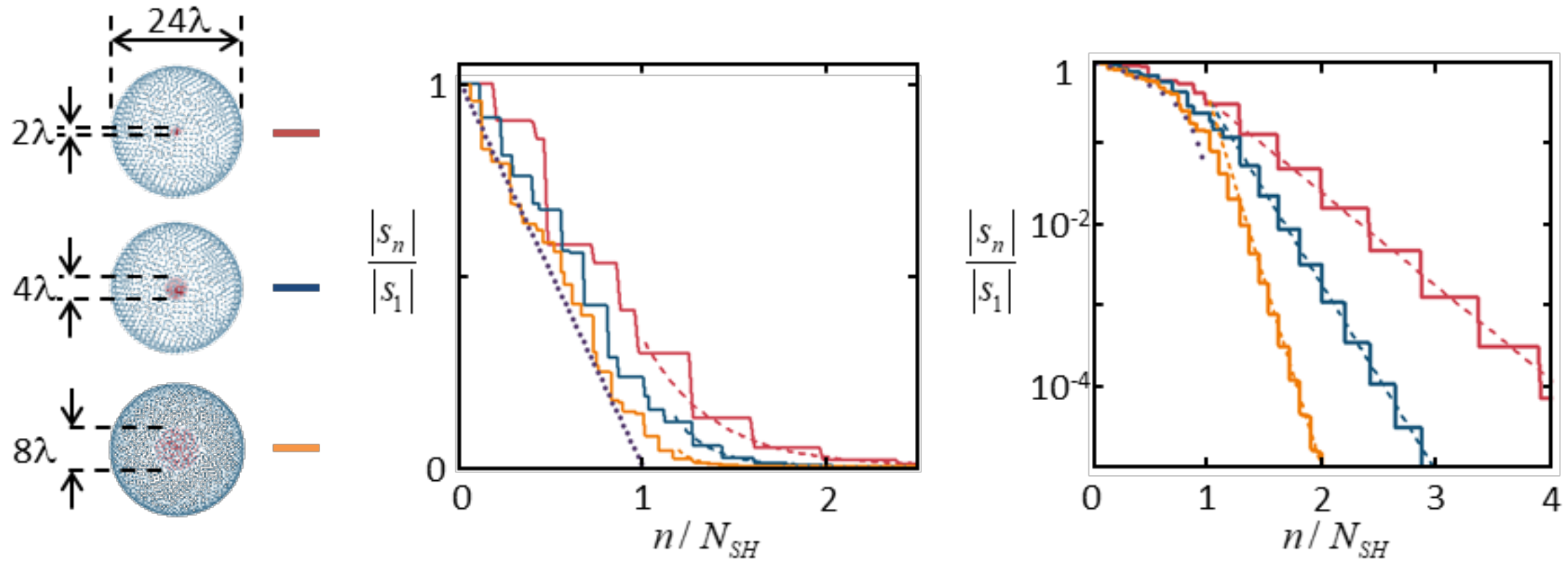

Fig. 21. Behavior of the magnitude of the singular values $|s_n|$, as a function of communication mode number *n*, relative to that of largest singular value $|s_1|$, for three different centered spherical "shell" source and receiver spaces. In each case, the receiving points are on the surface of a $24\lambda$ diameter sphere. The source points are on the surfaces of spheres of diameters $2\lambda$ (upper, red line), $4\lambda$ (middle, blue line), and $8\lambda$ (lower, orange line), respectively. The solid lines are drawn between the calculated values of $|s_n|/|s_1|$ in each case to guide the eye. The dashed and dotted lines are heuristic functions shown for comparison (see text). On the horizontal axis, the mode numbers for each curve are divided by the corresponding spherical heuristic numbers, which are $N_{SH2} \simeq 50.3$, $N_{SH4} \simeq 201$, and $N_{SH8} \simeq 804$ for the 2, 4, and $8\lambda$ diameter source spheres, respectively. 1600 source and receiver points are used for the $2\lambda$ and $4\lambda$ cases, and 2400 for the $8\lambda$ case, distributed approximately uniformly over the sphere surfaces [83].

Fig. 21 shows the calculated singular value magnitudes (relative to the strongest (first) one) for three different situations with increasing radius of the inner "source" spherical shell surface. The singular value magnitudes are plotted both on a linear scale (left graph) and a logarithmic scale (right graph). We see first from the left graph in Fig. 21 that this situation does not lead to anything like paraxial degeneracy of the "well-coupled" modes. Other than some "step" structure, there is no large "plateau" of approximately equal singular values. Indeed, intriguingly, as we increase the size of the "source" sphere, these "strongly coupled" singular values apparently asymptote towards a simple straight line (the dotted line in the left graph), with an intercept we call a *spherical heuristic number* of

$$N_{SH} = 16\pi r^2 / \lambda^2 \tag{74}$$

where *r* is the radius of the sphere of sources; that is, the line goes from 1 at $n = 1$ down to 0 at $n = N_{SH}$, or equivalently, a function $1 - n / N_{SH}$. We can justify this number also with a heuristic argument given in **Appendix C**. This number $N_{SH}$ corresponds to one mode for every square half wavelength (a surface area element $(\lambda/2)^2$) on the source sphere surface.

As we continue past this $N_{SH}$ in each case, we see an underlying exponential fall-off in the singular values on a "staircase" line in each case. The dashed lines in each case are given by the function

$$f_{SH}(n) = \frac{1}{3}\exp\left(-\frac{3}{8}\frac{(n - N_{SH})}{\sqrt{N_{SH}}}\right) \tag{75}$$

Note that this expression is heuristic; it is not derived, and the constants 1/3 and 3/8 are simply chosen to give exponentials that, by eye, approximately describe the apparent exponential fall of in the singular values as *n* begins to significantly exceed $N_{SH}$ in each case. It is worth noting, however, that we are able to use the same coefficients, 1/3 and 3/8, for all three cases shown. We can however argue for the $\sqrt{N_{SH}}$ factor in the denominator in the exponential by a similar rationalization to that above for the paraxial case.

Note that, despite there being nowhere for the wave go to avoid the receiving sphere, we still see an exponential fall-off in coupling the weakly coupled modes, rather similar in form to the two-dimensional paraxial case above. It is also true in this spherical case that increasing the radius of the "receiving" sphere makes essentially no difference to the form of the lines in Fig. 21. Again, we conclude that the weakly coupled modes therefore also correspond to propagating (not evanescent) modes, with inverse square behavior of their intensities, at least for receiving radii much larger than the source sphere.

The simplicity of the asymptotic behaviors of the singular values suggests some analytic solution may explain these. Indeed, we expect there may be analytic solutions in such a spherical case, likely involving spherical Bessel functions for the radial behavior and spherical harmonics for the angular behavior, for example.

## 5.5. Deducing sources to give a particular wave

The SVD approach gives a straightforward way to calculate just what source in the source space or volume is required to generate a specific wave in the receiving space or volume. Suppose we want a (normalized) wave (or vector of amplitudes) $|\phi_{Ro}\rangle$ in the receiving space. Because the (normalized) set of eigenfunctions or eigenvectors $\{|\phi_{Rj}\rangle\}$ is a complete set for the receiving space, then we can expand $|\phi_{Ro}\rangle$ in this set, as

$$|\phi_{Ro}\rangle = \sum_j a_j |\phi_{Rj}\rangle \tag{76}$$

where

$$a_j = \langle \phi_{Rj} | \phi_{Ro} \rangle \tag{77}$$

Suppose, for the moment, that we want to generate just one component, $a_q|\phi_{Rq}\rangle$ of this sum (Eq. (76)). Then because $\mathsf{G}_{SR}|\psi_{Sj}\rangle = s_j|\phi_{Rj}\rangle$ (Eq. (31)), to generate that component we need an amplitude of the (normalized) source function $|\psi_{Sq}\rangle$ of $a_q/s_q$. We can repeat this argument for each component, adding up the results. So, quite generally, the required source function $|\psi_{So}\rangle$ to generate $|\phi_{Ro}\rangle$ is

$$|\psi_{So}\rangle = \sum_j \frac{a_j}{s_j}|\psi_{Sj}\rangle \equiv \sum_j \frac{1}{s_j}\langle \phi_{Rj}|\phi_{Ro}\rangle |\psi_{Sj}\rangle \tag{78}$$

This point has been known, at least for the specific case of prolate spheroidal basis sets and correspondingly simple apertures, for some time [84].

### 5.5.1. Sources for an arbitrary combination of specific receiver modes

In the example of Fig. 22, we have deliberately chosen to try to create a set of amplitudes at the receiver positions that is a specific arbitrary (normalized) superposition [85] of the first 14 "receiver" communications modes, here of the approximately paraxial example as in Fig. 9, Fig. 10, and Fig. 12.

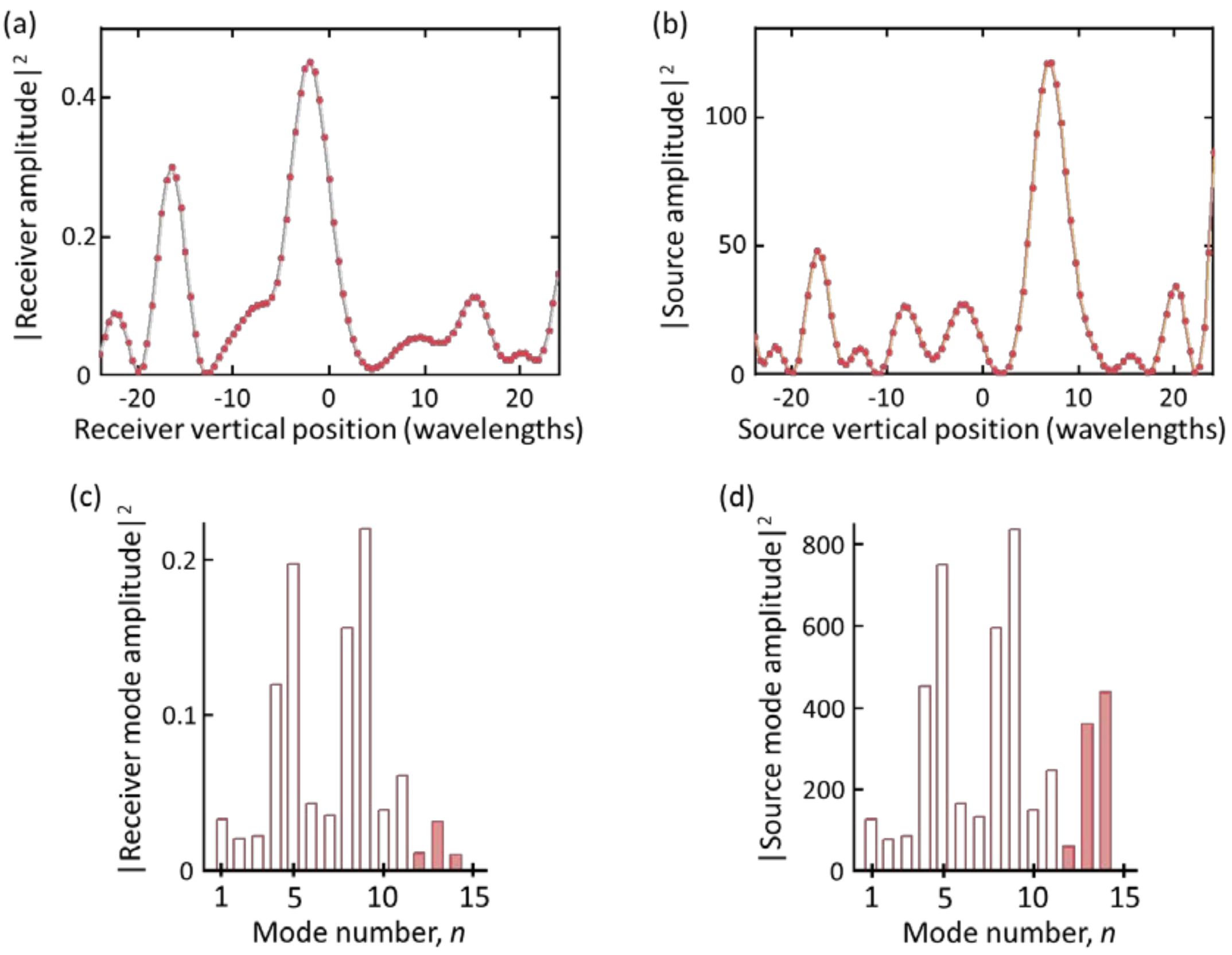

Fig. 22. Example of constructing the required sources to generate a specific received "wave" or set of received amplitudes. To avoid the additional graphic complication of handling phases, the values plotting in the graphs are for the squared magnitudes of the relevant quantities. The source and receiver points are as in Fig. 9, with the corresponding modes as in Fig. 10 and Fig. 12. (a) Desired receiver values (points) and the calculated actual values generated (line) using the calculated source values. (b ) Corresponding sources values (points, joined by lines for visual clarity). (c) Values for each receiver mode used to construct the desired values at the receiver points. (d) Corresponding required values for each source mode to generate the desired receiver values. Note in particular in (c) and (d) that, though the required source values for each mode largely track the required receiver values for each mode for the first 10 modes (which all have similar singular values), for Modes 12, 13, and 14 in particular (which are highlighted in (c) and (d)), the required values for the source modes have to rise because the singular values are becoming smaller.

The points in Fig. 22 (a) show the modulus squared of the desired amplitudes at the receiver points (so, the effective power desired at each of these points). In this case, because we restricted to waves that could be created just from these first 14 modes, we are able to generate exactly the wave we want by using Eq. (78) to set the corresponding amplitudes of the first 14 source functions. The grey curve that passes through the points in Fig. 22 (a) is not a line joining the points (as we have done in earlier graphics); rather it is the modulus squared of the calculated wave as a function of vertical position at the line of receivers. Obviously, this approach generates the form we wanted at the receiver points. The modulus squared of the corresponding amplitudes at the source points is shown in Fig. 22 (b).

Fig. 22 (c) shows the modulus squared of the corresponding receiver modal amplitudes using to make up the superposition. Note in particular that we have chosen finite amplitudes for Modes 12, 13, and 14 (highlighted as the solid red bars in Fig. 22 (c)) so that we can see the effect of including some modes with relatively small singular values. The singular values are all similar for the first 10 modes. In general, we see from Eq.(78) the modulus squared of the required amplitude in the corresponding source modes should be larger by a factor $1/|s_j|^2$ for the *j*th mode so as to generate the required receiver values. For the first 10 modes in this approximately paraxial problem, this source quantity (Fig. 22 (d)) varies only slightly from $1/|s_1|^2 \simeq 3752$ for Mode 1 to $1/|s_{10}|^2 \simeq 3803$ for Mode 10. Consequently, the modulus squared of the source amplitudes for these first 10 modes (in Fig. 22 (d)) looks essentially identical in form for these first 10 modes to the corresponding modulus squared of the receiver mode amplitudes (Fig. 22 (c)).

Note, though, that the required relative magnitudes of the source amplitudes grow substantially especially as we consider Modes 12, 13, and 14 (we have highlighted these also in Fig. 22 (d)). Once we get to Mode 14, $1/|s_{14}|^2 \simeq 44694$, and the required mode amplitude magnitude has grown accordingly. Hence we see explicitly in this example that, if our desired set of receiver amplitudes require the use of modes with small singular values, the amplitude of the corresponding source mode has to be increased accordingly. The reason why the shapes of the curves in Fig. 22 (a) and Fig. 22 (b) are different is because we need these relatively larger amplitudes of the weakly coupled modes.

.

Note again that, for this case where we made up the desired set of amplitudes at the receivers as a specific linear combination of a finite number of the (receiver) communications modes, we were guaranteed to be able to make up the required source amplitudes from a finite linear combination of the corresponding source communications modes. If we examine more general functions, we are not guaranteed any such finite linear combination, however, and we look at some such examples next.

### 5.5.2. Sources for a Gaussian spot – passing the diffraction limit

A Gaussian "beam" is a well-known example form with straightforward mathematical properties, and is one that also occurs in optics as a good approximation to the beam form from confocal laser cavities. Such a Gaussian distribution of desired amplitudes at the receiver points does not, however, correspond to any of the communications modes for finite source and receiver spaces, and mathematically generating that set of Gaussian amplitudes will require a linear combination of multiple receiver communications modes.

In the examples in Fig. 23 (which uses the same sets of sources and receivers as in Fig. 9, Fig. 10, Fig. 12, and Fig. 22), we have chosen to ask for receiver amplitudes of the form

$$\phi\left(y_{Rj}\right) = c_R\left(0, y_{Rj}\right)\exp\left(-\frac{\left(y_{Rj} - y_o\right)^2}{w^2}\right) \tag{79}$$

These amplitudes correspond to a Gaussian in the *y* direction (vertical on the graphs) with $1/e$ amplitude half width of *w* (which is also therefore $1/e^2$ "intensity" half width of the set of modulus squared amplitudes $|\phi_{Rj}|^2$), with unit amplitude at the center of the Gaussian (i.e., at the point $y_o$), and with confocally curved phase fronts from the factor $c_R\left(0, y_{Rj}\right)$ (as in Eq. (55)). For all the graphs except Fig. 23 (e), we choose $y_o = 0$, which corresponds to a Gaussian centered vertically. Fig. 23 (e) has the desired center shifted down (here, technically a positive shift) by 18λ.

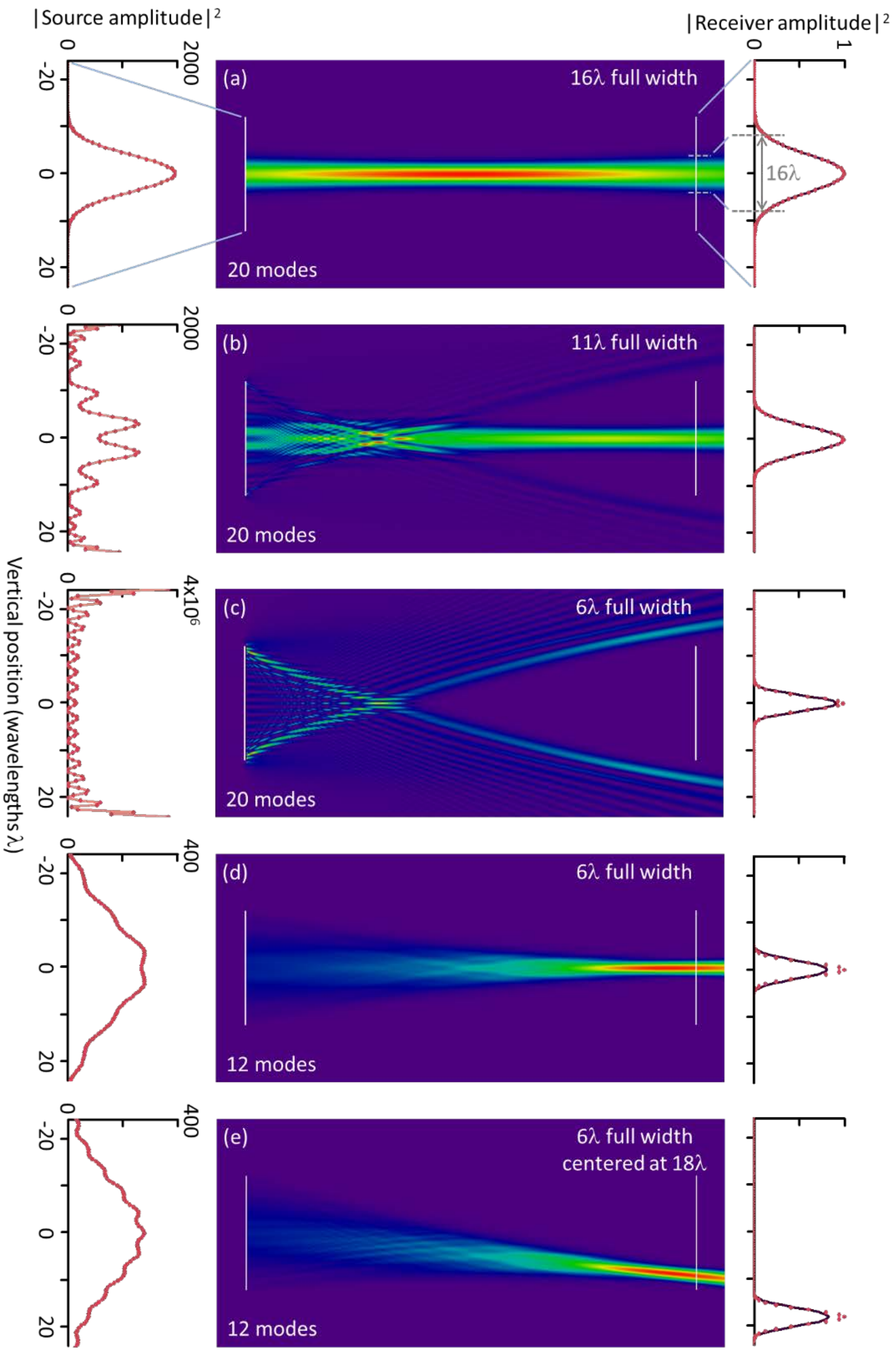

Fig. 23. Plots of the required sources (the red points, joined by red lines for visual clarity, are the modulus squared of the amplitudes of the "front" line of sources) on the graphs on the left to attempt to synthesize the desired receiver values, shown as the red points on the graphs on the right (the modulus squared of the desired receiver amplitude is plotted). The actual resulting values of the modulus squared of the receiver amplitude are shown as the black line in these graphs on the right . Beam intensity, multiplied by the distance from the sources on the left for graphic clarity, is shown in the middle pictures in false color. Results for three different desired Gaussian widths are shown in (a), (b) and (c) respectively, with calculations based on using the first 20 communications modes. For (d) and (e), the calculation is restricted to using only 12 modes, and for (e), the desired position is shifted down by 18λ.

We have confocally curved the desired phase front because such waves are “easier” to construct, especially for “off-center” desired beams (as in Fig. 23 (e)), because of the underlying confocal curvature of the beams with large numbers of “bumps”. Constructing “flat” phase fronts away from the center tends to require the use of significant amplitudes of more modes than we might expect from just the intensity shape of the beam.

In the superpositions, we only use the first 20 communications modes in our calculations for Fig. 23 (a), (b), and (c), and the first 12 for (d) and (e). 20 modes are apparently sufficient for (a) 16λ and (b) 11λ full widths; the resulting generated amplitudes (the black lines) correspond well with the desired values (red points), at least as seen by eye in these calculations. The source form for the 16λ case also looks to be a smooth and essentially Gaussian curve. (We can rationalize later why the source form here is also Gaussian.)

For the 11λ full-width case, the form of the source is more complicated, which essentially reflects the fact that we are close to violating what we typically regard as diffraction limits, which loosely here means we are starting to require the use of weakly coupled modes. As a result, we see significant amplitudes of some of the high numbered modes, which also shows in the emergence of parts of the beam that miss the receiver space.

For the 6λ case in Fig. 23 (c), large amounts of weakly coupled high-numbered modes are required. Note in particular that the vertical scale maximum ( $4\times10^6$ ) on the source graph in (c) is 2000 times as high as that in Fig. 23 (a) and (b), and that there are particularly large amplitudes of the sources at the two extreme ends of the source region. Now in the picture of the beam intensity, on this false color scale, the amplitude of the desired Gaussian beam does not even show up, and large parts of the beam miss the receiver region entirely (above and below). Nonetheless, an approximately Gaussian beam of approximately the correct width is generated along the line of receivers, as we see in the graph on the right of Fig. 23 (c). The very large source amplitudes and the substantial parts of the beam missing the receiver space entirely are because there are significant amplitudes of high-numbered and very weakly coupled communications modes.

The red bars in Fig. 24 (b) show the amount of the various modes (as the modulus squared of their desired amplitudes) that we require. Now we see that for this 6λ wide beam, there are small but significant received amplitudes required though Mode 23 on this graph. If we include 23 modes in the calculation, we can do somewhat better in the beam shape, as shown by the orange curve in Fig. 24 (a), though even then we still do not quite reach the desired shape.

If we look closely at the graph on the right in Fig. 23 (c), we see that the desired Gaussian beam is not quite correctly created; the peak is not quite high enough in the center. The reason for this discrepancy is that we do not have enough modes in our calculation. We have plotted the region near the center in more detail in Fig. 24. Now the calculation for the 20 mode case is the light blue line in Fig. 24 (a), which is below the desired peak (as given by the red points) in the center (and is also slightly wider near the base of the curve).

In principle, we could keep adding more, higher-numbered modes and continue to increase the accuracy of the created beam shape. However, the source amplitudes required for these higher-numbered modes are increasing very rapidly because of the very small singular values associated with these high-numbered communications modes. In fact, beyond Mode 23, the coupling is so weak that the calculation starts to have significant numerical errors in conventional 64 bit calculations, and further improvements are essentially beyond such standard calculations.

As we see in Fig. 24 (a), the shape calculated with just 12 modes is not much different from those with added higher-numbered modes. (In these calculations for a centered beam, only the odd numbered modes are actually required because of the symmetry, so we only plot values required for the source and receiver modes for the odd modes in Fig. 24 (b).) The first 12 modes here are all strongly coupled (see Fig. 9 (b)), with the first 10 being approximately equally well coupled. As we add in further modes (modes 13 and 15) for the 16 mode calculation, we see some improvement in the resulting shape (the pink curve in Fig. 24 (a)) with corresponding required additional source modes (pink bars in Fig. 24 (b)). The required source mode values for the 20 mode and 23 mode cases are shown as the blue and orange bars respectively in Fig. 24 (b).

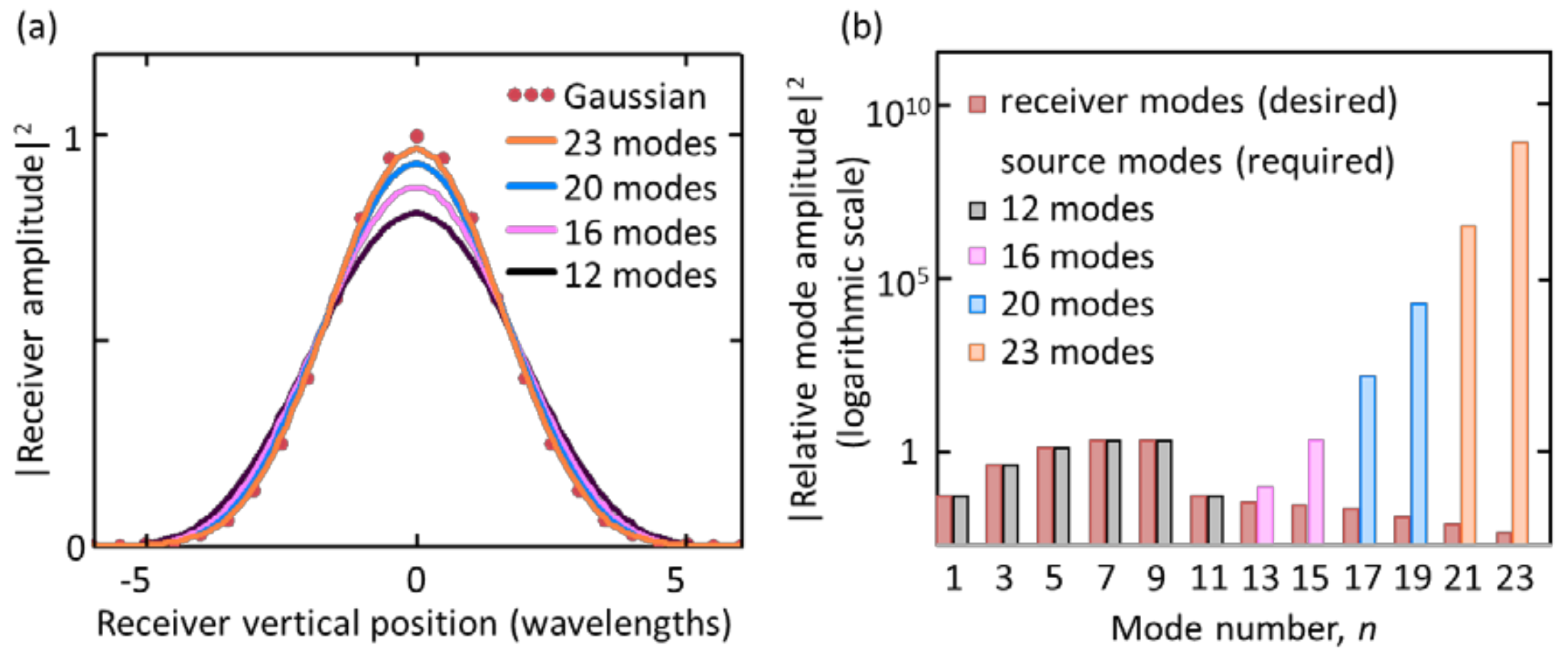

Fig. 24. Illustration of the effect of changing the number of modes used in trying to create the Gaussian amplitudes for the 6-wavelength wide Gaussian shape in Fig. 23 (c) (red points in (a) here). The red bars in (b) show the mode amplitudes (modulus squared) required for the receiver communications modes (up to Mode 23) to attempt to create the desired Gaussian shape. The grey bars show the corresponding relative strengths of the (modulus squared of) the source mode amplitudes up to Mode 12. (Only odd numbered modes occur in this problem because of the symmetry, and only those amplitudes are therefore plotted here.) The pink, blue, and orange colored bars show progressively the additional required (modulus squared) amplitudes for 16, 20, and 23 mode calculations. (The overall vertical position of the source mode (modulus squared) amplitudes on this logarithmic scale is adjusted to match the corresponding receiver mode (modulus squared) amplitude for the strongest coupled mode for easier comparison of relative magnitudes.) (a) shows that a sharper peak and a slightly narrrow shape do result from adding further modes, but (b) shows that the required amplitudes of the additional higher-numbered modes become enormous, illustrating the practical impossiblity of substantial exceeding diffraction limits.

In Fig. 24, we see that adding in higher-numbered modes makes only small improvements in the desired shape, and at the cost of extremely large amplitudes of the source modes. We are therefore illustrating the practical impossibility of substantially passing the diffraction limit; even extremely large amplitudes in the additional sources make relatively little improvement to the resulting shape.

In this case, if we restrict ourselves to using only the well-coupled modes (e.g., up to Mode 12), we can generate a reasonable shape of received beam (Fig. 23 (d)), even if it is not quite as narrow as we would have wanted, while avoiding any large source amplitudes and while also having the beam essentially all arriving in the receiver space.

Fig. 23 (e) illustrates what happens if we retain this 12 mode calculation but now ask for a beam displaced by 18 wavelengths from the center. We see that we are able to shift the beam while retaining what appears to be a similar shape, and we expect similar beam “scanning” behavior over the entire receiver space.

### 5.5.3. “Top-hat” function

As another example, we can attempt to generate a “top-hat” function – one that is constant within a range and zero elsewhere. We use the same configuration as in Fig. 23, and we ask in this example for a centered “top-hat” set of receiver amplitudes with a width of 18$\lambda$. The resulting actual receiver amplitudes and the relative mode amplitudes are plotted in Fig. 25. Except for the use of the top-hat function rather than the Gaussian, Fig. 25 is otherwise similar to Fig. 24.

The top-hat function is approximately created, but the edges are not abrupt, consistent with diffraction limitations, and additionally there is “ringing” – spatial oscillations – in the amplitude across the top of the "top-hat", reminiscent of the Gibbs phenomenon in the use of Fourier series to represent "square" functions. We also see similar behavior to that seen with the Gaussian in that the inclusion of further higher-numbered modes in the calculation leads to relatively little improvement in the form of the actual amplitudes at the receivers, and very large amplitudes of higher-numbered source modes are required even for these benefits.

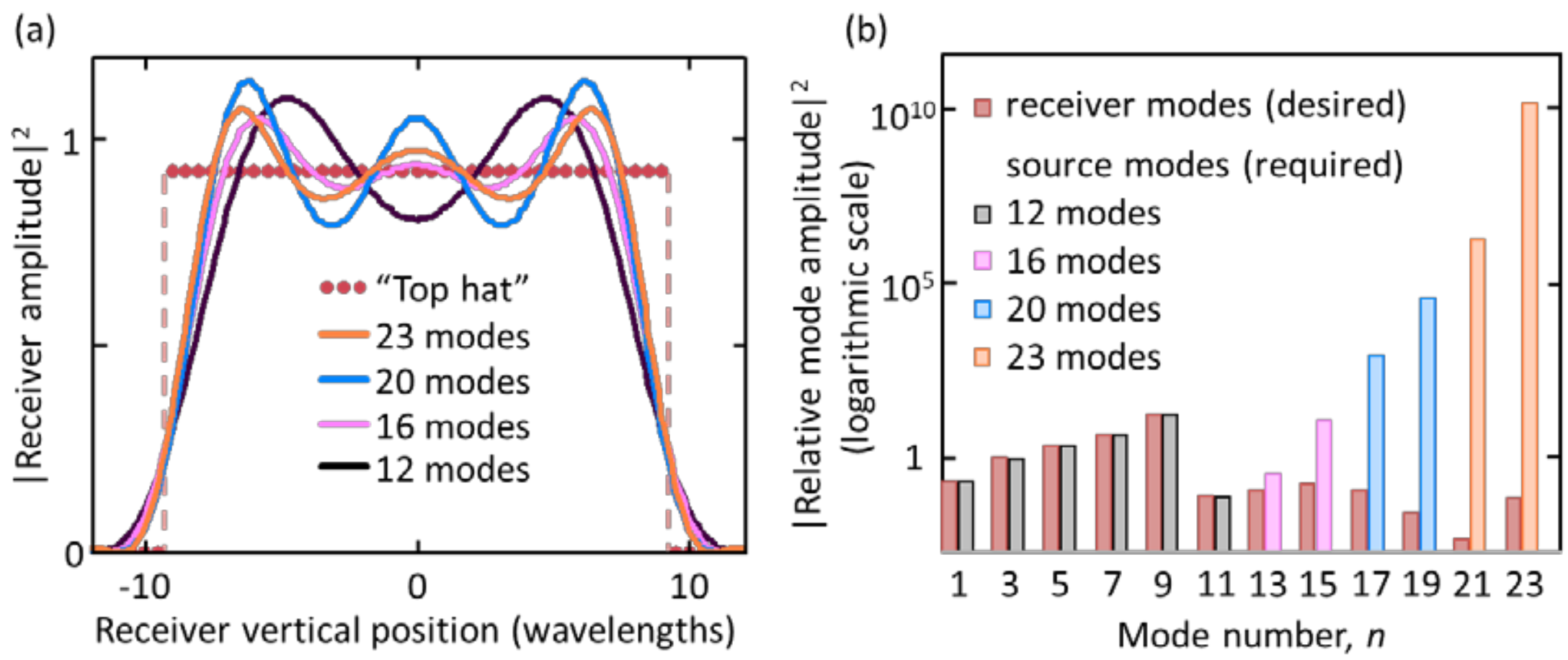

Fig. 25. (a) Receiver amplitude (modulus squared) for the desired "top-hat" function of width 18λ (red points and dashed light red lines to guide the eye) and the various actual amplitudes (modulus squared) for use of different numbers of modes, simlarly to Fig. 24 (a). (b) Relative amplitudes of the (modulus squared of) the receiver and transmitter modes, similalry to Fig. 24 (b).

### 5.5.4. Notes on passing the diffraction limit

Generally, these examples are illustrating that

> the reason why we cannot focus past conventional diffraction limits is that we must then use very weakly coupled communications modes, leading to very large source amplitudes.

There is no way of avoiding this for given source and receiver spaces. If we look back at Fig. 12, we see that, once we pass the "well-coupled" modes, the number of "bumps" in the subsequent modes only improves linearly as the coupling strength falls of exponentially. Equivalently

> linear improvements in resolution past the diffraction limit essentially require exponential increases in source amplitudes

Our discussion here is, of course, only an example to illustrate how this communications mode approach relates to resolution limits. Such limits are well understood and have been comprehensively reviewed [86, 87]. [88 - 90] are examples of using such a modal approach experimentally for super-resolution, based on the analytic prolate spheroidal functions and/or equivalent "sampling theory" approaches (see section 7.3 and [33]). [90] explicitly makes the same point we are making here that the rapid fall-off the singular values prevents effective super-resolution. These conclusions are also consistent with recent innovative approaches to sub-diffraction imaging [91], which indicate large amplitudes of sub-diffraction effective image sources are required to image such features into the far field.

# 6. Mathematics of continuous functions, operators, and vector spaces

Using finite collections of point sources and receivers, we have seen many of the behaviors of orthogonal wave channels between sources and receivers. But we have two problems:

1) Why is it that adding more sources and receivers (in given volumes) does not continue to increase the number of usable communications channels or "degrees of freedom"? We have some heuristic answers, but no general mathematical principle.
2) How can we transition mathematically continuous functions rather than discrete "point" sources and receivers? For example, we may actually have current densities on antennas or in continuous solids, and we may want to understand continuous waves in receiving volumes or even use them as effective continuous sources in diffraction problems.

Both of these questions are answered in this section. To address them, we need to step back mathematically from finite matrices, and set up an approach based on functional analysis. The results are very general and powerful.

To introduce functional analysis properly would take too much space. Unfortunately, though there are substantial texts in the field (e.g., [92 - 94]), their style and length can be forbidding. To help, I wrote a separate introduction [36] that, though shorter by about a factor of 10 than standard texts, does present all we need, including all proofs. With that as a reference, in sections 6.1 to 6.6, I introduce the main ideas and terminology, and give a path to the answers we need.

As we progress, we develop extended ideas of the inner product (section 6.6). We need this to handle electromagnetism properly. This development clarifies how to get back to a simpler "matrix-vector" algebra (in what we call an "algebraic shift"). We also complete the mathematics of SVD in section 6.7 and **Appendix D**.

One other key point is that we argue from the physics (sections 6.8 and 6.9) that the core operators we need (such as Green's functions) will be so-called *Hilbert-Schmidt operators*. That in turn allows us to use the particularly powerful mathematics of so-called *compact* operators, and also gives us the sum rule $S$ that underlies why we can only have finite numbers of usable communications channels.

## 6.1. Functions, vectors, numbers, and spaces

In this mathematics, we can think of functions as being mathematical "vectors", (e.g., column vectors of numbers), with possibly infinite numbers of elements; these might be the values of the function at each of a possibly infinite number of points, for example. If needed, we use the term "mathematical" vector to distinguish these from "geometrical vectors" such as a position vector $\mathbf{r}$, or an electric field $\mathbf{E}$. From here on, we use the terms "functions" and (mathematical) "vectors" interchangeably. Initially, we use Greek lower case italic letters, such as $\alpha$, to represent functions (or vectors), and Roman italic letters, such as $a$, for (complex) numbers or "scalars".

An (abstract) *space* is simply a set (here, of elements that are vectors) with some additional axiomatic properties (such as the inner product). So our vectors (or functions) will be elements in (or members of) what we call a *vector* (or function) *space*. Defining spaces is important; specifically, we will have different spaces for source functions and for received waves (or wave functions). Though these functions will generally also be in different physical volumes, by *space* we mean this abstract space, not the physical volume. Our "source" and "receiving" spaces may also differ in other ways – they may even also have different "inner products", for example.

## 6.2. Inner products

The most important axiomatic property we add to our space is the "inner product". This gives a well-defined effective "length" of a function (a norm), and a "distance" between two functions (a metric). It also defines "orthogonality" of the functions. The following defines an inner product, and anything with these properties is an inner product.

For all vectors $\alpha$, $\beta$ and $\gamma$ in a vector space, and all (complex) scalars *a*, we define an *inner product* $(\alpha,\beta)$, which is a (complex) scalar, through the following properties:

(IP1) $(\gamma,\alpha+\beta)=(\gamma,\alpha)+(\gamma,\beta)$

(IP2) $(\gamma,a\alpha)=a(\gamma,\alpha)$ (where $a\alpha$ is the vector or function in which all the values in the vector or function $\alpha$ are multiplied by the (complex) scalar *a*)

(IP3) $(\beta,\alpha)=(\alpha,\beta)^*$

(IP4) $(\alpha,\alpha)\geq 0$, with $(\alpha,\alpha)=0$ if and only if $\alpha=0$ (the zero vector)

(80)

The first two "linearity" properties, (IP1) and (IP2), are useful in describing linear systems (as in linear superposition of waves) [95]. (IP3) gives a useful algebra for working with complex quantities, and (IP4) means that the inner product gives us a positive real *norm*

$$\|\alpha\|=\sqrt{(\alpha,\alpha)} \tag{81}$$

A space with a norm is called a *normed space*. A norm expresses an idea of "length" for a function as some single, real number. The existence of a norm then allows us to define a *metric* (the metric "induced" by the norm), which then gives a real number that we can use as the "distance" between two functions or vectors. For two vectors or functions $\alpha$ and $\beta$, that metric would be defined as

$$d_P(\alpha,\beta)\equiv\|\alpha-\beta\|=\sqrt{(\alpha-\beta,\alpha-\beta)} \tag{82}$$

A space with a metric is called a *metric space*.

With ordinary geometric vectors, the norm is just the length of the geometric vector, and the metric is the distance between the "tips" of the two vectors if their other ends are "joined" at the same point. The usual dot product of geometric vectors is an inner product, satisfying all the properties (IP1) – (IP4), though geometric vectors have only real components, so (IP3) just corresponds to the geometric dot product being commutative. Note that the inner product is *not* however in general commutative because of the complex conjugate in (IP3). A vector or function space that has an inner product with these properties is called an *inner-product (vector) space*.

If some situation, such as waves in linear media, can be usefully described using vectors with such an inner product, then these properties of the norm and the metric mean that we can exploit much of the mathematics from real analysis; hence, we can use ideas of convergence of sequences of numbers, to discuss convergence of functions as well, and that idea is at the core of functional analysis.

The other very important use of an inner product is to define *orthogonality*. Specifically,

a non-zero element $\alpha$ of an inner product space is said to be *orthogonal* to a non-zero element $\beta$ of the same space if and only if $(\alpha,\beta)=0$

(83)

This is a generalization of the idea that the geometrical vector "dot" product is similarly used to define orthogonality (or "being at right angles") in geometric space. Note here that we extend that idea to allow for complex vector "components" and for arbitrary, even infinite, numbers of dimensions.

Note these essential properties of an inner product (IP1) to (IP4), listed in (80), leave considerable flexibility. When we set up a given vector space, we formally choose the inner product for that space; in different spaces we may make different choices. For a "receiving" space, we may want the inner product to correspond to the "energy" in a wave, and we can define such "energy" inner products. For the sources, on the other hand, we might just want a simple Cartesian inner product (as in ordinary vector multiplications), for example, for different current distributions. For a reason that will become clear later, we call such a choice the *underlying inner product* of the space [96]. We return to specific forms of inner products once we have defined the idea of "operators".

## 6.3. Sequences and convergence

We can obviously write a list of multiple elements in a set or space, but we need to distinguish two kinds of list. The first kind simply lists elements in the set, conventionally written by enclosing the list within curly brackets. So $\{1.7, 3.6, 2\}$ is a set that contains the three real numbers 1.7, 3.6, and 2. The order of the elements in this simple list does not matter, so $\{1.7, 2, 3.6\}$ means the same as $\{1.7, 3.6, 2\}$.

We often do care about the order of numbers, however. 1.7, 2, and 3.6 might be the values of some function at successive points, or they might be the *x*, *y*, and *z* coordinates of some point in space. The second kind of list, called a *sequence* gives the elements in a given order, and is conventionally written by enclosing the elements in ordinary braces [97]. For example, $(1.7, 3.6, 2)$ is a sequence of the real numbers 1.7, 3.6, and 2 in this order.

In functional analysis, by default, a sequence of elements is usually an infinitely long [98] list of elements in a particular order, as in some (infinitely long) sequence $(\alpha_1, \alpha_2, \alpha_3, \cdots)$ of vectors. A *subsequence* is just some of the elements of such a sequence, but retaining the relative order. So, one subsequence of $(\alpha_1, \alpha_2, \alpha_3, \cdots)$ would be $(\alpha_1, \alpha_3, \cdots)$, where we have missed out element $\alpha_2$, but have kept the order of the rest. (A subsequence is therefore also a sequence in its own right.)

Many of the proofs in functional analysis depend on convergence of sequences (or subsequences). Specifically, does a sequence converge so that, for every element $\alpha_j$ after some specific *n*th element $\alpha_n$ (so for $j > n$), each such element is closer and closer (in the sense of the metric) to some specific element $\beta$? If so, $\beta$ is the limit of the sequence. In real analysis (the analysis of numbers rather than functions), the sequence $(1, 1/2, 1/4, 1/8, 1/16, \cdots)$ converges towards 0, for example.

Another formal convergence is called Cauchy convergence, in which, essentially, the separation between elements gets closer and closer [99]. It can be proved that every convergent sequence in a metric space is such a *Cauchy sequence* of elements.

Mathematically, it can be important whether or not the limit of a convergent sequence lies within the space. So, in our example $(1, 1/2, 1/4, 1/8, 1/16, \cdots)$, it matters whether 0 is within the range of numbers allowed in the space or not. A (metric) space is said to be *complete* if every Cauchy sequence in the space converges to a limit that is also an element of the space. This completeness essentially means that we are not "missing out" specific functions from the space, and we are careful to include functions that might be at the "edges" in a mathematical sense of our space – i.e., that are the limits of sequences of functions that are otherwise within the space. These are all reasonable requirements in the physical problems of interest to us, and we presume such completeness in our spaces.

## 6.4. Hilbert spaces

With this background, we can now define a Hilbert (vector) space.

A Hilbert space is a complete inner-product space. (84)

Generally, we notate spaces using italic upper case letters such as *D*, *F*, *G*, and *R*, with subscripts to distinguish spaces if necessary, and if we use *H* (with or without a subscript) for a space, then it is certainly a Hilbert space. The mathematics of Hilbert spaces is powerful and very useful. One particularly powerful aspect is the idea of a basis set.

### 6.4.1. Orthogonal sets and basis sets in Hilbert spaces

An *orthogonal set* of elements (vectors) in a Hilbert space is a subset of the space whose (non-zero) elements are pairwise orthogonal – every member is orthogonal to every other member. So, for any two (non-zero) members $\alpha$ and $\gamma$ of this orthogonal set, $(\alpha,\gamma)=0$ unless $\alpha=\gamma$. Even more convenient is an *orthonormal set*, which is an orthogonal set in which every element is *normalized* to have a norm of 1, i.e., $(\alpha,\alpha)=\sqrt{(\alpha,\alpha)}=1$. We presume that we can index the members with an integer or natural number index *j* or *k*, for example. Then, using the Kronecker delta (Eq. (27)), for an orthonormal set

$$(\alpha_j,\alpha_k)=\delta_{jk} \tag{85}$$

A *linear combination* of vectors $\beta_1,\ldots,\beta_m$ of a vector space is an expression of the form $d_1\beta_1+d_2\beta_2+\ldots+d_m\beta_m$ for some set of scalars $\{d_1,d_2,\ldots,d_m\}$. We can choose to have a set of vectors defined as those vectors $\gamma$ that can be represented by a linear combination of the vectors in an orthonormal set $\{\alpha_1,\alpha_2,\ldots\}$, i.e., by the sum

$$\gamma=a_1\alpha_1+a_2\alpha_2+\cdots\equiv\sum_j a_j\alpha_j \tag{86}$$

We can also call such an expression the *expansion* of $\gamma$ in the basis $\alpha_j$ (i.e., in the set $\{\alpha_1,\alpha_2,\ldots\}$), and the numbers $a_j$ are called the *expansion coefficients*. A set of orthogonal (and, preferably, orthonormal) vectors that can be used to represent any vector in a space can be called an (orthogonal or orthonormal) *basis* for that space. The number of such functions in the basis – i.e., in the set $\{\alpha_1,\alpha_2,\ldots\}$ – is the *dimensionality* of the basis set and of the corresponding space. Depending on the space, this dimensionality could be finite or infinite. By definition, a basis, because it can represent any function in a given space, is also said to be a *complete set* of functions for the space [100]. The coefficient $a_j$ is easily extracted by forming the inner product with $\alpha_j$, i.e.,

$$(\alpha_j,\gamma)=a_j \tag{87}$$

Indeed, we this is the defining equation for the *expansion coefficients* $a_j$. Now we come to a key attribute of a Hilbert space (see, e.g., [36]):

There is always a basis for a Hilbert space (88)

That is, there is always some complete set of orthogonal (or orthonormal) functions $\{\alpha_1,\alpha_2,\ldots\}$ that forms a basis for any given Hilbert space. Importantly, this also applies to infinite-dimensional Hilbert spaces. Of course, if there is one basis set, then there are many possible basis sets – in fact an infinite number – because we can make them from orthogonal linear combinations of the set $\{\alpha_1,\alpha_2,\ldots\}$.

### 6.4.2. An "algebraic shift" to Dirac notation for vectors and inner products

We can now view this set of numbers $\{a_j\}$ as being the *representation* of the function $\gamma$ in the basis $\{\alpha_1, \alpha_2, \ldots\}$, and we can choose to write them as a column vector,

$$\gamma = \begin{bmatrix} a_1 \\ a_2 \\ \vdots \end{bmatrix} \equiv |\gamma\rangle \tag{89}$$

We can now use a Dirac "ket vector" notation $|\gamma\rangle$ as a short-hand for just such a column vector of such expansion coefficients and, indeed, for the function it represents. The usefulness of this notational shift to Dirac notation goes deeper, however. Consider an inner product of two functions $\eta$ and $\mu$ in a given Hilbert space. Expanding each function in this basis $\{\alpha_1, \alpha_2, \ldots\}$ gives

$$\eta = \sum_k r_k \alpha_k \text{ and } \mu = \sum_k t_k \alpha_k \tag{90}$$

where

$$r_k = (\alpha_k, \eta) \text{ and } t_k = (\alpha_k, \mu) \tag{91}$$

are inner products formed using the underlying inner product in the space. Using these expansions and the inner product "linearity" properties ((IP1) and IP2) in (80)), we have

$$(\mu, \eta) = \sum_{p,q} t_p^* r_q (\alpha_p, \alpha_q) = \sum_{p,q} t_p^* r_q \delta_{pq} = \sum_{p,q} t_p^* r_q \equiv [t_1^*, t_2^*, \ldots] \begin{bmatrix} r_1 \\ r_2 \\ \vdots \end{bmatrix} \tag{92}$$

Once we make an "algebraic shift" of regarding the ket vectors in Dirac notation $|\mu\rangle$ and $|\eta\rangle$ as being vectors of expansion coefficients constructed using the underlying inner product in the space, i.e., defining the Dirac ket vectors, using the underlying inner product, as in

$$|\eta\rangle \equiv \begin{bmatrix} (\alpha_1, \eta) \\ (\alpha_2, \eta) \\ \vdots \end{bmatrix} \text{ and } |\mu\rangle \equiv \begin{bmatrix} (\alpha_1, \mu) \\ (\alpha_2, \mu) \\ \vdots \end{bmatrix} \tag{93}$$

then the subsequent mathematics of the inner products is simply the Cartesian inner product

$$(\mu, \eta) \equiv \begin{bmatrix} (\alpha_1, \mu)^* & (\alpha_2, \mu)^* & \cdots \end{bmatrix} \begin{bmatrix} (\alpha_1, \eta) \\ (\alpha_2, \eta) \\ \vdots \end{bmatrix} \equiv \langle\mu||\eta\rangle \equiv \langle\mu|\eta\rangle \tag{94}$$

where we have now introduced the "bra" vector $\langle\mu|$ as being the row vector whose elements are the complex conjugates of the expansion coefficients, and the shorthand $\langle\mu||\eta\rangle \equiv \langle\mu|\eta\rangle$ for such an inner product. We will take a further step in this "algebraic shift" once we have defined linear operators and their matrix representation below.

## 6.5. Linear operators

In our mathematics, we also need operators. An operator mathematically turns one function into another, or, equivalently, generates a "new" function from an "old" one. It maps from functions in a "domain" space $D$ to a "range" space $R$ (which may be a different space). In our case, we are particularly interested in operators that generate a "wave" function in a receiving volume from a "source" function in a source volume, for example. We notate operators using an upper-case letter in a sans-serif font, as in $\mathsf{A}$. We can write the action of the operator $\mathsf{A}$ on any vector or function $\alpha$ in its domain $D$ to generate a vector or function $\gamma$ in its range $R$ as

$$\gamma = \mathsf{A}\alpha \tag{95}$$

### 6.5.1. Definition of linear operators

The operators of most interest to us are linear operators, defined as follows:

> For any two vectors or functions $\alpha$ and $\beta$ in its domain $D$, and any (possibly complex) scalar $c$, a *linear operator* $\mathsf{A}$ is required to have two properties:
>
> O1 $\quad \mathsf{A}(\alpha+\beta) = \mathsf{A}\alpha + \mathsf{A}\beta$
>
> O2 $\quad \mathsf{A}(c\alpha) = c\mathsf{A}\alpha$

(96)

All operators we consider will be linear. Note all matrices are linear operators in this sense.

### 6.5.2. Operator norms and bounded operators

We can now usefully introduce the idea of operator norms. We need this because it allows us to consider convergence, now of the functions in the range that result from operating on functions in the domain, and to define "boundedness". For an operator to be *bounded*, we require that, for any vector $\alpha$ in the domain $D$ and with a finite norm $\|\alpha\|_D = \sqrt{(\alpha,\alpha)_D}$ , the resulting vector $\mathsf{A}\alpha$ in the range $R$ must have a finite norm $\|\mathsf{A}\alpha\|_R \equiv \sqrt{(\mathsf{A}\alpha,\mathsf{A}\alpha)_R}$ . (Here the subscripts $D$ and $R$ are allowing for possibly different inner products in the domain and range.) With physical operators representing waves, we expect boundedness – finite inputs should give finite outputs. Formally, we can restate this as

$$\|\mathsf{A}\|_{sup} = \sup_{\substack{\alpha \text{ in } D \\ \alpha\neq 0}} \frac{\|\mathsf{A}\alpha\|_R}{\|\alpha\|_D} < \infty \tag{97}$$

The supremum ("sup") here essentially [101] means the largest value for any non-zero vector $\alpha$ in the domain $D$, so here the norm of the "largest" vector we could get in the range by operating with $\mathsf{A}$ on a normalized vector in the domain. This expression also formally defines the *supremum (operator) norm* $\|\mathsf{A}\|_{sup}$ for the operator, which can be used in various proofs. Later, we will define another operator norm (the Hilbert-Schmidt norm). Formally, the existence of such a norm implies the operator is bounded, and *vice versa*. Quite generally for operator norms, we can also write

$$\|\mathsf{A}\alpha\| \le \|\mathsf{A}\|\|\alpha\| \tag{98}$$

which is obvious from Eq. (97) for the supremum norm.

### 6.5.3. Matrix representation of linear operators and use of Dirac notation

Because any Hilbert space has some complete basis set, we can use this property and the underlying inner product to represent a linear operator as a matrix. Suppose, then, that we have two Hilbert spaces, $H_1$ and $H_2$, which may be different spaces and may have different underlying inner products. We presume orthonormal basis sets $\{\alpha_1,\alpha_2,\ldots\}$ in $H_1$ and $\{\beta_1,\beta_2,\ldots\}$ in $H_2$. Both Hilbert spaces may be infinite dimensional, and so these basis sets may also be infinite.

We presume that a bounded linear operator $\mathsf{A}_{21}$ maps from vectors in space $H_1$ to vectors in space $H_2$ – for example, mapping an vector $\eta$ in $H_1$ to some vector $\sigma$ in $H_2$

$$\sigma = \mathsf{A}_{21}\eta \tag{99}$$

Quite generally, we could construct the (underlying) inner product between this resulting vector and an arbitrary vector $\mu$ in $H_2$. Specifically, we would have

$$(\mu,\sigma)_2 \equiv (\mu, \mathsf{A}_{21}\eta)_2 \tag{100}$$

Here, since both vectors $\mu$ and $\sigma$ are in $H_2$, we are using the underlying inner product in $H_2$ and we use the subscript "2" to make this clear. Now we can define a *matrix element*, which is generally a complex number, as

$$a_{jk} = (\beta_j, \mathsf{A}_{21}\alpha_k)_2 \tag{101}$$

Now we expand the vectors $\eta$ and $\mu$ on their corresponding basis sets using the underlying inner product in each space. So, we have

$$\eta = \sum_k r_k \alpha_k \quad \text{and} \quad \mu = \sum_j t_j \beta_j \tag{102}$$

where the $r_k$ and the $t_j$ are complex numbers given by

$$r_k = (\alpha_k, \eta)_1 \quad \text{and} \quad t_j = (\beta_j, \mu)_2 \tag{103}$$

Then, we can rewrite Eq. (100) as

$$(\mu, \mathsf{A}_{21}\eta)_2 = \sum_j t_j^* \left(\beta_j, \mathsf{A}_{21}\left[\sum_k r_k \alpha_k\right]\right)_2 = \sum_{j,k} t_j^* r_k (\beta_j, \mathsf{A}_{21}\alpha_k)_2 = \sum_{j,k} t_j^* a_{jk} r_k \tag{104}$$

Equivalently, substituting back for $r_k$ and $t_j$ using (103), and noting that, therefore, by IP3, $t_j^* = (\mu, \beta_j)$ we could choose to write

$$\mathsf{A}_{21} \equiv \sum_{j,k} (\cdot, \beta_j) a_{jk} (\alpha_k, \cdot) \tag{105}$$

where we substitute the vector being operated on for the dot "$\cdot$" in each case, as in substituting $\mu$ for the dot in the left inner product and $\eta$ for the dot in the right inner product to evaluate the same result as in Eq. (104) [102].

Now we can to complete our an "algebraic shift" towards a matrix-vector algebra, written in Dirac notation. We can write the "ket" version of the vectors $\eta$ and $\mu$, though with the basis $\{\beta_1, \beta_2, \ldots\}$ for $\mu$ because it is in space $H_2$, i.e.,

$$|\eta\rangle \equiv \begin{bmatrix} r_1\left(=(\alpha_1,\eta)_1\right) \\ r_2\left(=(\alpha_2,\eta)_1\right) \\ \vdots \end{bmatrix} \quad \text{and} \quad |\mu\rangle \equiv \begin{bmatrix} t_1\left(=(\beta_1,\mu)_2\right) \\ t_2\left(=(\beta_2,\mu)_2\right) \\ \vdots \end{bmatrix} \tag{106}$$

Using Eq. (101) for the matrix elements, we can also write a matrix $\mathsf{A}_{21}$ as a matrix representation of the operator $\mathsf{A}_{21}$.

$$\mathsf{A}_{21} \equiv \begin{bmatrix} a_{11} & a_{12} & \cdots \\ a_{21} & a_{22} & \cdots \\ \vdots & \vdots & \ddots \end{bmatrix} \tag{107}$$

(Note we use the same notation for these two technically different things – the operator and its matrix representation – but the context will resolve the confusion: if we are using Dirac notation, then we are using the matrix representation of the operator.) Then the sum $\sum_{j,k} t_j^* a_{21} r_k$ can be interpreted as the vector-matrix-vector product

$$\sum_{j,k} t_j^* a_{jk} r_k \equiv \langle \mu | \mathsf{A}_{21} | \eta \rangle \tag{108}$$

This completes our "algebraic shift" to Dirac or matrix-vector notation. So, now,

we can use Dirac or matrix-vector notation as long as we presume

- The bra or ket vectors are considered to contain the expansion coefficients constructed using a basis and the underlying inner product for that space in which the corresponding function exists, as in Eq. (106)
- The operators are considered to be matrices with matrix elements as in Eq. (101), again are based on the use of the underlying inner products in the corresponding spaces.

Note in writing underlying inner products with operators, such as $(\mu,\sigma)_2 \equiv (\mu, \mathsf{A}_{21}\eta)_2$ in Eq. (100), we have only ever required the operator to operate to the "right". Indeed, for some operators, such as conventional derivative operator, for example, that may be a "legal" requirement. One particular benefit of this "algebraic shift" is that, though the actual operator may only be able to operate to the "right", in the matrix-vector/Dirac version, the matrix version of the operator can also "operate" to the left, which gives us the convenience of the usual associative laws of matrix-vector algebra. Explicitly, for example, we can "break up" $\langle\mu|\mathsf{A}_{21}|\eta\rangle$ as

$$\langle\mu|\mathsf{A}_{21}|\eta\rangle \equiv \langle\mu|\left(\mathsf{A}_{21}|\eta\rangle\right) \equiv \left(\langle\mu|\mathsf{A}_{21}\right)|\eta\rangle \tag{109}$$

So, summarizing these benefits, the use of Dirac notation in this way

- "hides" the underlying inner products for subsequent algebra, leaving just a Cartesian inner product for bra and ket vectors
- allows full use of associative laws as in matrix-vector multiplications

So, used in this way, Dirac notation can handle situations with different underlying inner products in different spaces. It gives a simple and convenient algebra for working within and between these more sophisticated Hilbert spaces [103].

We can usefully go one step further, writing the matrix $\mathsf{A}_{21}$ itself in terms of bra and ket vectors. For the moment, we will be explicit about what spaces the vector are in by using "1" and "2" subscripts on the vectors. Specifically, we write a *bilinear expansion* [104]

$$\mathsf{A}_{21} \equiv \sum_{j,k} a_{jk} \left|\beta_j\right\rangle_2 {}_1\langle\alpha_k| \tag{110}$$

This form results in the same matrix elements as in Eq. (101). Explicitly,

$$\begin{aligned} {}_2\langle\mu|\mathsf{A}_{21}|\eta\rangle_1 &\equiv {}_2\langle\mu|\left(\sum_{j,k} a_{jk}\left|\beta_j\right\rangle_2 {}_1\langle\alpha_k|\right)|\eta\rangle_1 \\ &= \sum_p t_p^* \; {}_2\langle\beta_p|\left(\sum_{j,k} a_{jk}\left|\beta_j\right\rangle_2 {}_1\langle\alpha_k|\right)\sum_q r_q\left|\alpha_q\right\rangle_1 = \sum_p t_p^* \sum_{j,k}\delta_{pj} a_{jk}\, {}_1\delta_{kq}\sum_q r_q = \sum_{j,k} t_j^* a_{jk} r_k \end{aligned} \tag{111}$$

in agreement with Eq. (104), so this approach for writing matrices works here also. Dropping the subscript notation on the vectors, instead of Eq. (110) we can just write

$$\mathsf{A}_{21} \equiv \sum_{j,k} a_{jk}\left|\beta_j\right\rangle\langle\alpha_k| \tag{112}$$

Note that a linear operator like $\mathsf{A}_{21}$ from one Hilbert space to another can be written in such an "outer-product" form as in Eq. (110) on any desired basis sets for each Hilbert space. Of course, the numbers $a_{jk}$ will be different depending on the basis sets chosen. From this point on, we will use either the "original" notation with explicit underlying inner products, or the Dirac notation, depending on which is the most convenient.

### 6.5.4. Adjoint operator

For an operator $\mathsf{A}$ that maps vectors in space $H_1$ to vectors in space $H_2$, the corresponding adjoint operator $\mathsf{A}^\dagger$ (in a notation that anticipates a key result here) is one that maps vectors from space $H_2$ to space $H_1$, and that, for any vectors $\eta$ in $H_1$ and $\mu$ in $H_2$, satisfies

$$\left(\mu, \mathsf{A}\eta\right)_2 = \left(\mathsf{A}^\dagger\mu, \eta\right)_1 \tag{113}$$

which is the defining equation for an adjoint operator. To relate this to matrices, we can consider orthonormal basis sets $\{\alpha_1, \alpha_2, \ldots\}$ in $H_1$ and $\{\beta_1, \beta_2, \ldots\}$ in $H_2$ and write the matrix elements of these two operators as

$$a_{jk} = \left(\beta_j, \mathsf{A}\alpha_k\right)_2 \text{ and } b_{kj} = \left(\alpha_k, \mathsf{A}^\dagger \beta_j\right)_1 \tag{114}$$

Then from Eqs. (113) and (114), and using the inner product property (IP3) from (80), we have

$$b_{kj} = \left(\alpha_k, \mathsf{A}^\dagger \beta_j\right)_1 = \left(\mathsf{A}^\dagger \beta_j, \alpha_k\right)_1^* = \left(\beta_j, \mathsf{A}\alpha_k\right)_2^* = a_{jk}^* \tag{115}$$

which means that, as anticipated in the notation, in matrix form, the adjoint operator $\mathsf{A}^\dagger$ is simply the matrix that is the Hermitian adjoint of the matrix version of $\mathsf{A}$. Generally we can use this superscript $\dagger$ to indicate the Hermitian adjoint operation. Note also that

$$\left(\mathsf{A}^\dagger\right)^\dagger = \mathsf{A} \tag{116}$$

### 6.5.5. Compact operators

Compact operators are a key category for our purposes. One definition is as follows:

> The operator $\mathsf{A}$ (from the normed space $F$ to the normed space $G$) is *compact* if and only if it maps every bounded sequence $(\alpha_m)$ of vectors in $F$ into a sequence in $G$ that has a convergent subsequence.

(117)

Such a mathematical definition contains just enough to allow various useful mathematical proofs, it is rather technical, and does not directly reveal what is so powerful and relevant about them. For our purposes, however, compactness is what in the end allows us to deduce that we only have finite numbers of useful channels in communication.

We can begin to see the point of these operators through an extreme example case. Consider an infinite-dimensional Hilbert space, with an orthonormal basis $\{\alpha_1, \alpha_2, \ldots\}$. Physically, we could think of these as representing orthogonal source functions, such as orthogonal current distributions, inside a source space, for example. For any two such basis vectors, the "distance" between them is defined by the metric

$$\begin{aligned} d\left(\alpha_j, \alpha_k\right) \equiv \sqrt{\left(\alpha_j - \alpha_k, \alpha_j - \alpha_k\right)} &= \sqrt{\left(\alpha_j, \alpha_j\right) + \left(\alpha_k, \alpha_k\right) - \left(\alpha_k, \alpha_j\right) - \left(\alpha_j, \alpha_k\right)} \\ &= \sqrt{1 + 1 - 0 - 0} = \sqrt{2} \end{aligned} \tag{118}$$

(This can be visualized as the distance between the "tips" of two unit vectors that are at right angles.) So, we can construct an infinite sequence that is just these basis vectors, each used exactly once, such as the sequence $(\alpha_1, \alpha_2, \ldots)$. This sequence does not converge, and has no convergent subsequences; every pair of elements in the sequence has a "distance" between them of $\sqrt{2}$ because they are all orthogonal.

A compact operator operating on that infinite (and non-converging) sequence of different basis vectors will create a sequence of vectors that will have some convergent subsequence [105]. This will ultimately lead to only finite numbers of usable channels in communication, even with infinite dimensional spaces.

### 6.5.6. Mathematical definition of Hilbert-Schmidt operators

Hilbert-Schmidt operators form a particularly important class of compact operators. This is because, as we will discuss later, the important operators we encounter for Green's functions for waves are such Hilbert-Schmidt operators, though we postpone that physical discussion. The mathematical definition of a Hilbert-Schmidt operator is as follows:

For a Hilbert space $H_1$ with an orthonormal basis $\{\alpha_1, \alpha_2, \ldots\}$ and a bounded operator $\mathsf{A}$ that maps from vectors in $H_1$ to vectors in a Hilbert space $H_2$, then $\mathsf{A}$ is a Hilbert-Schmidt operator if and only if $S = \sum_j \|\mathsf{A}\alpha_j\|^2 < \infty$ (119)

Because we use this specific sum elsewhere, we name it the *sum rule limit S*. The square root of this sum-rule limit $S$ can be called the *Hilbert-Schmidt norm* of the operator, i.e.,

$$\|\mathsf{A}\|_{HS} = \sqrt{S} \equiv \sqrt{\sum_j \|\mathsf{A}\alpha_j\|^2} \tag{120}$$

For any arbitrary complete basis sets $\{|\alpha_j\rangle\}$ and $\{|\beta_k\rangle\}$ in $H_1$, starting from this definition, we can prove three other equivalent expressions for $S$ [36].

$$\begin{aligned} S \equiv \|\mathsf{A}\|_{HS}^2 &= \sum_{j,k} |a_{kj}|^2 \\ &= \sum_j \langle \alpha_j | \mathsf{A}^\dagger \mathsf{A} | \alpha_j \rangle = \sum_k \langle \beta_k | \mathsf{A}^\dagger \mathsf{A} | \beta_k \rangle \\ &\equiv Tr(\mathsf{A}^\dagger \mathsf{A}) = Tr(\mathsf{A}\mathsf{A}^\dagger) \end{aligned} \tag{121}$$

One particularly important property, as mentioned above, is that [36]

Hilbert-Schmidt operators are compact. (122)

Other important results for our purposes are that [36]

if $\mathsf{A}$ is a Hilbert-Schmidt operator, so also are $\mathsf{A}^\dagger$, $\mathsf{A}^\dagger\mathsf{A}$ and $\mathsf{A}\mathsf{A}^\dagger$ (123)

### 6.5.7. Hermitian operators

The most general definition of a *Hermitian* or *self-adjoint* operator $\mathsf{A}$ is that, for all vectors or functions $\beta$ and $\gamma$ in the relevant Hilbert space or spaces,

$$(\beta, \mathsf{A}\gamma) = (\mathsf{A}\beta, \gamma) \tag{124}$$

If we compare this with the definition of the adjoint operator, Eq.(113), we see that this means a Hermitian operator is equal to its own adjoint,

$$\mathsf{A} = \mathsf{A}^\dagger \tag{125}$$

and for the matrix elements of the operator on some basis set(s), we have

$$a_{jk} = a_{kj}^* \tag{126}$$

Using the general mathematical relation for two operators or matrices $\mathsf{B}$ and $\mathsf{C}$, $(\mathsf{BC})^\dagger = \mathsf{C}^\dagger\mathsf{B}^\dagger$ (see Eq. (13), and easily proved in summation notations, for example), and Eq. (116), we see that both $\mathsf{A}^\dagger\mathsf{A}$ and $\mathsf{A}\mathsf{A}^\dagger$ are Hermitian, regardless of whether $\mathsf{A}$ is Hermitian. Hence, for a Hilbert-Schmidt operator $\mathsf{A}$ (which is not necessarily Hermitian), the operators $\mathsf{A}^\dagger\mathsf{A}$ and $\mathsf{A}\mathsf{A}^\dagger$ are both compact and Hermitian

(127)

This is relevant because in general the Green's function operator $\mathsf{G}$, coupling sources in one volume to generate waves in another volume, is not generally Hermitian, though it will be a Hilbert-Schmidt operator; however, the operators $\mathsf{G}^\dagger\mathsf{G}$ and $\mathsf{G}\mathsf{G}^\dagger$ will be Hilbert-Schmidt, compact, and Hermitian.

Quite generally, for an operator $\mathsf{A}$, some vector $\alpha$ is an *eigenvector* (or *eigenfunction*) of $\mathsf{A}$ if and only if, for some number $c$

$$\mathsf{A}\alpha = c\alpha \tag{128}$$

in which case *c* is the corresponding eigenvalue.

With these definitions, we can write down several important and useful properties of Hermitian operators (see [36] for derivations and proofs).

- for any Hermitian operator $\mathsf{A}$, $(\beta, \mathsf{A}\beta)$ is a real number
- all eigenvalues of Hermitian operators are necessarily real.
- for a Hermitian operator, eigenvectors for different eigenvalues are orthogonal
- if a non-zero eigenvalue of compact Hermitian operator has some number $n > 1$ of orthogonal corresponding eigenvectors (known as *degenerate eigenvectors*), then this number *n* (known as the *degeneracy* or *multiplicity*) is finite
- if a compact Hermitian operator is operating on an infinite dimensional space, then the sequence of eigenvalues $(c_p)$ must tend to zero as $p \to \infty$.

## 6.5.8. The spectral theorem for compact Hermitian operators

The *spectral theorem* is a particularly important and powerful theorem for the eigenfunctions of a compact Hermitian operator, and can be stated as follows:

> For a compact Hermitian operator $\mathsf{A}$ mapping from a Hilbert space *H* onto itself, the set of eigenfunctions $\{\beta_j\}$ of $\mathsf{A}$ is complete for describing any vector $\phi$ that can be generated by the action of the operator on an arbitrary vector $\psi$ in the space *H*, i.e., any vector $\phi = \mathsf{A}\psi$. If all the eigenvalues of $\mathsf{A}$ are non-zero, then the set $\{\alpha_j\}$ will be complete for the Hilbert space *H*; if not, then we can extend the set by Gram-Schmidt orthogonalization to form a complete set for *H*.

(129)

A consequence is that we can write any such compact Hermitian operator in terms of its eigenfunctions and corresponding eigenvalues as

$$\mathsf{A} = \sum_{j=1}^{\infty} r_j \beta_j (\beta_j, \cdot) \tag{130}$$

(where we substitute the vector being operated on for the dot "$\cdot$") or, in Dirac notation

$$\mathsf{A} = \sum_{j=1}^{\infty} r_j |\beta_j\rangle\langle\beta_j| \tag{131}$$

Here, the eigenvalues $r_j$ are whatever ones are associated with the corresponding eigenvector $\beta_j$. (Note here that for the case of degenerate eigenvalues, we presume that we have written an orthogonal set of eigenvectors for each such degenerate eigenvalue (which we are always free to do) and for indexing purposes for an *p*-fold degenerate eigenvalue, we simply repeat the eigenvalue *p* times in this sum, once for each of the corresponding orthogonal degenerate eigenvectors.)

Another important property is the following [36]:

> The eigenvectors $\beta_j$ of a compact Hermitian operator can be found by a progressive variational technique, finding the largest possible result for $\|\mathsf{A}\beta_j\|$ where $\beta_j$ is constrained to be orthogonal to all the previous eigenvectors. This will also give a corresponding set of eigenvalues $r_j$ in descending order of their magnitude.

(132)

This means, physically, that the eigenfunctions are essentially the "best" functions we can choose if we are trying to maximize performance in specific ways (such as maximizing power coupling between sources and the resulting waves), and we could even find them physically just by looking for the best such performance.

### 6.5.9. Positive operators

A *positive operator* $\mathsf{C}$ [106] is one for which, for any vector $\beta$ in relevant space(s),

$$(\beta, \mathsf{C}\beta) \geq 0 \tag{133}$$

In particular, we can prove [36] that

any operator that can be written in the form $\mathsf{C} = \mathsf{B}^\dagger \mathsf{B}$ where $\mathsf{B}$ is a linear operator, is a positive operator (134)

The property (133) automatically implies that

any eigenvalues $c$ of a positive operator are positive (non-negative), i.e., $c \geq 0$ (135)

Note then that, following from (127), (134) and (135)

if $\mathsf{A}$ is a Hilbert-Schmidt operator, then the operators $\mathsf{A}^\dagger \mathsf{A}$ and $\mathsf{A}\mathsf{A}^\dagger$ (in addition to being Hermitian Hilbert-Schmidt (compact) operators) are positive operators and therefore any eigenvalues of either of them are necessarily positive (technically, $\geq 0$) (136)

## 6.6. Inner products involving operators

We have mentioned that there can be many ways of choosing the inner product for a Hilbert space, and the choice we make may depend on the problem. Positive Hermitian operators give a particularly broad class of ways we can set up inner products.

### 6.6.1. Operator-weighted inner product

Suppose we have a positive Hermitian operator $\mathsf{W}$ that acts on functions such as $\alpha$, $\beta$, and $\gamma$, and suppose we already have defined an inner product of the form $(\beta, \alpha)$ with all the properties IP1 to IP4 as in (80). Now, the action of $\mathsf{W}$ on a vector $\gamma$ is to generate another vector $\alpha$ as in $\alpha = \mathsf{W}\gamma$. So, we can form the inner product $(\beta, \alpha) \equiv (\beta, \mathsf{W}\gamma)$. Now, from the Hermiticity of $\mathsf{W}$, we know that $(\beta, \mathsf{W}\gamma) = (\mathsf{W}\beta, \gamma)$, as in Eq. (124), and by (IP3), we know that $(\beta, \mathsf{W}\gamma) = (\mathsf{W}\gamma, \beta)^*$. So, let us define a new entity, which we could call an *operator-weighted inner product* [107],

$$(\beta, \gamma)_{\mathsf{W}} \equiv (\beta, \mathsf{W}\gamma) \tag{137}$$

Then, using first Eq. (124) and then IP3, we have

$$(\beta, \gamma)_{\mathsf{W}} = (\beta, \mathsf{W}\gamma) = (\mathsf{W}\beta, \gamma) = (\gamma, \mathsf{W}\beta)^* \equiv (\gamma, \beta)_{\mathsf{W}}^* \tag{138}$$

Hence this new entity, based on a positive Hermitian operator $\mathsf{W}$, also satisfies the property IP3 of an inner product. It is straight forward to show that, because $\mathsf{W}$ is linear, this entity also satisfies (IP1), as in

$$\begin{aligned}(\gamma, \alpha + \beta)_{\mathsf{W}} &\equiv (\gamma, \mathsf{W}(\alpha + \beta)) = (\gamma, \mathsf{W}\alpha + \mathsf{W}\beta) = (\gamma, \mathsf{W}\alpha) + (\gamma, \mathsf{W}\beta) \\ &= (\gamma, \alpha)_{\mathsf{W}} + (\gamma, \beta)_{\mathsf{W}}\end{aligned} \tag{139}$$

and (IP2), as in

$$(\gamma, a\alpha)_{\mathsf{W}} \equiv (\gamma, \mathsf{W}a\alpha) = (\gamma, a\mathsf{W}\alpha) = a(\gamma, \mathsf{W}\alpha) \equiv a(\gamma, \alpha)_{\mathsf{W}} \tag{140}$$

As for (IP4), because $\mathsf{W}$ is by choice a positive operator we already know by Eq. (133) that any entity $(\beta, \mathsf{W}\beta)$ is a positive real number, and hence $(\beta, \gamma)_{\mathsf{W}}$ satisfies (IP4). So, for any positive (linear) Hermitian operator $\mathsf{W}$, we can construct an (operator-weighted) inner product of the form given by Eq. (137).

One simple example of such an inner product is what we could call a (simple) *weighted inner product*. If we have a non-zero positive real "weighting" function $w(x)$, then we could use an expression

$$(\alpha, \beta)_w = \int w(x)\alpha^*(x)\beta(x)dx \tag{141}$$

to define a "legal" inner product. In this case, the weight function can be viewed as a diagonal, positive Hermitian operator on a "position" basis.

### 6.6.2. Transformed inner product

With a positive operator that can be written as in (134) in the form

$$\mathsf{W} = \mathsf{B}^\dagger \mathsf{B} \tag{142}$$

where $\mathsf{B}$ is a linear operator, we can take an additional step that opens another sub-class of inner products. Specifically, we could define what we could call a *transformed inner product* [107]. To do this, we first write the operator-weighted inner product with $\mathsf{W}$ as in Eq. (137), with this form (142)

$$(\beta, \gamma)_{\mathsf{W}} \equiv (\beta, \mathsf{W}\gamma) = (\beta, \mathsf{B}^\dagger \mathsf{B}\gamma) = (\mathsf{B}\beta, \mathsf{B}\gamma) \tag{143}$$

where in the last step we have used the definition of an adjoint operator as Eq. (113) and the property (116).

We can regard the operator $\mathsf{B}$ as transforming a vector $\beta$ - after all, $\mathsf{B}$ operating on $\beta$ is just a linear transform [108] acting on $\beta$ – and we could write generally

$$(\beta, \gamma)_{T\mathsf{B}} \equiv (\mathsf{B}\beta, \mathsf{B}\gamma) \tag{144}$$

where the subscript "$T\mathsf{B}$" indicates this inner product with respect to the transformation $\mathsf{B}$ of the vectors in the inner product. Because this is just a rewriting of an operator-weighted inner product, as in Eq. (143), we already know it is a "legal inner product. Note we can use any (bounded) linear operator $\mathsf{B}$ to construct such a transformed inner product because it can construct a (bounded) positive operator using Eq. (142). We will encounter just such an inner product as an energy inner product for the electromagnetic field.

## 6.7. Singular value decomposition

We have already discussed singular value decomposition (SVD) for finite matrices, where it is well known. With the mathematics of functional analysis, however, we can extend this idea more generally to any compact operators (so, even with infinite dimensional spaces). We can at the same time complete the formal proofs of the various statements, Eqs. (29) - (35), even for infinite dimensional spaces, and these statements now apply for $\mathsf{G}_{SR}$ as any compact operator. The results for finite matrices are then just a special case. Since this is now a purely mathematical argument, we defer it to **Appendix D**.

## 6.8. Physical coupling operators as Hilbert-Schmidt operators

We see that having coupling operators or Green's functions that are Hilbert-Schmidt operators opens powerful mathematical tools for infinite-dimensional spaces, as required for continuous functions. We can, however, now argue, based on physical presumptions, that the free-space Green's functions associated with wave equations quite generally are Hilbert-Schmidt operators, as also are a wide range of physical coupling operators.

We start with monochromatic scalar Green's function as in Eq. (4), now written as being between points $\mathbf{r}_S$ and $\mathbf{r}_R$ in the source and receiving volumes, $V_S$ and $V_R$, respectively.

$$G_\omega(\mathbf{r}_R;\mathbf{r}_S) = -\frac{1}{4\pi}\frac{\exp(ik|\mathbf{r}_R - \mathbf{r}_S|)}{|\mathbf{r}_R - \mathbf{r}_S|} \tag{145}$$

We follow an approach similar to [5], though using the notations and results from the general mathematics we developed above for Hilbert spaces. Presuming the source volume $V_S$ is finite, and that the source and receiver volumes are separate (i.e., not overlapping) [109, 110], with some minimum distance $r_{min}$ between them,

$$\int_{V_S} |G_\omega(\mathbf{r}_R;\mathbf{r}_S)|^2 d^3\mathbf{r}_S = \frac{1}{16\pi^2}\int_{V_S}\frac{1}{|\mathbf{r}_R - \mathbf{r}_S|^2} d^3\mathbf{r}_S \le C = \frac{V_S}{16\pi^2 r_{min}^2} \tag{146}$$

where, with these assumptions, $C$ is finite. Hence, integrating this finite quantity over a finite volume $V_R$, also gives a finite result we can call $S$, given by

$$S = \int_{V_R}\int_{V_S} |G_\omega(\mathbf{r}_R;\mathbf{r}_S)|^2 d^3\mathbf{r}_S d^3\mathbf{r}_R \le \frac{V_R V_S}{16\pi^2 r_{min}^2} \tag{147}$$

Indeed, with this specific Green's function we can even let the receiving volume $V_R$ be a spherical shell of some finite thickness $w$ but of arbitrarily large radius $r_a$ and still get a finite, limiting result for $S$, i.e., because as $r_a \to \infty$, $V_R \to 4\pi r_a^2 w$ and $r_{min} \to r_a$

$$S \to \frac{V_S w}{4\pi} \text{ as } r_a \to \infty \tag{148}$$

which we can interpret as resulting from the "inverse square" behavior characteristic of power or energy in waves.

With this finite $S$, we can examine the Hilbert-Schmidt nature of this $G_\omega$. Since by presumption our source and receiver spaces are Hilbert spaces, we have complete orthonormal basis sets in each, which we can write formally as $\{\alpha_{Sq}\} \equiv \{\alpha_{S1}, \alpha_{S2}, \cdots\}$ and $\{\alpha_{Rp}\} \equiv \{\alpha_{R1}, \alpha_{R2}, \cdots\}$, respectively. Since these are just spatial functions, we can also write them in the form $\alpha_{Sq}(\mathbf{r}_S)$ and $\alpha_{Rp}(\mathbf{r}_R)$ respectively. For the moment, we use simple Cartesian inner products in integral form for the underlying inner products i.e.,

$$(\mu_S, \eta_S) \equiv \int_{V_S} \mu_S^*(\mathbf{r}_S)\eta_S(\mathbf{r}_S) d^3\mathbf{r}_S \text{ and } (\mu_R, \eta_R) \equiv \int_{V_R} \mu_R^*(\mathbf{r}_R)\eta_R(\mathbf{r}_R) d^3\mathbf{r}_R \tag{149}$$

where $\mu_S$, $\eta_S$, $\mu_R$, and $\eta_R$ are arbitrary functions in their spaces.

Using the form Eq.(105) for expanding an operator on two basis sets, we can rewrite Eq. (147) using these basis sets [111]

$$G_\omega(\mathbf{r}_R;\mathbf{r}_S) \equiv \sum_{p,q} g_{pq}(\cdot, \alpha_{Rp})(\alpha_{Sq}, \cdot) \equiv \sum_{p,q} g_{pq}\alpha_{Rp}(\mathbf{r}_R)\alpha_{Sq}^*(\mathbf{r}_S) \tag{150}$$

Completing the inner products with basis functions $\alpha_{Ri}(\mathbf{r}_R)$ on the left and $\alpha_{Sj}(\mathbf{r}_S)$ on the right, and using the orthonormality of the respective basis sets, we obtain

$$g_{ij} = \int_{V_S}\int_{V_R} \alpha_{Ri}^*(\mathbf{r}_R) G_\omega(\mathbf{r}_R;\mathbf{r}_S)\alpha_{Sj}(\mathbf{r}_S) d^3\mathbf{r}_R d^3\mathbf{r}_S \tag{151}$$

Now, using (150), we can write

$$\left|G_\omega\left(\mathbf{r}_R;\mathbf{r}_S\right)\right|^2 \equiv \left[\sum_{p,q} g_{pq}\alpha_{Rp}\left(\mathbf{r}_R\right)\alpha_{Sq}^*\left(\mathbf{r}_S\right)\right]\left[\sum_{m,n} g_{mn}\alpha_{Rm}\left(\mathbf{r}_R\right)\alpha_{Sn}^*\left(\mathbf{r}_S\right)\right]^* \\ = \sum_{p,q,m,n} g_{pq}g_{mn}^*\alpha_{Rp}\left(\mathbf{r}_R\right)\alpha_{Rm}^*\left(\mathbf{r}_R\right)\alpha_{Sq}^*\left(\mathbf{r}_S\right)\alpha_{Sn}\left(\mathbf{r}_S\right) \tag{152}$$

So

$$\int_{V_S}\int_{V_R}\left|G_\omega\left(\mathbf{r}_R;\mathbf{r}_S\right)\right|^2 d^3\mathbf{r}_R d^3\mathbf{r}_S \\ = \sum_{p,q,m,n} g_{pq}g_{mn}^*\int_{V_R}\alpha_{Rp}\left(\mathbf{r}_R\right)\alpha_{Rm}^*\left(\mathbf{r}_m\right)d^3\mathbf{r}_R\int_{V_S}\alpha_{Sq}^*\left(\mathbf{r}_S\right)\alpha_{Sn}\left(\mathbf{r}_S\right)d^3\mathbf{r}_S \\ = \sum_{p,q,m,n} g_{pq}g_{mn}^*\delta_{pm}\delta_{qn} = \sum_{p,q}\left|g_{pq}\right|^2 \tag{153}$$

Putting this result together with Eq. (147), then, we have a key result

$$S = \int_{V_S}\int_{V_R}\left|G_\omega\left(\mathbf{r}_R;\mathbf{r}_S\right)\right|^2 d^3\mathbf{r}_R d^3\mathbf{r}_S = \sum_{p,q}\left|g_{pq}\right|^2 \tag{154}$$

Taking this result together with the definition of a Hilbert-Schmidt operator, Eq. (121), and the result (148), we therefore conclude that

> the scalar Green's function $G_\omega\left(\mathbf{r}_R;\mathbf{r}_S\right) = -\dfrac{1}{4\pi}\dfrac{\exp\left(ik\left|\mathbf{r}_R-\mathbf{r}_S\right|\right)}{\left|\mathbf{r}_R-\mathbf{r}_S\right|}$ operating from a finite source volume $V_S$ to a receiver volume $V_R$ that is either finite or is a spherical shell of arbitrarily large radius, is a Hilbert-Schmidt operator. (155)

Note in particular that, by evaluating the integral in Eq. (147)

> we can establish the sum rule $S$ without solving the SVD problem for the eigenfunctions and eigenvalues (156)

We can see from the structure of this proof that even broader classes of physical operators between source and receiver spaces with also be Hilbert-Schmidt operators. First, note that

> any finite coupling operator $D\left(\mathbf{r}_R;\mathbf{r}_S\right)$ between finite volumes $V_S$ and $V_R$ is a Hilbert-Schmidt operator (157)

To see this, we can repeat the formal integration as in Eq. (147) with $D$ instead of $G_\omega$, which will give a finite result because of the finiteness of $F$ and the finiteness of the volumes. Then we can follow through similar algebra as in Eqs. (150) to (153) to prove a result as in Eq. (154) with $D$ instead of $G_\omega$.

We can also extend the result to operator-weighted inner products, as in Eq. (137), at least if those operators are "local" – that is the action of this (bounded) (positive Hermitian) "weighting" operator $\mathsf{W}$ on a function at a point $\mathbf{r}$ in a given space only depends on the value and/or the spatial derivatives of the function at a given point, and so it can be written as $\mathsf{W}(\mathbf{r})$. This may be reasonably obvious but we give the full proof in **Appendix E**. So

> any finite coupling operator $D\left(\mathbf{r}_R;\mathbf{r}_S\right)$ between finite volumes $V_S$ and $V_R$ and for which any finite functions in the associated Hilbert spaces lead to finite operator-weighted inner products, is a Hilbert-Schmidt operator with respect to those inner products (158)

This is then a very broad category of physical coupling operators. If the operator $D(\mathbf{r}_R;\mathbf{r}_S)$ is such that its magnitude falls off as $\sim 1/r$ with distance *r* from the source, as is common in wave operators so that the energy in the wave does not grow with distance from the source, then the receiving volume can be a spherical shell of arbitrary radius, and the operator will still be a Hilbert-Schmidt operator.

Thus far, we have only formally considered scalar operators, but the generalization to operators for vector fields (such as the electromagnetic field), is straightforward, and we complete this in section 8.6 once we have introduced the necessary notations. The operators are then formally tensor or dyadic, but as long as they are bounded, they will also be Hilbert-Schmidt operators.

Generally, we see that the Hilbert-Schmidt nature of these physical operators follows immediately if the operators give finite results in the receiving volume from finite sources in the source volume and if the source and receiver volumes are both finite. With wave operators, we expect this extends even to spherical shell receiving volumes of arbitrary radius. Hence, following from these Hilbert-Schmidt properties of the coupling operators, we have a key result that we can state informally but correctly as follows:

> For all operators that give waves in one volume from sources in another, the sum of the squares of the coupling strengths between orthogonal sources in one volume and orthogonal waves in the other is a finite "sum rule" number *S* that can be calculated by integration using the operator, without otherwise solving the problem. This is true even if the orthogonal sets are infinite.

## 6.9. Diffraction operators

So far, we have discussed actual sources in one space generating waves in another. Another very common class of problems uses wavefronts as effective sources to describe diffraction and beam propagation. So, if we know the wave on one surface, then we can hope to calculate it on another. This approach goes back to Huygens [112]. We could call the operator that relates the resulting wave on a second surface to the wave amplitudes on first surface the *diffraction operator*. In the simplest "Huygens" approach, then, we have a set of effective point sources on the first surface whose density is proportional to the wave amplitude on that surface. In that case, the diffraction operator for scalar waves is just the same Green's function, Eq. (4), we have used when considering "actual" point sources, so in that approximation all our results using this Green's function carry over to diffraction problems.

The simple point-source approach to diffraction operators does have known problems; notoriously, it leads to non-existent "backwards" waves. The effective source also should not radiate uniformly in angle even in the forward direction; one solution, taken by Fresnel, introduces an *ad hoc* angle-dependent "obliquity factor" to the Green's function [68, 113], (and see [114] for a general discussion of diffraction).

Generally, diffraction operators can be constructed using such integrations of the wave equation (Kirchhoff for scalar waves (e.g., [68, 114]), and Stratton and Chu [115, 116] for electromagnetic waves). Such integrations show that the waves generated from sources inside some closed surface can be emulated outside the surface by effective sources on the surface. Though technically only valid if we consider the entire closed surface, typically we pretend that we can consider effective sources only over some finite surface corresponding to the aperture of the system. See discussions of communications modes for scalar [117] and vector [118] diffraction, and [89] for the Debye-Wolf vector wave approach to a diffraction operator.

We will not provide further details of such diffraction operators here. However, a key point is that, for the same reason that the actual Green's functions between sources and receiving points are Hilbert-Schmidt operators, all such diffraction operators will be Hilbert-Schmidt as well, as in (158) . The waves they give are finite, and the source area or volume is finite. So certainly the resulting Hilbert-Schmidt integral, as in Eq. (154) or Eq. (226), will be finite for any finite receiving volume, and also, we expect, for any spherical shell receiving volume of arbitrary radius because we also expect the resulting waves to fall off as $\sim 1/r$ or faster. Hence, we can apply the general results of our analysis also to diffraction operators, or, equivalently,

> within the limitations of diffraction operators, communications modes based on diffraction operators correctly model the channels for waves between surfaces.

## 6.10. Using the sum rule to validate practical, finite basis sets

The formal mathematics above for continuous functions generally leads to basis sets with infinite numbers of elements. In practice, we prefer finite sets that are "complete enough" for our problems. Using the sum rule of Eq. (154) (or, more generally, Eq. (226)), we can establish a simple criterion for knowing when to stop adding elements to our sets.

First, we evaluate the sum rule $S$ by performing the required Green's function or coupling operator integral over the source and receiver volumes (as in Eq. (154) or Eq. (226)). Then we choose any convenient finite sets of orthonormal functions – $N_S$ functions $\beta_{NSq}(\mathbf{r}_S)$ in the source space and $N_R$ functions $\beta_{NRp}(\mathbf{r}_R)$ in the receiving space. Then we evaluate coupling matrix elements $b_{pq}$ as in Eq. (151) or Eq. (225), but with our sets $\beta_{NSq}(\mathbf{r}_S)$ and $\beta_{NRp}(\mathbf{r}_R)$ instead of the complete basis sets $\alpha_{Sq}(\mathbf{r}_S)$ and $\alpha_{Rp}(\mathbf{r}_R)$. Then we compare

$$S_b = \sum_{p=1}^{N_R}\sum_{q=1}^{N_S}\left|b_{pq}\right|^2 \tag{159}$$

to $S$. If $S_b$ is close enough for our purposes to $S$ then we can stop adding functions to our basis. The most strongly coupled channel that we could be missing would be one with power" coupling strength $S - S_b$. Once that power coupling strength becomes too small to be useful, we can stop, and work with the finite sets we now have.

As a simple example, we can consider "uniform patch" basis functions for two surfaces, approximated by point sources at the center of the patch, as in section 5.3.5 above. We already have a hint as to how many such points we would require in each space from the heuristic results above in section 5.3.5; there we deduced desirable minimum spacings of such points (Eqs. (198) and (70)) if they are to be good approximations to uniform "patches" of sources and/or receivers. Consider, for example, the sets of 9 source and receiver points as in Fig. 6 above. We will consider these as approximating continuous lines of length $h = 9\times\lambda/2 = 4.5\lambda$, equivalent to assigning each "point" source to a linear "patch" of size (length) $\lambda/2$. As in Fig. 6, these are separated by a distance $L = 5\lambda$. In this case, the sum rule integrals become, with a scalar Green's function as in Eq.(4),

$$\begin{aligned} S &= \int_{y_R=-h/2}^{h/2}\int_{y_S=-h/2}^{h/2}\left|G_\omega\left(y_R;y_S\right)\right|^2 dy_S\,dy_R \\ &= \frac{1}{(4\pi)^2}\int_{y_R=-h/2}^{h/2}\int_{y_S=-h/2}^{h/2}\frac{1}{L^2+\left(y_S-y_R\right)^2}dy_S\,dy_R \simeq \frac{0.726}{(4\pi)^2} \end{aligned} \tag{160}$$

For a "patch" source or receiver of length $\lambda/2$, so that the source or receiver function is normalized on integrating over the patch, the linear source or receiver density in the patch should be $\sqrt{2/\lambda}$. For the equivalent point source, that entire density over a $\lambda/2$ length is concentrated now in a "point", so an effective point source amplitude of $(\lambda/2)\times\sqrt{2/\lambda} = \sqrt{\lambda/2}$, and similarly for the equivalent point receiver. So, with this normalization, we should use an expression

$$\left|g_{ij}\right| \equiv \sqrt{\lambda/2}\times\sqrt{\lambda/2}\,\frac{1}{4\pi\left(\sqrt{L^2+\left(y_{Sj}-y_{Ri}\right)^2}\right)} \tag{161}$$

where $y_{Sj}$ and $y_{Ri}$ are the 9 source and receiver vertical positions. Summing over all 9 source and receiver points gives

$$S_b = \sum_{i=1}^{9} \sum_{j=1}^{9} \left| g_{ij} \right|^2 \simeq \frac{0.726}{(4\pi)^2} \tag{162}$$

which is identical to the result in Eq. (160) within numerical error. The near exact equality here between the results in Eqs. (160) and (161) is somewhat accidental, because the point sources are already an approximation to line-segment sources. The agreement nonetheless confirms that, in this case, no additional points should be required for a useful model here, and the point sources and receivers effectively capture all the strongly coupled channels. So, some further subdivision into smaller patches, hence adding more basis functions, would essentially make no difference; it would find no further strongly coupled channels.

# 7. Communications modes and common families of functions

So far, we evaluated communications modes numerically for several representative "toy" problems, and have justified that, with point sources chosen sufficiently close, such results are valid approximations to continuous source distributions and effective sources in diffraction problems. We also know, however, that under analytic paraxial approximations, useful families of continuous functions emerge, and we consider these here.

## 7.1. Prolate spheroidal functions and relation to Hermite-Gaussian and Laguerre-Gaussian approximations

Paraxial analysis of laser resonators without considering the finite size of the mirrors [119] leads to Hermite-Gaussian functions in rectangular coordinate systems and to Laguerre-Gaussians in cylindrical coordinates. With finite mirror sizes in laser resonators [120], and in work with a diffraction operator for waves between finite apertures [33], the so-called prolate spheroidal functions emerge; with finite rectangular apertures, we obtain the linear prolate functions, and finite circular apertures lead to circular prolate spheroidal functions.

Hermite-Gaussian and Laguerre-Gaussian families are discussed extensively in connection with laser resonators (e.g., [2]), and Laguerre-Gaussians have received much recent interest because of their so-called "orbital" angular momentum (OAM) [30, 31] (see section 7.2 below). Prolate spheroidal functions are less well known, possibly because we cannot express them in simple formulas (though we can calculate them [121 - 124]), but they have some important mathematical properties.

Prolate spheroidal functions (see [33] for a discussion) arose for solving some quite unrelated problems in prolate spheroidal coordinates [125], but the linear versions became better known for being the eigenfunctions of the finite Fourier transform. (The circular version correspondingly gives the eigenfunctions of the finite Hankel transform.) They are thus of some interest in signal processing problems.

We do not have space here to discuss the mathematics of these functions in detail, but we can show the relation to our approach. First, generally, these functions can emerge in scalar communications mode analysis if we take a paraxial approximation, which means (i) that we approximate the distance $|\mathbf{r}_S - \mathbf{r}_R|$ in the denominator of the Green's function (Eq. (4)) by $L$, the separation between the surfaces, and (ii) in the complex exponential $\exp(ik|\mathbf{r}_R - \mathbf{r}_S|)$ we approximate

$$|\mathbf{r}_S - \mathbf{r}_R| \simeq L\left(1 + \frac{(x_R - x_S)^2}{2L} + \frac{(y_R - y_S)^2}{2L}\right) \tag{163}$$

Then, for rectangular source and receiver apertures of equal size, for example, linear prolate spheroidal functions (with confocal phase curvatures) emerge as the communications mode functions in each aperture [5, 126]. Notably, the eigenvalues (or singular values) clearly show the same "paraxial degeneracy" we saw in our numerical examples above – nearly constant up to the paraxial heuristic number, and then falling off rapidly. Similarly, for circular apertures, the circular prolate spheroidal functions are the communications mode functions in each (circular) aperture.

Hermite-Gaussians and Laguerre-Gaussians are only approximate communications modes functions to the extent that they are approximations to the corresponding prolate spheroidal functions. Once significant field amplitudes approach the boundaries of the rectangular or circular apertures, the corresponding prolate spheroidal functions correctly incorporate the effects of those boundaries, but the Hermite-Gaussians and Laguerre-Gaussians do not. Hermite-Gaussians and Laguerre-Gaussians can be derived as solutions to differential equations under conditions without boundaries [119], but they do not usefully emerge as solutions to our integral equation approach [127]. The absence of boundaries can also lead to apparently infinite sets of communications modes with finite coupling strengths, but this is unphysical, and is cut off when the problem is solved correctly with boundaries on the volumes or surfaces. A failure to understand this point can lead to confusion on available channels, especially for Laguerre-Gaussians as OAM beams.

## 7.2. Orbital angular momentum beams and degrees of freedom in communications

Given the considerable recent interest in OAM beams and modes [30, 31], it is important here to make three points explicitly. In systems in which positive and negative angular momentum "modes" are degenerate – that is, they have the same coupling strength (which is usually the case in systems without some explicit helicity or nonreciprocity), we can make the simple statement that

> the use of orbital angular momentum does not increase the number of usable channels beyond those already available without the use of orbital angular momentum

and, indeed

> we can obtain the same number of usable channels when using beams of zero orbital angular momentum

The proof of these statements is straightforward [128]. Furthermore, for reasons associated with the Hilbert-Schmidt nature of the Green's functions in wave problems,

> the use of orbital angular momentum or any other form of spatial multiplexing does not allow for infinite numbers of usable channels in communicating between finite volumes

A communications mode analysis removes any such confusion about the available modes. See also these critiques of OAM modes and comparisons to other approaches [18, 22, 23, 129 - 136].

## 7.3. Paraxial degeneracy, sets of functions, and Fourier optics

When we have paraxial (approximate) degeneracy, up to the paraxial heuristic number $N_P$ we have some reasonable flexibility in the choice of our sets of communications mode functions. If this paraxial degeneracy were perfect, mathematically any orthogonal linear combination of those degenerate functions would be an equally good choice, at least for the first $N_P$ functions. If the degeneracy is not perfect, then such orthogonal linear combinations of source functions will lead to receiver functions that are in general not quite orthogonal. Then we would have some crosstalk, but this would be a matter of degree, and we would therefore have some practical freedom.

This paraxial (approximate) degeneracy also allows us to connect to other descriptions of optics, especially as $N_P$ becomes large, as it may be in conventional optical systems. For example, we can ask, in an approximately paraxial system like Fig. 9 or Fig. 15 (d), for the sets of source functions that would generate sinusoidal-shaped transverse field patterns (with confocal curvatures), with integer numbers of half periods fitting within the receiver surface. If we do that, we would find numerically that, up to some number somewhat smaller than the appropriate paraxial heuristic number (in each direction for Fig. 15(d)), we would be able to create such sinusoidal patterns. The mathematics of the problem also looks numerically similar to a Fourier transform between the source surface and the receiving surface. Hence, we can find a correspondence to the common Fourier optics approach [114] to optical systems, at least as the paraxial systems become large. This is only approximate, of course, and plane waves or simple interfering combinations of them are not generally the true communications modes of the system, but this paraxial degeneracy allows us to link approximately to those kinds of descriptions.

# 8. Extending to electromagnetic waves

So far in our explicit simulations and discussions we have considered just the simplest case of scalar waves and the corresponding Green's functions, and a simple "Cartesian" inner product to enable the results of functional analysis. For electromagnetic waves, however, we need to go beyond this, both in the way we describe the wave and in the form of the inner product we need in some situations. Fortunately, we can derive relatively simple results for both the Green's function and an energy inner product. In **Appendix F**, we derive these results in detail for a uniform isotropic medium. Here we summarize key results.

## 8.1. How many independent fields?

A first question is how many effective "fields" do we need to count? We know that electromagnetism can be described using the electric field **E** and the magnetic field **B**. Each of these is a vector with three components at each point in space, so naively we might think we need 6 scalar fields to describe them. Of course, Maxwell's equations relate **E** and **B** to each other and to the charge density $\rho$ and the current density **J**, and $\rho$ and **J** are themselves related by conservation of charge. One way of reducing the number of scalar fields is to change to a description in terms of a (magnetic) vector potential **A** and a scalar potential $\Phi$, thereby reducing to 4 scalar fields. But there is still arbitrariness here, which formally appears as the freedom to choose the "gauge" or, equivalently, set $\nabla \cdot \mathbf{A}$ .

In communications problems, we are not interested in any static fields; communicating information requires changing fields. In **Appendix F**, we show that, if we separate out any static fields, then we can set up a new gauge – the M-gauge.

This new M-gauge allows us to express the (changing) electromagnetic field in terms only of the vector potential in this gauge, $\mathbf{A}_M$ , and using only the current density **J** as the source of the fields.

So, we can conclude that

> we only need consider 3 independent scalar fields in counting modes and "degrees of freedom" of the electromagnetic field for communication problems.

For such a gauge, if we know $\mathbf{A}_M$ , we can get back to the electric and magnetic fields (neglecting any static fields) using

$$\mathbf{E}_M = -\frac{\partial \mathbf{A}_M}{\partial t} \tag{164}$$

$$\mathbf{B}_M = \nabla \times \mathbf{A}_M \tag{165}$$

## 8.2. A vector wave equation for electromagnetic fields

In this M-gauge, we find we can write a wave equation for **A**, driven by the current density distribution **J**. For monochromatic waves at (angular) frequency $\omega$, we find

$$\nabla\times\nabla\times\mathbf{A}_{\omega M}-k^2\mathbf{A}_{\omega M}=\mu\mathbf{J}_\omega \tag{166}$$

with

$$k^2=\omega^2\varepsilon\mu\equiv\omega^2/v^2 \tag{167}$$

with dielectric constant $\varepsilon$, magnetic permeability $\mu$, wave (phase) velocity

$$v=\sqrt{1/\varepsilon\mu} \tag{168}$$

and the monochromatic driving current density

$$\mathbf{J}(\mathbf{r},t)=\mathbf{J}_\omega(\mathbf{r})\exp(-i\omega t)+c.c. \tag{169}$$

This wave equation uses the $\nabla\times\nabla\times$ operator rather than the $\nabla^2$ of the scalar wave equation, but this allows just one wave equation for the entire vector field, and driven just by the current density. (Full time-dependent versions are given in **Appendix F**.)

## 8.3. Green's functions for electromagnetic waves

The resulting Green's function for Eq. (166) is more complicated than the scalar Green's function for two reasons. First, needs to embody the necessary vector attributes. Since in general the resulting wave vector field at a point may not be in the same direction as the vector current density at some other point that generates the wave, the Green's function has a corresponding "dyadic" character (signified by a "double line" notation (as in Eq. (170) below). (Dyadics and their algebra are introduced and described in **Appendix H**.) Second, though the resulting Green's function has far-field "propagating" parts that fall off as $\sim 1/R$ (where $R$ is the distance from the source), just like the scalar Green's function, it has additional near field parts falling off as $\sim 1/R^2$ and $\sim 1/R^3$. The full dyadic Green's function with all these terms is derived in **Appendix F**, for both monochromatic and full time-dependent cases.

In communications we are likely most interested in the propagating ("$\sim 1/R$") waves. If $\mathbf{R}=\mathbf{r}-\mathbf{r}'$ is the vector separation between a source point $\mathbf{r}'$ and the point of interest **r**, (so these points are separated by a distance $R$), then in this M-gauge, the far-field (propagating) monochromatic dyadic Green's function for the vector potential is

$$\overline{\overline{G}}_{\omega MP}(\mathbf{r};\mathbf{r}')=-(\hat{\mathbf{e}}_1\hat{\mathbf{e}}_1+\hat{\mathbf{e}}_2\hat{\mathbf{e}}_2)G_\omega(\mathbf{r};\mathbf{r}') \tag{170}$$

where $G_\omega(\mathbf{r};\mathbf{r}')$ is the scalar Green's function as before Eq. (4), and the unit vectors $\hat{\mathbf{e}}_1$ and $\hat{\mathbf{e}}_2$ are perpendicular to each other and to **R** (Fig. 26). (We are otherwise free to choose the directions of $\hat{\mathbf{e}}_1$ and $\hat{\mathbf{e}}_2$.)

We have had to use a dyadic notation in Eq. (170). Briefly, a "dyad" such as $\hat{\mathbf{e}}_a\hat{\mathbf{e}}_b$ can be viewed as a pair of vectors, one, here $\hat{\mathbf{e}}_a$, "waiting to be operated on" by a vector from the left, and the other ($\hat{\mathbf{e}}_b$) similarly "waiting to operate" on by a vector on the right. (See **Appendix H** for an extended discussion of dyads.) So, for example, we could operate mathematically on the left and right of $\hat{\mathbf{e}}_a\hat{\mathbf{e}}_b$ using the vector dot products with, say, a vector potential **A** and a current density **J** to obtain

$$\mathbf{A}\cdot\hat{\mathbf{e}}_a\hat{\mathbf{e}}_b\cdot\mathbf{J}\equiv(\mathbf{A}\cdot\hat{\mathbf{e}}_a)(\hat{\mathbf{e}}_b\cdot\mathbf{J})\equiv A_aJ_b \tag{171}$$

which is the product of the component $J_b$ of **J** in the $\hat{\mathbf{e}}_b$ direction and the component $A_a$ of **A** in the $\hat{\mathbf{e}}_a$ direction.

The specific dyads $\hat{\mathbf{e}}_1\hat{\mathbf{e}}_1$ and $\hat{\mathbf{e}}_2\hat{\mathbf{e}}_2$ have a simple physical meaning in Eq. (170): the resulting propagating vector potential field at $\mathbf{r}$ has two polarization components (see Fig. 26), in perpendicular directions $\hat{\mathbf{e}}_1$ and $\hat{\mathbf{e}}_2$, that are transverse to the "direction of propagation" (the vector $\mathbf{R}$), and that are each driven by the corresponding component of the current density at the "point" source at $\mathbf{r}'$. (Note that the electric field $\mathbf{E}_M$, from Eq. (164), is parallel to $\mathbf{A}_M$, so $\mathbf{E}_M$ and $\mathbf{A}_M$ have the same polarization directions.)

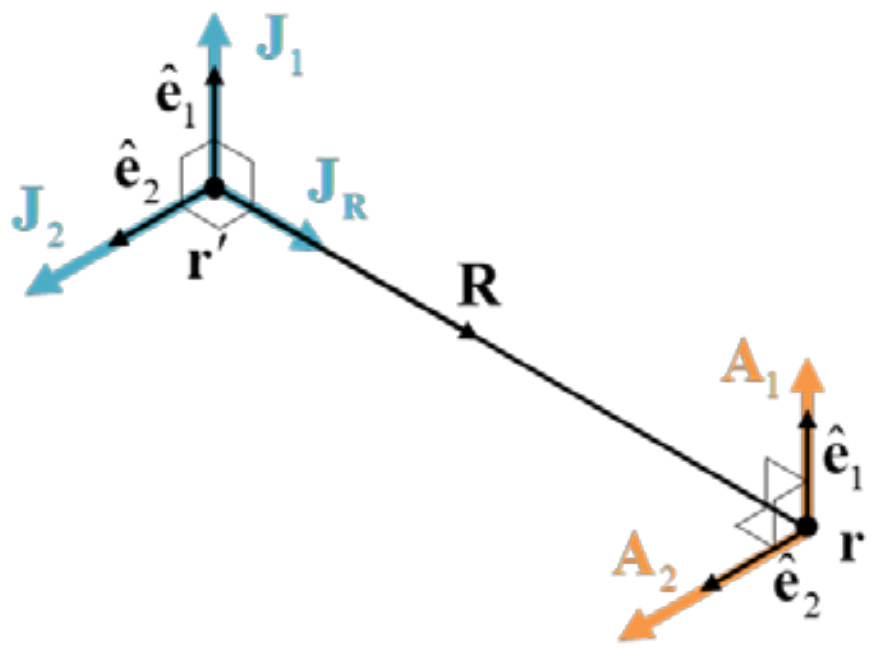

Fig. 26. Illustration of current density elements $\mathbf{J}_1$ and $\mathbf{J}_2$ at $\mathbf{r}'$, in two directions $\hat{\mathbf{e}}_1$ and $\hat{\mathbf{e}}_2$, both perpendicular to each other and to the vector $\mathbf{R} = \mathbf{r} - \mathbf{r}'$, generating corresponding vector potential components $\mathbf{A}_1$ and $\mathbf{A}_2$ in those same directions at $\mathbf{r}$. Note that, for propagating waves, a current density $\mathbf{J}_\mathbf{R}$ in the direction of $\mathbf{R}$ does not generate a corresponding "longitudinal" vector potential component in that direction at $\mathbf{r}$, though near-field ("non-propagating") terms can do so.

This might seem almost a trivially obvious answer from our normal understanding of polarization in optics, but note that we have constructed this limiting case from a novel full treatment of the electromagnetic field, and a rigorous counting of the independent field components.

We also have rigorous results that extend beyond this limiting case. The full monochromatic Green's function is the sum

$$\overline{\overline{G}}_{\omega M} = \overline{\overline{G}}_{\omega MP} + \overline{\overline{G}}_{\omega MN} \tag{172}$$

of this "propagating" Green's function $\overline{\overline{G}}_{\omega MP}(\mathbf{r};\mathbf{r}')$ and the near-field ("non-propagating") Green's function

$$\overline{\overline{G}}_{\omega MN} = \left[\frac{1}{kR}\left(i - \frac{1}{kR}\right)\left(2\hat{\mathbf{R}}\hat{\mathbf{R}} - \hat{\mathbf{e}}_1\hat{\mathbf{e}}_1 - \hat{\mathbf{e}}_2\hat{\mathbf{e}}_2\right)\right] G_\omega(R) \tag{173}$$

which also gives "longitudinal" fields – i.e., vector components that are not perpendicular to $\mathbf{R}$ (or the corresponding unit length vector $\hat{\mathbf{R}}$). The full dyadic time-dependent Green's functions are also give below in Eqs. (289), (290), and (291).

In a paraxial approximation, we might also approximate the vector directions $\hat{\mathbf{e}}_1$, $\hat{\mathbf{e}}_2$, and $\hat{\mathbf{R}}$, which generally depend on the choices of the points $\mathbf{r}'$ in the source plane or volume and $\mathbf{r}$ in the receiving plane or volume, with the fixed coordinate directions $\hat{\mathbf{x}}$, $\hat{\mathbf{y}}$, and $\hat{\mathbf{z}}$ respectively. In that case, Eq. (170) would separate into two scalar Green's functions, one for the $\hat{\mathbf{x}}$-polarized field, driven by the $\hat{\mathbf{x}}$ component ($J_x$) of $\mathbf{J}$, and the second for the $\hat{\mathbf{y}}$-polarized field, driven by the $\hat{\mathbf{y}}$ component ($J_y$) of $\mathbf{J}$, for two planes or volumes separated in the $\hat{\mathbf{z}}$ direction. Generally, however, we do not have to make that paraxial approximation, and we should note that these coordinate directions $\hat{\mathbf{e}}_1$ and $\hat{\mathbf{e}}_2$ change depending on $\mathbf{r}'$ and $\mathbf{r}$.

Finally, we can note for the specific case of monochromatic fields at (angular) frequency $\omega$, for the amplitudes $\mathbf{A}_{\omega M}(\mathbf{r})$ and $\mathbf{E}_{\omega M}(\mathbf{r})$ of the "positive frequency" (i.e., $\propto \exp(-i\omega t)$) parts of the vector potential and the electric field in a standard complex notation (see Eqs. (299) and (305) below), then Eq. (164) becomes

$$\mathbf{E}_{\omega M} = i\omega\mathbf{A}_{\omega M} \tag{174}$$

So, within this constant factor $i\omega$, the electric field and the magnetic vector potential are identical, and conceptually we can work just with the electric field, deducing the magnetic field if necessary just using the monochromatic version of the corresponding Maxwell equation (see below, Eq. (232)), or, equivalently, from Eq. (165), with $\mathbf{B}_{\omega M}$ as in Eq. (300)

$$\mathbf{B}_{\omega M} = (1/i\omega)\nabla\times\mathbf{E}_{\omega M} \tag{175}$$

For full time-dependent cases, as for pulsed fields, however, we should continue working with the magnetic vector potential to derive the full electromagnetic behavior.

## 8.4. Inner products for electromagnetic quantities and fields

### 8.4.1. Cartesian inner product for sets of sources or receivers

We could have some set of $N_S$ sources, driven by some (complex) mathematical vector $|\psi_S\rangle$ of $N_S$ amplitudes, such as voltages or currents from some output amplifiers. Similarly we could have some set of $N_R$ receivers that give a (complex) mathematical vector of received amplitudes $|\phi_R\rangle$ with $N_R$ elements, which might also be voltages or currents. As long as the generated wave is linear in the source amplitudes and the received signals are linear in the wave amplitude, we can choose to perform simple Cartesian inner products in the space of source (mathematical) vectors $|\psi_S\rangle$ and in the space of received (mathematical) vectors $|\phi_R\rangle$. In this case, we are essentially defining our ideas of orthogonality in these mathematical spaces before transmission and after reception. It makes no difference in principle to such inner products if the sources have particular vector directions and the receivers detect waves with particular vector directions. Of course, such vector aspects will come into the matrix of coupling coefficients between sources and receivers. Such simple Cartesian inner products are likely to be useful in situations, as in acoustics or radio-frequency electromagnetics, where we have corresponding separate source and receiver elements like sets of loudspeakers, microphones, and antennas.

### 8.4.2. Cartesian inner product for vector fields

We can extend the idea of a simple Cartesian inner product, as in Eq. (149), to vector fields using the ordinary dot product between the vectors. For two vector fields $\mu(\mathbf{r})$ and $\eta(\mathbf{r})$ in some volume *V*, we could write such a Cartesian (vector) inner product as

$$(\mu,\eta) \equiv \int_V \mu^*(\mathbf{r})\cdot\eta(\mathbf{r})\,d^3\mathbf{r} \tag{176}$$

which is valid because it would satisfy all the mathematical criteria as in Eq. (80).

Just what inner product we want to use in a given situation might depend on the physical system in a given space. For example, for some currents out of a set of amplifiers each driving wires as radiators, we might use such a simple Cartesian inner product of the currents (or current densities) to define orthogonality between different overall "transmitting" outputs, so such a Cartesian inner product like Eq. (176) might be appropriate for current density **J** in a source volume.

### 8.4.3. Electromagnetic mode example

We show an electromagnetic example in Fig. 27 (based on the Cartesian inner products of both source and receiver amplitude vectors). Here we presume point sources and receivers, spaced as in the earlier scalar wave example in Fig. 6; here, though, these sources contain three (geometrically) orthogonal vector current sources, each of which can be set separately, and similarly each receiver detects field amplitude (vector potential, or more realistically, electric field) separately in three orthogonal directions. Now, instead of 9 mathematical sources and receivers in this example, we have 27 of each, in vector groups of 3 at each point. Using the dyadic Green's function, Eq. (172), we now construct a $27 \times 27$ matrix coupling these sources and receivers, and perform the singular value decomposition to establish the current source modes, which are now 27-element vectors. In Fig. 27, we show the mode coupling strengths (the modulus squared of the singular values) in decreasing order for all 27 of the resulting modes.

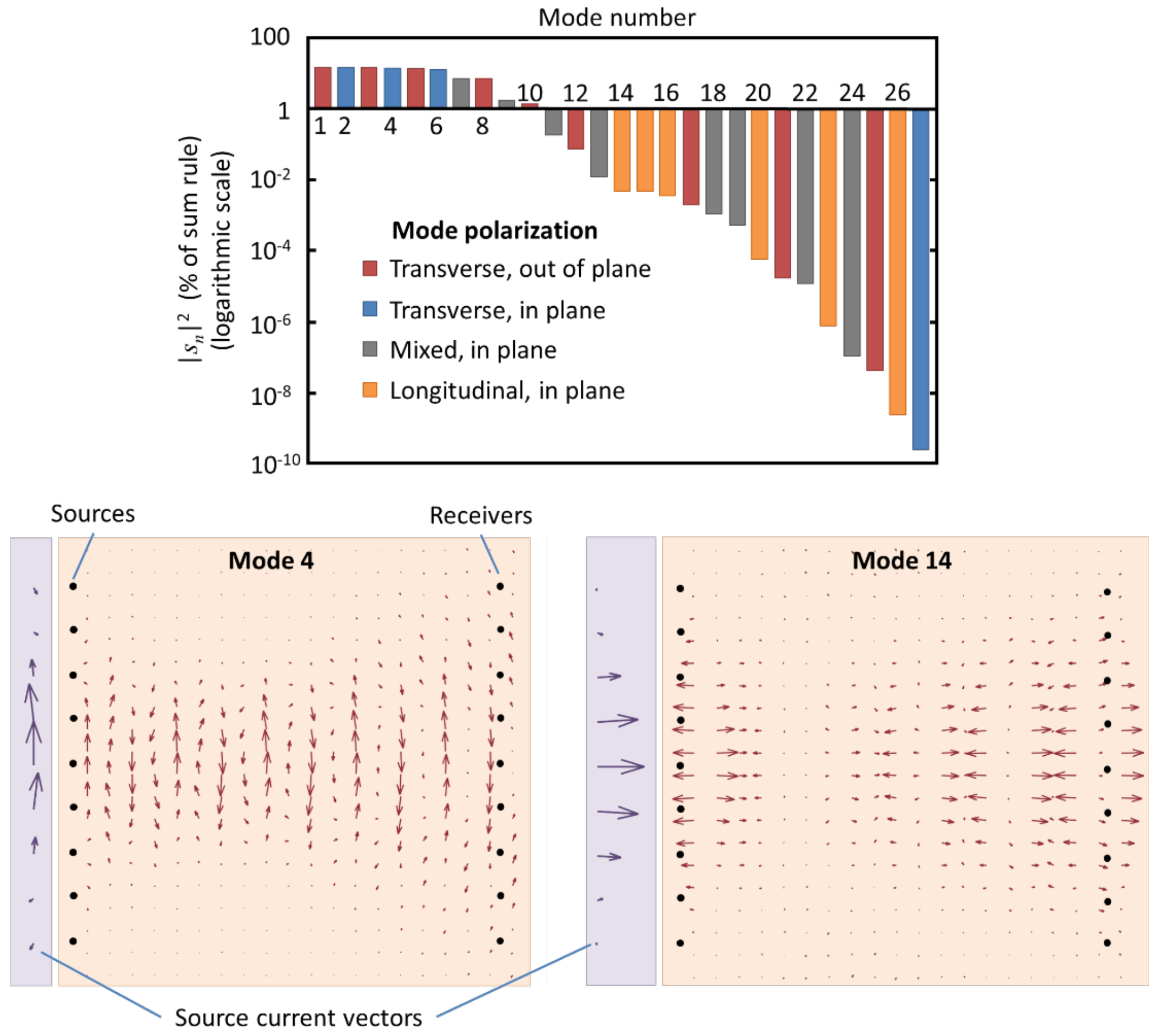

Fig. 27. Mode coupling strengths $|s_n|^2$ and vector source current and (vector potential) fields for two example modes for an electromagnetic system with lines of point vector sources and receivers, spaced vertically by $\lambda/2$ in lines $5\lambda$ apart, as in the earlier scalar wave example of Fig. 6. Mode 4 is a well-coupled mode whose polarization, both in the current sources and the resulting waves is in the plane of the paper and is also substantially tranverse to the propagation direction from left to right. The vector field plots and the source current vectors show only the real part of the complex values, so are essentially "snapshots" of the current and field vectors. Some of the modes have source and wave polarization entirely in the direction out of the plane, and transverse to the propagation direction in this plane (red bars). All other modes have polarizations of currents and waves in the plane. Some of those are substantially transverse (blue bars); others of those are mixed between transverse and longitudinal (i.e., with polarization in the horizontal $z$ direction) (grey bars). Some are almost entirely longitudinal (orange bars), with Mode 14 being the strongest of these longitudinal modes. For graphic clarity, the amplitude of the wave is multiplied by a factor depending on the horizontal distance $z$ from the sources. For Mode 4, this factor is $\sqrt{z}$ to compensate for the expansion of the wave in the directions out of the plane. For Mode 14, we also need to compensate for the additional "$1/R$" fall off because the longitudinal wave Green's function falls off at least as fast as $1/R^2$. To prevent the particularly large near-field amplitudes from dominating the drawn vector lengths, the factor used is $(z-0.2\lambda)^{3/2}$ rather than just $z^{3/2}$. Though the amplitude changes from left to right are therefore artificial, the relative amplitudes within a vertical column of vectors, and the directions of the vectors, are, however, correct.

Because of the geometry of this example, there are 9 resulting modes for which the sources and receivers are entirely polarized out of the plane, and the wave field in the cross-section that is the plane of the paper is also polarized this way. The polarization for the sources and receivers of the remaining 18 modes lies entirely in the plane, as does the wave field polarization in the plane plotted. The 9 "out of plane" modes (red bars) have behavior quite similar to the scalar results in Fig. 6, with 3 or 4 strongly coupled modes (here Modes 1, 3, 5,and 8). The first 3 or 4 other strongly coupled modes are substantially (Modes 2, 4, and 6 (blue bars)) or mostly (Mode 7 (grey bar)) transverse in the plane. The vector field plot for Mode 4 is shown in Fig. 27; this is a "single bump" beam. We can reasonably interpret these results as showing that, using propagating vector waves, we have twice as many well-coupled channels as in the scalar case because we can now use two polarizations.

The various other modes here are not strongly coupled, and hence might be practically uninteresting. Various other of the in-plane modes have a somewhat mixed character, with some significant longitudinal components in the polarization. It is, however, worth noting that we do also acquire additional modes strongly associated with the longitudinal polarization. There is a group of 3 of these (Modes 14, 15, and 16), with approximately equal coupling strength; Mode 14 is plotted, which shows dominant longitudinal polarization and a "single bump" character. Modes 15 and 16 have two- and three-bump character. These three modes can be seen as longitudinal analogs of the sets of transverse polarized and strongly coupled modes. They are weakly coupled because they result from terms in the Green's function that fall off as $1/R^2$ or faster; however, we see that longitudinally polarized electromagnetic waves show diffraction behavior similar to that of transverse waves.

We can understand this because, though their wave amplitude falls off faster with $R$, the wave still incorporates the underlying scalar Green's function that determines the interference behavior underlying diffraction. See [118] for other vector field examples.

### 8.4.4. Energy inner product for the electromagnetic field

We might be passing the wave in a receiving volume into some lossless optical network to separate out the power or energy to different output ports. Then we might want the inner product to define orthogonality so that adding up the energy or power in the different orthogonal channels or ports gave the total energy or power in the field. In general, for an electromagnetic field, a simple Cartesian inner product as in Eq. (176) does not separate out orthogonal parts of the field whose energies we can add to get the total, and this inner product $(\mathbf{A}_M, \mathbf{A}_M)$ does not reliably give a measure of the energy in the field $\mathbf{A}_M$.

In such a case, when considering inner products and orthogonality directly in the field itself, not in the sources or receivers, we would like an "energy" inner product; specifically, we want the property that the sum of the energies in the orthogonal components of the field is the total energy of the field. Such an energy inner product is also useful for quantizing the electromagnetic field (section 9). Using the M-gauge, we derive an energy inner product for the electromagnetic field in Appendix F, which we summarize here.

Since the entire properties of the electromagnetic field (at least for communications) can be described using the vector potential in the M-gauge, we set up the inner product using that. Briefly, we use an operator that we call $\sqrt{\mathsf{U}}$ to mathematically generate the $\mathbf{E}_M$ and $\mathbf{B}_M$ electric and magnetic fields from the vector potential $\mathbf{A}_M$. Then we can formally calculate the energy density $u$ in the (time-varying) field based on a standard expression (see Eq. (295) and associated discussion)

$$u = \frac{1}{2}\left(\varepsilon \mathbf{E}\cdot\mathbf{E} + \frac{1}{\mu}\mathbf{B}\cdot\mathbf{B}\right) \tag{177}$$

The transformed inner product (Eqs. (143) and (144)) with respect to $\sqrt{\mathsf{U}}$ is formally

$$(\boldsymbol{\mu},\boldsymbol{\eta})_{T\sqrt{\mathsf{U}}} \equiv \left(\sqrt{\mathsf{U}}\boldsymbol{\mu},\sqrt{\mathsf{U}}\boldsymbol{\eta}\right) \tag{178}$$

where now $\mu(\mathbf{r})$ and $\eta(\mathbf{r})$ are vector potential fields (e.g., in the "receiving" volume). We derive the full time-dependent form of (178) below (see Eqs. (310) to (313)). From this we can deduce the simpler monochromatic case, for which we have the operator

$$\sqrt{\mathsf{U}_\omega} = \begin{bmatrix} i\omega\sqrt{\varepsilon} & 0 & 0 \\ 0 & i\omega\sqrt{\varepsilon} & 0 \\ 0 & 0 & i\omega\sqrt{\varepsilon} \\ 0 & \frac{-1}{\sqrt{\mu}}\frac{\partial}{\partial x_3} & \frac{1}{\sqrt{\mu}}\frac{\partial}{\partial x_2} \\ \frac{1}{\sqrt{\mu}}\frac{\partial}{\partial x_3} & 0 & \frac{-1}{\sqrt{\mu}}\frac{\partial}{\partial x_1} \\ \frac{-1}{\sqrt{\mu}}\frac{\partial}{\partial x_2} & \frac{1}{\sqrt{\mu}}\frac{\partial}{\partial x_1} & 0 \end{bmatrix} \tag{179}$$

(using dashed lines to separate the matrix elements). So, with three orthogonal coordinate directions $\hat{\mathbf{x}}_1$, $\hat{\mathbf{x}}_2$, and $\hat{\mathbf{x}}_3$, and writing $\mathbf{A}_{\omega M}(\mathbf{r}) = A_{\omega M1}(\mathbf{r})\hat{\mathbf{x}}_1 + A_{\omega M2}(\mathbf{r})\hat{\mathbf{x}}_2 + A_{\omega M3}(\mathbf{r})\hat{\mathbf{x}}_3$

$$\sqrt{\mathsf{U}_\omega}\mathbf{A}_{\omega M}(\mathbf{r}) \equiv \sqrt{\mathsf{U}_\omega}\begin{bmatrix} A_{\omega M1}(\mathbf{r}) \\ A_{\omega M2}(\mathbf{r}) \\ A_{\omega M3}(\mathbf{r}) \end{bmatrix} \equiv \begin{bmatrix} \sqrt{\varepsilon}E_{\omega M1}(\mathbf{r}) \\ \sqrt{\varepsilon}E_{\omega M2}(\mathbf{r}) \\ \sqrt{\varepsilon}E_{\omega M3}(\mathbf{r}) \\ \left(1/\sqrt{\mu}\right)B_{\omega M1}(\mathbf{r}) \\ \left(1/\sqrt{\mu}\right)B_{\omega M2}(\mathbf{r}) \\ \left(1/\sqrt{\mu}\right)B_{\omega M3}(\mathbf{r}) \end{bmatrix} \equiv \left[\sqrt{\mathsf{U}_\omega}\mathbf{A}_{\omega M}(\mathbf{r})\right] \tag{180}$$

where by the notation $\left[\sqrt{\mathsf{U}_\omega}\mathbf{A}_{\omega M}(\mathbf{r})\right]$ we mean the 6-element column vector version. Note here that have explicitly clarified that this operator essentially generates the electric and magnetic field components, appropriately weighted for constructing the total energy density. (The factor ½ in Eq. (177) disappears in Eq. (180) because of the way we define monochromatic field amplitudes.) To complete the construction of the inner product as in Eq. (178), we formally construct the 6-element mathematical column vectors  and $\left[\sqrt{\mathsf{U}_\omega}\boldsymbol{\eta}(\mathbf{r})\right]$, which are both functions of space. Then we take the Hermitian adjoint $\left[\sqrt{\mathsf{U}_\omega}\boldsymbol{\mu}(\mathbf{r})\right]^\dagger$, which is a 6-element row vector, and form the product $\left[\sqrt{\mathsf{U}_\omega}\boldsymbol{\mu}(\mathbf{r})\right]^\dagger\left[\sqrt{\mathsf{U}_\omega}\boldsymbol{\eta}(\mathbf{r})\right]$, which is a scalar function of $\mathbf{r}$. Then we integrate that over the volume $V$ of interest (e.g., the receiving volume), as in

$$(\boldsymbol{\mu},\boldsymbol{\eta})_{T\sqrt{\mathsf{U}_\omega}} \equiv \int_V \left[\sqrt{\mathsf{U}_\omega}\boldsymbol{\mu}(\mathbf{r})\right]^\dagger\left[\sqrt{\mathsf{U}_\omega}\boldsymbol{\eta}(\mathbf{r})\right]d^3\mathbf{r} \tag{181}$$

For a monochromatic vector potential $\mathbf{A}_{\omega M}(\mathbf{r})$, the total energy of the field in $V$ is then

$$U = \left(\mathbf{A}_{\omega M},\mathbf{A}_{\omega M}\right)_{T\sqrt{\mathsf{U}_\omega}} \tag{182}$$

## 8.5. Energy-orthogonal modes for arbitrary volumes

Typically, in considering "modes" of the electromagnetic field in free space, we pretend we have some resonator, and we presume that its resonant modes will be orthogonal in some suitable sense. For free-space, for example, it is common to pretend we have some cuboidal box, with perfectly reflecting walls. This box will therefore have standing plane wave resonant oscillation modes, which we can check satisfy Maxwell's equations, and which we can then count. Possibly, we will take some limit as we make the box large, to get some density of modes per unit box volume. We may also just presume that we can using traveling waves instead of the standing ones, possibly with periodic boundary conditions (though those would not be resonant modes). Generally, the justification for such choices is weak; we do not have such a box, nor is free space periodic. A rationalization is that these fictitious approaches give results that ultimately seem to agree with experiments, such as for the thermal radiation of light fields in arguments like Planck's radiation law and Einstein's A&B coefficients.

Now that we have an energy inner product, in the monochromatic (Eq. (182)) or full time-dependent (Eq. (314)) version, we can directly define electromagnetic fields that are orthogonal in energy. And, we can do this in *any* volume, without the fiction of a resonator or spatial periodicity. The total energy of a field that is a linear combination of these functions is the sum of the energies in these orthogonal components.

There could be many ways of setting up such (energy) orthogonal waves in some arbitrary "receiving" volume $V_R$, which we also require to be solutions of Maxwell's equations. Note first that we can guarantee the waves are Maxwell equation solutions by formally generating them using current sources in some other "source" volume $V_S$. We could then proceed as follows. (We will give this for the monochromatic case, though we could use a full time-dependent case instead.)

1) Choose a current source function $J_{\omega b1}(\mathbf{r}_S)$ in $V_S$. (Here the subscript *b* indicates that this is going to be a basis function for the source space.) For later convenience, we can choose this to be a function that is normalized (using a simple Cartesian inner product in $V_S$ as in Eq. (176)).
2) Calculate the resulting vector potential wave (in the M-gauge) $\mathbf{A}_{\omega M1}(\mathbf{r}_R)$ in $V_R$ using the dyadic Green's function $\overline{\overline{G}}_{\omega M}(\mathbf{r}_R;\mathbf{r}_S)$ as in Eq. (172) using the corresponding Green's function integral
$$\mathbf{A}_{\omega M1}(\mathbf{r}_R) = \mu \int_V \overline{\overline{G}}_{\omega M}(\mathbf{r}_R;\mathbf{r}_S) \cdot \mathbf{J}_{\omega b1}(\mathbf{r}_S)\, d^3\mathbf{r}_S \tag{183}$$
and construct a normalized version $\mathbf{A}_{\omega Mb1}(\mathbf{r}_R)$ (using the energy inner product in $V_R$ as in Eq. (181)).
3) Choose a second (normalized) current source function $\mathbf{J}_{\omega b2}(\mathbf{r}_S)$ in $V_S$, orthogonal to $\mathbf{J}_{\omega b1}(\mathbf{r}_S)$ [137, 138]. Now calculate the corresponding wave $\mathbf{A}_{\omega M2}(\mathbf{r}_R)$ as in Eq. (183). Now retain only the part $\mathbf{A}_{\omega\perp}(\mathbf{r}_S)$ that is orthogonal to $\mathbf{A}_{\omega Mb1}(\mathbf{r}_R)$, using the energy inner product Eq. (181) in $V_R$, which we can do by formally, "projecting out" the component of $\mathbf{A}_{\omega Mb1}(\mathbf{r}_R)$
$$\mathbf{A}_{\omega\perp}(\mathbf{r}_R) = \mathbf{A}_{\omega M2}(\mathbf{r}_R) - \left(\mathbf{A}_{\omega Mb1}, \mathbf{A}_{\omega M2}\right)_{T\sqrt{\mathrm{U}_\omega}} \mathbf{A}_{\omega Mb1}(\mathbf{r}_R) \tag{184}$$

Now normalize $\mathbf{A}_{\omega\perp}(\mathbf{r}_S)$ to give the second wave function $\mathbf{A}_{\omega Mb2}(\mathbf{r}_R)$.

We then proceed similarly, choosing a subsequent source functions $\mathbf{J}_{\omega bn}(\mathbf{r}_S)$, orthogonal to all preceding ones $\mathbf{J}_{\omega bm}(\mathbf{r}_S)$. Then, in the resulting vector potential field in $V_R$, using an extended "projecting out" of the components in all previous functions, we retaining only the (normalized) part $\mathbf{A}_{\omega Mbn}(\mathbf{r}_R)$ of the resulting wave in each case, which will be orthogonal all previous such functions $\mathbf{A}_{\omega Mbm}(\mathbf{r}_R)$. This process is just a version of a Gram-Schmidt orthogonalization, and it will generate a set of waves $\mathbf{A}_{\omega Mbn}(\mathbf{r}_R)$ in $V_R$ that are orthogonal with respect to the energy inner product in $V_R$.

We have also incidentally generated an orthogonal set of source functions $\mathbf{J}_{\omega bn}(\mathbf{r}_S)$ during this process. So, now we have basis sets, $\mathbf{J}_{\omega bn}(\mathbf{r}_S)$ and $\mathbf{A}_{\omega Mbn}(\mathbf{r}_R)$, each orthonormal with respect to the underlying inner products in their respective spaces (Cartesian for the source space, energy for the receiving space).

To emphasize, with such a process, and noting explicitly that orthogonality here is in the sense of the entire vector functions being orthogonal, whether or not the geometrical vector components of the fields are at right angles,

> we can construct orthogonal sets of electromagnetic waves for any shape of volume, avoiding fictitious boxes or resonators, with the energy of any superposition being the sum of the energies of the orthogonal components.

If we wanted to make sure, for example, that we had a set that corresponded to any possible propagating electromagnetic waves for that volume $V_R$, for $V_S$ we could use a large spherical shell surrounding $V_R$.

## 8.6. Sum rule and communications modes for electromagnetic fields

We briefly discussed the concept of the sum rule for the electromagnetic field case above in section 6.8. Now that we have the necessary notation and definitions we can now explicitly write the integral for the sum rule $S$ in this electromagnetic case, using the energy inner product, Eq. (181), in $V_R$, as

$$S = \int_{V_S}\int_{V_R}\sum_{p=1}^{3}\sum_{q=1}^{3}\hat{\mathbf{x}}_p \cdot \left[\sqrt{\mathsf{U}_\omega}\,\overline{\overline{G}}_\omega(\mathbf{r}_R;\mathbf{r}_S)\right]^\dagger \left[\sqrt{\mathsf{U}_\omega}\,\overline{\overline{G}}_\omega(\mathbf{r}_R;\mathbf{r}_S)\right]\cdot \hat{\mathbf{x}}_q d^3\mathbf{r}_R d^3\mathbf{r}_S \quad (185)$$

Now that we have constructed basis sets $\mathbf{J}_{\omega bn}(\mathbf{r}_S)$ and $\mathbf{A}_{\omega Mbn}(\mathbf{r}_R)$ that are orthogonal with respect to the underlying inner products in their respective spaces, we can now generate a coupling matrix between these orthogonal sets. With these sets, the matrix elements would be

$$g_{ij} = \int_{V_S}\int_{V_R}\left[\sqrt{\mathsf{U}_\omega}\,\mathbf{A}_{\omega Mbi}(\mathbf{r}_R)\right]^\dagger \left[\sqrt{\mathsf{U}_\omega}\,\overline{\overline{G}}_\omega(\mathbf{r}_R;\mathbf{r}_S)\right]\cdot \mathbf{J}_{\omega bj}(\mathbf{r}_S)\, d^3\mathbf{r}_R d^3\mathbf{r}_S \quad (186)$$

We could also keep on adding functions until $\sum_{i,j}\left|g_{ij}\right|^2$ was sufficiently close to $S$, as discussed in section 6.9, and then we would have a finite matrix that we could practically call $\mathsf{G}_{SR}$ for this problem. Then we could perform the singular value decomposition of this matrix to construct the singular values $s_j$ and the corresponding sets of communications mode functions $\{|\psi_j\rangle\}$ and $\{|\phi_j\rangle\}$. The $\{|\psi_j\rangle\}$ would now be mathematical vectors whose elements were the expansion coefficients $h_{jq}$ on the basis $\mathbf{J}_{\omega bq}(\mathbf{r}_S)$, and similarly $\{|\phi_j\rangle\}$ would be vectors of expansion coefficients $f_{jq}$ on the $\mathbf{A}_{\omega Mbn}(\mathbf{r}_R)$ basis.

For subsequent work, we might well want to use the current density and vector potential functions corresponding to these communications modes, i.e.,

$$\mathbf{J}_{\omega CMj}(\mathbf{r}_S) = \sum_q h_{jq}\mathbf{J}_{\omega bq}(\mathbf{r}_S) \text{ and } \mathbf{A}_{\omega CMj}(\mathbf{r}_S) = \sum_p f_{jp}\mathbf{A}_{\omega Mbp}(\mathbf{r}_R) \quad (187)$$

as appropriate orthogonal basis sets since they represent the “best” choices of basis functions.

# 9. Quantizing the electromagnetic field using the M-gauge

Typical standard approaches [9, 139 - 141] to quantization of the electromagnetic field are based on the Coulomb gauge and "transverse" vector potentials, with monochromatic modes chosen either arbitrarily as plane waves or as modes of some resonator. Now that we know that we can generate some complete (or complete enough) set of (energy) orthogonal functions $\mathbf{A}_{\omega Mj}(\mathbf{r}_R)$ for representing any electromagnetic wave in $V_R$, we can use these to quantize the electromagnetic field in the M-gauge. As a result,

> we need no fictitious resonator or "box", allowing quantization for any volume and we avoid the formal problems of the Coulomb gauge [142, 143]. The resulting quantization then proceeds otherwise essentially similarly to those standard approaches. (As in typical standard approaches, we will consider waves only in a uniform isotropic background medium.) Note also that our approach can also be based on the full dyadic Green's function, including all near field terms if we wish; we would simply include those in the evaluation of the $\mathbf{A}_{\omega Mj}(\mathbf{r}_R)$.

To start, we formally expand any (classical) electromagnetic wave in $V_R$ in the basis $\mathbf{A}_{\omega Mj}(\mathbf{r}_R)$. We formally choose expansion coefficients $a_j(t)$ that explicitly include the time-dependent factor $\exp(-i\omega t)$ (and that is their only time dependence). To make the $a_j(t)$ dimensionless, we can introduce a multiplying factor with dimensions of the square root of energy, choosing $\sqrt{\hbar\omega}$ (cunningly anticipating a later result). So, formally,

$$\mathbf{A}_{\omega M}(\mathbf{r}_R,t)=\sqrt{\hbar\omega}\sum_j\left[a_j(t)\mathbf{A}_{\omega Mj}(\mathbf{r}_R)+a_j^*(t)\mathbf{A}_{\omega Mj}^*(\mathbf{r}_R)\right] \tag{188}$$

Now we can formally evaluate the energy, or equivalently the (classical) Hamiltonian in this field using the energy inner product in the receiving volume. That becomes [144]

$$\begin{aligned} H &=\left(\mathbf{A}_R(\mathbf{r}_R,t),\mathbf{A}_R(\mathbf{r}_R,t)\right)_{T\sqrt{\mathrm{U}_\omega}}=\hbar\omega\sum_{m,n}\left(a_m(t)\mathbf{A}_{\omega Mm}(\mathbf{r}_R),a_n(t)\mathbf{A}_{\omega Mn}(\mathbf{r}_R)\right)_{T\sqrt{\mathrm{U}_\omega}} \\ &=\hbar\omega\sum_{m,n}a_m^*(t)a_n(t)\delta_{mn}=\sum_j\hbar\omega a_j^*(t)a_j(t) \end{aligned} \tag{189}$$

where we have used the (energy) orthonormality of the $\mathbf{A}_{\omega Mj}(\mathbf{r}_R)$. We see, not surprisingly, that the resulting (classical) Hamiltonian is just the sum of the Hamiltonians

$$H_j=\hbar\omega a_j^*(t)a_j(t) \tag{190}$$

We give the detailed steps in the resulting quantization of such Hamiltonians below for completeness in **Appendix I**, but the results are straightforward. We obtain the familiar quantum-mechanical Hamiltonian form

$$\hat{H}_j=\hbar\omega\left(\hat{a}_j^\dagger\hat{a}_j+\frac{1}{2}\right) \tag{191}$$

for the *j*th mode, where the annihilation and creation operators $\hat{a}_j$ and $\hat{a}_j^\dagger$ obey the usual commutation relations and other algebraic properties (see Eqs. (373) - (375)). We can also write appropriate field operators for the vector potential and the electric and magnetic fields (see Eqs. (370) - (372)).

# 10. Linear scatterers and optical devices

As we have set up the physics and mathematics of our SVD approach to linear wave systems, we have mostly used the example of a free-space or "uniform medium" Green's function as the coupling operator between the source space and the receiving space. However, as we discussed in section 6.8, as in the statement (157), any finite coupling operator $D(\mathbf{r}_R;\mathbf{r}_S)$ between finite volumes $V_S$ and $V_R$ is a Hilbert-Schmidt operator (and including even dyadic operators for electromagnetic fields). Any fixed physical linear system can be described by such an operator, which we could call a scattering operator or a "device" operator. We can model any fixed linear scatterers or linear optical devices, or more generally linear "objects", in this way – free space propagation, complex multiple scatterers, waveguide channels, sophisticated linear optical systems or devices, or any object behaving linearly with respect to the incident field (including those that absorb radiation). Of course, the detailed analysis could be complicated, especially for systems involving multiple scattering. However, we can draw some conclusions here that apply even without knowing the details of the scattering or device operator.

## 10.1. Existence of orthogonal functions and channels

First, we note that all the mathematics of the SVD, following from the Hilbert-Schmidt nature of any such system, applies also to any finite linear scattering or device operator. The SVD can always be performed on any such compact operator (see section 6.7), leading to the existence of orthogonal sets of functions in the source space and the receiver space. So, we can formally conclude, as mentioned in section 1, that any linear optical "device" can be viewed as a mode converter, converting from specific sets of functions in the input space one-by-one to specific corresponding functions in the output space. Also, as stated in section 1.4, there is a set of independent channels through any linear scatterer; that is, there is a set $\{|\psi_j\rangle\}$ of orthogonal source functions $|\psi_j\rangle$ that couple, one by one, to corresponding members $|\phi_j\rangle$ of the orthogonal set $\{|\phi_j\rangle\}$ of wave functions in the receiving space (even if we do not know what these functions are).

These mode-converter basis sets $\{|\psi_j\rangle\}$ and $\{|\phi_j\rangle\}$ and the associated singular values $s_j$ tell us everything that can be known about this scatterer or device based on waves generated from the source volume and detected in the receiving volume; they are a complete description as far as these sources and detectors are concerned. They are sufficient to reconstruct the matrix $\mathsf{D}$ corresponding to some scattering or device operator $D$ for these source and receiver spaces. Note explicitly, now that we are considering wave systems other than just free-space propagation, this approach is valid even if the system is lossy, as long as that is "linear" loss – that is, the "output" field amplitude is linearly proportional to the "input" field amplitude – and for systems with finite linear gain.

## 10.2. Establishing the orthogonal channels through any linear scatterer

Though we might not know the operator $D$ or its matrix $\mathsf{D}$, there are at least two ways to establish it for any scatterer. One way is simply to measure the scattering matrix from some set of sources to some set of receiving points (see, e.g., [145 - 148]) in some interferometric optical experiment. Of course for a complex scatterer, this could take some time, but it is possible, and it is then also possible to perform the mathematical SVD of that matrix to establish the functions $|\psi_j\rangle$, $|\phi_j\rangle$, and the singular values $s_j$.

A second approach [12] is to have a physical system find the SVD by a sequential maximization process. In this case, we can in principle use two meshes of interferometers as in Fig. 5(a), but with the scatterer in the space between the two meshes, as in Fig. 28 [149]. By an iterative procedure back and forward between one side and the other, and based only on single-parameter power optimizations, it is possible for the meshes to set themselves up so that the one on the left generates the $|\psi_j\rangle$ from the corresponding single-mode inputs on the left, and couples the resulting $|\phi_j\rangle$ into corresponding single-mode outputs on the right, automatically establishing the best channels through the scatterer. Effectively, the physical system has performed the SVD of the optical system [150] and has embedded its results in the settings of the interferometers.

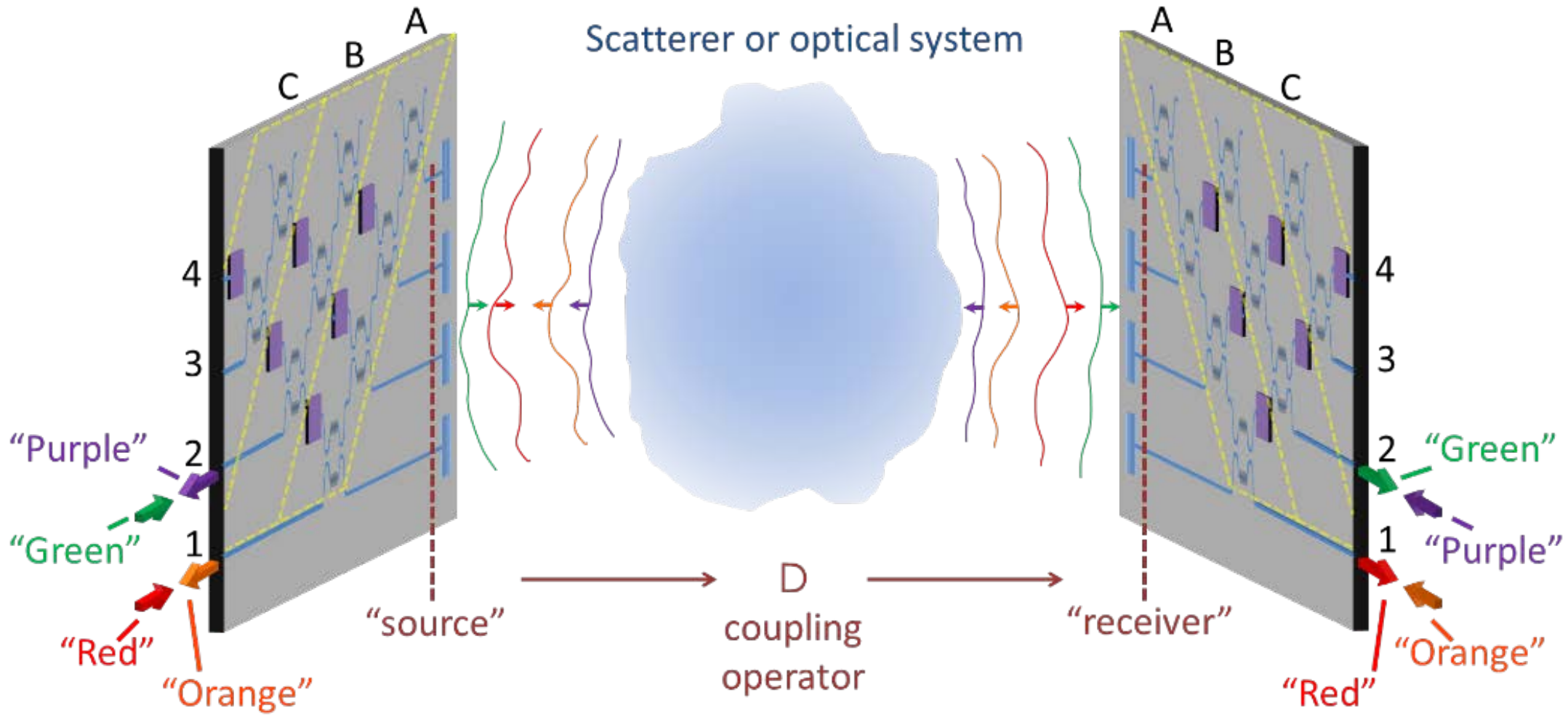

Fig. 28. Conceptual apparatus for finding the best orthogonal channels through any reciprocal linear scatterer or optical system at a given frequency (after [12]), nominally described here by some coupling operator $D$ from left to right, using two interferometer meshes on either side of the scatterer. To find the most strongly coupled channel, we shine light into the "red" input waveguide 1 on the left, and adjust the interferometers in row A on the right to maximize the output "red" power in the output waveguide 1 on the right. Then we run in reverse, shining the "orange" power (actually, at the same wavelength as the "red" power – colors here are for graphic clarity only) backwards into the "output" waveguide 1 on the right, and adjust the row A interferometers on the left to maximize the "orange" power backwards out of "input" waveguide 1 on the left. We repeat this "red"/"orange" process forwards and backwards until the system converges, having found the most strongly coupled channel through the system. Then, leaving row A on both sides set, we repeat a similar process with the "green" and "purple" beams, now in waveguide 2 on both sides. This will find the second most strongly coupled channel. We can then repeat for the waveguides 3. (No final "waveguides 4" process is required because it is automatically configured as the only remaining orthogonal channel). The process has found the 4 most strongly coupled channels in this system. Technically, this process has effectively found the singular value decomposition of the optical system between the waveguide amplitudes at the "source" dashed line on the left and those at the "receiving" dashed line on the right, effectively embedding the unitary matrices $U^\dagger$ and $V$ of the SVD of $D = VD_{diag}U^\dagger$ in the interferometer settings in the meshes on the left and the right.

## 10.3. Bounding the dimensionalities of the spaces

If the scatterer or device has some finite volume $V_D$, we could think of two separate problems, especially if we know what are the coupling operators $G_{SD}$, from the source space to the device volume $V_D$, and $G_{DR}$, from the device volume to the receiving space (see Fig. 29.).

These might, for example, just be simple free-space Green's functions. Then solving for the communications modes of $G_{SD}$ (from sources to the device) and $G_{DR}$ (from the device to the receiving volume) would give us maximum numbers of usable channels into the device and out of it.

Immediately, for example, such an approach would tell us how many channels we need to block or emulate to make some object appear "invisible" as seen from some receiving volume, based on sources in some source volume, limiting the necessary complexity of any active "cloaking" device [151] around the scattering object.

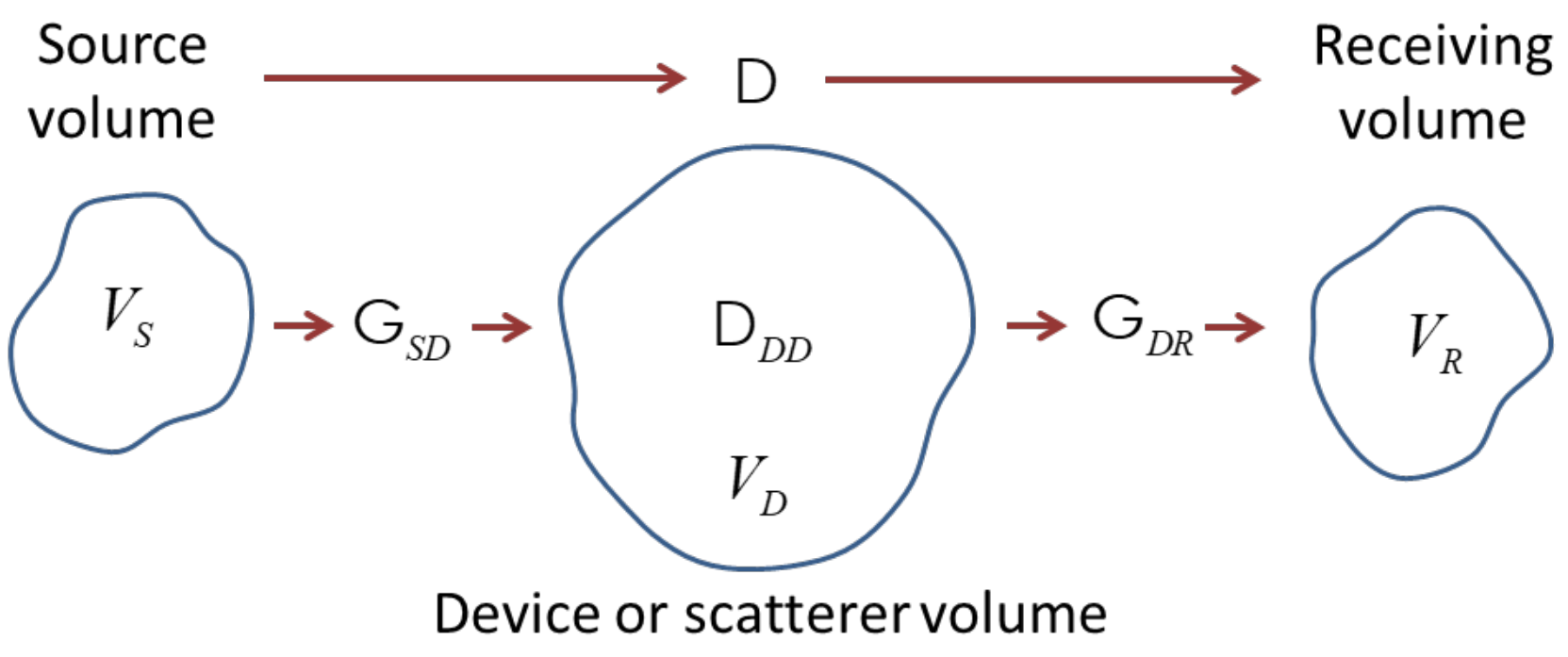

Fig. 29. Configurations to consider either the full coupling operator $D$ from source to receiving volumes, or a succession of three operators, including an "internal" scattering operator $D_{DD}$ that generates effective sources for waves to the receiving volume from the incoming waves from the source volume).

We could also choose to look at an "internal" device operator $D_{DD}$ as one that maps from the input waves into the device to the effective sources inside the device that generate the resulting "external" scattered waves. In that case, we could view the overall operator coupling $D$ from the original source volume $V_S$ to the ultimate receiving volume $V_R$ as being the product

$$D = G_{DR} D_{DD} G_{SD} \tag{192}$$

Such an approach could then begin to link to approaches to limits to optical devices, as in [152, 153], and also to discussions of the necessary complexity of optical devices, as in [154], though detailed discussions of these topics are beyond the present work.

## 10.4. Emulating an arbitrary linear optical device and proving any such device is possible – arbitrary matrix-vector multiplication

We have discussed above that we can approximate linear optical systems by using sufficiently many "patches" of sources (or waves acting as sources through diffraction operators) and corresponding "patches" of receivers, or equivalently any appropriate and sufficient large orthogonal basis sets of sources or input waves and of output waves. Explicit discussions above were for simple "free-space" propagation, but the same concepts and limits apply as we want to approximate optical devices.

Suppose, then, that we make some optical apparatus that effectively samples a "source" light field, such as with grating couplers acting as appropriate "patch" collecting devices, and delivers the output of each such patch into a single mode waveguide to pass into some optical waveguide system, and similarly couples output waveguides into similar output "patches" to generate output waves. Then, if we can make the waveguide system in the middle so that it can implement arbitrary linear mappings between the input waveguides and the output waveguides, then, within the approximation of the input and output light fields by these patches, we can make an arbitrary linear optical component.

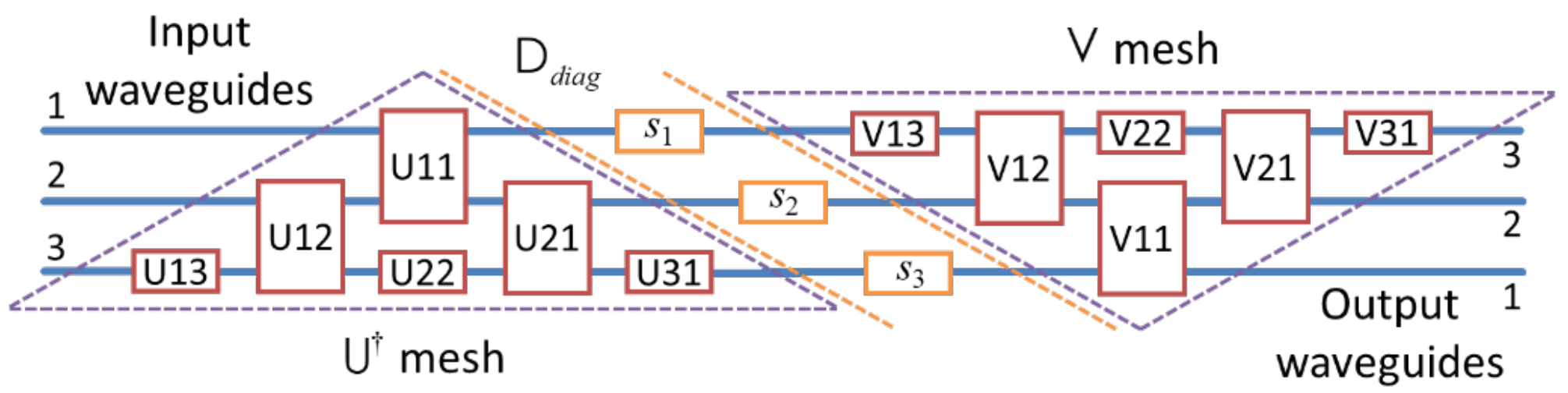

Fig. 30. Singular value decomposition architecture for constructing any matrix (here $3\times3$) in optics between a sets of waveguide input and output amplitudes [25].

Fig. 30 shows one way of making such an arbitrary linear mapping between input and output waveguides. Here we exploit the idea of SVD, but in this case using it as a way to construct a matrix physically [25]. We know from Eq. (221) that any compact matrix or operator can be written in the form $\mathsf{D}=\mathsf{V}\mathsf{D}_{diag}\mathsf{U}^\dagger$. So if we can emulate an arbitrary unitary matrix $\mathsf{U}^\dagger$, a diagonal matrix $\mathsf{D}_{diag}$, and another unitary matrix $\mathsf{V}$, in the correct sequence in some waveguide system, then we physically create a system to give the effect of an arbitrary matrix or operator $\mathsf{D}$, at least up to the dimensionality of this waveguide system, and within the approximation of the actual light fields by "patches" or some other form of sampling or transformation.

In Fig. 30, an input unitary mesh of interferometers (see Fig. 5) implements any $\mathsf{U}^\dagger$ (at least up to a dimensionality of three in this simple example). A line of modulators (with controllable phase and amplitude implements the singular values (the diagonal elements of $\mathsf{D}_{diag}$), with gain if needed for singular values greater than 1. Another unitary mesh implements $\mathsf{V}$. Hence, we can emulate any matrix, up to the dimensionality of the mesh (here $3\times3$).

This approach may be practically useful for various linear processing functions [25, 57, 58]. Note, incidentally, that, unlike some previous approaches to optical "matrix-vector" multipliers [155], this approach does not have fundamental "splitting" loss; unless we want loss so we can implement non-unitary matrices, this approach has no loss other than background loss in the components. This class of architectures can also be trained directly and progressively using the mode-converter basis functions of interest to implement the corresponding matrix [25].

Because it gives a constructive proof that any linear optical component is possible in principle, this approach also gives us an apparatus for thought experiments, as in the derivation of new radiation laws [7], which we discuss below.

# 11. Mode-converter basis sets as fundamental optical descriptions

Note that, for any linear scatterer, device or "object" with the mode-converter basis functions $|\psi_j\rangle$ and $|\phi_j\rangle$

$$\begin{array}{l}\text{the only radiation scattered into an "output" wave } |\phi_j\rangle \\ \text{is from the corresponding input wave or source } |\psi_j\rangle\end{array} \tag{193}$$

and

the input wave or source $|\psi_j\rangle$ only scatters
into the corresponding output wave $|\phi_j\rangle$.
(194)

These statements follow automatically from the orthogonality of the sets $|\psi_j\rangle$ and $|\phi_j\rangle$ and the "pairing" in the communications mode or mode-converter description. These sets therefore acquire a "fundamental" status for describing any linear scatterers, devices or objects, and we can use them to derive some quite basic (and novel) results.

## 11.1. Radiation laws

Kirchhoff's radiation law states that the "absorptivity" and the "emissivity" of an object must be equal so that the object can come to thermal equilibrium with other bodies just by exchanging radiation (see, e.g., [7]). For the total radiation from a body at a given wavelength, this is straightforward. It is sometimes extended to a "directional" radiation law, though the proof of that neglects diffraction, and none of these laws applies to non-reciprocal objects. Using a thought experiment that exploits an arbitrary linear optical system and the properties of the mode-converter basis sets, new laws can be derived, including the effects of diffraction and non-reciprocity [7].

Suppose that the object $O$ in Fig. 31 has a specific mode-converter pair of input and output functions $|\psi_1\rangle$ and $|\phi_1\rangle$. Suppose also that the optical machine in Fig. 31 is set so that any input light at its input port 1 generates such a wave $|\psi_1\rangle$ at the outputs of the waveguide to free-space converters; we imagine we have enough of these "patch" generators to do a sufficiently good job of synthesizing all the possibly well-connected input mode-converter functions. Similarly, any light in the corresponding mode-converter output function $|\phi_1\rangle$ is collected and appears at output port 1 of the optical machine.

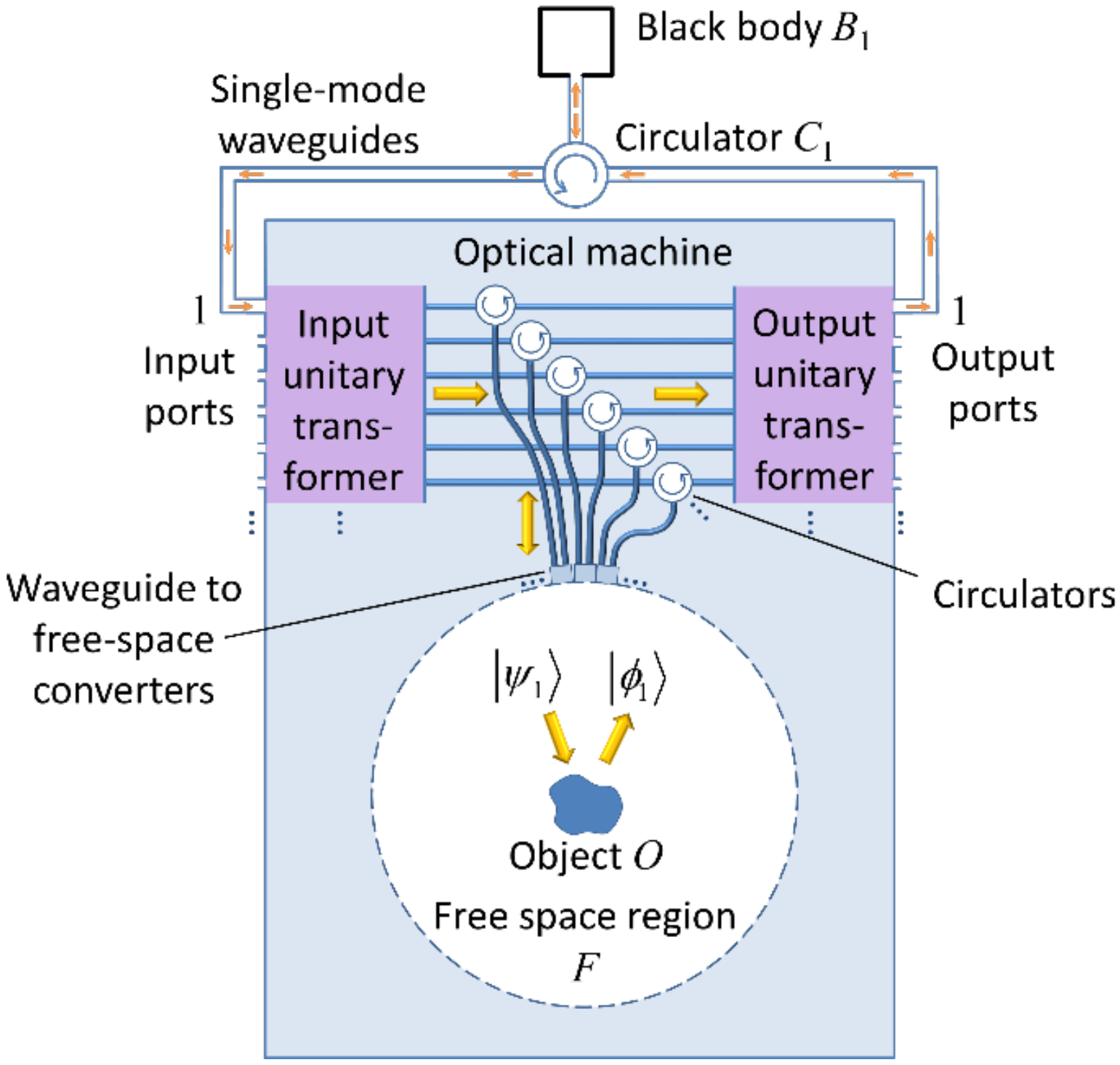

Fig. 31. Thought-experiment apparatus for establishing a modal thermal radiation law (after [7]).

Note that no light into any of the other ports leads to any scattering into this mode $|\phi_1\rangle$ by (193), and none of the light from $|\psi_1\rangle$ is scattered into any other output mode by (194). Indeed, we can simply connect the other output ports of the machine back round to the corresponding input ports. There is no interaction between these channels through scattering.

Now we presume that the black body $B_1$ emits and absorbs light only through a single-mode waveguide. A circulator separates forward and backward light. So, any light emitted by $B_1$ appears as light in mode $|\psi_1\rangle$. Some of that light may then be absorbed by the object *O*. Any light that is not absorbed is then scattered into $|\phi_1\rangle$ and sent back to $B_1$. But to establish thermal equilibrium between the black body and the object, we therefore require that an amount of light is emitted from the object into $|\phi_1\rangle$ that is equal to the amount absorbed from $|\psi_1\rangle$. Hence

> the absorptivity of mode-converter input mode $|\psi_1\rangle$ is equal to the
> emissivity into corresponding mode-converter output mode $|\phi_1\rangle$ (195)

establishing a previously unknown "modal" radiation law (Law 1 of [7]). Note this law includes all effects of diffraction, and it is also valid even if the object *O* is non-reciprocal.

From this modal law, we can algebraically derive three other laws of thermal radiation, one of which (Law 3) is essentially the original law for total absorptivity and emissivity, and another (Law 4) is a correct (and much more general) version of the "directional" law:

> for reciprocal objects, the absorptivity of any input beam is equal to the emissivity back into that same beam.

## 11.2. A modal "A and B coefficient" argument – the *M* coefficient for emission and absorption

Following an approach related to that used for the above radiation laws, we can construct an argument similar to Einstein's classic "A and B" coefficient argument (see, e.g., [139]), relating absorption, spontaneous emission, and stimulated emission based on thermodynamic and statistical mechanics arguments; now, however, we construct it directly for a mode-converter pair. We give the detail of this argument in **Appendix J**. Now we presume we have some quantum "two-level" system inside some otherwise-lossless optical environment, which can be anything we want – e.g., free space, a resonator, or a waveguide – and we establish the corresponding mode-converter basis sets $\{|\psi_j\rangle\}$ and $\{|\phi_j\rangle\}$, or at least some pair or pairs of interest. Specifically, we conclude the following:

> If a quantum system has probabilities $P_2$ and $P_1$ of being in its upper and lower states respectively, and if the probability per unit time that a photon in the mode-converter input mode $|\psi_j\rangle$ is absorbed by the quantum system is $MP_1$, then if there are $n_p$ photons in input mode $|\psi_j\rangle$ the probability per unit time that a photon is emitted into the corresponding mode-converter output mode $|\phi_j\rangle$ is $(n_p+1)MP_2$. (196)

Now, instead of the A and B coefficients, we only have one coefficient, *M*, for a given mode-converter pair [156]. This simple result works mode by mode. We have avoided using the free-space density of states and any fictitious boxes or resonators we might use to construct that. The modal basis is derived entirely based on the actual optical configuration of quantum system and any otherwise loss-less optical system in which it is embedded, thereby giving a more general result.

## 11.3. Mode-converter basis sets as physical properties of a system

Necessarily, much of the discussion so far has centered on the mathematical process of SVD for establishing the communications modes and mode-converter basis sets. However, an important point is that these can be established entirely physically, at least in principle, without any mathematics. We already made this point in section 10.2, showing how to establish these mode pairs by maximization of transmission through a scatterer. With the radiation laws, we have another option in principle: we could establish by some iterative process just what input mode leads to the largest absorption in some (partially) absorbing object, thereby establishing $|\psi_1\rangle$. If necessary, then we can find the corresponding $|\phi_1\rangle$ by looking at the thermal emission and establishing the best possible mode power collection (e.g., into a single mode fiber). We could use similar apparatus as in Fig. 28, and that would then let us progress to finding the next such pair, using the second waveguide on each side, and so on. This is not necessarily a very practical approach, but it emphasizes that these modes are physical properties of the systems that can be established by measurement.

# 12. Conclusions

The idea of considering modes as pairs of functions determined from singular value decomposition is a useful and powerful approach for understanding waves in communications and in linear scattering and optics more broadly. It is supported by rigorous mathematics and by deep physical concepts in waves generally, including for full vector electromagnetic waves. This approach is applicable in acoustics, in wireless communications, and in optics, and provides a unified framework for such problems.

This approach clarifies the number and specific nature of orthogonal channels in communication with waves – the communications modes – including rigorous counting of these channels, and shows how to establish them – mathematically, in calculations, and in practice. It introduces and rationalizes many "heuristic" behaviors of numbers of usable channels. It provides the "best" modal basis sets for describing any optical component or wave scatterer. Those mode-converter basis set pairs are the most economical ones for describing and analyzing any such component or scatterer, and they lead to fundamental physical laws that apply one-by-one to these mode pairs, including new and more general radiation laws and a new compact version of the fundamental argument relating absorption, spontaneous emission and stimulated emission of quantum systems.

These approaches are valid for any size scale of object or system, from large imaging optics to radio antennas and nanophotonic devices. As such they complement and complete the many existing approaches – from conventional ray-tracing imaging optics, through Fourier optics and standard families of functions and beams, to direct solutions of wave equations – that are useful separately at different scales and complexities of systems. We might expect these communications mode and mode-converter basis set approaches to be particularly useful for systems that are too large for convenient direct wave calculations and too small for the simpler Fourier and ray-tracing approaches. At any scale, however, they can offer the most economical description in terms of the modes that matter most.

We can hope that the ultimate simplicity, clarity and rigor of these approaches will be practically useful, fundamentally significant, and stimulating to new insights into waves and how we can use them.

# Appendix A Approximating uniform line or patch sources with point sources

Suppose we divide a source surface or line into "patches" of area $A_p$ (for a surface) or width $d_p$ (for a line). Our question is whether there is much difference between considering a point source of amplitude $h$ in the middle of the patch or a uniformly distributed source over the entire area or length of the patch, with source (areal) density $\eta_A = h / A_p$, or source (linear) density $\eta_d = h / d_p$.

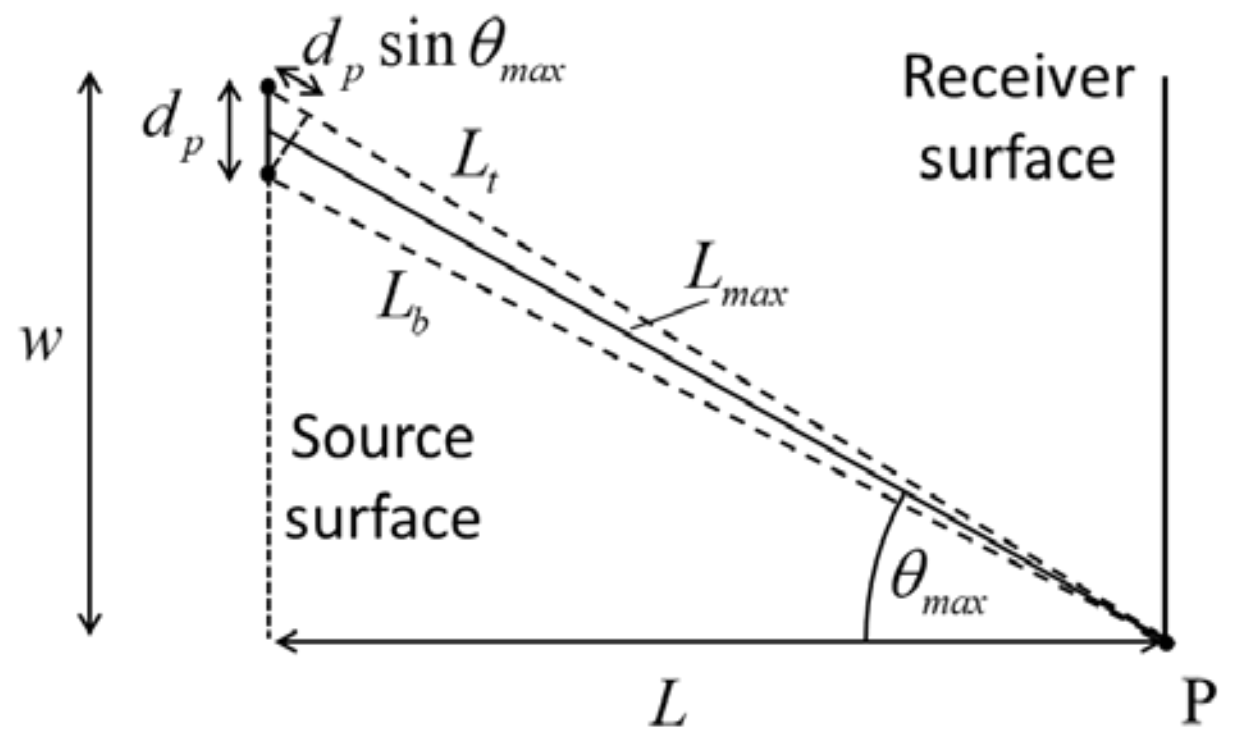

Fig. 32. Schematic for estimating the largest size of source "patch" for which a uniform source density across the patch can be approximated by a point source in the middle of the patch.

We can consider the most distant point $P$ from this source on the receiving surface, at some distance $L_{max}$ from the (center of) the patch (see Fig. 32). If there is no difference in the wave amplitude at $P$ between a source at one extreme end of the patch and one at the other, then we can consider that it makes no difference whether we consider the point source or the equivalent source density spread over the entire area or width of the source patch.

We presume that the linear dimension of these patches is much smaller than the separation $L$ between the source and receiver surfaces, so there is negligible difference in the "$1/r$" factor in the Green's function between these two extreme points. However, we do have to consider the phase difference between these two extreme source points for waves arriving at $P$. In Fig. 32, we illustrate the case of a "line" source patch. We can consider that our approximation will start to break down if the difference in the two lengths $L_t$ from the "top" point and $L_b$ from the "bottom" point reaches $\sim \lambda / 2$, because then we will have destructive interference between the waves from these two extreme sources.

From Fig. 32, $L_t = L_{max} + \left(d_p / 2\right)\sin\theta_{max}$ and $L_b = L_{max} - \left(d_p / 2\right)\sin\theta_{max}$ so

$$L_t - L_b = d_p \sin\theta_{max} \tag{197}$$

So, keeping this below $\lambda / 2$ requires

$$d_p < \lambda / 2\sin\theta_{max} \tag{198}$$

The situation for an area "patch" is slightly different in that the most extreme distance difference would be between the diagonally opposite corners of the patch, so possibly up to $\sqrt{2}$ larger than the largest linear dimension. However, for this simple heuristic we neglect that minor difference, and conclude that the spacing of our point sources (in either direction for area patches) should satisfy the approximate limit of Eq. (198) if they are reasonably to approximate uniform patches.

If the source and receiving surfaces are approximately the same size (with a linear dimension *w*) and are centered on a common axis, then $\tan\theta_{max} = w/L$. For paraxial situations, so where $w \ll L$, then $\sin(\theta_{max}) \simeq \tan(\theta_{max})$, so we have $d_p < \lambda L / 2w$, which is the result Eq. (70) in the main text.

# Appendix B Longitudinal heuristic angle

To understand the effective "diffraction angle" from a longitudinal line of sources, we can construct an argument based on two point sources, A and B, spaced a distance $2\Delta z$ apart in the *z* direction, as in Fig. 33.

Suppose the relative phase of the two sources is such that they add constructively along the *z* direction (i.e., the direction from A or B to P). As we move away from P in a direction perpendicular to *z*, such as by some amount $\delta y$ in the *y* direction to some point Q, the relative distance to the points A and B will change. Specifically, the distance from A to Q is $s_{AQ} = \left[(z_o + \Delta z)^2 + \delta y^2\right]^{1/2}$ and from B to Q, $s_{BQ} = \left[(z_o - \Delta z)^2 + \delta y^2\right]^{1/2}$.

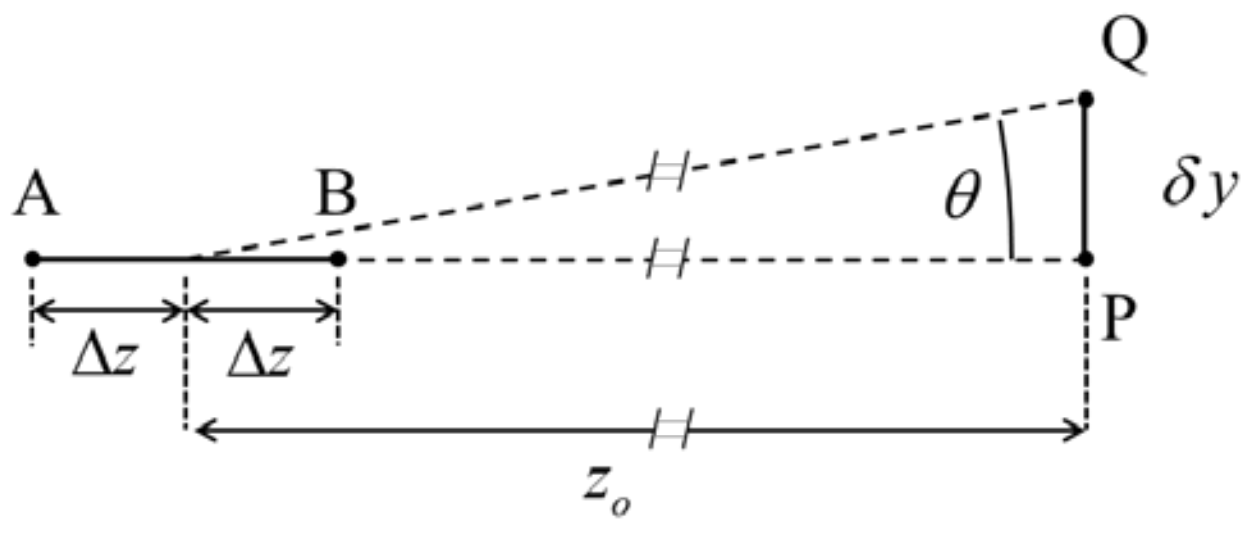

Fig. 33. Two point sources A and B spaced $2\Delta z$ apart along the *z* axis, at a very large distance $z_o$ from a point *P*. Point Q is spaced a relatively small distance $\delta y$ away from P in the *y* direction. $\theta$ is the angle subtended by the line segment PQ relative to the midpoint between A and B. (Not to scale; $z_o \gg 2\Delta z$.)

We can also think of the separation between P and Q as an angle *θ*. Presuming $\delta y \ll z_o$, then $\theta \simeq \delta y / z_o$, or equivalently $\delta y = z_o\theta$. Writing

$$\Delta s = s_{BQ} - s_{AQ} \tag{199}$$

and dropping terms $\propto (\Delta z / z_o)^2$ as being relatively too small as we let $z_o$ become arbitrarily large, we have

$$\Delta s \simeq z_o\left\{\left[1-(2\Delta z / z_o)+\theta^2\right]^{1/2} - \left[1+(2\Delta z / z_o)+\theta^2\right]^{1/2}\right\} \tag{200}$$

Using [157] $\sqrt{1+\varepsilon} \simeq 1 + (\varepsilon/2) - (\varepsilon^2/8)$, after some algebra, $\Delta s \simeq -2\Delta z + \Delta z\,\theta^2$. But $s_{BP} - s_{AP} = -2\Delta z$. So the change in the relative distance from points A and B as we move from P to Q is

$$\delta s = (s_{BQ} - s_{AQ}) - (s_{BP} - s_{AP}) \simeq \Delta z\,\theta^2 \tag{201}$$

We are interested specifically in the angle $\theta_L$ by which this relative distance has changed by $\lambda/2$, corresponding from a change from constructive to destructive interference, and hence a minimum in the intensity. So, for $\delta s = \lambda/2$ we have what we call the *longitudinal heuristic angle* $\theta_L = \sqrt{\lambda / 2\Delta z}$, which is the desired result (Eq. (71) in the main text).

# Appendix C Spherical heuristic number

We can rationalize the approximate effective number of channels for a spherical source as follows. Consider some small lateral line AB of length $d$ on an outer "receiving" sphere of a very large radius $R$ (Fig. 34), and consider two point sources P and Q on the extreme sides of the source sphere and in the same plane as AB and on a line parallel to AB.

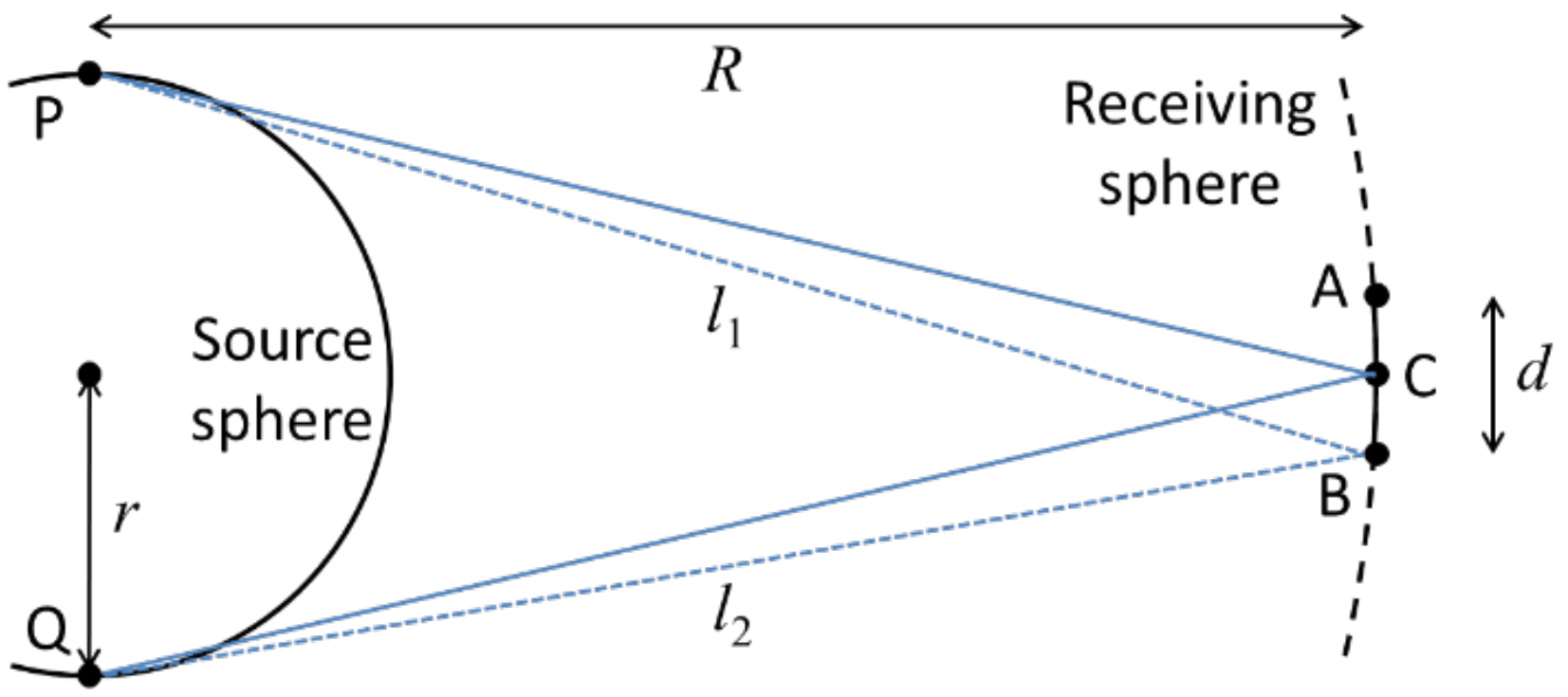

Fig. 34. Construction for deducing an effective number of degrees of freedom from a spherical source.

We ask that $d$ is small enough such that the distance difference $\Delta l = l_1 - l_2$ from the two point sources changes by just $\lambda / 2$ as we move from the middle (point C) to one end (point B) of this line, so we could move from constructive interference to cancellation. So

$$\begin{aligned}\Delta l &= \sqrt{\left(r + d/2\right)^2 + R^2} - \sqrt{\left(r - d/2\right)^2 + R^2} \\ &\simeq R\left(1 + \frac{1}{2}\left(\frac{r + d/2}{R}\right)^2 - 1 - \frac{1}{2}\left(\frac{r - d/2}{R}\right)^2\right) \simeq \frac{rd}{R} = \frac{\lambda}{2}\end{aligned} \qquad (202)$$

So $d = R\lambda / 2r$ . We can reason similarly in the other direction (out of the "paper") leading to a patch of area $d^2 = R^2\lambda^2 / 4r^2$ . The total number of such patches on the surface area $4\pi R^2$ of the receiving sphere is therefore $N_{SH} = 4\pi R^2 / d^2 = 16\pi r^2 / \lambda^2$ , which is the result Eq. (74) in the main text. Note that this result depends only on the radius $r$ of the source sphere. [158] derives a number equivalent to this by a "sampling" argument.

# Appendix D Singular value decomposition of compact operators

For a compact (but not necessarily Hermitian) operator $\mathsf{A}$ that maps from a Hilbert space $H_S$ to a Hilbert space $H_R$ (which will be the source and receiver spaces in our physical problems), the operator $\mathsf{A}^\dagger\mathsf{A}$ maps from $H_S$ back into $H_S$, and the operator $\mathsf{A}\mathsf{A}^\dagger$ maps from $H_R$ back into $H_R$ . We know that both of these operators are then compact and Hermitian (see (127)), so by the spectral theorem for compact Hermitian operators (129), the set of eigenfunctions $\{|\psi_j\rangle\}$ of $\mathsf{A}^\dagger\mathsf{A}$ is a complete set for $H_S$, and the set of eigenfunctions $\{|\phi_j\rangle\}$ of $\mathsf{A}\mathsf{A}^\dagger$ is a complete set for $H_R$, and we will choose these sets to be orthonormal in their spaces. In SVD, we are interested in both of these sets of eigenfunctions, and the corresponding eigenvalues.

Both $\mathsf{A}^\dagger\mathsf{A}$ and $\mathsf{A}\mathsf{A}^\dagger$ are also positive operators by (134) (or (136)). So, by (136), their eigenvalues are also positive (non-negative). So, we can choose to write the eigenvalues of $\mathsf{A}^\dagger\mathsf{A}$ as $c_j = |s_j|^2$. The eigen equation for $\mathsf{A}^\dagger\mathsf{A}$ is then

$$\mathsf{A}^\dagger\mathsf{A}|\psi_j\rangle = |s_j|^2 |\psi_j\rangle \tag{203}$$

So, using the expansion of the form Eq. (131) for $\mathsf{A}^\dagger\mathsf{A}$, we have

$$\mathsf{A}^\dagger\mathsf{A} = \sum_{j=1}^{\infty} |s_j|^2 |\psi_j\rangle\langle\psi_j| \tag{204}$$

So,

$$\|\mathsf{A}\psi_n\|^2 = \langle\psi_n|\mathsf{A}^\dagger\mathsf{A}|\psi_n\rangle = |s_n|^2 \tag{205}$$

Then

$$\|\mathsf{A}\psi_n\| = |s_n| \tag{206}$$

So for all non-zero eigenvalues $s_n$ we can construct a set of functions $\{|\phi_n\rangle\}$ in $H_R$ (where we have used a notation that anticipates the answer – we have not yet proved these are also the eigenfunctions of $\mathsf{A}\mathsf{A}^\dagger$), that we define as

$$|\phi_n\rangle = \frac{1}{s_n}\mathsf{A}|\psi_n\rangle \tag{207}$$

This set of functions is, first, normalized; that is

$$\langle\phi_n|\phi_n\rangle = \frac{1}{s_n^* s_n}\langle\psi_n|\mathsf{A}^\dagger\mathsf{A}|\psi_n\rangle = \frac{|s_n|^2}{s_n^* s_n} = 1 \tag{208}$$

and we have

$$\begin{aligned}
\langle\phi_m|\phi_n\rangle &= \frac{1}{s_m^* s_n}\langle\psi_m|\mathsf{A}^\dagger\mathsf{A}|\psi_n\rangle = \frac{1}{s_m^* s_n}\langle\psi_m|\left(\sum_{j=1}^{\infty}|s_j|^2|\psi_j\rangle\langle\psi_j|\right)|\psi_n\rangle \\
&= \frac{1}{s_m^* s_n}\sum_{j=1}^{\infty}|s_j|^2\langle\psi_m|\psi_j\rangle\langle\psi_j|\psi_n\rangle = \frac{1}{s_m^* s_n}\sum_{j=1}^{\infty}|s_j|^2\langle\psi_m|\psi_j\rangle\delta_{jn} \\
&= \frac{|s_n|^2}{s_m^* s_n}\langle\psi_m|\psi_n\rangle = \frac{|s_n|^2}{s_m^* s_n}\delta_{mn} = \delta_{mn}
\end{aligned} \tag{209}$$

so this set $\{|\phi_n\rangle\}$ is also orthonormal.

Now suppose we consider an arbitrary function $|\psi\rangle$ in $H_S$. Then we can expand it in the orthonormal set $\{|\psi_j\rangle\}$, as in Eqs. (86) and (87), to obtain

$$|\psi\rangle = \sum_j \langle\psi_j|\psi\rangle|\psi_j\rangle \tag{210}$$

So, using (207), and the associative law (109) for products in Dirac notation

$$\begin{aligned}
\mathsf{A}|\psi\rangle &= \mathsf{A}\sum_j\langle\psi_j|\psi\rangle|\psi_j\rangle = \sum_j\langle\psi_j|\psi\rangle\mathsf{A}|\psi_j\rangle = \sum_j s_j\langle\psi_j|\psi\rangle|\phi_j\rangle \\
&= \sum_j s_j|\phi_j\rangle\langle\psi_j|\psi\rangle = \left(\sum_j s_j|\phi_j\rangle\langle\psi_j|\right)|\psi\rangle
\end{aligned} \tag{211}$$

Since $\psi$ was arbitrary, we can therefore write

$$\mathsf{A} \equiv \sum_j s_j \left|\phi_j\right\rangle\left\langle\psi_j\right| \tag{212}$$

which is the *singular value decomposition* (SVD) of the operator $\mathsf{A}$ from a space $H_S$ to a possibly different space $H_R$. The complex numbers $s_j$ are the *singular values*. Note this definition can be for infinite-dimensional spaces.

Finally, it remains only to verify that the $\left\{\left|\phi_j\right\rangle\right\}$ are the eigenfunctions of $\mathsf{A}\mathsf{A}^\dagger$. From (212) and the general algebraic identity for Hermitian adjoints of vector-vector products [159],

$$\mathsf{A}^\dagger \equiv \sum_k s_k^* \left|\psi_k\right\rangle\left\langle\phi_k\right| \tag{213}$$

So from (212)

$$\begin{aligned}\mathsf{A}\mathsf{A}^\dagger &\equiv \sum_{j,k}\left(s_j\left|\phi_j\right\rangle\left\langle\psi_j\right|\right)\left(s_k^*\left|\psi_k\right\rangle\left\langle\phi_k\right|\right) = \sum_{j,k} s_j s_k^* \left|\phi_j\right\rangle\left\langle\psi_j\middle|\psi_k\right\rangle\left\langle\phi_k\right| \\ &= \sum_{j,k} s_j s_k^* \left|\phi_j\right\rangle \delta_{jk} \left\langle\phi_k\right| = \sum_j \left|s_j\right|^2 \left|\phi_j\right\rangle\left\langle\phi_j\right|\end{aligned} \tag{214}$$

Hence

$$\mathsf{A}\mathsf{A}^\dagger \left|\phi_j\right\rangle = \left|s_j\right|^2 \left|\phi_j\right\rangle \tag{215}$$

So the $\left\{\left|\phi_j\right\rangle\right\}$ are indeed the eigenfunctions of $\mathsf{A}\mathsf{A}^\dagger$, and note that they have the same eigenvalues as $\mathsf{A}^\dagger\mathsf{A}$ [160]. Note, finally, from (212),

$$\mathsf{A}\left|\psi_k\right\rangle = \sum_j s_j \left|\phi_j\right\rangle\left\langle\psi_j\middle|\psi_k\right\rangle = \sum_j s_j \left|\phi_j\right\rangle \delta_{jk} = s_k \left|\phi_k\right\rangle \tag{216}$$

So the operator $\mathsf{A}$ maps one-by-one from the set of orthogonal functions $\left\{\left|\psi_j\right\rangle\right\}$ in $H_S$ to the corresponding member of the set of orthogonal functions $\left\{\left|\phi_j\right\rangle\right\}$ in $H_R$ with coupling amplitude $s_j$ in each case. There is also a complementary mathematical relation that we can similarly deduce from Eq. (213)

$$\mathsf{A}^\dagger \left|\phi_k\right\rangle = s_k^* \left|\psi_k\right\rangle \tag{217}$$

We can also formally rewrite the SVD, Eq. (212), in matrix form. To do so, we can write a matrix that is diagonal on some basis $\left\{\left|\gamma_j\right\rangle\right\}$ as

$$\mathsf{D}_{diag} = \sum_j s_j \left|\gamma_j\right\rangle\left\langle\gamma_j\right| \tag{218}$$

where $s_j$ are the diagonal elements, and we can define two matrices

$$\mathsf{U} = \sum_p \left|\psi_p\right\rangle\left\langle\gamma_p\right| \text{ and } \mathsf{V} = \sum_q \left|\phi_q\right\rangle\left\langle\gamma_q\right| \tag{219}$$

both of which are technically unitary [161]. Then

$$\begin{aligned}\mathsf{V}\mathsf{D}_{diag}\mathsf{U}^\dagger &= \left(\sum_q \left|\phi_q\right\rangle\left\langle\gamma_q\right|\right)\left(\sum_j s_j \left|\gamma_j\right\rangle\left\langle\gamma_j\right|\right)\left(\sum_p \left|\gamma_p\right\rangle\left\langle\psi_p\right|\right) \\ &= \sum_{q,j,p} \left|\phi_q\right\rangle \delta_{qj} s_j \delta_{jp} \left\langle\psi_p\right| = \sum_j s_j \left|\phi_j\right\rangle\left\langle\psi_j\right| = \mathsf{A}\end{aligned} \tag{220}$$

Hence, an equivalent form for writing singular value decomposition is the "matrix" version

$$\mathsf{A} = \mathsf{V}\mathsf{D}_{diag}\mathsf{U}^\dagger \tag{221}$$

with $\mathsf{U}$ and $\mathsf{V}$ as in (219). In this case, we can view the $\left|\phi_j\right\rangle$ as being the columns of $\mathsf{V}$ and $\left\langle\psi_j\right|$ as being the rows of $\mathsf{U}^\dagger$ (or $\left|\psi_j\right\rangle$ as being the columns of $\mathsf{U}$).

# Appendix E Hilbert-Schmidt operators with weighted inner products

Suppose we have two possibly-different weighting operators $\mathsf{W}_R(\mathbf{r}_R)$ and $\mathsf{W}_S(\mathbf{r}_S)$ in source and receiving spaces respectively. We can now follow through an argument similar to that in section 6.8. Instead of (149), we can write the operator-weighted inner products

$$(\mu_S,\eta_S)_{\mathsf{W}_S} \equiv \int_{V_S} \mu_S^*(\mathbf{r}_S)\mathsf{W}_S(\mathbf{r}_S)\eta_S(\mathbf{r}_S)d^3\mathbf{r}_S \tag{222}$$

$$(\mu_R,\eta_R)_{\mathsf{W}_R} \equiv \int_{V_R} \mu_R^*(\mathbf{r}_R)\mathsf{W}_R(\mathbf{r}_R)\eta_R(\mathbf{r}_R)d^3\mathbf{r}_R \tag{223}$$

We also formally presume that, for finite functions $\mu_S$, $\eta_S$, $\mu_R$, and $\eta_R$, these inner products are finite. So, for some coupling operator (which may be a Green's function)

$$\mathsf{D}(\mathbf{r}_R;\mathbf{r}_S) = \sum_{p,q} d_{pq}(\cdot,\alpha_{Rp})_{\mathsf{W}_R}(\alpha_{Sq},\cdot)_{\mathsf{W}_S} \equiv \sum_{p,q} d_{pq}\alpha_{Rp}(\mathbf{r}_R)\alpha_{Sq}^*(\mathbf{r}_S) \tag{224}$$

We can find the $d_{ij}$ as usual by premultiplying by $\alpha_{Ri}^*(\mathbf{r}_R)$, postmultiplying by $\alpha_{Sj}(\mathbf{r}_S)$, we can integrate with the weighting operators as in

$$d_{ij} = \int_{V_S}\int_{V_R} \alpha_{Ri}^*(\mathbf{r}_R)\left[\mathsf{W}_R(\mathbf{r}_R)\mathsf{D}(\mathbf{r}_R;\mathbf{r}_S)\right]\left[\mathsf{W}_S(\mathbf{r}_S)\alpha_{Sj}(\mathbf{r}_S)\right]d^3\mathbf{r}_R d^3\mathbf{r}_S \tag{225}$$

Here we have used square brackets $[\cdots]$ to indicate explicitly that the weighting operators only have to operate on the function to their immediate right. Now consider the integral

$$S_D = \int_{V_S}\int_{V_R} \left[\mathsf{W}_S(\mathbf{r}_S)\mathsf{D}^\dagger(\mathbf{r}_S;\mathbf{r}_R)\right]\left[\mathsf{W}_R(\mathbf{r}_R)\mathsf{D}(\mathbf{r}_R;\mathbf{r}_S)\right]d^3\mathbf{r}_R d^3\mathbf{r}_S \tag{226}$$

where

$$\mathsf{D}^\dagger(\mathbf{r}_S;\mathbf{r}_R) = \sum_{m,n} d_{mn}^*\alpha_{Rm}^*(\mathbf{r}_R)\alpha_{Sn}(\mathbf{r}_S) \equiv \mathsf{D}^*(\mathbf{r}_R;\mathbf{r}_S) \tag{227}$$

By presumption, $\mathsf{W}_R(\mathbf{r}_R)\mathsf{D}(\mathbf{r}_R;\mathbf{r}_S)$ and $\mathsf{W}_S(\mathbf{r}_S)\mathsf{D}^\dagger(\mathbf{r}_S;\mathbf{r}_R)$ are finite. So the integral (226) of finite functions over finite volumes is finite. Then

$$\begin{aligned}
&\left[\mathsf{W}_S(\mathbf{r}_S)\mathsf{D}^\dagger(\mathbf{r}_S;\mathbf{r}_R)\right]\left[\mathsf{W}_R(\mathbf{r}_R)\mathsf{D}(\mathbf{r}_R;\mathbf{r}_S)\right] \\
&= \left\{\sum_{m,n} d_{mn}^*\left[\mathsf{W}_S(\mathbf{r}_S)\alpha_{Sn}(\mathbf{r}_S)\right]\alpha_{Rm}^*(\mathbf{r}_R)\right\}\left\{\sum_{p,q} d_{pq}\left[\mathsf{W}_R(\mathbf{r}_R)\alpha_{Rp}(\mathbf{r}_R)\right]\alpha_{Sq}^*(\mathbf{r}_S)\right\} \\
&= \sum_{m,n,p,q} d_{mn}^* d_{pq}\alpha_{Sq}^*(\mathbf{r}_S)\left[\mathsf{W}_S(\mathbf{r}_S)\alpha_{Sn}(\mathbf{r}_S)\right]\alpha_{Rm}^*(\mathbf{r}_R)\left[\mathsf{W}_R(\mathbf{r}_R)\alpha_{Rp}(\mathbf{r}_R)\right]
\end{aligned} \tag{228}$$

Formally integrating (228) over the two volumes, as in (226), gives

$$S_D = \sum_{m,n,p,q} d_{mn}^* d_{pq}\delta_{qn}\delta_{mp} = \sum_{p,q}\left|d_{pq}\right|^2 \tag{229}$$

which, since we know $S_D$ is finite, proves this is a Hilbert-Schmidt operator.

# Appendix F Electromagnetic Gauge, Green's functions and energy inner product

Here we derive Green's functions and energy inner products for electromagnetic fields in uniform media. The electromagnetism background is relatively standard (see, e.g., [41, 110, 116, 162 – 164]. However, we need a new "gauge" – the "M-gauge". This gauge lets us write all fields of interest just using the magnetic vector potential and clarifies that there are altogether only three independent field components for communications. That then allows a single, novel dyadic vector potential Green's function and a novel "energy" inner product for the electromagnetic field.

## F.1. Background electromagnetism

We presume an isotropic uniform medium, so with constant, scalar permeability $\mu$ and permittivity $\varepsilon$. We consider a free charge density $\rho$ and a corresponding "conduction" current density $\mathbf{J}$ of that charge. Then, we can write Maxwell's equations as

$$\text{(M1)} \qquad \nabla \cdot \varepsilon \mathbf{E} = \rho \tag{230}$$

$$\text{(M2)} \qquad \nabla \cdot \mathbf{B} = 0 \tag{231}$$

$$\text{(M3)} \qquad \nabla \times \mathbf{E} = -\frac{\partial \mathbf{B}}{\partial t} \tag{232}$$

$$\text{(M4)} \qquad \nabla \times (\mu)^{-1} \mathbf{B} = \mathbf{J} + \varepsilon \frac{\partial \mathbf{E}}{\partial t} \tag{233}$$

We can also explicitly write out the charge conservation condition

$$\nabla \cdot \mathbf{J} = -\frac{\partial \rho}{\partial t} \tag{234}$$

We want to know how many independent field components we need to describe communication. To reduce from the nominal 6 different scalar field quantities required to write the vector components of $\mathbf{E}$ and $\mathbf{B}$, we change first a description using the magnetic vector potential $\mathbf{A}$ and a scalar (electric) potential $\Phi$. To relate these to $\mathbf{E}$ and $\mathbf{B}$, we follow several standard steps. Since for any vector field $\mathbf{F}$

$$\nabla \cdot (\nabla \times \mathbf{F}) = 0 \tag{235}$$

then, since $\nabla \cdot \mathbf{B} = 0$ (M3) (Eq. (231)), we can write

$$\mathbf{B} = \nabla \times \mathbf{A} \tag{236}$$

where $\mathbf{A}$ is the *(magnetic) vector potential.* Next, from (M3) (Eq. (232), we now have

$$\nabla \times \mathbf{E} = -\frac{\partial}{\partial t} (\nabla \times \mathbf{A})$$

so, presuming we can interchange the order of differentiations, we can write

$$\nabla \times \left( \mathbf{E} + \frac{\partial}{\partial t} \mathbf{A} \right) = 0 \tag{237}$$

Now we also have the vector calculus identity

$$\nabla \times (\nabla F) = 0 \tag{238}$$

for any scalar field $F$. So, we can argue that any such field $\mathbf{E} + \partial \mathbf{A} / \partial t$ whose curl is zero can always be written therefore as the gradient of some other scalar function, i.e.,

$$\mathbf{E} + \frac{\partial}{\partial t} \mathbf{A} = -\nabla \Phi \tag{239}$$

for some scalar field $\Phi$, which we call the *scalar potential*. Rewriting Eq. (239), we have

$$\mathbf{E} = -\nabla \Phi - \frac{\partial}{\partial t} \mathbf{A} \tag{240}$$

These two equations (236) and (240) describe the magnetic and electric fields in terms of these two potentials. Using Eq. (240) in (M1) (Eq.(230)) gives (with constant isotropic $\varepsilon$)

$$\nabla\cdot\left(\nabla\Phi+\frac{\partial}{\partial t}\mathbf{A}\right)=-\frac{\rho}{\varepsilon} \tag{241}$$

Noting that $\nabla\cdot\left(\nabla F\right)$ is just another notation for $\nabla^2 F$ for any scalar field *F*, and interchanging the order of the derivatives, gives

$$\nabla^2\Phi+\frac{\partial}{\partial t}\left(\nabla\cdot\mathbf{A}\right)=-\frac{\rho}{\varepsilon} \tag{242}$$

From (M4) (Eq. (233)), and using Eq. (236) for **B** and Eq. (240) for **E**, we have

$$\nabla\times\left(\mu\right)^{-1}\nabla\times\mathbf{A}=\mathbf{J}+\varepsilon\frac{\partial}{\partial t}\left(-\nabla\Phi-\frac{\partial}{\partial t}\mathbf{A}\right) \tag{243}$$

Interchanging the order of the derivatives and rearranging gives

$$\nabla\times\left(\mu\right)^{-1}\nabla\times\mathbf{A}+\varepsilon\frac{\partial^2\mathbf{A}}{\partial t^2}+\varepsilon\frac{\partial\nabla\Phi}{\partial t}=\mathbf{J} \tag{244}$$

Since we are presuming that $\varepsilon$ and $\mu$ are simply constants, then Eq. (244) becomes

$$\nabla\times\nabla\times\mathbf{A}+\varepsilon\mu\frac{\partial^2\mathbf{A}}{\partial t^2}+\varepsilon\mu\frac{\partial\nabla\Phi}{\partial t}=\mu\mathbf{J} \tag{245}$$

Using the vector identity

$$\nabla\times\left(\nabla\times\mathbf{F}\right)\equiv\nabla\left(\nabla\cdot\mathbf{F}\right)-\nabla^2\mathbf{F} \tag{246}$$

it is common to rearrange Eq. (245) to obtain

$$\nabla^2\mathbf{A}-\varepsilon\mu\frac{\partial^2\mathbf{A}}{\partial t^2}-\nabla\left(\nabla\cdot\mathbf{A}+\varepsilon\mu\frac{\partial\Phi}{\partial t}\right)=-\mu\mathbf{J} \tag{247}$$

though we will deliberately *not* take this approach below.

These equations (242) and (245) (or (247)) are still coupled between the two potentials **A** and Φ. To simplify further, we need to choose a *gauge* for these potentials – i.e., the specific choice of **A** and Φ for given **E** and **B** fields. To understand why we need to make a choice, note, first, that we could represent a specific **B** field using a first or "old" choice $\mathbf{A}_{old}$, with $\mathbf{B}=\nabla\times\mathbf{A}_{old}$ as in Eq. (236). Because of the vector calculus identity $\nabla\times\left(\nabla F\right)=0$ (Eq. (238)), we could add the gradient of some scalar function $\Psi$, called the *gauge function*, to $\mathbf{A}_{old}$ to create a new vector potential

$$\mathbf{A}_{new}=\mathbf{A}_{old}+\nabla\Psi \tag{248}$$

without making any change to the magnetic field **B**; specifically,

$$\mathbf{B}=\nabla\times\mathbf{A}_{old}=\nabla\times\mathbf{A}_{new} \tag{249}$$

However, if we make no further changes, we see from Eq. (240) that the new vector potential would add a term $-\partial\nabla\Psi/\partial t$ to the electric field **E**. To avoid this, therefore, we can add a term $-\partial\Psi/\partial t$ to the potential $\Phi$; that is, we write

$$\Phi_{new}=\Phi_{old}-\frac{\partial\Psi}{\partial t} \tag{250}$$

We can check this explicitly by calculating **E** using $\Phi_{new}$ and $\mathbf{A}_{new}$; that is,

$$\begin{aligned}\mathbf{E}&=-\nabla\Phi_{new}-\frac{\partial}{\partial t}\mathbf{A}_{new}=-\nabla\left(\Phi_{old}-\frac{\partial\Psi}{\partial t}\right)-\frac{\partial}{\partial t}\left(\mathbf{A}_{old}+\nabla\Psi\right)\\&=-\nabla\Phi_{old}-\frac{\partial}{\partial t}\mathbf{A}_{old}-\nabla\frac{\partial\Psi}{\partial t}+\frac{\partial}{\partial t}\nabla\Psi=-\nabla\Phi_{old}-\frac{\partial}{\partial t}\mathbf{A}_{old}\end{aligned} \tag{251}$$

where we presume we can interchange the order of temporal and spatial derivatives. So, Eq. (251) shows explicitly that we are free to choose the gauge function $\Psi$ as long as we use the transformation rules Eqs. (248) and (250). A new choice of gauge function $\Psi$ gives a *gauge transformation*.

## F.2. Choosing a gauge for communications problems

We want to choose a gauge that leaves only the minimum number of field components so that we can count modes properly. The Coulomb and Lorentz gauges are particularly common (see, e.g., [116, 162, 163]). Typically, a gauge is set in practice by some choice for $\nabla \cdot \mathbf{A}$. For the Coulomb gauge (with subscript "*C*"), that choice is

$$\nabla \cdot \mathbf{A}_C = 0 \tag{252}$$

and for the Lorentz gauge (with subscript "*L*")

$$\nabla \cdot \mathbf{A}_L = -\frac{1}{c^2}\frac{\partial \Phi_L}{\partial t} \tag{253}$$

Many other choices are possible (see, e.g., [165, 166]). Using each such choice in the wave equation (247) leads to four "scalar" equations or their equivalent, one wave equation for each vector component of **A**, and another equation for the scalar potential $\Phi$. In the Coulomb and the Lorenz gauges, the driving source terms are the charge density $\rho$ and the current density **J**. Immediately, this tells us that there are no more than 4 independent functions required to specify any field that we would create as a result – one for $\rho$ and one each for the three vector components of **J**. However, the charge density $\rho$ and the current density **J** are linked by conservation of charge, Eq. (234), so even these 4 may be too many. There are also technical problems with the Coulomb gauge in particular [142, 143] as we describe waves coming from sources.

### F.2.1. A gauge for communications – the M-gauge

We are interested here in sending changing fields from a source. We can, without significant restriction, suppose that before some time $t_o$, all electric and magnetic fields have been constant, and that we have a known charge density $\rho_o(\mathbf{r})$ that has been fixed up to this point. (We can also presume all effective magnetic currents $\mathbf{J}_m(\mathbf{r})$ have been stable and fixed, and that we have had other fixed "solenoidal" current densities $\mathbf{J}_o(\mathbf{r})$, i.e., ones for which $\nabla \cdot \mathbf{J}_o = 0$). So before time $t_o$, we have some electrostatic field $\mathbf{E}_o(\mathbf{r})$ that we could obtain by a solution of Maxwell's first equation (M1) with this charge density $\rho_o(\mathbf{r})$, as well as possibly some magnetostatic field $\mathbf{B}_o(\mathbf{r})$ from Maxwell's fourth equation (M4), with some (fixed) $\mathbf{J}_o(\mathbf{r})$ and $\mathbf{J}_m(\mathbf{r})$.

Now we presume that after time $t_o$, we have some new additional current density $\mathbf{J}(\mathbf{r},t)$. These new currents give changes $\Delta\rho(\mathbf{r},t)$ in charge density that necessarily and only result from such currents, so we should not need an additional independent driving term corresponding to the change in charge density. The changes in the electromagnetic field, and any propagating components of that, should result only from these currents $\mathbf{J}(\mathbf{r},t)$.

Now our task is to construct a new gauge, which, using the "m" from "coMmunications", we call the "M" gauge [167], with the only time-dependent driving terms being from the three vector components of the current density $\mathbf{J}(\mathbf{r},t)$.The key to this gauge is to choose the scalar potential as being associated just with the original fixed charge density $\rho_o(\mathbf{r})$, so it obeys a simple, fixed Poisson equation

$$\nabla^2 \Phi_M(\mathbf{r}) = -\frac{\rho_o(\mathbf{r})}{\varepsilon} \tag{254}$$

which, within an arbitrary additive constant, has the solution

$$\Phi_M(\mathbf{r}) = \frac{1}{4\pi\varepsilon}\int \frac{\rho_o(\mathbf{r}')}{|\mathbf{r}-\mathbf{r}'|}d^3\mathbf{r}' \tag{255}$$

Because neither $\rho_o$ or $\Phi_M$ has any time dependence, we therefore have

$$\frac{\partial \Phi_M}{\partial t} = 0 \tag{256}$$

and, trivially,

$$\nabla \frac{\partial \Phi_M}{\partial t} = 0 \tag{257}$$

We have, as usual in any gauge (Eq. (240))

$$\mathbf{E} = -\nabla\Phi_M - \frac{\partial \mathbf{A}_M}{\partial t} \tag{258}$$

With the fixed electrostatic field

$$\mathbf{E}_o = -\nabla\Phi_M \tag{259}$$

we could write

$$\mathbf{E} = \mathbf{E}_o + \mathbf{E}_M \tag{260}$$

with $\mathbf{E}_M = -\partial \mathbf{A}_M / \partial t$ (Eq. (164) with $\mathbf{E}_M$ containing all the time-varying and propagating electric fields [168]. Similarly, we could write

$$\mathbf{B} = \mathbf{B}_M + \mathbf{B}_o \tag{261}$$

with $\mathbf{B}_M = \nabla \times \mathbf{A}_M$ (Eq. (165)) as usual for gauge potentials.

Generally, a gauge is practically defined by the choice of $\nabla \cdot \mathbf{A}$, and we need this result later. We give the formal derivation of this below in **Appendix G**, with the result

$$\nabla \cdot \mathbf{A}_M(\mathbf{r},t) = \frac{1}{\varepsilon}\int_{t'=t_o}^{t}\int_{t''=t_o}^{t'} \nabla \cdot \mathbf{J}(\mathbf{r},t'')\,dt''dt' \tag{262}$$

### F.2.2. Wave equations in the M-gauge

Now returning to Eq. (245) and using Eq. (257) to eliminate the term $\varepsilon\mu(\partial\nabla\Phi/\partial t)$ because it is zero in this gauge, we obtain the wave equation

$$\nabla \times \nabla \times \mathbf{A}_M + \varepsilon\mu\frac{\partial^2 \mathbf{A}_M}{\partial t^2} = \mu\mathbf{J} \tag{263}$$

which lets us define what will be the phase velocity $v = \sqrt{1/\varepsilon\mu}$ (Eq. (168)).

Often we are most interested in monochromatic fields – that is, fields whose time dependence can be written in the form $\sin(\omega t)$, $\cos(\omega t)$, $\exp(i\omega t)$, or $\exp(-i\omega t)$, or linear combinations of these, with some specific choice of the angular frequency $\omega$. In this case,

$$\frac{\partial^2 \mathbf{A}_M}{\partial t^2} = -\omega^2 \mathbf{A}_M \tag{264}$$

For the definiteness and simplicity, for such monochromatic fields we can use the form

$$\mathbf{A}_M(\mathbf{r},\mathbf{t}) = \mathbf{A}_{\omega M}(\mathbf{r})\exp(-i\omega t) + \mathbf{A}^*_{\omega M}(\mathbf{r})\exp(i\omega t) \equiv \mathbf{A}_{\omega M}(\mathbf{r})\exp(-i\omega t) + c.c. \tag{265}$$

where "*c. c.*" stands for "complex conjugate. The addition of the complex conjugate ensures that the field $\mathbf{A}_M(\mathbf{r},t)$ is real. In Eq. (265) $\mathbf{A}_{\omega M}(\mathbf{r})$ is in general a complex amplitude function (which therefore holds any phase information for the fields). As is common, we perform the algebra for the field $\mathbf{A}_{\omega M}(\mathbf{r})\exp(-i\omega t)$. If necessary to get back to real fields, we can formally repeat the calculation with the complex conjugate of the form in Eq. (265), and add the two. In the monochromatic case, therefore, instead of Eq. (263) we have $\nabla \times \nabla \times \mathbf{A}_{\omega M} - k^2 \mathbf{A}_{\omega M} = \mu \mathbf{J}_{\omega}$ (Eq. (166)), with $k^2 = \omega^2 \varepsilon\mu \equiv \omega^2 / v^2$ (Eq. (167), and $\mathbf{J}(\mathbf{r},t) = \mathbf{J}_{\omega}(\mathbf{r})\exp(-i\omega t) + c.c.$(Eq. (169)).

So, the answer to the number of independent functions required to specify an electromagnetic field is that there are three such functions of space and time for all communications purposes, plus other functions of space only to specify background static fields. Our choice of gauge has separated these out, as desired. The consequent price is that is that the resulting wave equation (263) has a $\nabla \times \nabla \times$ spatial derivative rather than a more common $\nabla^2$ one [169]. However, this is manageable and even useful, as we show below.

## F.3. Dyadic Green's function for the vector potential in the M-gauge

Approaches to the mathematical Green's function solutions for **A** for equations of the form of (263) or (166) can be repurposed from discussions of "dyadic" Green's functions that can be derived directly for **E** and **B** in electromagnetism [41, 110, 170 - 173]. A dyadic Green's function expresses that fact that the resulting vector field or vector potential function may not be parallel to the vector source that generates it [174]. We give a brief tutorial introduction to dyadics [175], together with some necessary identities, in **Appendix H** below.

### F.3.1. Derivation of general form for monochromatic waves

In part because the vector and dyadic calculus becomes somewhat involved here, we will consider the monochromatic version of the wave equation, Eq.(166), first. We now propose a dyadic Green's function $\overline{\overline{G}}_{\omega M}(\mathbf{r};\mathbf{r}')$ that, starting from some vector "point source" element at position $\mathbf{r}'$ will lead to some vector wave (not necessarily in the same vector direction) at position **r**. Following the usual approach with Green's functions, we write in this dyadic case

$$\mathbf{A}_{\omega M}(\mathbf{r}) = \mu \int_V \overline{\overline{G}}_{\omega M}(\mathbf{r};\mathbf{r}') \cdot \mathbf{J}_{\omega}(\mathbf{r}') d^3\mathbf{r}' \tag{266}$$

and we formally write the corresponding wave equation for the dyadic Green's function as

$$\nabla \times \nabla \times \overline{\overline{G}}_{\omega M} - k^2 \overline{\overline{G}}_{\omega M} = \overline{\overline{I}}\delta(\mathbf{r}-\mathbf{r}') \tag{267}$$

Note the presence of the idem factor (unit dyadic) $\overline{\overline{I}}$ on the right in Eq. (267), as required to make the right hand side a dyadic entity to match the left hand side. Taking the divergence of both sides of Eq. (267), and noting that the divergence of the curl of a function is necessarily zero (a result that also works in dyadic form [170]), we have

$$-k^2 \nabla \cdot \overline{\overline{G}}_{\omega M} = \nabla \cdot \left(\overline{\overline{I}}\delta(\mathbf{r}-\mathbf{r}')\right) \tag{268}$$

So, using the identity Eq. (345) and rearranging, we have [176]

$$\nabla \cdot \overline{\overline{G}}_{\omega M} = -\frac{1}{k^2} \nabla \delta(\mathbf{r}-\mathbf{r}') \tag{269}$$

Now, using the identity Eq.(246), which also works in dyadic form [170], Eq. (267) becomes

$$\nabla^2 \overline{\overline{G}}_{\omega M} - \nabla\left(\nabla \cdot \overline{\overline{G}}_{\omega M}\right) + k^2 \overline{\overline{G}}_{\omega M} = -\overline{\overline{I}}\delta(\mathbf{r}-\mathbf{r}') \tag{270}$$

and we can substitute using Eq. (269) for $\nabla \cdot \overline{\overline{G}}_{\omega M}$ to obtain, after rearrangement,

$$(\nabla^2 + k^2)\overline{\overline{G}}_{\omega M} = -\left(\overline{\overline{I}} + \frac{1}{k^2}\nabla\nabla\right)\delta(\mathbf{r}-\mathbf{r}') \tag{271}$$

Now let us propose a form for $\overline{\overline{G}}_{\omega M}$ , and check to see whether it can be a solution to Eq. (271), and hence to (267). Specifically, we propose

$$\overline{\overline{G}}_{\omega M}\left(\mathbf{r};\mathbf{r}'\right)=-\left(\overline{\overline{I}}+\frac{1}{k^2}\nabla\nabla\right)g\left(\mathbf{r};\mathbf{r}'\right) \tag{272}$$

where $g\left(\mathbf{r};\mathbf{r}'\right)$ is a scalar function to be determined. Substituting this into Eq. (271) gives

$$\left(\nabla^2+k^2\right)\left(\overline{\overline{I}}+\frac{1}{k^2}\nabla\nabla\right)g\left(\mathbf{r};\mathbf{r}'\right)=\left(\overline{\overline{I}}+\frac{1}{k^2}\nabla\nabla\right)\delta\left(\mathbf{r}-\mathbf{r}'\right) \tag{273}$$

Exchanging order of terms and derivatives on the left gives

$$\left(\overline{\overline{I}}+\frac{1}{k^2}\nabla\nabla\right)\left[\left(\nabla^2+k^2\right)g\left(\mathbf{r};\mathbf{r}'\right)\right]=\left(\overline{\overline{I}}+\frac{1}{k^2}\nabla\nabla\right)\delta\left(\mathbf{r}-\mathbf{r}'\right) \tag{274}$$

This equation can be satisfied if

$$\left(\nabla^2+k^2\right)g\left(\mathbf{r};\mathbf{r}'\right)=\delta\left(\mathbf{r}-\mathbf{r}'\right) \tag{275}$$

But this is just the Green's function equation for a scalar wave equation, as in Eq. (4), so

$$g\left(\mathbf{r};\mathbf{r}'\right)\equiv G_\omega\left(\mathbf{r};\mathbf{r}'\right)=-\frac{1}{4\pi}\frac{\exp\left(ik\left|\mathbf{r}-\mathbf{r}'\right|\right)}{\left|\mathbf{r}-\mathbf{r}'\right|} \tag{276}$$

So, substituting back into Eq. (272) we have

$$\overline{\overline{G}}_{\omega M}\left(\mathbf{r};\mathbf{r}'\right)=-\left(\overline{\overline{I}}+\frac{1}{k^2}\nabla\nabla\right)G_\omega\left(\mathbf{r};\mathbf{r}'\right)=\frac{1}{4\pi}\left(\overline{\overline{I}}+\frac{1}{k^2}\nabla\nabla\right)\frac{\exp\left(ik\left|\mathbf{r}-\mathbf{r}'\right|\right)}{\left|\mathbf{r}-\mathbf{r}'\right|} \tag{277}$$

Hence, we now have a relatively straightforward expression for the Green's function of the "$\nabla\times\nabla\times$" wave equation (166).

### F.3.2. Explicit form for the dyadic Green's function for monochromatic waves

In the Green's function in Eq. (277) the only variable is $R=\left|\mathbf{r}-\mathbf{r}'\right|$, which we can regard as a radius variable in spherical coordinates around the fixed point $\mathbf{r}'$. So, with no dependence on the $\theta$ and $\phi$ coordinates

$$\nabla G_\omega\left(\mathbf{r};\mathbf{r}'\right)\equiv\left[\frac{d}{dR}G_\omega\left(R\right)\right]\nabla R \tag{278}$$

Now

$$\frac{d}{dR}G_\omega\left(R\right)=-\frac{1}{4\pi}\frac{d}{dR}\left[\frac{\exp\left(ikR\right)}{R}\right]=-\frac{1}{4\pi}\left(\frac{ik}{R}-\frac{1}{R^2}\right)\exp\left(ikR\right)=\left(ik-\frac{1}{R}\right)G_\omega\left(R\right) \tag{279}$$

So, from Eq. (278), and using the identity $\nabla R=\hat{\mathbf{R}}$ (Eq. (350))

$$\nabla G_\omega\left(\mathbf{r};\mathbf{r}'\right)=\left(ik-\frac{1}{R}\right)G_\omega\left(R\right)\hat{\mathbf{R}} \tag{280}$$

Continuing,

$$\nabla\nabla G_\omega\left(\mathbf{r};\mathbf{r}'\right)=\nabla\left[\left(ik-\frac{1}{R}\right)G_\omega\left(R\right)\right]\hat{\mathbf{R}}+\left(ik-\frac{1}{R}\right)G_\omega\left(R\right)\nabla\hat{\mathbf{R}} \tag{281}$$

where we use $\nabla\mathbf{R} = \overline{\overline{I}}$ (see Eq. (351)). We can now progressively work out the remaining parts of this expression (281).

$$\begin{aligned}\nabla\left[\left(ik - \frac{1}{R}\right)G_\omega(R)\right] &= -G_\omega(R)\nabla\left(\frac{1}{R}\right) + \left(ik - \frac{1}{R}\right)\nabla G_\omega(R) \\ &= G_\omega(R)\frac{\hat{\mathbf{R}}}{R^2} + \left(ik - \frac{1}{R}\right)^2 G_\omega(R)\hat{\mathbf{R}} = \left[\left(ik - \frac{1}{R}\right)^2 + \frac{1}{R^2}\right]G_\omega(R)\hat{\mathbf{R}}\end{aligned} \quad (282)$$

where we have used the result $\nabla(1/R) == -\hat{\mathbf{R}}/R^2$ (Eq. (352)). Substituting using Eq. (353) ( $\nabla\hat{\mathbf{R}} = (1/R)\left(\overline{\overline{I}} - \hat{\mathbf{R}}\hat{\mathbf{R}}\right)$ ) and (282), we can therefore rewrite Eq. (281) as

$$\begin{aligned}\nabla\nabla G_\omega(\mathbf{r};\mathbf{r}') &= \left\{\left[\left(ik - \frac{1}{R}\right)^2 + \frac{1}{R^2}\right]\hat{\mathbf{R}}\hat{\mathbf{R}} + \left(ik - \frac{1}{R}\right)\frac{1}{R}\left(\overline{\overline{I}} - \hat{\mathbf{R}}\hat{\mathbf{R}}\right)\right\}G_\omega(R) \\ &= \left[-k^2\hat{\mathbf{R}}\hat{\mathbf{R}} + \frac{ik}{R}\left(\overline{\overline{I}} - 3\hat{\mathbf{R}}\hat{\mathbf{R}}\right) - \frac{1}{R^2}\left(\overline{\overline{I}} - 3\hat{\mathbf{R}}\hat{\mathbf{R}}\right)\right]G_\omega(R)\end{aligned} \quad (283)$$

So, finally, using Eq. (283) in Eq. (277) gives

$$\begin{aligned}\overline{\overline{G}}_{\omega M}(\mathbf{r};\mathbf{r}') &= -\left[\overline{\overline{I}} - \hat{\mathbf{R}}\hat{\mathbf{R}} + \frac{i}{kR}\left(\overline{\overline{I}} - 3\hat{\mathbf{R}}\hat{\mathbf{R}}\right) - \frac{1}{k^2R^2}\left(\overline{\overline{I}} - 3\hat{\mathbf{R}}\hat{\mathbf{R}}\right)\right]G_\omega(R) \\ &= -\left[\overline{\overline{I}} - \hat{\mathbf{R}}\hat{\mathbf{R}} + \frac{1}{kR}\left(i - \frac{1}{kR}\right)\left(\overline{\overline{I}} - 3\hat{\mathbf{R}}\hat{\mathbf{R}}\right)\right]G_\omega(R)\end{aligned} \quad (284)$$

We see from Eq. (284) that we have two different kinds of terms. Specifically, writing $\overline{\overline{G}}_{\omega M} = \overline{\overline{G}}_{\omega MP} + \overline{\overline{G}}_{\omega MN}$ (Eq. (172)) we can define a "propagating" Green's function

$$\overline{\overline{G}}_{\omega MP} = -\left(\overline{\overline{I}} - \hat{\mathbf{R}}\hat{\mathbf{R}}\right)G_\omega(R) \quad (285)$$

and a "near-field" Green's function

$$\overline{\overline{G}}_{\omega MN} = -\left[\frac{1}{kR}\left(i - \frac{1}{kR}\right)\left(\overline{\overline{I}} - 3\hat{\mathbf{R}}\hat{\mathbf{R}}\right)\right]G_\omega(R) \quad (286)$$

The magnitude of the propagating Green's function $\overline{\overline{G}}_{\omega MP}$ falls off as $1/R$, from the $1/R$ dependence of the scalar Green's function $G_\omega(R)$, and its behavior is characteristic of a propagating wave. The "near-field" Green's function, by contrast, consists only of terms whose magnitude is falling off as $1/R^2$ or $1/R^3$, which are characteristic of near-field fields, not propagating ones.

We can usefully rewrite these Green's functions Eqs. (285) and (286) by rewriting the unit dyadic $\overline{\overline{I}}$ using a coordinate direction $\hat{\mathbf{R}}$ and two other directions, given by unit vectors $\hat{\mathbf{e}}_1$ and $\hat{\mathbf{e}}_2$, which are perpendicular to each other and to $\hat{\mathbf{R}}$, giving (as in Eq. (338))

$$\overline{\overline{I}} = \hat{\mathbf{R}}\hat{\mathbf{R}} + \hat{\mathbf{e}}_1\hat{\mathbf{e}}_1 + \hat{\mathbf{e}}_2\hat{\mathbf{e}}_2 \quad (287)$$

so, substituting Eq. (287) in Eq. (285) gives $\overline{\overline{G}}_{\omega MP} = -\left(\hat{\mathbf{e}}_1\hat{\mathbf{e}}_1 + \hat{\mathbf{e}}_2\hat{\mathbf{e}}_2\right)G_\omega(R)$ (Eq. (170)).

So, in this propagating Green's function there is never any component of the vector potential in the direction of propagation (the "radial" direction $\hat{\mathbf{R}}$) (i.e., there is no "longitudinal" propagating vector potential wave); all the propagating vector potential waves are "transverse".

By contrast, the "near-field" Green's function actually has a larger magnitude in the "longitudinal" radial direction $\hat{\mathbf{R}}$ than in each of the other two transverse directions $\hat{\mathbf{e}}_1$ and $\hat{\mathbf{e}}_2$; specifically, substituting using (287) in Eq. (286), gives (Eq. (173))

$$\overline{\overline{G}}_{\omega MN} = \left[\frac{1}{kR}\left(i - \frac{1}{kR}\right)\left(2\hat{\mathbf{R}}\hat{\mathbf{R}} - \hat{\mathbf{e}}_1\hat{\mathbf{e}}_1 - \hat{\mathbf{e}}_2\hat{\mathbf{e}}_2\right)\right] G_\omega(R)$$

### F.3.3. Green's functions for general time-dependent waves

We can follow through a similar analysis for the full time-dependent case, based on a scalar retarded Green's function

$$G(\mathbf{r},t;\mathbf{r}',t') = -\frac{1}{4\pi}\frac{\delta\left(t - t' - |\mathbf{r}-\mathbf{r}'|/v\right)}{|\mathbf{r}-\mathbf{r}'|} \tag{288}$$

For reasons of space, we omit the detailed derivation here, but the dyadic aspects are all similar. The resulting Green's function of the full time-dependent wave equation Eq. (263) also can usefully be written (using the notation also of (287)) as a sum

$$\overline{\overline{G}}_M = \overline{\overline{G}}_{MP} + \overline{\overline{G}}_{MN} \tag{289}$$

of a propagating term

$$\begin{aligned}\overline{\overline{G}}_{MP}(\mathbf{r},t;\mathbf{r}',t') &= (1/4\pi R)\left(\overline{\overline{I}} - \hat{\mathbf{R}}\hat{\mathbf{R}}\right)\delta(t - t' - R/v) = -\left(\overline{\overline{I}} - \hat{\mathbf{R}}\hat{\mathbf{R}}\right)G(\mathbf{r},t;\mathbf{r}',t') \\ &= -\left(\hat{\mathbf{e}}_1\hat{\mathbf{e}}_1 + \hat{\mathbf{e}}_2\hat{\mathbf{e}}_2\right)G(\mathbf{r},t;\mathbf{r}',t')\end{aligned} \tag{290}$$

and a near- field term

$$\begin{aligned}\overline{\overline{G}}_{MN}(\mathbf{r},t;\mathbf{r}',t') &= \frac{v}{4\pi R^2}\left(\overline{\overline{I}} - 3\hat{\mathbf{R}}\hat{\mathbf{R}}\right)\left[\Theta\left(t - t' - |\mathbf{r}-\mathbf{r}'|/v\right) + \frac{v}{R}\Xi\left(t - t' - |\mathbf{r}-\mathbf{r}'|/v\right)\right] \\ &= -\frac{v}{4\pi R^2}\left(\left(2\hat{\mathbf{R}}\hat{\mathbf{R}} - \hat{\mathbf{e}}_1\hat{\mathbf{e}}_1 - \hat{\mathbf{e}}_2\hat{\mathbf{e}}_2\right)\right)\left[\Theta\left(t - t' - |\mathbf{r}-\mathbf{r}'|/v\right) + \frac{v}{R}\Xi\left(t - t' - |\mathbf{r}-\mathbf{r}'|/v\right)\right]\end{aligned} \tag{291}$$

where $\Theta(a)$ and $\Xi(a)$ are the Heaviside and "ramp" functions, respectively, defined as

$$\Theta(a) = \begin{cases}1, & a \geq 0 \\ 0, & a < 0\end{cases} \quad \text{and} \quad \Xi(a) = \begin{cases}a, & a \geq 0 \\ 0, & a < 0\end{cases} \tag{292}$$

and we use the formal results

$$\int_{t_a = t_o}^{t_b} \delta(t_a - t' - R/v)\,dt_a = \Theta(t_b - t' - R/v) \tag{293}$$

$$\int_{t_b = t_o}^{t}\int_{t_a = t_o}^{t_b} \delta(t_a - t' - R/v)\,dt_a dt_b = \int_{t_b = t_o}^{t} \Theta(t_b - t' - R/v)\,dt_b = \Xi(t - t' - R/v) \tag{294}$$

Just as for the monochromatic case, the propagating term gives "transverse" vector potential waves.

### F.3.4. Green's functions for the electric and magnetic fields

The Green's function formalism we have set up so far for the vector potential in the M-gauge is complete for describing the electromagnetic field resulting from (time-varying) current sources (and any corresponding changing charge distributions). It can therefore be used to derive dyadic Green's functions for the electric and magnetic fields, based on the relations Eqs (164) and (165) that give the electric and magnetic fields from the vector potential in this gauge. For reasons of space, we omit this here. Also, these can be derived directly from Maxwell's equations (see, e.g., [41, 170]). As we would expect, the choice of gauge does not make any difference to the result for the electric and magnetic fields. The point of our approach with the M-gauge is to clarify how many independent variables and fields there are and to give an inner product form that is directly suitable for use in functional analysis, which we derive next.

## F.4. Energy inner product for the vector potential

Using the M-gauge, we can conveniently set up an inner product for the electromagnetic field, and this can be an "energy" inner product. With this inner product to define orthogonality and energy, the total energy of a field is the sum of the energies of its orthogonal components.

### F.4.1. Expressions for energy density in electromagnetic fields

#### F.4.1.1. General time-dependent form

A standard expression for the energy density in an electromagnetic field is ([162], p. 259)

$$u=(1/2)(\mathbf{E}\cdot\mathbf{D}+\mathbf{B}\cdot\mathbf{H}) \tag{295}$$

In a lossless, uniform, isotropic medium with dielectric constant $\varepsilon$ and magnetic permeability $\mu$, this can be rewritten as $u=(1/2)(\varepsilon\mathbf{E}\cdot\mathbf{E}+\mu^{-1}\mathbf{B}\cdot\mathbf{B})$ (Eq. (177)).

#### F.4.1.2. Monochromatic form

If we are considering a monochromatic field at angular frequency $\omega$, we can write

$$\mathbf{E}(\mathbf{r},t)=\mathbf{E}_{o\omega}(\mathbf{r})\cos(\omega t+\theta_e) \tag{296}$$

$$\mathbf{B}(\mathbf{r},t)=\mathbf{B}_{o\omega}(\mathbf{r})\cos(\omega t+\theta_m) \tag{297}$$

where $\theta_e$ and $\theta_m$ are phase angles [177], and where both $\mathbf{E}_{o\omega}(\mathbf{r})$ and $\mathbf{B}_{o\omega}(\mathbf{r})$ are real [178]. Then the energy density of Eq. (177), now considered to be averaged over a cycle (which introduces a factor of ½) [179], can be written

$$u=(1/4)(\varepsilon\mathbf{E}_{o\omega}\cdot\mathbf{E}_{o\omega}+\mu^{-1}\mathbf{B}_{o\omega}\cdot\mathbf{B}_{o\omega}) \tag{298}$$

However, we also can conveniently write [180] (e.g., for work in quantum mechanics)

$$\mathbf{E}(\mathbf{r},t)=\mathbf{E}_{\omega}(\mathbf{r})\exp(-i\omega t)+\mathbf{E}_{\omega}^{*}(\mathbf{r})\exp(i\omega t)\equiv\mathbf{E}_{\omega}(\mathbf{r})\exp(-i\omega t)+c.c. \tag{299}$$

$$\mathbf{B}(\mathbf{r},t)=\mathbf{B}_{\omega}(\mathbf{r})\exp(-i\omega t)+c.c. \tag{300}$$

where $\mathbf{E}_{\omega}$ and $\mathbf{B}_{\omega}$ are complex numbers, incorporating the phase shifts, i.e.,

$$\mathbf{E}_{\omega}(\mathbf{r})=(\mathbf{E}_{o\omega}(\mathbf{r})/2)\exp(-i\theta_e) \tag{301}$$

$$\mathbf{B}_{\omega}(\mathbf{r})=(\mathbf{B}_{o\omega}(\mathbf{r})/2)\exp(-i\theta_m) \tag{302}$$

We will then get the same answer as in Eq. (298) for the (time-averaged) energy densities if we write with our new $\mathbf{E}_{\omega}$ and $\mathbf{B}_{\omega}$, instead of Eq. (298),

$$u=\varepsilon\mathbf{E}_{\omega}^{*}\cdot\mathbf{E}_{\omega}+\mu^{-1}\mathbf{B}_{\omega}^{*}\cdot\mathbf{B}_{\omega} \tag{303}$$

Below we will need the monochromatic forms of the vector potential. We make similar definitions to those for the electric and magnetic fields. So, with a real monochromatic vector potential that we could write, analogously to Eqs. (296) and (297), as

$$\mathbf{A}_M(\mathbf{r},t)=\mathbf{A}_{o\omega M}(\mathbf{r})\cos(\omega t+\theta_M) \tag{304}$$

for a phase angle $\theta_M$ (which may also be a function of position) and, analogously to Eqs. (299) and (300), we can write Eq. (265) ($\mathbf{A}_M(\mathbf{r},\mathbf{t})=\mathbf{A}_{\omega M}(\mathbf{r})\exp(-i\omega t)+c.c.$), where, analogously to Eqs. (301) and (302)

$$A_{\omega M}(\mathbf{r}) = \left(\mathbf{A}_{o\omega M}(\mathbf{r})/2\right)\exp\left(-i\theta_M\right) \tag{305}$$

## F.4.2. Inner product form

Our goal here is to construct an inner product that works with the electromagnetic field expressed through the vector potential. Using the vector potential in the M-gauge, we can deduce the entire electromagnetic field responsible for any communications, as expressed by $\mathbf{E}_M$ and $\mathbf{B}_M$ .

### F.4.2.1. General time-dependent form

Now, immediately from Eq. (177), and using the expressions for the electric field, Eq. (164), and the magnetic field, Eq. (165), in the M-gauge, we have

$$u = \frac{1}{2}\left(\varepsilon\left[-\frac{\partial \mathbf{A}_M^*}{\partial t}\right]\cdot\left[-\frac{\partial \mathbf{A}_M}{\partial t}\right] + \frac{1}{\mu}\left[\nabla\times\mathbf{A}_M^*\right]\cdot\left[\nabla\times\mathbf{A}_M\right]\right) \tag{306}$$

Now, in component form, with

$$\mathbf{A}_M \equiv \hat{\mathbf{x}}_1 A_{M1} + \hat{\mathbf{x}}_2 A_{M2} + \hat{\mathbf{x}}_3 A_{M3} \tag{307}$$

then

$$-\frac{\partial \mathbf{A}_M}{\partial t} = -\hat{\mathbf{x}}_1\frac{\partial A_{M1}}{\partial t} - \hat{\mathbf{x}}_2\frac{\partial A_{M2}}{\partial t} - \hat{\mathbf{x}}_3\frac{\partial A_{M3}}{\partial t} \tag{308}$$

and

$$\nabla\times\mathbf{A}_M = \hat{\mathbf{x}}_1\left(\frac{\partial A_{M3}}{\partial x_2} - \frac{\partial A_{M2}}{\partial x_3}\right) + \hat{\mathbf{x}}_2\left(\frac{\partial A_{M1}}{\partial x_3} - \frac{\partial A_{M3}}{\partial x_1}\right) + \hat{\mathbf{x}}_3\left(\frac{\partial A_{M2}}{\partial x_1} - \frac{\partial A_{M1}}{\partial x_2}\right) \tag{309}$$

So, now we can construct a $6\times 3$ matrix, which we will call $\sqrt{\mathsf{U}}$ for a reason that will become apparent later. The idea of this matrix is that it will operate on the three vector components of $\mathbf{A}_M$ , expressed as a 3-element column vector, to generate a six-element column vector whose first three elements are the three vector components of $\mathbf{E}$ (weighted by $\sqrt{\varepsilon/2}$ ) and whose second three elements are the three vector components of $\mathbf{B}$ (weighted by $\sqrt{1/2\mu}$ ). This matrix then becomes (inserting dashed lines to separate elements explicitly in the matrix)

$$\sqrt{\mathsf{U}} = \frac{1}{\sqrt{2}}\begin{bmatrix} -\sqrt{\varepsilon}\frac{\partial}{\partial t} & 0 & 0 \\ 0 & -\sqrt{\varepsilon}\frac{\partial}{\partial t} & 0 \\ 0 & 0 & -\sqrt{\varepsilon}\frac{\partial}{\partial t} \\ 0 & \frac{-1}{\sqrt{\mu}}\frac{\partial}{\partial x_3} & \frac{1}{\sqrt{\mu}}\frac{\partial}{\partial x_2} \\ \frac{1}{\sqrt{\mu}}\frac{\partial}{\partial x_3} & 0 & \frac{-1}{\sqrt{\mu}}\frac{\partial}{\partial x_1} \\ \frac{-1}{\sqrt{\mu}}\frac{\partial}{\partial x_2} & \frac{1}{\sqrt{\mu}}\frac{\partial}{\partial x_1} & 0 \end{bmatrix} \tag{310}$$

Then, with

$$\mathbf{E}_M \equiv \hat{\mathbf{x}}_1 E_{M1} + \hat{\mathbf{x}}_2 E_{M2} + \hat{\mathbf{x}}_3 E_{M3} \tag{311}$$

$$\mathbf{B}_M \equiv \hat{\mathbf{x}}_1 B_{M1} + \hat{\mathbf{x}}_2 B_{M2} + \hat{\mathbf{x}}_3 B_{M3} \tag{312}$$

we have

$$\sqrt{\mathsf{U}}\begin{bmatrix} A_{M1} \\ A_{M2} \\ A_{M3} \end{bmatrix} = \frac{1}{\sqrt{2}}\begin{bmatrix} \sqrt{\varepsilon}E_{M1} \\ \sqrt{\varepsilon}E_{M2} \\ \sqrt{\varepsilon}E_{M3} \\ \left(1/\sqrt{\mu}\right)B_{M1} \\ \left(1/\sqrt{\mu}\right)B_{M2} \\ \left(1/\sqrt{\mu}\right)B_{M3} \end{bmatrix} \equiv \left[\sqrt{\mathsf{U}}\mathbf{A}_M\right] \tag{313}$$

Here we use a notation $\left[\sqrt{\mathsf{U}}\mathbf{A}_M\right]$ with the square brackets to indicate that this entity should be thought of as mathematical 6-element column vector. Now, we can construct a transformed inner product with respect to this operator $\sqrt{\mathsf{U}}$, as discussed in section 6.6.2. For two non-zero vector potential fields, $\boldsymbol{\mu}(\mathbf{r},t)$ and $\boldsymbol{\eta}(\mathbf{r},t)$, this inner product is written

$$(\boldsymbol{\mu},\boldsymbol{\eta})_{T\sqrt{\mathsf{U}}} \equiv \left(\sqrt{\mathsf{U}}\boldsymbol{\mu},\sqrt{\mathsf{U}}\boldsymbol{\eta}\right) \tag{314}$$

and this defines an energy inner product for electromagnetic fields. For two (non-zero) electromagnetic fields, the inner product will be zero if and only if the two fields are orthogonal with respect to this energy inner product. See the discussion given above (around Eqs. (180) to (182)) explicitly for the monochromatic case above for how to complete the construction of this inner product as in integral over space. The only differences in the full time-dependent case are that the inner product $(\boldsymbol{\mu},\boldsymbol{\eta})_{T\sqrt{\mathsf{U}}}$ is formally completed at some time *t*, and that we use $\sqrt{\mathsf{U}}$ rather than the monochromatic version $\sqrt{\mathsf{U}_\omega}$ .

When formed between a vector potential field $\mathbf{A}_M$ and itself, we obtain the total energy *U* at some time *t* of the electromagnetic field (neglecting static fields) in this volume *V*

$$U = \left(\mathbf{A}_M,\mathbf{A}_M\right)_{T\sqrt{\mathsf{U}}} \equiv \left(\sqrt{\mathsf{U}}\mathbf{A}_M,\sqrt{\mathsf{U}}\mathbf{A}_M\right) \tag{315}$$

Expressed as a transformed inner product [181], as in (314), we have our desired energy inner product for the electromagnetic field in a volume *V* at a time *t*.

#### F.4.2.2. Monochromatic form

For monochromatic fields at angular frequency $\omega$ (so with implicit time-dependence of the form $\exp(-i\omega t)$ in ), $\partial/\partial t \to -i\omega$. We also need to work with the energy density as in Eq. (303), which leads to the elimination of the factor of ½ compared to the full time-dependent version, so the monochromatic version of the operator becomes as in Eq. (179) above.

# Appendix G Divergence of the vector potential in the M-gauge

Generally, a choice of gauge formally results from a choice of the divergence of the vector potential, and we establish this here for $\nabla\cdot\mathbf{A}_M$. Formally, we establish the gauge function $\Psi_{CM}$ that transforms from the Coulomb gauge to the M-gauge. In the Coulomb gauge, by choice (Eq. (252)) $\nabla\cdot\mathbf{A}_C = 0$. Now the wave equation (242) becomes the Poisson equation

$$\nabla^2\Phi_C = -\frac{\rho}{\varepsilon} \tag{316}$$

That is, the potential $\Phi_C$ is simply the ("instantaneous") electrostatic potential associated with the charge density $\rho$ (hence the name "Coulomb" gauge). Presuming we know $\rho$ and its behavior in space and time, then we can solve Eq. (316) to obtain

$$\Phi_C(\mathbf{r},t)=\frac{1}{4\pi\varepsilon}\int\frac{\rho(\mathbf{r}',t)}{|\mathbf{r}-\mathbf{r}'|}d^3\mathbf{r}' \tag{317}$$

where the integral is over all the volume containing any charge density. We can check this solution Eq. (317) by taking $\nabla^2$ of both sides (noting that this is with respect to the non-primed position variables, such as $\mathbf{r}$.)

Now, we formally define the difference $\Delta\rho(\mathbf{r},t)$ between the actual (free) charge density at any given place and time, $\rho(\mathbf{r},t)$ (which can change after time $t_o$), and the original, (fixed) charge density, $\rho_o(\mathbf{r})$

$$\Delta\rho=\rho-\rho_o \tag{318}$$

Explicitly, then, using the form as in Eq. (250), we can write

$$\Phi_M=\Phi_C-\frac{\partial\Psi_{CM}}{\partial t} \tag{319}$$

and using Eqs. (255), (316), and (317), we can therefore write

$$\frac{\partial\Psi_{CM}}{\partial t}=\Phi_C-\Phi_M=\frac{1}{4\pi\varepsilon}\int\frac{\rho(\mathbf{r}',t)}{|\mathbf{r}-\mathbf{r}'|}d^3\mathbf{r}'-\frac{1}{4\pi\varepsilon}\int\frac{\rho_o(\mathbf{r}')}{|\mathbf{r}-\mathbf{r}'|}d^3\mathbf{r}'=\frac{1}{4\pi\varepsilon}\int\frac{\Delta\rho(\mathbf{r}',t)}{|\mathbf{r}-\mathbf{r}'|}d^3\mathbf{r}' \tag{320}$$

The charge density $\Delta\rho(\mathbf{r},t)$ results entirely from the currents $\mathbf{J}(\mathbf{r},t)$ that flow after time $t_o$, and so the charge conservation equation (234) becomes

$$\nabla\cdot\mathbf{J}=-\frac{\partial\Delta\rho}{\partial t} \tag{321}$$

So, from Eq. (320) we can write

$$\frac{\partial^2\Psi_{CM}}{\partial t^2}=\frac{1}{4\pi\varepsilon}\int\frac{\partial\Delta\rho(\mathbf{r}',t)/\partial t}{|\mathbf{r}-\mathbf{r}'|}d^3\mathbf{r}'=-\frac{1}{4\pi\varepsilon}\int\frac{\nabla'\cdot\mathbf{J}(\mathbf{r}',t)}{|\mathbf{r}-\mathbf{r}'|}d^3\mathbf{r}' \tag{322}$$

Integrating once with respect to time therefore gives us

$$\frac{\partial\Psi_{CM}(\mathbf{r},t')}{\partial t}=-\frac{1}{4\pi\varepsilon}\int_{t''=t_o}^{t'}\left[\int_V\frac{\nabla'\cdot\mathbf{J}(\mathbf{r}',t'')}{|\mathbf{r}-\mathbf{r}'|}d^3\mathbf{r}'\right]dt'' \tag{323}$$

where for clarity now we are also explicitly showing the spatial integral is over the volume $V$ containing currents. Integrating a second time with respect to time gives us

$$\Psi_{CM}(\mathbf{r},t)=-\frac{1}{4\pi\varepsilon}\int_{t'=t_o}^{t}\int_{t''=t_o}^{t'}\left[\int_V\frac{\nabla'\cdot\mathbf{J}(\mathbf{r}',t'')}{|\mathbf{r}-\mathbf{r}'|}d^3\mathbf{r}'\right]dt''dt' \tag{324}$$

which is therefore the gauge function that transforms from the Coulomb gauge to the M-gauge. Now Eq. (248) becomes

$$\mathbf{A}_M=\mathbf{A}_C+\nabla\Psi_{CM} \tag{325}$$

Since $\nabla\cdot\mathbf{A}_C=0$ (Eq. (252)), we therefore have

$$\nabla\cdot\mathbf{A}_M=\nabla^2\Psi_{CM} \tag{326}$$

Hence, from Eq. (324), and using the identity

$$\delta(\mathbf{r}-\mathbf{r}') \equiv -\frac{1}{4\pi}\nabla^2 \frac{1}{|\mathbf{r}-\mathbf{r}'|} \tag{327}$$

we obtain

$$\begin{aligned}
\nabla\cdot\mathbf{A}_M(\mathbf{r},t) &= -\frac{1}{4\pi\varepsilon}\nabla^2 \int_{t'=t_o}^{t}\int_{t''=t_o}^{t'}\left[\int_V \frac{\nabla'\cdot\mathbf{J}(\mathbf{r}',t'')}{|\mathbf{r}-\mathbf{r}'|}d^3\mathbf{r}'\right]dt''dt' \\
&= -\frac{1}{4\pi\varepsilon}\int_{t'=t_o}^{t}\int_{t''=t_o}^{t'}\left[\int_V \nabla'\cdot\mathbf{J}(\mathbf{r}',t'')\nabla^2\frac{1}{|\mathbf{r}-\mathbf{r}'|}d^3\mathbf{r}'\right]dt''dt' \\
&= \frac{1}{\varepsilon}\int_{t'=t_o}^{t}\int_{t''=t_o}^{t'}\left[\int_V \nabla'\cdot\mathbf{J}(\mathbf{r}',t'')\delta(\mathbf{r}-\mathbf{r}')d^3\mathbf{r}'\right]dt''dt' = \frac{1}{\varepsilon}\int_{t'=t_o}^{t}\int_{t''=t_o}^{t'}\nabla\cdot\mathbf{J}(\mathbf{r},t'')dt''dt'
\end{aligned} \tag{328}$$

which is the result quoted above (Eq. (262)).

# Appendix H Dyadic notation and useful identities for Green's functions

A *dyad* is pair of vectors, written as $\mathbf{ab}$, with no symbol between the vectors **a** and **b**. **a** is a vector that is "waiting to operate on (or be operated on by)" a vector from the left, and similarly **b** is a vector that is "waiting to operate on" a vector on the right. So, for example, with two vectors **c** and **d**, and with $\mathbf{c}\cdot\mathbf{a}=s$ and $\mathbf{b}\cdot\mathbf{d}=w$, where *s* and *w* are scalars, then

$$\mathbf{c}\cdot(\mathbf{ab})\cdot\mathbf{d} = (\mathbf{c}\cdot\mathbf{a})(\mathbf{b}\cdot\mathbf{d}) = sw \tag{329}$$

Now we see why there is no symbol inserted between **a** and **b** in the dyadic; once these dot product operations have been performed to the left and the right, we have a simple scalar multiplication in the middle. For a vector operation to the left, such as a cross product, then we are left with a dyad $\mathbf{fg}$ in this case, i.e., with $\mathbf{f}=\mathbf{c}\times\mathbf{a}$ and $\mathbf{g}=\mathbf{b}\times\mathbf{d}$,

$$\mathbf{c}\times(\mathbf{ab})\times\mathbf{d} = (\mathbf{c}\times\mathbf{a})(\mathbf{b}\times\mathbf{d}) = \mathbf{fg} \tag{330}$$

A *dyadic* is an extension to sums of dyads, such as the sum of three dyads, typically written using the unit vectors in the coordinate system. So with three orthogonal coordinate directions with corresponding unit vectors $\hat{\mathbf{x}}_1$, $\hat{\mathbf{x}}_2$, and $\hat{\mathbf{x}}_2$ then we could write three vectors (or vector functions), one in each coordinate direction.

$$\mathbf{F}_j = \sum_{i=1}^{3} f_{ij}\hat{\mathbf{x}}_i \tag{331}$$

So, with $j=1,2,3$, we could write one dyadic

$$\overline{\overline{F}} = \sum_{j=1}^{3}\mathbf{F}_j\hat{\mathbf{x}}_j = \sum_{i=1}^{3}\sum_{j=1}^{3} f_{ij}\hat{\mathbf{x}}_i\hat{\mathbf{x}}_j \tag{332}$$

where conventionally the dyadic is notated with a double line above it. Formally, we can define appropriate products for dyadics. For some vector **a**, we can define the following:

*anterior scalar product*
$$\mathbf{a}\cdot\overline{\overline{F}} = \sum_{j=1}^{3}(\mathbf{a}\cdot\mathbf{F}_j)\hat{\mathbf{x}}_j = \sum_{i=1}^{3}\sum_{j=1}^{3} a_i f_{ij}\hat{\mathbf{x}}_j \tag{333}$$

*posterior scalar product*
$$\overline{\overline{F}}\cdot\mathbf{a} = \sum_{j=1}^{3}\mathbf{F}_j(\hat{\mathbf{x}}_j\cdot\mathbf{a}) = \sum_{i=1}^{3}\sum_{j=1}^{3} a_j f_{ij}\hat{\mathbf{x}}_i \tag{334}$$

*anterior vector product*
$$\mathbf{a}\times\overline{\overline{F}} = \sum_{j=1}^{3}(\mathbf{a}\times\mathbf{F}_j)\hat{\mathbf{x}}_j \tag{335}$$

*posterior vector product*
$$\overline{\overline{F}} \times \mathbf{a} = \sum_{j=1}^{3} \mathbf{F}_j \left( \hat{\mathbf{x}}_j \times \mathbf{a} \right) \tag{336}$$

The result of the scalar products is a vector in both cases, and the result of the vector product is a dyadic in both cases.

The unit dyadic or *idem factor* $\overline{\overline{I}}$ leaves a vector unchanged in either scalar product, i.e.,

$$\mathbf{a} \cdot \overline{\overline{I}} = \overline{\overline{I}} \cdot \mathbf{a} = \mathbf{a} \tag{337}$$

so we can think of it as the dyadic "identity" operator. We could also write

$$\overline{\overline{I}} = \sum_{j=1}^{3} \hat{\mathbf{x}}_j \hat{\mathbf{x}}_j \equiv \hat{\mathbf{x}}_1 \hat{\mathbf{x}}_1 + \hat{\mathbf{x}}_2 \hat{\mathbf{x}}_2 + \hat{\mathbf{x}}_3 \hat{\mathbf{x}}_3 \tag{338}$$

which is therefore the sum of the three dyads $\hat{\mathbf{x}}_1 \hat{\mathbf{x}}_1$, $\hat{\mathbf{x}}_2 \hat{\mathbf{x}}_2$, and $\hat{\mathbf{x}}_3 \hat{\mathbf{x}}_3$.

## H.1. Vector calculus extended to dyadics

To extend vector calculus to dyadics, we build on the anterior scalar and vector products. Specifically, the divergence of a dyadic function can be defined, using Eq. (333), by

$$\nabla \cdot \overline{\overline{F}} = \sum_{j=1}^{3} \left( \nabla \cdot \mathbf{F}_j \right) \hat{\mathbf{x}}_j = \sum_{i=1}^{3} \sum_{j=1}^{3} \frac{\partial f_{ij}}{\partial x_i} \hat{\mathbf{x}}_j \tag{339}$$

and the curl can be defined, following from Eq. (335)

$$\nabla \times \overline{\overline{F}} = \sum_{j=1}^{3} \left( \nabla \times \mathbf{F}_j \right) \hat{\mathbf{x}}_j = \sum_{i=1}^{3} \sum_{j=1}^{3} \left( \nabla f_{ij} \times \hat{\mathbf{x}}_i \right) \hat{\mathbf{x}}_j \tag{340}$$

where we have used the vector calculus identity (for a fixed vector $\hat{\mathbf{x}}$)

$$\nabla \times \left( f(\mathbf{r}) \hat{\mathbf{x}} \right) = \nabla f(\mathbf{r}) \times \hat{\mathbf{x}} \tag{341}$$

We can also usefully define the gradient of a vector function. Specifically, for some vector function $\mathbf{F}(\mathbf{r})$ with components $f_j(\mathbf{r})$, i.e.,

$$\mathbf{F}(\mathbf{r}) = \sum_{j=1}^{3} f_j(\mathbf{r}) \hat{\mathbf{x}}_j \tag{342}$$

we can choose to write the dyadic

$$\nabla \mathbf{F}(\mathbf{r}) = \sum_{j=1}^{3} \left( \nabla f_j(\mathbf{r}) \right) \hat{\mathbf{x}}_j = \sum_{i=1}^{3} \sum_{j=1}^{3} \frac{\partial f_j(\mathbf{r})}{\partial x_i} \hat{\mathbf{x}}_i \hat{\mathbf{x}}_j \tag{343}$$

Note that, just as the divergence of a vector function gives a scalar function, the divergence of a dyadic gives a vector. Similarly, just as the curl of a vector function gives a vector function, then the curl of a dyadic gives a dyadic, and just as the gradient of a scalar function gives a vector function, the gradient of a vector function gives a dyadic.

We can also introduce here two other relations we need. First, for a dyadic of the form

$$\overline{\overline{F}} = f(\mathbf{r}) \overline{\overline{I}} \tag{344}$$

then

$$\nabla \cdot \overline{\overline{F}} = \nabla \cdot \left( f(\mathbf{r}) \overline{\overline{I}} \right) = \sum_{j=1}^{3} \nabla \cdot \left( f(\mathbf{r}) \hat{\mathbf{x}}_j \right) \hat{\mathbf{x}}_j = \sum_{j=1}^{3} \frac{\partial f(\mathbf{r})}{\partial x_j} \hat{\mathbf{x}}_j = \nabla f \tag{345}$$

Second, for this same form of dyadic, Eq. (344), we can consider $\nabla \nabla \cdot \overline{\overline{F}}$. From Eq. (345)

$$\nabla\nabla\cdot\overline{\overline{F}}=\sum_{j=1}^{3}\nabla\left(\frac{\partial f\left(\mathbf{r}\right)}{\partial x_j}\hat{\mathbf{x}}_j\right)=\sum_{j=1}^{3}\frac{\partial^2 f\left(\mathbf{r}\right)}{\partial x_i\partial x_j}\hat{\mathbf{x}}_i\hat{\mathbf{x}}_j=\nabla\nabla f\left(\mathbf{r}\right) \tag{346}$$

which is effectively introducing and defining a new operator, "$\nabla\nabla$". Viewed as an operator in its own right (so, without the "dot" in "$\nabla\nabla\cdot$"), $\nabla\nabla$ is an operator that takes a scalar function, here $f\left(\mathbf{r}\right)$, and creates a dyadic. Note that we must *not* notate it as $\nabla^2$, which is the Laplacian (equivalent to $\nabla\cdot\nabla$).

## H.2. Useful derivatives for dyadics and Green's functions

Several derivatives are useful in working with dyadics. First, with

$$\mathbf{R}=\mathbf{r}-\mathbf{r}' \tag{347}$$

then

$$R=\left|\mathbf{r}-\mathbf{r}'\right| \tag{348}$$

and we can write $\mathbf{R}$ in terms of Cartesian coordinates as

$$\mathbf{R}=x_1\hat{\mathbf{x}}_1+x_2\hat{\mathbf{x}}_2+x_3\hat{\mathbf{x}}_3 \tag{349}$$

(Note that we are now using $x_1$, $x_2$, and $x_3$ as the components of $\mathbf{R}$, not of $\mathbf{r}$.)

The gradient of $R$, which also obviously depends only on $R$ (and not on spherical coordinates $\theta$ and $\phi$), is simply (by definition) $\hat{\mathbf{R}}$ - the unit vector in the $\mathbf{r}-\mathbf{r}'$ radial direction of interest, and we can write this formally as

$$\nabla R=\hat{\mathbf{R}}\equiv\frac{\mathbf{r}-\mathbf{r}'}{\left|\mathbf{r}-\mathbf{r}'\right|}\equiv\frac{\mathbf{R}}{R} \tag{350}$$

Then from Eq. (343) and the definition Eq. (338)

$$\nabla\mathbf{R}=\sum_{i=1}^{3}\sum_{j=1}^{3}\frac{\partial x_j}{\partial x_i}\hat{\mathbf{x}}_i\hat{\mathbf{x}}_j=\sum_{j=1}^{3}\hat{\mathbf{x}}_j\hat{\mathbf{x}}_j=\overline{\overline{I}} \tag{351}$$

(which we see, incidentally, gives us another expression for $\overline{\overline{I}}$). We also have

$$\nabla\left(\frac{1}{R}\right)=-\frac{1}{R^2}\nabla R=-\frac{\hat{\mathbf{R}}}{R^2} \tag{352}$$

So,

$$\nabla\hat{\mathbf{R}}=\nabla\left(\frac{\mathbf{R}}{R}\right)=\frac{\nabla\mathbf{R}}{R}+\mathbf{R}\nabla\left(\frac{1}{R}\right)=\frac{\overline{\overline{I}}}{R}-\frac{\mathbf{R}}{R^2}\nabla\mathbf{R}=\frac{1}{R}\left(\overline{\overline{I}}-\hat{\mathbf{R}}\hat{\mathbf{R}}\right) \tag{353}$$

Note carefully that Eq. (351) is for the vector $\mathbf{R}$, whereas Eq. (353) is for the unit vector $\hat{\mathbf{R}}$. Expressions (351) and (353) are particularly useful in working with dyadic Green's functions. We also need three other related derivatives. First,

$$\nabla\left(\frac{\hat{\mathbf{R}}}{R}\right)=\frac{1}{R}\nabla\hat{\mathbf{R}}+\hat{\mathbf{R}}\nabla\left(\frac{1}{R}\right)=\frac{1}{R^2}\left(\overline{\overline{I}}-\hat{\mathbf{R}}\hat{\mathbf{R}}\right)-\hat{\mathbf{R}}\frac{\hat{\mathbf{R}}}{R^2}=\frac{1}{R^2}\left(\overline{\overline{I}}-2\hat{\mathbf{R}}\hat{\mathbf{R}}\right) \tag{354}$$

and noting, second, that

$$\nabla\left(\frac{1}{R^2}\right)=-\frac{2}{R^3}\nabla R=-\frac{2\hat{\mathbf{R}}}{R^3} \tag{355}$$

we have our third required derivative

$$\nabla\left(\frac{\hat{\mathbf{R}}}{R^2}\right)=\frac{1}{R^2}\nabla\hat{\mathbf{R}}+\hat{\mathbf{R}}\nabla\left(\frac{1}{R^2}\right)=\frac{1}{R^3}\left(\bar{\bar{I}}-\hat{\mathbf{R}}\hat{\mathbf{R}}\right)-2\hat{\mathbf{R}}\frac{\hat{\mathbf{R}}}{R^3}=\frac{1}{R^3}\left(\bar{\bar{I}}-3\hat{\mathbf{R}}\hat{\mathbf{R}}\right) \quad (356)$$

# Appendix I Quantization of the electromagnetic field in the M-gauge

The algebraic steps in the quantization of the field in the M-gauge are similar to those in the common "transverse" Coulomb gauge approaches (see, e.g., [139, 140]), but we avoid fictitious resonators or boxes, the formal problems of the Coulomb gauge [142, 143], and any separation into "longitudinal" and "transverse" fields. Our approach can proceed for arbitrary volumes and can include all near-field terms if we wish.

Our basis functions or "modes" are any (energy) orthogonal set of monochromatic vector potential fields $\mathbf{A}_{\omega Mj}(\mathbf{r}_R)$ in the volume $V_R$, and we can expand a monochromatic field in them, as in Eq. (188), with expansion coefficients $a_j(t)$ that explicitly include the time-dependent factor $\exp(-i\omega t)$ (and that is their only time dependence), and we can write a classical Hamiltonian Eq. (189) for the field. Still in a classical view, for one such "mode", we formally propose a pair of "canonical" variables – a generalized "position", $q_j$ and a generalized "momentum" $p_j$ given by

$$q_j(t)=\sqrt{\hbar/2\omega}\left[a_j(t)+a_j^*(t)\right] \text{ and } p_j(t)=-i\sqrt{\hbar\omega/2}\left[a_j(t)-a_j^*(t)\right] \quad (357)$$

We note in passing that these expressions are readily inverted to give

$$a_j=\sqrt{1/2\hbar\omega}\left[\omega q_j+ip_j\right] \text{ and } a_j^*=\sqrt{1/2\hbar\omega}\left[\omega q_j-ip_j\right] \quad (358)$$

Because of the $\exp(-i\omega t)$ time-dependence of $a_j(t)$, we have

$$\partial q_j(t)/\partial t=p_j(t) \text{ and } \partial p_j(t)/\partial t=-\omega^2 q_j(t) \quad (359)$$

We can formally rewrite the Hamiltonian Eq. (189) as

$$H=(1/2)\sum_j\left[p_j^2(t)+\omega^2 q_j^2(t)\right] \quad (360)$$

These variables then satisfy Hamilton's equations

$$\frac{\partial H}{\partial p_j}=\frac{\partial q_j}{\partial t} \text{ and } \frac{\partial H}{\partial q_j}=-\frac{\partial p_j}{\partial t} \quad (361)$$

So, we can take a typical approach in quantum mechanics and quantize the "oscillator" by postulating we can replace the variables $p$ and $q$ with operators $\hat{p}$ and $\hat{q}$ that will in turn lead to appropriate commutation relations, which we postulate to be

$$\left[\hat{q}_m,\hat{p}_n\right]\equiv\hat{q}_m\hat{p}_n-\hat{p}_n\hat{q}_m=i\hbar\delta_{mn} \quad (362)$$

$$\left[\hat{q}_m,\hat{q}_n\right]=0 \text{ and } \left[\hat{p}_m,\hat{p}_n\right]=0 \quad (363)$$

(Note, incidentally, that we will use a "hat", as in $\hat{q}$ and $\hat{p}$, to indicate a quantum mechanical operator as distinct from its corresponding classical quantity.) These commutation relations are consistent with a typical presumption of replacing $p_j$ with $\hat{p}_j=-i\hbar\partial/\partial q_j$ and using $q_j$ as its own operator $\hat{q}_j$, as is appropriate in a "position" representation of quantum mechanics, for some (generalized) position $q_j$. The corresponding proposed Hamiltonian becomes, from Eq. (360), the operator

$$\hat{H}=(1/2)\sum_j\left[\hat{p}_j^2(t)+\omega^2\hat{q}_j^2(t)\right] \quad (364)$$

We can define, the *annihilation operator* $\hat{a}_j$ and the corresponding *creation operator* $\hat{a}_j^\dagger$

$$\hat{a}_j = \sqrt{1/2\hbar\omega}\left[\omega\hat{q}_j + i\hat{p}_j\right] \text{ and } \hat{a}_j^\dagger = \sqrt{1/2\hbar\omega}\left[\omega\hat{q}_j - i\hat{p}_j\right] \quad (365)$$

for "mode" *j*, where these two operators are Hermitian adjoints of one another, as the notation suggests. Note that these expressions in Eqs. (365) are operator analogs of those in Eqs. (358). Correspondingly, then, we can directly write the inversions of these as

$$\hat{q}_j = \sqrt{\hbar/2\omega}\left[\hat{a}_j + \hat{a}_j^\dagger\right] \text{ and } \hat{p}_j = i\sqrt{\hbar\omega/2}\left[\hat{a}_j^\dagger - \hat{a}_j\right] \quad (366)$$

and the commutation relations for the annihilation and creation operators then become

$$\left[\hat{a}_m, \hat{a}_n^\dagger\right] \equiv \hat{a}_m\hat{a}_n^\dagger - \hat{a}_n^\dagger\hat{a}_m = \delta_{mn} \quad (367)$$

$$\left[\hat{a}_m, \hat{a}_n\right] = 0 \text{ and } \left[\hat{a}_m^\dagger, \hat{a}_n^\dagger\right] = 0 \quad (368)$$

Rewriting Eq. (364) using this notation gives $\hat{H} = (\hbar\omega/2)\sum_j\left[\hat{a}_j\hat{a}_j^\dagger + \hat{a}_j^\dagger\hat{a}_j\right]$, which we can write using the commutation relation Eq. (367) in the more familiar form [182] as

$$\hat{H} = \sum_j \hbar\omega\left(\hat{a}_j^\dagger\hat{a}_j + (1/2)\right) \quad (369)$$

Note that, though we have proceeded here based on the M-gauge description of electromagnetism, and without the usual assumptions of the Coulomb gauge approach (i.e., "transverse" sources, fields and potentials, and "pretend" resonators) we have come to a familiar result. We can now develop the consequences of this approach further.

Because of the analogy between $a_j$ and $\hat{a}_j$ and between $a_j^*$ and $\hat{a}_j^\dagger$, we merely need to substitute the operator $\hat{a}_j$ for $a_j$ and the operator $\hat{a}_j^\dagger$ for $a_j^*$ in Eq. (188) to obtain the corresponding vector potential field operator

$$\hat{\mathbf{A}}_{\omega M}(\mathbf{r}_R, t) = \sqrt{\hbar\omega}\sum_j\left[\hat{a}_j(0)\mathbf{A}_{\omega Mj}(\mathbf{r}_R)\exp(-i\omega t) + h.c.\right] \quad (370)$$

where we have written $\hat{a}_j(t) = \hat{a}_j(0)\exp(-i\omega t)$ and we use the terminology "h. c." to stand for "Hermitian conjugate" (which is the same as Hermitian adjoint), and we note that the Hermitian adjoint of a scalar function is just the complex conjugate of that scalar function. Using Eq. (164) ( $\mathbf{E}_M = -\partial\mathbf{A}_M/\partial t$ ) and Eq. (165) ( $\mathbf{B}_M = \nabla\times\mathbf{A}_M$ ), we can write the analogous electric and magnetic field operators as, respectively,

$$\hat{\mathbf{E}}_{\omega M}(\mathbf{r}_R, t) = i\omega\sqrt{\hbar\omega}\sum_j\left[\hat{a}_j(0)\mathbf{A}_{\omega Mj}(\mathbf{r}_R)\exp(-i\omega t) - h.c.\right] \quad (371)$$

(note the "-" sign before the "*h.c.*") and

$$\hat{\mathbf{B}}_{\omega M}(\mathbf{r}_R, t) = \sqrt{\hbar\omega}\sum_j\left[\hat{a}_j(0)\left[\nabla\times\mathbf{A}_{\omega Mj}(\mathbf{r}_R)\right]\exp(-i\omega t) + h.c.\right] \quad (372)$$

Obviously, from Eq. (369), we can divide the Hamiltonian into a sum of Hamiltonians, one, $\hat{H}_j$, for each "mode" or basis function $\mathbf{A}_{\omega Mj}(\mathbf{r}_R)$. Specifically $\hat{H}_j = \hbar\omega\left(\hat{a}_j^\dagger\hat{a}_j + (1/2)\right)$ (Eq. (191) and we have the usual properties of the annihilation and creation operators for the mode (see, e.g., [9]). Specifically, the eigenstates of these individual mode Hamiltonians are the number states or Fock states $|n_j\rangle$, where conventionally $n_j$ is interpreted as the number of photons in the mode, and the energy eigen equation is

$$\hat{H}|n_j\rangle = \hbar\omega\left(n_j + (1/2)\right)|n_j\rangle \quad (373)$$

where we can also if we wish define a number operator

$$\hat{N}_j = \hat{a}_j^\dagger \hat{a}_j \text{ with } \hat{N}_j |n_j\rangle = n_j |n_j\rangle \quad (374)$$

The commutation relation Eq. (367) gives the standard "raising" and "lowering" properties

$$\hat{a}_j^\dagger |n_j\rangle = \sqrt{n_j + 1} |n_j + 1\rangle \text{ and } \hat{a}_j |n_j\rangle = \sqrt{n_j} |n_j - 1\rangle \quad (375)$$

with $\hat{a}_j |0\rangle = 0$.

# Appendix J Modal "A and B" Coefficient Argument

We presume we have a quantum mechanical system that has a probability $P_1$ of being in a lower state, and a probability $P_2$ of being in an upper state. We presume this system (or an ensemble of identical systems) is sitting in some optical environment that is otherwise lossless, such as some resonator, waveguide, or other dielectric environment.

In thermal equilibrium with a heat reservoir with which the system can exchange energy, as usual the ratio of these probabilities $P_2$ and $P_1$ is given by the Boltzmann factor

$$\frac{P_2}{P_1} = \exp\left(-\frac{E_{21}}{k_B T}\right) \quad (376)$$

where $E_{21}$ is the (positive) energy separation of the states, $T$ is the temperature and $k_B$ is Boltzmann's constant.

Now we consider a specific one $|\psi_j\rangle$ of the mode-converter input modes [6, 7] for this optical system, which will have a corresponding mode-converter output mode $|\phi_j\rangle$. We presume the photon energy in this mode is approximately $\hbar\omega \simeq E_{21}$.

We presume that the probability per unit time that a photon in this mode-converter input mode $|\psi_j\rangle$ is absorbed by the quantum system can be written as

$$R_{12j} = M_{12j} P_1 \quad (377)$$

where $M_{12j}$ is some constant that is characteristic of the quantum system and its interaction with light in $|\psi_j\rangle$ (and therefore at frequency $\omega$); specifically, we presume this constant does not itself depend on temperature.

Using the Planck distribution, the number of photons in this mode in thermal equilibrium at some temperature $T$ is

$$n_j = \frac{1}{\exp(E_{21}/k_B T) - 1} \quad (378)$$

So the total absorption rate of photons from this mode in thermal equilibrium is

$$W_{12j} = n_j R_{12j} = n_j M_{12j} P_1 \quad (379)$$

For emission into the mode-converter output mode $j$, we presume a rate

$$W_{21j} = L_{21j} P_2 + n_j M_{21j} P_2 \quad (380)$$

Here we are proposing what will be a spontaneous emission term ($L_{21j} P_2$), which is independent of the number of photons in the mode, and what will be a stimulated emission term ($n_j M_{21j} P_2$) that is proportional to the number of photons in the mode. Again, we presume that $L_{21j}$ and $M_{21j}$ are constants that are characteristic of the quantum system and its interaction with light in this input mode $j$ and that do not themselves depend on temperature.

This argument is closely analogous to Einstein's A and B coefficient argument (see, e.g., [139]). The $M_{12}$ and $M_{21}$ coefficients are close analogs to the $B_{12}$ and $B_{21}$ coefficients in Einstein's argument. Because we have avoided having to define densities of modes in free space, we do not have a direct analog to the $A$ coefficient in Einstein's argument (which assumed free space modes), but the coefficient $L_{21}$ is taking on the analogous role in the argument for the mode of interest.

Now, as noted above ((194)), any power not absorbed from $\left|\psi_j\right\rangle$ is scattered into the corresponding mode-converter output mode $\left|\phi_j\right\rangle$. So, if the arrival rate of photons in the mode-converter input mode $\left|\psi_j\right\rangle$ in thermal equilibrium is $Q_j$, then the scattering rate into the mode-converter output mode $\left|\phi_j\right\rangle$ is

$$S_j = Q_j - W_{12j} \tag{381}$$

So, the total number of photons per unit time emitted and scattered into the mode-converter output mode $\left|\phi_j\right\rangle$ is

$$W_{TOTj} = S_j + W_{21j} = Q_j - W_{12j} + W_{21j} \tag{382}$$

Now, as discussed above in section **11.1** in the derivation of Law 1 in [7], in thermal equilibrium, the total number of photons arriving at the system in mode-converter input mode $\left|\psi_j\right\rangle$ must equal the total number emerging (so, the sum of emitted and scattered photons) from the system into mode-converter output mode $\left|\phi_j\right\rangle$. This is because we can construct an optical machine that couples these, and only these, as the output and input light for a single-mode black body, with which we much be able to come to thermal equilibrium. Notably, in this approach, no light in any other orthogonal input modes is coupled by scattering into the mode-converter output mode $\left|\phi_j\right\rangle$ (see (193)), so we have accounted for all possible output light here. Therefore, in thermal equilibrium,

$$W_{TOTj} = Q_j \tag{383}$$

So, from Eq. (382),

$$W_{12j} = W_{21j} \tag{384}$$

So, from Eqs. (379) and (380)

$$n_j M_{12j} P_1 = L_{21j} P_2 + n_j M_{21j} P_2 \tag{385}$$

Then, using Eq. (378)

$$M_{12j} = \left[\frac{L_{21}}{n_j} + M_{21j}\right]\frac{P_2}{P_1} = \left[L_{21j}\left\{\exp\left(\frac{E_{21j}}{k_B T}\right) - 1\right\} + M_{21j}\right]\frac{P_2}{P_1} \tag{386}$$

Now using Eq. (376)

$$M_{12j} = L_{21j} + \left(M_{21j} - L_{21j}\right)\exp\left(\frac{-E_{21}}{k_B T}\right) \tag{387}$$

Now, by assumption, all of $M_{12j}$, $L_{21j}$, and $M_{12j}$ are independent of temperature. Hence, the only way the expression on the right hand side can be independent of temperature is if

$$M_{21j} = L_{21j} \tag{388}$$

in which case, from Eq. (387) we are left with

$$M_{12j} = L_{21j} \tag{389}$$

Finally, we can summarize the overall result, which is that

$$M_{12j} = M_{21j} = L_{21j} \equiv M \tag{390}$$

In other words, these coefficients for absorption, stimulated emission and spontaneous emission are identical for any mode-converter pair of input and output functions. We can therefore use one coefficient for all three processes. We can restate this as above ((196)).

# Appendix K History and literature review of communications modes and related concepts

## K.1. Early history of degrees of freedom in optics and waves

By the 1950's, the idea of quantifying communication and information was developing rapidly [183], including the idea of the sampling theorem, as discussed, e.g., by Shannon [183 - 185]. Following on from the sampling theorem, the idea that there should be some bound on degrees of freedom in optics was proposed by both Gabor [74, 186] and Toraldo di Francia [187]. Gabor, based on a Fourier optics view, gives a heuristic approach based on the number of approximately non-overlapping Gaussian spots one could form on a second surface from source or waves on a first surface. Toraldo di Francia [187] works directly from the sampling theorem. Both these approaches lead to results essentially equivalent to our paraxial heuristic numbers (Eqs. (59) and (64)). Sampling theorem approaches have continued in the literature, with other useful results (e.g., for spherical surfaces and bounding volumes [158, 188]).

Use of the sampling theorem in optics and in waves generally is based first on some physical argument on a maximum "transverse" component to the wavevector that would be encountered or supported in some configuration of source and receiver spaces (in our notation); this therefore gives a a maximum transverse spatial frequency, which becomes the analog of bandwidth. Second, these approaches presume some maximum physical extent of an aperture or apertures (which becomes the analog of the time window).

Such sampling theory approaches are useful in paraxial situations with regular apertures (usually one-dimensional or rectangular) and where Fourier optics [114] is already a good approximation, and in some other far-field situations (e.g., spherical [158]). They are not going to work well in situations with irregular shaped surfaces and with most volumes – there is then no simple way to choose the "aperture". With small surfaces or volumes, there is no obvious clear "cut-off" spatial frequency; for sources in particular, a small source needs to be described by continuous functions, with continuous spatial-frequency spectra. Such problems are well known mathematically [184] and in the context of optics [33, 189].

## K.2. Eigenfunctions for wave problems with regular apertures

Getting past these problems of the sampling theorem requires a change in the mathematics. A key step was the mathematical realization that an obscure but known set of functions – the prolate spheroidals – had remarkable "eigenfunction" properties relevant to these problems with finite "windows" [190], with specific consequences for optics [33, 120, 189]. The first optical realization here was that these were the correct description of the modes of confocal laser resonators [120] (in the paper that proposed those resonators); they correctly give the beam shapes on the mirrors. Later laser resonator work (see, e.g., [111]) would replace those with Hermite-Gaussian and Laguerre-Gaussian approximations, which are much easier to work with mathematically. For lasers, this approximation is generally valid because only the modes with vanishingly small amplitudes at the edges of the mirrors are going to have low enough loss to oscillate, and in those cases there is little or no difference compared to the (correct) prolate spheroidal functions. Another property of these Hermite-Gaussian and Laguerre-Gaussian beams is that their shape does not change as they propagate, other than for changes in size; that can unfortunately lead to the false inference that modes are generally beams with this "constant shape" property, which we have seen is not the case in the actual communications modes in, e.g., section 5.

A key mathematical property of the prolate spheroidals is that they are eigenfunctions of the finite Fourier transform for the "linear" prolate spheroidals (which is relevant for linear and rectangular apertures) and of the finite Hankel transform for the "circular" prolate spheroidals (for circular apertures) (see, e.g., [33]). So, in a Fourier-optics paraxial approximation, they are the input wavefunctions or sources in a finite rectangular or circular aperture that will produce waves of the same shape in a corresponding aperture in the Fourier domain; extending to a second (or inverse) Fourier transform (and such a pair of transforms is essentially one way of looking at imaging), the same shape will be reproduced (inverted) at the image plane also. They are therefore also eigenfunctions of imaging viewed in this way. So, even allowing for the effects of finite apertures, these are functions that will be imaged perfectly through a paraxial system. See also [84] for continued discussion of such imaging.

Though this early work does not use the terminology of SVD, these prolate spheroidal functions can also be viewed as SVD functions in the sense we discuss in this paper, and the "coupling strength" eigenvalue associated with these eigenfunctions is then a singular value. These prolate spheroidals are a special case of SVD in two senses: (1) as communications modes, the source function and the received wave are identical (which is not generally the case for communications modes); (2) these solutions only work for source and receive surfaces of the same specific type of shape (rectangular or circular).

A key result, though, of this move to an eigenfunction problem, even if only for some special cases, is that, essentially for the first time, it removes the mathematical problems mentioned above of a sampling theorem approach. There are indeed now infinite sets of functions, both for sources and received waves, not finite ones, but the effective number of degrees of freedom comes from the behavior of the eigenvalues: up to some effective number, they are all essentially the same (which is our "paraxial degeneracy"), and then they drop off so rapidly that, below some practical threshold, they can be neglected. As we have also illustrated more generally, this effective number, at least in these simple cases of large, regular apertures in a paraxial approximation, agrees with the heuristic results of the sampling theory approach.

It was likely known to these early authors that they were dealing with Hilbert-Schmidt operators, which in turn are compact and therefore have "good" eigenfunction properties, Surprisingly, however, the idea that these singular values were also obeying a sum rule (which is in the end the Hilbert-Schmidt norm) does not seem to have been used or exploited until the much later, and more general, work on eigenfunction and SVD approaches [5, 191, 192] as communications modes.

## K.3. Emergence of communications modes

The idea of communications modes, as presented here, as the general answers to the orthogonal "best" channels, including continuous sources and with volumes as well as surfaces, is introduced first in 1998 in [191] and extended in 2000, first in the scalar wave case [5], and then for electromagnetic waves in [192]. The electromagnetic analysis in the present work formally supersedes [192] by reducing the problem to one only requiring the magnetic vector potential, and hence removing any remaining ambiguity about how many independent fields are required for communications problems.

### K.3.1. Wireless communications

Ideas of SVD, especially for finite matrices, have been routine as mathematical techniques for many decades, at least back to the 1960's, with fundamental work from the late 19$^{th}$ century. The use of SVD is common in signal processing and statistics. In wireless, as MIMO emerged in the 1990's, the channel matrix H between different spatial sources and receivers becomes one on which SVD could be performed to evaluate the best channels. There are certainly references to the corresponding operator $HH^{\dagger}$ and its eigenvalues by 1998 [193]. Normally, though, the full channel matrix may not be known to both the transmitter and the receiver, and the use here is to calculate the channel capacity [197, 198], not the optimum transmit and receive modes. In wireless systems, the matrix is usually a finite one, based on finite numbers of transmit and receive antennas.

Following [5, 191, 192], a body of work emerges in the wireless literature either explicitly using communications mode concepts or related results on degrees of freedom in spatial channels and/or spatial multiplexing (see, e.g., [13, 134, 194 - 206] ). The text [13] refers to what we call the communications modes as "eigenmodes of the channel" or "eigenchannels". Beam forming for MIMO can use SVD of the channel matrix to form the modes, which would then formally be communications modes in our notation. Optical techniques based on optical modes could also be used to set MIMO modes in beam forming [207].

### K.3.2. Electromagnetic scattering and imaging

The ideas of communications modes have been used extensively in scattering of radiation, including [208 - 218]. The kinds of behaviors of singular values illustrated in section 5 are also seen extensively in much of this work. Recent work analyzes imaging with electromagnetic beams with SVD [219], and such techniques have also been proposed for near-field scanning [220].

### K.3.3. Optics

Communications modes have been applied directly to the analysis of many optical systems [117, 118, 221 - 227]. [228] re-establishes communications modes based on an optimization approach. [229] analyzes x-ray waveguides using them. The optical case is also extended into noise-limited systems in a modern analysis using communications modes [40]. [18] compares communications modes and other beam forms for free-space communications. The concept of communications modes can be extended to partially coherent fields [226].

A body of work usefully analyzing optical systems of various kinds uses the terminology of "optical eigenmodes" [133, 230 - 233]. These seem to correspond to the communications modes in the present paper, or, for more complex optical systems, the mode-converter basis sets. For example, the "optical eigenmode" input and output functions in [133] are orthogonal sets in the input and output spaces, coupled with "intensity" strengths associated with an eigenvalue that appears to be the modulus squared of the communications mode or mode-converter singular values in our notation. Since such sets are unique (within geometric degeneracies), these are the same sets as those in our terminology. This work goes on to count the number of degrees of freedom by counting these sets (e.g., in [133]), though the authors do not appear to link their work to the communications mode formalism or to the earlier work on degrees of freedom discussed above. The general statement of these optical eigenmodes includes the explicit extension to what we describe as weighted or operator-weighted inner products (see [230] Eqs. (1) and (2)).

## K.4. Complex optics, matrix representations, and mode-converter basis sets

There has been growing interest in being able to characterize and/or controllably propagate through complex optical systems. These include strongly scattering media [233 - 248], and multiple-mode optical fibers for communications [249, 250] and imaging [251 - 256]. Such work typically employs spatial light modulators (SLMs) to characterize the matrix that relates input modes to output modes, or at least to establish one or more strong channels through the scatterer. The matrix in such work is often called a "transmission" matrix or, sometimes, a "transfer matrix" [250]. In the terminology of the present paper, we would call this a "device" matrix or, generally a "scattering" matrix. All such systems in which the matrix is fully characterized can be analyzed using the SVD approach (our "mode-converter basis sets") to establish the best channels. [257] similarly uses the SVD approach to analyze arbitrary polarization components.

If we want to be able to use more than one such channel at a time without incurring simple beamsplitting losses, we need to go beyond simple single SLMs to "multiple layered" approaches to generate and/or separate overlapping but orthogonal modes of the system. There are two known controllable approaches, one using multiple planes of SMLs or diffractive optics elements [258 - 260], and the other using meshes of integrated Mach-Zehnder modulators [12, 24 – 29, 57 – 65, 261, 262]. Such meshes can also be used to synthesize any linear optical component or matrix up to the dimensionality of the mesh; the general scheme to accomplish that relies on the SVD architecture, so the device is directly emulating the SVD of the desired matrix [25].

# Appendix L Novel results in this work

Though our primary goal here is a tutorial introduction and review, to give a complete picture, we have included some apparently novel work. It would be unethical to pass off either existing work as novel or new work as established fact. To avoid both errors, we list here what may be novel. Electromagnetism in particular is a subject that has been investigated for many years by many researchers, and we ask the reader's forgiveness if we have missed priority by others.

## L.1. Minor extensions of prior work and introduction of new terminology

The mathematics in section 3 is standard for the algebra of finite matrices. The emphasis on the sum rule, Eq. (36), is less common in that algebra, but the concept is implicit. The discussion on the constraint on the choice of coupling strengths in section 3.9 is mathematically obvious, but may not be generally understood yet in waves and optics.

All the explicit numerical examples in sections 4, 5, and 8 were performed for this paper. The mathematical technique used there for point sources and receivers of establishing the optimum (communications mode) channels by performing the SVD of the resulting finite coupling matrix is likely obvious to many working in radio frequency wireless theory, though may not be obvious in optics or acoustics. Working out the actual communications modes and their behavior and the resulting beams (other than for the special case of prolate spheroidal wavefunctions in paraxial examples) is less common, though we have discussed this before [5, 23, 192]. The explicit behaviors of the weakly coupled modes, as shown, for example, in Fig. 10 and Fig. 12, do not appear to have been presented before.

That the singular values are essentially constant up to a specific number is well known for the specific case of the prolate spheroidal solutions for rectangular or circular apertures, but the clarification that this is a general property of paraxial problems for surfaces or volumes of uniform thickness is a new generalization that also justifies our introduction of the term "paraxial degeneracy" to describe this.

Heuristic arguments based on an approximate cut-off positions or angles (by which interference from extreme points in a source area or volume transitions from originally constructive interference to being destructive) are not new in themselves – we have used them before [5], for example. The specific versions of these for terms we have introduced here – the paraxial heuristic number(s) (Eqs. (59) and (64) in section 5.3.4), the longitudinal heuristic angle (Eq. (71) in section 5.4.1), and the spherical heuristic number (Eq. (74) in section 5.4.2) – are new for this work.

The discussion, in general terms of communications modes, of the difficulty of passing the diffraction limit is somewhat novel, though the core idea is well known to some, at least in the special case of prolate spheroidal functions (e.g., [90]).

The mathematics of functions, operators and vectors in section 6, is all technically standard in functional analysis. The explicit notion that we would have different "underlying inner products" (a term we have introduced here) in different Hilbert spaces in the problem (in our case, source and receiving spaces) is not common, even though nothing has to be added to the functional analysis mathematics to support this idea (other than some notation to distinguish these inner products). As a result, the explicit discussion of making what we call an "algebraic shift" to Dirac notation using these underlying inner products may be novel, though again it is all implicit in the underlying mathematics.

The discussion of inner products involving operators in section 6.6 involves no actual new mathematics, but the explicit discussion is not common, and we have introduced the terms "operator-weighted inner product" and "transformed inner product" to clarify specific classes here.

The issue that "orbital" angular momentum beams do not generally introduce new degrees of freedom for communications is further clarified here (section 7.2).

The term "M-gauge" is also introduced in this work, though the concept and its consequence may be more substantial than a minor extension, so we discuss this further below.

## L.2. Novel observations

It is well known that many of the Green's functions used with waves are Hilbert-Schmidt operators (see, e.g., [110])). The general statements (157) and (158) may go beyond such discussions, however.

The clarification of the size scales for the transition in directionality from being from longitudinal extent to being from transverse extent (Eq. (73) and section 5.4.1) – a transition from a longitudinal "antenna" view to an "optics" view – may be novel.

The observations for spherical shell source and receiver spaces (section 5.4.2) that the "well-coupled" communications mode singular values asymptote to a straight line, and that line intersects the axis at the spherical heuristic number, are both new, and await an analytic explanation.

It is known for the specific case of prolate spheroidal functions (which are necessarily in a paraxial approximation) that the singular values past the paraxial heuristic number drop off essentially exponentially (see [80 - 82], and section 5.3.4.2). The generality of this exponential behavior, however, as seen in paraxial and non-paraxial cases, and for many different shapes of source and receiver volumes, is a new observation that also awaits a clear explanation. The observation that, in the far field, the corresponding waves are not evanescent but propagating (section 5.3.4.2) may also be novel.

That transverse-polarized electromagnetic waves show simple diffraction behavior similar to that of scalar waves is, of course, well known. The observation (section 8.4.3) that such diffraction behavior also holds for longitudinally polarized (and hence "non-propagating") electromagnetic waves may be surprising and novel.

## L.3. Substantial new concepts and results

### L.3.1. Introduction of the M-gauge for electromagnetism

The M-gauge (section 8 and **Appendix F**) and its consequences are significant new concepts introduced here. First, this resolves that there are only 3 independent vector components for the electromagnetic field for all problems involving changes in the field (as in communications), which can then be completely represented by the magnetic vector potential **A** in this gauge. The resulting vector wave equation is then solved with a Green's function that is a mathematical analog of the conventional dyadic Green's function of the electric field **E**, but applied here, apparently in a novel manner, to **A**. (This approach avoids the conventional separation into so-called "longitudinal" and "transverse" field components that have various formal problems.) This use of **A** then allows a novel energy inner product for the electromagnetic field (section 8.4.4 and **Appendix F.4**), which then allows the full power of functional analysis to be exploited, including the construction of orthogonal sets of waves without any restriction to specific volumes (like cuboidal boxes).

### L.3.2. Novel quantization of the electromagnetic field

The possibility of constructing "energy orthogonal" sets of electromagnetic fields for any volume in turn allows a novel quantization of the electromagnetic field (section **9** and **Appendix I**), now on a rigorous basis of functions in any volume of interest, avoiding the "longitudinal/transverse" field separation and any restriction to "plane-wave" modes.

### L.3.3. Novel "M-coefficient" modal alternate to Einstein's "A&B" coefficient argument

With the use of the mode-converter basis functions, the novel "M-coefficient" argument (section **11.2** and **Appendix J**) can replace Einstein's "A&B" coefficient argument with a simple result that applies mode-by-mode and for a quantum system in any otherwise loss-less dielectric environment, not just in free space.

# 13. Acknowledgements

This material is based upon work supported by the Air Force Office of Scientific Research under award numbers FA9550-15-1-0335 and FA9550-17-1-0002. I am pleased to acknowledge many stimulating discussions with Shanhui Fan.

then the “knee” in the curves here moves closer to $N_{Hy}$ - that is, the factor that here is 0.985 moves closer to 1. The form of the curve, explicitly including the exponential decay rate, does not change, however, with the singular values falling off exponentially with the same exponent.

[79] In this rationalization, we presume we can approximately “factorize” the modes into a product of “horizontal” and “vertical” mode forms, like those seen with “line” sources and receivers. Up to $n \simeq N_H$, both the horizontal and vertical forms are for modes below the corresponding $N_{Hx}$ and $N_{Hy}$ limits. However, for $n > N_H$, one or other of the horizontal or vertical forms must exceed its corresponding $N_{Hx}$ and $N_{Hy}$ limits. So, there will be a set of $N_{Hx}$ “horizontal” modes that correspond to the first “vertical” mode past the limit, and similarly a set of $N_{Hy}$ modes that correspond to the first “horizontal” mode past the limit. So we expect to see a “step” with $\approx N_{Hx} + N_{Hy}$ modes with approximately equal singular values. A similar argument for successive weaker modes in one or other direction leads to a subsequent step, and so on. Because there is a number of such modes on each step that is therefore proportional (in this square case) to $N_{Hx} = N_{Hy} = \sqrt{N_H}$, we divide by $\sqrt{N_H}$ in the exponential. Of course, this is not quite a complete counting of all the possible weakly coupled modes, because there will also be modes in which both the “horizontal” and “vertical” modes are both “weak”, so this rationalization is not a complete description, but it does give some sense as to why we can see “steps” and the $\sqrt{N_H}$ factor in the denominator in the approximate exponential.

[80] The general behavior of singular values for prolate spheroidal functions is well known, and [81] expands this discussion for the weakly coupled values in a general “Fourier transform” approach, showing that the number of “degrees of freedom” increases only logarithmically as the minimum acceptable singular value is decreased ([81], Eq. (2)), which is consistent with an exponentialy decaying strength of the singular values. [82] extends this to more dimensions. Insofar as these Fourier transform approaches are valid, which may hold in the limit of large structures separated by even larger distances, they give some explanation for this phenomenon.

[83] We used a “spiral” approach to obtain an approximately uniform distribution of points on the spherical surface. See E. B. Saff and A. B. J. Kuijlaars, “Distributing many points on a sphere,” The Mathematical Intelligencer, **19**, No. 1, 5 – 11 (1997), in our case using a “golden section” ratio angular increment $\pi(3-\sqrt{5})$. A pseudo-code version of this algorithm is as follows for distributing Nsp points on a sphere of unit radius, returning arrays x, y, and z of the corresponding *x*, *y*, and *z* values for each point. Here, “;” is a statement separator, and $x_k$ is the kth element of the array x.

g := 0; dz := 2/Nsp; s :=1-dz/2; dg := $\pi\left(3-\sqrt{5}\right)$;

For k := 1 to Nsp

r:= $\sqrt{1-s^2}$ ; $x_k$ := r cos(g); $y_k$ := r sin(g); $z_k$ := s; s := s – dz; g := g + dg;

[85] The modal amplitudes used were chosen somewhat arbitrarily, but with suitable values of for Modes 12, 13, and 14, to illustrate the key points without also showing extreme behavior (such as very large source amplitudes at specific sources). The actual modal amplitudes in this example for the desired received field are (to 3 significant figures), in order for Modes 1 to 14, 0.183, 0.143, 0.148, 0.346, 0.445, 0.207, 0.188, 0.395, 0.469, 0.198, 0.247, 0.104, 0.178, 0.099. The phases of the modes were chosen randomly with a uniform distribution over all phases from -$\pi$ to $\pi$; explicitly, those phases, for Modes 1 to 14, are $\pi$ times the following values: 0.823, 0.912, -0.668, -0.235, -0.141, -0.353, 0.678, 0.749, -0.919, 0.711, -0.098, -0.170, 0.935, 0.325. (All numbers are quoted to three significant figures, though the actual values used had higher precision.)

[95] In (IP2), for good reason, we choose a notation convention here that is the other way round from most (but not all) mathematics texts. Common mathematical notation for (IP2) would have $(a\gamma,\alpha)=a(\gamma,\alpha)$, which, with (IP3) would give $(\gamma,a\alpha)=a^*(\gamma,\alpha)$. Our choice corresponds better with the order we encounter in our "algebraic shift" to Dirac notation, and gives a natural form of the associative property of multiplication as in matrix-vector notation.

[96] This term "underlying inner product" is one that I am introducing here for clarity.

[97] Note this is technically a re-use of a notation; we already used $(\alpha,\beta)$ with such ordinary braces for the inner product. Such re-use is unfortunately rather common in mathematical texts.

[98] Unfortunately, this "infinite-long" aspect of a given sequence may well not be stated clearly or explicitly in functional analysis.

[99] In a metric space with a metric $d(\alpha,\beta)$, a sequence $(\alpha_n)$ is said to be Cauchy (or to be a Cauchy sequence) if for every real number $\varepsilon>0$ (no matter how small) there is a number $N$ (a positive integer or natural number) such that, for every $m,n>N$, $d(\alpha_m,\alpha_n)<\varepsilon$.

[100] This use of "complete" in a "complete set" is different from the idea of a "complete" space; this confusion is unfortunate, but is unavoidable because of common usage.

[101] Technically, the supremum is the smallest number that is greater than or equal to all the numbers being considered.

[102] In a notation like this with this "dot", it is best to view these inner product operations as "waiting to happen"; just how much of the inner product operation we are effectively writing down here can be somewhat vague in mathematics texts. However, we will take the approach that both any "operator weighting" and any integral for the inner product are "waiting to be applied" and in that sense are not yet part of this expression.

[103] In its more common use in quantum mechanics, Dirac notation is not required to deal with the sophistication of different underlying inner products, though we see here that, with careful definitions, it can handle this extension.

[104] Quite generally, a form like $\left|\beta_j\right\rangle_2 {}_1\!\left\langle\alpha_k\right|$ is an outer product. In contrast to the inner product, which produces a complex number, and which necessarily only involves vectors in the same Hilbert space, the outer product generates a matrix from the multiplication in "column vector – row vector" order, and can involve vectors in different Hilbert spaces.

[105] The same problem does not arise in finite-dimensional spaces; if we construct an infinitely long sequence made up from just the finite number of basis vectors in the space, we will have to repeat at least one of the basis vectors an infinite number of times, which gives us at least one convergent subsequence – the (sub)sequence consisting of just that basis vector repeated an infinite number of times. In fact, we can prove [36] that it is sufficient that an operator has finite-dimensional range for it to be compact. A corollary is that operators described by finite matrices are compact.

[106] Note that there is some variation in notation in mathematics texts. Kreyszig [92] uses this definition for a positive operator, for example, and if the "$\geq$" sign is replaced by a ">" sign in (133), he would then call the operator positive-definite. Others, however, such as [93], would give (133) as defining a non-negative operator, using "positive operator" only if the "$\geq$" sign is replaced by a ">" sign.

[107] Both "operator-weighted inner product" and "transformed inner product" are terms we are adding here; I know of no other standard names for these concepts.

[108] Incidentally, note that, unlike many basis transformation, there is no requirement here that this transform is unitary.

[109] In fact, it is not even necessary with such a "$1/r^2$" integrand that the volumes do not overlap; the result of such an integral will be finite even if the resulting "$1/r^2$" singularity is included. See the discussion in [110], p. 140 and p. 173.

[111] Here, as noted in the discussion after Eq. (105), we have made the notational choice to leave the integrals over over $\mathbf{r}_R$ and $\mathbf{r}_S$ out of this part of the mathematics, including them later when we perform the actual inner products.

[113] The full solution for scalar waves requires two kinds of sources, which can be written as point sources and spatial dipoles, but can also be written as spatio-temporal dipoles [68]; only one kind of such spatio-temporal dipole is typically required, however, for "slowly varying" wavefronts, allowing a return to a simple view of effective sources on the wavefront, and hence a simple Green's function with an obliquity factor included.

[126] Using a notation after [125], the *n*th such eigenfunction solution in the *y* direction in a given source or receiver aperture is of the form $S_{0n}(c_L, \sqrt{ky/L})$, where the parameter $c_L = (\pi/2) N_{Py}$, with $N_{Py}$ being the paraxial heuristic number in the *y* direction. A similar set of solutions will exist in the *x* direction.

[127] If we try to derive solutions with transverse boundaries at infinity using integral equations (as in our Green's function SVD approaches), the corresponding singular values are all identical (there is no "loss" in the propagation because no wave "misses" the receiver space), so the eigensolutions are completely degenerate and any orthogonal basis is equally good, so the solution becomes mathematically trivial. Hermite-Gaussians will be solutions in one direction, for example, but so will any other set of complete functions. There is also a physical contradiction in such solutions without finite boundaries because we are violating the paraxial approximation by allowing the boundaries to extend arbitrarily in the transverse direction.

[128] Beams with a specific "orbital" angular momentum correspond to a phase variation in azimuthal angle $\phi$ with an integer "quantum number" $m$ in the form $\exp(im\phi)$. If the solutions with $\exp(im\phi)$ and $\exp(-im\phi)$ are degenerate, then we are free to construct the linear combinations with phase variations of the form $\cos(m\phi) = (1/2)[\exp(im\phi) + \exp(-im\phi)]$ and $\sin(m\phi) = (1/2i)[\exp(im\phi) - \exp(-im\phi)]$. These new solutions, each being equal sums of solutions with equal but opposite "orbital" angular